\documentclass[10pt]{article}
\usepackage{graphics,graphicx}
\renewcommand{\cal}{\mathcal}

\begin{document}
\marginparwidth 3cm
%

%%% \include{dtp3_rosetta}
%
%-- ROSETTA.TEX
%
%================
% General utility
%================
%
\newcommand{\nl}{\nonumber\\}
\newcommand{\nn}{\nonumber}
\newcommand{\ds}{\displaystyle}
\newcommand{\mpar}[1]{{\marginpar{\hbadness10000%
                      \sloppy\hfuzz10pt\boldmath\bf#1}}%
                      \typeout{marginpar: #1}\ignorespaces}
\def\mnew{\mpar{\hfil NEW \hfil}\ignorespaces}
\newcommand{\lpar}{\left(}                            % bracketing
\newcommand{\rpar}{\right)} 
\newcommand{\lrbr}{\left[}
\newcommand{\rrbr}{\right]}
\newcommand{\lcbr}{\left\{}
\newcommand{\rcbr}{\right\}} 
\newcommand{\rbrak}[1]{\lrbr#1\rrbr}
\newcommand{\bq}{\begin{equation}}                    % equationing
\newcommand{\eq}{\end{equation}}
\newcommand{\bqa}{\begin{eqnarray}}
\newcommand{\eqa}{\end{eqnarray}}
\newcommand{\ba}[1]{\begin{array}{#1}}
\newcommand{\ea}{\end{array}}
\newcommand{\ben}{\begin{enumerate}}
\newcommand{\een}{\end{enumerate}}
\newcommand{\bei}{\begin{itemize}}
\newcommand{\eei}{\end{itemize}}
\newcommand{\eqn}[1]{Eq.(\ref{#1})}
\newcommand{\eqns}[2]{Eqs.(\ref{#1}--\ref{#2})}
\newcommand{\eqnss}[1]{Eqs.(\ref{#1})}
\newcommand{\eqnsc}[2]{Eqs.(\ref{#1},~\ref{#2})}
\newcommand{\tbn}[1]{Tab.~\ref{#1}}
\newcommand{\tbns}[2]{Tabs.~\ref{#1}--\ref{#2}}
\newcommand{\tbnsc}[2]{Tabs.~\ref{#1},~\ref{#2}}
\newcommand{\fig}[1]{Fig.~\ref{#1}}
\newcommand{\figs}[2]{Figs.~\ref{#1}--\ref{#2}}
\newcommand{\sect}[1]{Sect.~\ref{#1}}
\newcommand{\subsect}[1]{Sub-Sect.~\ref{#1}}
%
% Miscellanea of symbols:
%========================
%
\newcommand{\TeV}{\;\mathrm{TeV}}                     
\newcommand{\GeV}{\;\mathrm{GeV}}
\newcommand{\MeV}{\;\mathrm{MeV}}
\newcommand{\nb}{\;\mathrm{nb}}
\newcommand{\pb}{\;\mathrm{pb}}
\newcommand{\fb}{\;\mathrm{fb}}
\def\Re{\mathop{\operator@font Re}\nolimits}
\def\Im{\mathop{\operator@font Im}\nolimits}
\newcommand{\ord}[1]{{\cal O}\lpar#1\rpar}
\newcommand{\group}{SU(2)\otimes U(1)}
\newcommand{\ib}{i}
\newcommand{\asums}[1]{\sum_{#1}}
\newcommand{\asumt}[2]{\sum_{#1}^{#2}}
\newcommand{\asum}[3]{\sum_{#1=#2}^{#3}}
%
% Powers of 10:
%==============
%
\newcommand{\tmi}{\times 10^{-1}}
\newcommand{\tmii}{\times 10^{-2}}
\newcommand{\tmiii}{\times 10^{-3}}
\newcommand{\tmiv}{\times 10^{-4}}
\newcommand{\tmfv}{\times 10^{-5}}
\newcommand{\tmfvi}{\times 10^{-6}}
\newcommand{\tmfvii}{\times 10^{-7}}
\newcommand{\tmfviii}{\times 10^{-8}}
\newcommand{\tmfix}{\times 10^{-9}}
\newcommand{\tmfx}{\times 10^{-10}}
%
% Fields:
%========
%
\newcommand{\fer}{{\rm{fer}}}
\newcommand{\bos}{{\rm{bos}}}
\newcommand{\lep}{{l}}
\newcommand{\had}{{h}}
\newcommand{\gen}{\rm{g}}
\newcommand{\dbl}{\rm{d}}
\newcommand{\philone}{\phi}
\newcommand{\philoneb}{\phi_{0}}
\newcommand{\phiind}[1]{\phi_{#1}}
\newcommand{\gBi}[2]{B_{#1}^{#2}}
\newcommand{\gBn}[1]{B_{#1}}
%
% vector-bosons
%--------------
%
\newcommand{\ph}{\gamma}
\newcommand{\ab}{A}
\newcommand{\abr}{A^r}
\newcommand{\abb}{A^{0}}
\newcommand{\abi}[1]{A_{#1}}
\newcommand{\abri}[1]{A^r_{#1}}
\newcommand{\abbi}[1]{A^{0}_{#1}}
\newcommand{\wb}{W}            
\newcommand{\wbi}[1]{W_{#1}}           
\newcommand{\wbp}{W^{+}}
\newcommand{\wbm}{W^{-}}
\newcommand{\wbpm}{W^{\pm}}
\newcommand{\wbpi}[1]{W^{+}_{#1}}
\newcommand{\wbmi}[1]{W^{-}_{#1}}
\newcommand{\wbpmi}[1]{W^{\pm}_{#1}}
\newcommand{\wbli}[1]{W^{[+}_{#1}}
\newcommand{\wbri}[1]{W^{-]}_{#1}}
\newcommand{\zb}{Z}
\newcommand{\zbi}[1]{Z_{#1}}
\newcommand{\vb}{V}
\newcommand{\vbi}[1]{V_{#1}}      
\newcommand{\vbiv}[1]{V^{*}_{#1}}      
\newcommand{\Pb}{P}
\newcommand{\Sb}{S}
\newcommand{\Bb}{B}
%
% Higgs-Kibble ghosts
%--------------------
%
\newcommand{\hk}{K}
\newcommand{\hKi}[1]{K_{#1}}
\newcommand{\hkg}{\phi}
\newcommand{\hkn}{\phi^{0}}                 
\newcommand{\hkp}{\phi^{+}}
\newcommand{\hkm}{\phi^{-}}
\newcommand{\hkpm}{\phi^{\pm}}
\newcommand{\hkmp}{\phi^{\mp}}
\newcommand{\hki}[1]{\phi^{#1}}
\newcommand{\hb}{H}
\newcommand{\hbi}[1]{H_{#1}}
\newcommand{\hkl}{\phi^{[+\cgfi\cgfi}}
\newcommand{\hkr}{\phi^{-]}}
%
% FP-ghosts
%----------
%
\newcommand{\fpx}{X}
\newcommand{\fpy}{Y}
\newcommand{\fpxp}{X^+}
\newcommand{\fpxm}{X^-}
\newcommand{\fpxpm}{X^{\pm}}
\newcommand{\fpxi}[1]{X^{#1}}
\newcommand{\fpyZ}{Y^{\ssZ}}
\newcommand{\fpyA}{Y^{\ssA}}
\newcommand{\fpyZA}{Y_{\ssZ,\ssA}}
\newcommand{\fpbxi}[1]{{\overline{X}}^{#1}}
\newcommand{\fpbyZ}{{\overline{Y}}^{\ssZ}}
\newcommand{\fpbyA}{{\overline{Y}}^{\ssA}}
\newcommand{\fpbyZA}{{\overline{Y}}^{\ssZ,\ssA}}
%
% Fermionic fields
%-----------------
%
\newcommand{\Flone}{F}
\newcommand{\fpsi}{\psi}
\newcommand{\fpsii}[1]{\psi^{#1}}
\newcommand{\fpsib}{\psi^{0}}
\newcommand{\fpsir}{\psi^r}
\newcommand{\fpsiL}{\psi_{_L}}
\newcommand{\fpsiR}{\psi_{_R}}
\newcommand{\fpsiLi}[1]{\psi_{_L}^{#1}}
\newcommand{\fpsiRi}[1]{\psi_{_R}^{#1}}
\newcommand{\fpsiLbi}[1]{\psi_{_{0L}}^{#1}}
\newcommand{\fpsiRbi}[1]{\psi_{_{0R}}^{#1}}
\newcommand{\fpsiLR}{\psi_{_{L,R}}}
\newcommand{\fbpsi}{{\overline{\psi}}}
\newcommand{\fbpsii}[1]{{\overline{\psi}}^{#1}}
\newcommand{\fbpsir}{{\overline{\psi}}^r}
\newcommand{\fbpsiL}{{\overline{\psi}}_{_L}}
\newcommand{\fbpsiR}{{\overline{\psi}}_{_R}}
\newcommand{\fbpsiLi}[1]{\overline{\psi_{_L}}^{#1}}
\newcommand{\fbpsiRi}[1]{\overline{\psi_{_R}}^{#1}}
\newcommand{\fe}{e}
\newcommand{\ff}{f}
\newcommand{\fep}{e^{+}}
\newcommand{\fem}{e^{-}}
\newcommand{\fepm}{e^{\pm}}
\newcommand{\fp}{f^{+}}
\newcommand{\fm}{f^{-}}
\newcommand{\fhp}{h^{+}}
\newcommand{\fhm}{h^{-}}
\newcommand{\fh}{h}
\newcommand{\flm}{\mu}
\newcommand{\flmp}{\mu^{+}}
\newcommand{\flmm}{\mu^{-}}
\newcommand{\fll}{l}
\newcommand{\fllp}{l^{+}}
\newcommand{\fllm}{l^{-}}
\newcommand{\flt}{\tau}
\newcommand{\fltp}{\tau^{+}}
\newcommand{\fltm}{\tau^{-}}
\newcommand{\fq}{q}
\newcommand{\fqi}[1]{\fq_{#1}}
\newcommand{\bfqi}[1]{\barq_{#1}}
\newcommand{\ffQ}{Q}
\newcommand{\fu}{u}
\newcommand{\fd}{d}
\newcommand{\fc}{c}
\newcommand{\fs}{s}
\newcommand{\fqp}{q'}
\newcommand{\fup}{u'}
\newcommand{\fdp}{d'}
\newcommand{\fcp}{c'}
\newcommand{\fsp}{s'}
\newcommand{\fdpp}{d''}
\newcommand{\ffi}[1]{f_{#1}}
\newcommand{\bffi}[1]{{\overline{f}}_{#1}}
\newcommand{\ffpi}[1]{f'_{#1}}
\newcommand{\bffpi}[1]{{\overline{f}}'_{#1}}
\newcommand{\ft}{t}
\newcommand{\ffb}{b}
\newcommand{\ffp}{f'}
\newcommand{\fft}{{\tilde{f}}}
\newcommand{\fl}{l}
\newcommand{\fli}[1]{\fl_{#1}}
\newcommand{\fnu}{\nu}
\newcommand{\fU}{U}
\newcommand{\fD}{D}
\newcommand{\fUc}{\overline{U}}
\newcommand{\fDc}{\overline{D}}
\newcommand{\fnul}{\nu_l}
\newcommand{\fnue}{\nu_e}
\newcommand{\fnum}{\nu_{\mu}}
\newcommand{\fnut}{\nu_{\tau}}
\newcommand{\fbe}{{\overline{e}}}
\newcommand{\fbu}{{\overline{u}}}
\newcommand{\fbd}{{\overline{d}}}
\newcommand{\fbf}{{\overline{f}}}
\newcommand{\fbfp}{{\overline{f}}'}
\newcommand{\fbl}{{\overline{l}}}
\newcommand{\fbnu}{{\overline{\nu}}}
\newcommand{\fbnul}{{\overline{\nu}}_{\fl}}
\newcommand{\fbnue}{{\overline{\nu}}_{\fe}}
\newcommand{\fbnum}{{\overline{\nu}}_{\flm}}
\newcommand{\fbnut}{{\overline{\nu}}_{\flt}}
\newcommand{\fuL}{u_{_L}}
\newcommand{\fdL}{d_{_L}}
\newcommand{\ffL}{f_{_L}}
\newcommand{\fbuL}{{\overline{u}}_{_L}}
\newcommand{\fbdL}{{\overline{d}}_{_L}}
\newcommand{\fbfL}{{\overline{f}}_{_L}}
\newcommand{\fuR}{u_{_R}}
\newcommand{\fdR}{d_{_R}}
\newcommand{\ffR}{f_{_R}}
\newcommand{\fbuR}{{\overline{u}}_{_R}}
\newcommand{\fbdR}{{\overline{d}}_{_R}}
\newcommand{\fbfR}{{\overline{f}}_{_R}}
%
% anti-fermions, GP's realization
%--------------------------------
%
\newcommand{\barf}{\overline f}                
\newcommand{\barl}{\overline l}
\newcommand{\barq}{\overline q}
\newcommand{\barqp}{\overline{q}'}
\newcommand{\barb}{\overline b}
\newcommand{\bart}{\overline t}
\newcommand{\barc}{\overline c}
\newcommand{\baru}{\overline u}
\newcommand{\bard}{\overline d}
\newcommand{\bars}{\overline s}
\newcommand{\barv}{\overline v}
\newcommand{\barnu}{\overline{\nu}}
\newcommand{\barne}{\overline{\nu}_{\fe}}
\newcommand{\barnm}{\overline{\nu}_{\flm}}
\newcommand{\barnt}{\overline{\nu}_{\flt}}
%
% gluon
%------
%
\newcommand{\glu}{g}
%
% (anti)proton
%-------------
%
\newcommand{\prot}{p}
\newcommand{\aprot}{{\bar{p}}}
\newcommand{\Nucln}{N}
%
% Vector resonances
%------------------
%
\newcommand{\tM}{{\tilde M}}
\newcommand{\tMs}{{\tilde M}^2}
\newcommand{\tW}{{\tilde \Gamma}}
\newcommand{\tWs}{{\tilde\Gamma}^2}
\newcommand{\fphi}{\phi}
\newcommand{\fJpsi}{J/\psi}
\newcommand{\fgpsi}{\psi}
\newcommand{\Glone}{\Gamma}
\newcommand{\Gloni}[1]{\Gamma_{#1}}
\newcommand{\Glones}{\Gamma^2}
\newcommand{\Glonec}{\Gamma^3}
\newcommand{\glone}{\gamma}
\newcommand{\glones}{\gamma^2}
\newcommand{\gloneq}{\gamma^4}
\newcommand{\gloni}[1]{\gamma_{#1}}
\newcommand{\glonis}[1]{\gamma^2_{#1}}
\newcommand{\Grest}[2]{\Gamma_{#1}^{#2}}
\newcommand{\grest}[2]{\gamma_{#1}^{#2}}
\newcommand{\resampl}{A_{_R}}
\newcommand{\resasyi}[1]{{\cal{A}}_{#1}}
\newcommand{\sSrest}[1]{\sigma_{#1}}
\newcommand{\Srest}[2]{\sigma_{#1}\lpar{#2}\rpar}
\newcommand{\Gdist}[1]{{\cal{G}}\lpar{#1}\rpar}
\newcommand{\sGdist}{{\cal{G}}}
\newcommand{\Aarea}{A_{0}}
\newcommand{\Aareai}[1]{{\cal{A}}\lpar{#1}\rpar}
\newcommand{\sAarea}{{\cal{A}}}
\newcommand{\resolw}{\sigma_{\Energ}}
\newcommand{\chizer}{\chi_{_0}}
\newcommand{\ini}{\rm{in}}
\newcommand{\fin}{\rm{fin}}
\newcommand{\ifi}{\rm{if}}
\newcommand{\ipf}{\rm{i+f}}
\newcommand{\tot}{\rm{tot}}
\newcommand{\Bac}{Q}
\newcommand{\Res}{R}
\newcommand{\Int}{I}
\newcommand{\NRe}{NR}
\newcommand{\ratoe}{\delta}
\newcommand{\ratoes}{\delta^2}
%
% QED-boxes
%----------
%
\newcommand{\Fbox}[2]{f^{\rm{box}}_{#1}\lpar{#2}\rpar}
\newcommand{\Dbox}[2]{\delta^{\rm{box}}_{#1}\lpar{#2}\rpar}
\newcommand{\Bbox}[3]{{\cal{B}}_{#1}^{#2}\lpar{#3}\rpar}
%
% Masses:
%========
%
\newcommand{\phm}{\lambda}
\newcommand{\phms}{\lambda^2}
\newcommand{\mV}{M_{_V}}
\newcommand{\mw}{M_{_W}}
\newcommand{\mX}{M_{_X}}
\newcommand{\mY}{M_{_Y}}
\newcommand{\LM}{M}
\newcommand{\mz}{M_{_Z}}
\newcommand{\bzm}{M_{_0}}
\newcommand{\mh}{M_{_H}}
\newcommand{\bhm}{M_{_{0H}}}
\newcommand{\mf}{m_f}
\newcommand{\mfp}{m_{f'}}
\newcommand{\mfh}{m_{h}}
\newcommand{\mt}{m_t}
\newcommand{\me}{m_e}
\newcommand{\mm}{m_{\mu}}
\newcommand{\mtau}{m_{\tau}}
\newcommand{\muq}{m_u}
\newcommand{\md}{m_d}
\newcommand{\muqp}{m'_u}
\newcommand{\mdqp}{m'_d}
\newcommand{\mc}{m_c}
\newcommand{\ms}{m_s}
\newcommand{\mb}{m_b}
\newcommand{\mup}{M_u}                              % pole masses
\newcommand{\mdp}{M_d}
\newcommand{\mcp}{M_c}
\newcommand{\msp}{M_s}
\newcommand{\mbp}{M_b}
%
% Masses squared, some cubed
%---------------------------
%
\newcommand{\mls}{m^2_l}
\newcommand{\mVs}{M^2_{_V}}
\newcommand{\mws}{M^2_{_W}}
\newcommand{\mwc}{M^3_{_W}}
\newcommand{\LMs}{M^2}
\newcommand{\LMc}{M^3}
\newcommand{\mzs}{M^2_{_Z}}
\newcommand{\mzc}{M^3_{_Z}}
\newcommand{\bzms}{M^2_{_0}}
\newcommand{\bzmc}{M^3_{_0}}
\newcommand{\bhms}{M^2_{_{0H}}}
\newcommand{\mhs}{M^2_{_H}}
\newcommand{\mfs}{m^2_f}
\newcommand{\mfc}{m^3_f}
\newcommand{\mfps}{m^2_{f'}}
\newcommand{\mfhs}{m^2_{h}}
\newcommand{\mfpc}{m^3_{f'}}
\newcommand{\mts}{m^2_t}
\newcommand{\mes}{m^2_e}
\newcommand{\mms}{m^2_{\mu}}
\newcommand{\mmc}{m^3_{\mu}}
\newcommand{\mmfour}{m^4_{\mu}}
\newcommand{\mmf}{m^5_{\mu}}
\newcommand{\mmfive}{m^5_{\mu}}
\newcommand{\mmsix}{m^6_{\mu}}
\newcommand{\mminv}{\frac{1}{m_{\mu}}}
\newcommand{\mtaus}{m^2_{\tau}}
\newcommand{\mus}{m^2_u}
\newcommand{\mds}{m^2_d}
\newcommand{\muqps}{m'^2_u}
\newcommand{\mdqps}{m'^2_d}
\newcommand{\mcs}{m^2_c}
\newcommand{\mss}{m^2_s}
\newcommand{\mbs}{m^2_b}
\newcommand{\mups}{M^2_u}
\newcommand{\mdps}{M^2_d}
\newcommand{\mcps}{M^2_c}
\newcommand{\msps}{M^2_s}
\newcommand{\mbps}{M^2_b}
%
% Some ratios
%------------
%
\newcommand{\muf}{\mu_{\ff}}
\newcommand{\mufs}{\mu^2_{\ff}}
\newcommand{\mufq}{\mu^4_{\ff}}
\newcommand{\mufx}{\mu^6_{\ff}}
\newcommand{\muz}{\mu_{_{\zb}}}
\newcommand{\muw}{\mu_{_{\wb}}}
\newcommand{\mut}{\mu_{\ft}}
\newcommand{\muzs}{\mu^2_{_{\zb}}}
\newcommand{\muws}{\mu^2_{_{\wb}}}
\newcommand{\muts}{\mu^2_{\ft}}
\newcommand{\muSW}{\mu^2_{_{\wb}}}
\newcommand{\muwq}{\mu^4_{_{\wb}}}
\newcommand{\muwsx}{\mu^6_{_{\wb}}}
\newcommand{\muwms}{\mu^{-2}_{_{\wb}}}
\newcommand{\muhs}{\mu^2_{_{\hb}}}
\newcommand{\muhq}{\mu^4_{_{\hb}}}
\newcommand{\muhsx}{\mu^6_{_{\hb}}}
\newcommand{\mutq}{\mu^4_{_{\hb}}}   % bardinworry to be checked by grep 
\newcommand{\mutsx}{\mu^6_{_{\hb}}}  %     -/-
\newcommand{\muL}{\mu}
\newcommand{\muS}{\mu^2}
\newcommand{\muQ}{\mu^4}
\newcommand{\muizs}{\mu^2_{0}}
\newcommand{\muizq}{\mu^4_{0}}
\newcommand{\muis}{\mu^2_{1}}
\newcommand{\muiis}{\mu^2_{2}}
\newcommand{\muiiis}{\mu^2_{3}}
\newcommand{\muii}[1]{\mu_{#1}}
\newcommand{\muisi}[1]{\mu^2_{#1}}
\newcommand{\muiqi}[1]{\mu^4_{#1}}
\newcommand{\muixi}[1]{\mu^6_{#1}}
\newcommand{\zm}{z_m}
\newcommand{\ri}[1]{r_{#1}}
\newcommand{\xw}{x_w}
\newcommand{\xws}{x^2_w}
\newcommand{\xwc}{x^3_w}
\newcommand{\xth}{x_t}
\newcommand{\xths}{x^2_t}
\newcommand{\xthc}{x^3_t}
\newcommand{\xthf}{x^4_t}
\newcommand{\xthv}{x^5_t}
\newcommand{\xthx}{x^6_t}
\newcommand{\xh}{x_h}
\newcommand{\xhs}{x^2_h}
\newcommand{\xhc}{x^3_h}
\newcommand{\Rl}{R_{\fl}}
\newcommand{\Rb}{R_{\ffb}}
\newcommand{\Rc}{R_{\fc}}
%
% Masses quartic
%---------------
%
\newcommand{\mwq}{M^4_{_\wb}}
\newcommand{\mwf}{M^4_{_\wb}}
\newcommand{\LMq}{M^4}
\newcommand{\mzq}{M^4_{_Z}}
\newcommand{\bzmq}{M^4_{_0}}
\newcommand{\mhq}{M^4_{_H}}
\newcommand{\mfq}{m^4_f}
\newcommand{\mfpq}{m^4_{f'}}
\newcommand{\mtq}{m^4_t}
\newcommand{\meq}{m^4_e}
\newcommand{\mmq}{m^4_{\mu}}
\newcommand{\mtauq}{m^4_{\tau}}
\newcommand{\muqq}{m^4_u}
\newcommand{\mdq}{m^4_d}
\newcommand{\mcq}{m^4_c}
\newcommand{\msq}{m^4_s}
\newcommand{\mbq}{m^4_b}
\newcommand{\mupq}{M^4_u}
\newcommand{\mdpq}{M^4_d}
\newcommand{\mcpq}{M^4_c}
\newcommand{\mspq}{M^4_s}
\newcommand{\mbpq}{M^4_b}
%
% Masses sixtupled
%-----------------
%
\newcommand{\mwx}{M^6_{_W}}
\newcommand{\mzx}{M^6_{_Z}}
\newcommand{\mfx}{m^6_f}
\newcommand{\mfpx}{m^6_{f'}}
\newcommand{\LMx}{M^6}
%
% More masses
%------------
%
\newcommand{\mer}{m_{er}}
\newcommand{\mlep}{m_l}
\newcommand{\mleps}{m^2_l}
\newcommand{\mone}{m_1}
\newcommand{\mtwo}{m_2}
\newcommand{\mtre}{m_3}
\newcommand{\mfor}{m_4}
\newcommand{\mlone}{m}
\newcommand{\mloneb}{\bar{m}}
\newcommand{\mind}[1]{m_{#1}}
\newcommand{\mones}{m^2_1}
\newcommand{\mtwos}{m^2_2}
\newcommand{\mtres}{m^2_3}
\newcommand{\mfors}{m^2_4}
\newcommand{\mlones}{m^2}
\newcommand{\minds}[1]{m^2_{#1}}
\newcommand{\moneq}{m^4_1}
\newcommand{\mtwoq}{m^4_2}
\newcommand{\mtreq}{m^4_3}
\newcommand{\mforq}{m^4_4}
\newcommand{\mloneq}{m^4}
\newcommand{\mindq}[1]{m^4_{#1}}
\newcommand{\mlonev}{m^5}
\newcommand{\mindv}[1]{m^5_{#1}}
\newcommand{\monex}{m^6_1}
\newcommand{\mtwox}{m^6_2}
\newcommand{\mtrex}{m^6_3}
\newcommand{\mforx}{m^6_4}
\newcommand{\mlonex}{m^6}
\newcommand{\mindx}[1]{m^6_{#1}}
\newcommand{\Mone}{M_1}
\newcommand{\Mtwo}{M_2}
\newcommand{\Mtre}{M_3}
\newcommand{\Mfor}{M_4}
\newcommand{\Mlone}{M}
\newcommand{\Mlonep}{M'}
\newcommand{\Miind}{M_i}
\newcommand{\Mind}[1]{M_{#1}}
\newcommand{\Minds}[1]{M^2_{#1}}
\newcommand{\Mindc}[1]{M^3_{#1}}
\newcommand{\Mindf}[1]{M^4_{#1}}
\newcommand{\Mones}{M^2_1}
\newcommand{\Mtwos}{M^2_2}
\newcommand{\Mtres}{M^2_3}
\newcommand{\Mfors}{M^2_4}
\newcommand{\Mlones}{M^2}
\newcommand{\Mloneps}{M'^2}
\newcommand{\Miinds}{M^2_i}
\newcommand{\Mlonec}{M^3}
\newcommand{\Monec}{M^3_1}
\newcommand{\Mtwoc}{M^3_2}
\newcommand{\Moneq}{M^4_1}
\newcommand{\Mtwoq}{M^4_2}
\newcommand{\Mtreq}{M^4_3}
\newcommand{\Mforq}{M^4_4}
\newcommand{\Mloneq}{M^4}
\newcommand{\Miindq}{M^4_i}
\newcommand{\Monex}{M^6_1}
\newcommand{\Mtwox}{M^6_2}
\newcommand{\Mtrex}{M^6_3}
\newcommand{\Mforx}{M^6_4}
\newcommand{\Mlonex}{M^6}
\newcommand{\Miindx}{M^6_i}
\newcommand{\meb}{m_0}
\newcommand{\mebs}{m^2_0}
%
% Pole masses again
%------------------
%
\newcommand{\Mq }{M_q  }
\newcommand{\MqS}{M^2_q}
\newcommand{\Ms }{M_s  }
\newcommand{\MsS}{M^2_s}
\newcommand{\Mc }{M_c  }
\newcommand{\McS}{M^2_c}
\newcommand{\Mb }{M_b  }
\newcommand{\MbS}{M^2_b}
\newcommand{\Mt }{M_t  }
\newcommand{\MtS}{M^2_t}
%
% Some quark masses
%------------------
%
\newcommand{\mq}{m_q}
\newcommand{\mqs}{m^2_q}
\newcommand{\mqS}{m^2_q}
\newcommand{\mqQ}{m^4_q}
\newcommand{\mqX}{m^6_q}
\newcommand{\mqp}{m'_q }
\newcommand{\mqpS}{m'^2_q}
\newcommand{\mqpQ}{m'^4_q}
%
% Some logs of mass ratios
%=========================
%
\newcommand{\lL}{l}
\newcommand{\ls}{l^2}
\newcommand{\LL}{L}
\newcommand{\LcalL}{\cal{L}}
\newcommand{\LS}{L^2}
\newcommand{\LC}{L^3}
\newcommand{\LQ}{L^4}
\newcommand{\lw}{l_w}
\newcommand{\Lw}{L_w}
\newcommand{\Lws}{L^2_w}
\newcommand{\Lz}{L_z}
\newcommand{\Lzs}{L^2_z}
\newcommand{\Li}[1]{L_{#1}}
\newcommand{\Lis}[1]{L^2_{#1}}
\newcommand{\Lic}[1]{L^3_{#1}}
%
% Mandelstam variables
%=====================
%
\newcommand{\sman}{s}
\newcommand{\tman}{t}
\newcommand{\uman}{u}
\newcommand{\smani}[1]{s_{#1}}
\newcommand{\bsmani}[1]{{\bar{s}}_{#1}}
\newcommand{\smans}{s^2}
\newcommand{\tmans}{t^2}
\newcommand{\umans}{u^2}
\newcommand{\shat}{{\hat s}}
\newcommand{\that}{{\hat t}}
\newcommand{\uhat}{{\hat u}}
%
% More invariant variables
%-------------------------
%
\newcommand{\smanp}{s'}
\newcommand{\smanpi}[1]{s'_{#1}}
\newcommand{\tmanp}{t'}
\newcommand{\umanp}{u'}
\newcommand{\kappi}[1]{\kappa_{#1}}
\newcommand{\zetai}[1]{\zeta_{#1}}
%
% QED
%====
%
% Phase space and QED-varia
%--------------------------
%
\newcommand{\Phaspi}[1]{\Gamma_{#1}}
\newcommand{\rbetai}[1]{\beta_{#1}}
\newcommand{\ralphai}[1]{\alpha_{#1}}
\newcommand{\rbetais}[1]{\beta^2_{#1}}
\newcommand{\Lambdi}[1]{\Lambda_{#1}}
\newcommand{\Nomini}[1]{N_{#1}}
\newcommand{\smlone}{\frac{-\sman-\ib\ep}{\mlones}}
%
% Angles in bremsstrahlung
%-------------------------
%
\newcommand{\theti}[1]{\theta_{#1}}
\newcommand{\delti}[1]{\delta_{#1}}
\newcommand{\phigi}[1]{\phi_{#1}}
\newcommand{\acoli}[1]{\xi_{#1}}
\newcommand{\scats}{s}
\newcommand{\scatss}{s^2}
\newcommand{\scatsi}[1]{s_{#1}}
\newcommand{\scatsis}[1]{s^2_{#1}}
\newcommand{\scatst}[2]{s_{#1}^{#2}}
\newcommand{\scatc}{c}
\newcommand{\scatcs}{c^2}
\newcommand{\scatci}[1]{c_{#1}}
\newcommand{\scatcis}[1]{c^2_{#1}}
\newcommand{\scatct}[2]{c_{#1}^{#2}}
\newcommand{\angamt}[2]{\gamma_{#1}^{#2}}
%
% More about bremsstrahlung
%--------------------------
%
\newcommand{\Regia}{{\cal{R}}}
\newcommand{\Iconi}[2]{{\cal{I}}_{#1}\lpar{#2}\rpar}
\newcommand{\sIcon}[1]{{\cal{I}}_{#1}}
\newcommand{\betaf}{\beta_{\ff}}
\newcommand{\betafs}{\beta^2_{\ff}}
\newcommand{\Kfact}[2]{{\cal{K}}_{#1}\lpar{#2}\rpar}
%
% Structure and flux functions
%-----------------------------
%
\newcommand{\Struf}[4]{{\cal D}^{#1}_{#2}\lpar{#3;#4}\rpar}
\newcommand{\sStruf}[2]{{\cal D}^{#1}_{#2}}
\newcommand{\Fluxf}[2]{H\lpar{#1;#2}\rpar}
\newcommand{\Fluxfi}[4]{H_{#1}^{#2}\lpar{#3;#4}\rpar}
\newcommand{\sFluxf}{H}
\newcommand{\Bflux}[2]{{\cal{B}}_{#1}\lpar{#2}\rpar}
\newcommand{\bflux}[2]{{\cal{B}}_{#1}\lpar{#2}\rpar}
\newcommand{\Fluxd}[2]{D_{#1}\lpar{#2}\rpar}
\newcommand{\fluxd}[2]{C_{#1}\lpar{#2}\rpar}
\newcommand{\Fluxh}[4]{{\cal{H}}_{#1}^{#2}\lpar{#3;#4}\rpar}
\newcommand{\Sluxh}[4]{{\cal{S}}_{#1}^{#2}\lpar{#3;#4}\rpar}
\newcommand{\Fluxhb}[4]{{\overline{{\cal{H}}}}_{#1}^{#2}\lpar{#3;#4}\rpar}
\newcommand{\sFluxhb}{{\overline{{\cal{H}}}}}
\newcommand{\Sluxhb}[4]{{\overline{{\cal{S}}}}_{#1}^{#2}\lpar{#3;#4}\rpar}
\newcommand{\sSluxhb}[2]{{\overline{{\cal{S}}}}_{#1}^{#2}}
\newcommand{\fluxh}[4]{h_{#1}^{#2}\lpar{#3;#4}\rpar}
\newcommand{\fluxhs}[3]{h_{#1}^{#2}\lpar{#3}\rpar}
\newcommand{\sfluxhs}[2]{h_{#1}^{#2}}
\newcommand{\fluxhb}[4]{{\overline{h}}_{#1}^{#2}\lpar{#3;#4}\rpar}
\newcommand{\Strufd}[2]{D\lpar{#1;#2}\rpar}
%
% Mass and momenta squared ratios
%================================
%
\newcommand{\rMQ}[1]{r^2_{#1}}
\newcommand{\rMQs}[1]{r^4_{#1}}
\newcommand{\rf}{w_{\ff}}
\newcommand{\zf}{z_{\ff}}
\newcommand{\rfs}{w^2_{\ff}}
\newcommand{\zfs}{z^2_{\ff}}
\newcommand{\rfc}{w^3_{\ff}}
\newcommand{\zfc}{z^3_{\ff}}
\newcommand{\df}{d_{\ff}}
\newcommand{\rfp}{w_{\ffp}}
\newcommand{\rfps}{w^2_{\ffp}}
\newcommand{\rfpc}{w^3_{\ffp}}
\newcommand{\rt}{w_{\ft}}
\newcommand{\rts}{w^2_{\ft}}
\newcommand{\dt}{d_{\ft}}
\newcommand{\dts}{d^2_{\ft}}
\newcommand{\rh}{r_{h}}
\newcommand{\Lnrt}{\ln{\rt}}
\newcommand{\Rw}{R_{_{\wb}}}
\newcommand{\Rws}{R^2_{_{\wb}}}
\newcommand{\Rz}{R_{_{\zb}}}
\newcommand{\Rzp}{R^{+}_{_{\zb}}}
\newcommand{\Rzm}{R^{-}_{_{\zb}}}
\newcommand{\Rzs}{R^2_{_{\zb}}}
\newcommand{\Rzc}{R^3_{_{\zb}}}
\newcommand{\Rv}{R_{_{\vb}}}
\newcommand{\rhw}{r_{_{\wb}}}
\newcommand{\rhz}{r_{_{\zb}}}
\newcommand{\rhws}{r^2_{_{\wb}}}
\newcommand{\rhzs}{r^2_{_{\zb}}}
%
% More ratios
%------------
%
\newcommand{\vqrato}{z}
\newcommand{\vqrats}{w}
\newcommand{\vqratq}{w^2}
\newcommand{\seyrat}{z}
\newcommand{\sexrat}{w}
\newcommand{\sehrat}{h}
\newcommand{\sewrat}{w}
\newcommand{\sezrat}{z}
\newcommand{\zetav}{\zeta}
\newcommand{\zetavi}[1]{\zeta_{#1}}
\newcommand{\bpo}{\beta^2}
\newcommand{\bpos}{\beta^4}
\newcommand{\bpt}{{\tilde\beta}^2}
\newcommand{\lap}{\kappa}
\newcommand{\hw}{h_{_{\wb}}}
\newcommand{\hz}{h_{_{\zb}}}
%
% Couplings
%==========
%
\newcommand{\ec}{e}
\newcommand{\ecs}{e^2}
\newcommand{\ect}{e^3}
\newcommand{\ecq}{e^4}
\newcommand{\ecb}{e_{_0}}
\newcommand{\ecbs}{e^2_{_0}}
\newcommand{\ecbq}{e^4_{_0}}
\newcommand{\eci}[1]{e_{#1}}
\newcommand{\ecis}[1]{e^2_{#1}}
\newcommand{\hate}{{\hat e}}
\newcommand{\gss}{g_{_S}}
\newcommand{\gsss}{g^2_{_S}}
\newcommand{\gssb}{g^2_{_{S_0}}}
\newcommand{\als}{\alpha_{_S}}
\newcommand{\as}{a_{_S}}
\newcommand{\ass}{a^2_{_S}}
\newcommand{\gf}{G_{\ssF}}
\newcommand{\gfs}{G^2_{\ssF}}
\newcommand{\gb}{g} 
\newcommand{\gbi}[1]{g_{#1}}
\newcommand{\gbb}{g_{0}}
\newcommand{\gbs}{g^2}
\newcommand{\gbc}{g^3}
\newcommand{\gbf}{g^4}
\newcommand{\gpb}{g'}
\newcommand{\gpbs}{g'^2}
\newcommand{\vc}[1]{v_{#1}}
\newcommand{\ac}[1]{a_{#1}}
\newcommand{\vcc}[1]{v^*_{#1}}
\newcommand{\acc}[1]{a^*_{#1}}
\newcommand{\hatv}[1]{{\hat v}_{#1}}
\newcommand{\vcs}[1]{v^2_{#1}}
\newcommand{\acs}[1]{a^2_{#1}}
\newcommand{\gcv}[1]{g^{#1}_{\ssV}}
\newcommand{\gca}[1]{g^{#1}_{\ssA}}
\newcommand{\gcp}[1]{g^{+}_{#1}}
\newcommand{\gcm}[1]{g^{-}_{#1}}
\newcommand{\gcpm}[1]{g^{\pm}_{#1}}
\newcommand{\vci}[2]{v^{#2}_{#1}}
\newcommand{\aci}[2]{a^{#2}_{#1}}
\newcommand{\vceff}[1]{v^{#1}_{\rm{eff}}}
\newcommand{\hvc}[1]{\hat{v}_{#1}}
\newcommand{\hvcs}[1]{\hat{v}^2_{#1}}
\newcommand{\Vc}[1]{V_{#1}}
\newcommand{\Ac}[1]{A_{#1}}
\newcommand{\Vcs}[1]{V^2_{#1}}
\newcommand{\Acs}[1]{A^2_{#1}}
\newcommand{\vpa}[2]{\sigma_{#1}^{#2}}
\newcommand{\vma}[2]{\delta_{#1}^{#2}}
\newcommand{\vfw}{\sigma^{a}_{\ff}}
\newcommand{\vfpw}{\sigma^{a}_{\ffp}}
\newcommand{\vfwi}[1]{\sigma^{a}_{#1}}
\newcommand{\vfwsi}[1]{\lpar\sigma^{a}_{#1}\rpar^2}
\newcommand{\vvfw}{\sigma^{a}_{\ff}}
\newcommand{\vvew}{\sigma^{a}_{\fe}}
%-------------------------------------------------> g,G - couplings 
\newcommand{\gv}{g_{_V}}
\newcommand{\ga}{g_{_A}}
\newcommand{\gve}{g^{\fe}_{_{V}}}
\newcommand{\gae}{g^{\fe}_{_{A}}}
\newcommand{\gvf}{g^{\ff}_{_{V}}}
\newcommand{\gaf}{g^{\ff}_{_{A}}}
\newcommand{\gva}{g_{_{V,A}}}
\newcommand{\gvae}{g^{\fe}_{_{V,A}}}
\newcommand{\gvaf}{g^{\ff}_{_{V,A}}}
\newcommand{\sGv}{{\cal{G}}_{_V}}
\newcommand{\cGa}{{\cal{G}}^{*}_{_A}}
\newcommand{\cGv}{{\cal{G}}^{*}_{_V}}
\newcommand{\sGa}{{\cal{G}}_{_A}}
\newcommand{\Gvf}{{\cal{G}}^{\ff}_{_{V}}}
\newcommand{\Gaf}{{\cal{G}}^{\ff}_{_{A}}}
\newcommand{\Gvaf}{{\cal{G}}^{\ff}_{_{V,A}}}
\newcommand{\Gve}{{\cal{G}}^{\fe}_{_{V}}}
\newcommand{\Gae}{{\cal{G}}^{\fe}_{_{A}}}
\newcommand{\Gvae}{{\cal{G}}^{\fe}_{_{V,A}}}
\newcommand{\gvl}{g^{\fl}_{_{V}}}
\newcommand{\gal}{g^{\fl}_{_{A}}}
\newcommand{\gval}{g^{\fl}_{_{V,A}}}
\newcommand{\gvb}{g^{\ffb}_{_{V}}}
\newcommand{\gab}{g^{\ffb}_{_{A}}}
\newcommand{\fvf}{F_{_V}^{\ff}}
\newcommand{\faf}{F_{_A}^{\ff}}
\newcommand{\fvl}{F_{_V}^{\fl}}
\newcommand{\fal}{F_{_A}^{\fl}}
\newcommand{\corat}{\kappa}
\newcommand{\corats}{\kappa^2}
%
% Deltology-rhoology-kappaology
%==============================
%                              
\newcommand{\dr}{\Delta r}
\newcommand{\drl}{\Delta r_{_L}}
\newcommand{\drh}{\Delta{\hat r}}
\newcommand{\drhw}{\Delta{\hat r}_{_W}}
\newcommand{\rhou}{\rho_{_U}}
\newcommand{\rhoz}{\rho_{_\zb}}
\newcommand{\rZ}{\rho_{_\zb}}
\newcommand{\rhob}{\rho_{_0}}
\newcommand{\rZf}{\rho^{\ff}_{_\zb}}
\newcommand{\rhoe}{\rho_{\fe}}
\newcommand{\rhof}{\rho_{\ff}}
\newcommand{\rhoi}[1]{\rho_{#1}}
\newcommand{\kZf}{\kappa^{\ff}_{_\zb}}
\newcommand{\rWf}{\rho^{\ff}_{_\wb}}
\newcommand{\brWf}{{\bar{\rho}}^{\ff}_{_\wb}}
\newcommand{\rHf}{\rho^{\ff}_{_\hb}}
\newcommand{\brHf}{{\bar{\rho}}^{\ff}_{_\hb}}
\newcommand{\rhoR}{\rho^R_{_{\zb}}}
\newcommand{\hatrh}{{\hat\rho}}
\newcommand{\ku}{\kappa_{_U}}
\newcommand{\rZdf}[1]{\rho^{#1}_{_\zb}}
\newcommand{\kZdf}[1]{\kappa^{#1}_{_\zb}}
\newcommand{\rdfL}[1]{\rho^{#1}_{_L}}
\newcommand{\kdfL}[1]{\kappa^{#1}_{_L}}
\newcommand{\rdfR}[1]{\rho^{#1}_{\rm{rem}}}
\newcommand{\kdfR}[1]{\kappa^{#1}_{\rm{rem}}}
\newcommand{\bark}{\overline\kappa}
%
% Weak mixing angles
%===================
%
\newcommand{\stw}{s_{\theta}}             % bare, Lagrangian parameters
\newcommand{\ctw}{c_{\theta}}
\newcommand{\stws}{s_{\theta}^2}
\newcommand{\stwc}{s_{\theta}^3}
\newcommand{\stwf}{s_{\theta}^4}
\newcommand{\stwx}{s_{\theta}^6}
\newcommand{\ctws}{c_{\theta}^2}
\newcommand{\ctwc}{c_{\theta}^3}
\newcommand{\ctwf}{c_{\theta}^4}
\newcommand{\ctwx}{c_{\theta}^6}
\newcommand{\stwfiv}{s_{\theta}^5}
\newcommand{\ctwfiv}{c_{\theta}^5}
\newcommand{\stwsix}{s_{\theta}^6}
\newcommand{\ctwsix}{c_{\theta}^6}
%
% on-shell sines
%---------------
%
\newcommand{\siw}{s_{_W}}           
\newcommand{\cow}{c_{_W}}
\newcommand{\siws}{s^2_{_W}}
\newcommand{\cows}{c^2_{_W}}
\newcommand{\siwc}{s^3_{_W}}
\newcommand{\cowc}{c^3_{_W}}
\newcommand{\siwf}{s^4_{_W}}
\newcommand{\cowf}{c^4_{_W}}
\newcommand{\siwx}{s^6_{_W}}
\newcommand{\cowx}{c^6_{_W}}
\newcommand{\sons}{s_{_W}}
\newcommand{\sonss}{s^2_{_W}}
\newcommand{\cons}{c_{_W}}
\newcommand{\cooss}{c^2_{_W}}
%
% effective and other weak mixing angles
%---------------------------------------
%
\newcommand{\szs}{{\overline s}^2}
\newcommand{\szq}{{\overline s}^4}
\newcommand{\czs}{{\overline c}^2}
\newcommand{\sbs}{s_{_0}^2}
\newcommand{\cbs}{c_{_0}^2}
\newcommand{\dss}{\Delta s^2}
\newcommand{\snes}{s_{\nu e}^2}
\newcommand{\cnes}{c_{\nu e}^2}
\newcommand{\shs}{{\hat s}^2}
\newcommand{\chs}{{\hat c}^2}
\newcommand{\chl}{{\hat c}}
\newcommand{\seffs}{s^2_{\rm{eff}}}
\newcommand{\seffsf}[1]{\sin^2\theta^{#1}_{\rm{eff}}}
\newcommand{\sress}{s^2_{\rm res}}                
\newcommand{\sR}{s_{_R}}
\newcommand{\sRs}{s^2_{_R}}
\newcommand{\ctwe}{c_{\theta}^6}
\newcommand{\sany}{s}
\newcommand{\cany}{c}
\newcommand{\sanys}{s^2}
\newcommand{\canys}{c^2}
%
% Spinology etc.
%===============
%
\newcommand{\sip}{u}                             % incoming particle
\newcommand{\siap}{{\bar{v}}}                    %    "     anti-p
\newcommand{\sop}{{\bar{u}}}                     % outgoing p
\newcommand{\soap}{v}                            %    "     anti-p
\newcommand{\ip}[1]{u\lpar{#1}\rpar}             % incoming particle
\newcommand{\iap}[1]{{\bar{v}}\lpar{#1}\rpar}    %    "     anti-p
\newcommand{\op}[1]{{\bar{u}}\lpar{#1}\rpar}     % outgoing p
\newcommand{\oap}[1]{v\lpar{#1}\rpar}            %    "     anti-p
%
% With polarization
%------------------
%
\newcommand{\ipp}[2]{u\lpar{#1,#2}\rpar}         % incoming particle
\newcommand{\ipap}[2]{{\bar v}\lpar{#1,#2}\rpar} %    "     anti-p
\newcommand{\opp}[2]{{\bar u}\lpar{#1,#2}\rpar}  % outgoing p
\newcommand{\opap}[2]{v\lpar{#1,#2}\rpar}        %    "     anti-p
\newcommand{\upspi}[1]{u\lpar{#1}\rpar}
\newcommand{\vpspi}[1]{v\lpar{#1}\rpar}
\newcommand{\wpspi}[1]{w\lpar{#1}\rpar}
\newcommand{\ubpspi}[1]{{\bar{u}}\lpar{#1}\rpar}
\newcommand{\vbpspi}[1]{{\bar{v}}\lpar{#1}\rpar}
\newcommand{\wbpspi}[1]{{\bar{w}}\lpar{#1}\rpar}
\newcommand{\udpspi}[1]{u^{\dagger}\lpar{#1}\rpar}
\newcommand{\vdpspi}[1]{v^{\dagger}\lpar{#1}\rpar}
\newcommand{\wdpspi}[1]{w^{\dagger}\lpar{#1}\rpar}
\newcommand{\Ubilin}[1]{U\lpar{#1}\rpar}
\newcommand{\Vbilin}[1]{V\lpar{#1}\rpar}
\newcommand{\Xbilin}[1]{X\lpar{#1}\rpar}
\newcommand{\Ybilin}[1]{Y\lpar{#1}\rpar}
\newcommand{\up}[2]{u_{#1}\lpar #2\rpar}
\newcommand{\vp}[2]{v_{#1}\lpar #2\rpar}
\newcommand{\ubp}[2]{{\overline u}_{#1}\lpar #2\rpar}
\newcommand{\vbp}[2]{{\overline v}_{#1}\lpar #2\rpar}
\newcommand{\Pje}[1]{\frac{1}{2}\lpar 1 + #1\,\gfd\rpar}
\newcommand{\Pj}[1]{\Pi_{#1}}
\newcommand{\trace}{\mbox{Tr}}
%
% polarization operators and related things
%==========================================
%
\newcommand{\Poper}[2]{P_{#1}\lpar{#2}\rpar}
\newcommand{\Loper}[2]{\Lambda_{#1}\lpar{#2}\rpar}
\newcommand{\proj}[3]{P_{#1}\lpar{#2,#3}\rpar}
\newcommand{\sproj}[1]{P_{#1}}
\newcommand{\Nden}[3]{N_{#1}^{#2}\lpar{#3}\rpar}
\newcommand{\sNden}[1]{N_{#1}}
\newcommand{\nden}[2]{n_{#1}^{#2}}
%
% Wave functions
%===============
%
\newcommand{\vwf}[2]{e_{#1}\lpar#2\rpar}             % vector wave funct.
\newcommand{\vwfb}[2]{{\overline e}_{#1}\lpar#2\rpar}
\newcommand{\pwf}[2]{\epsilon_{#1}\lpar#2\rpar}      % photon wave funct.
\newcommand{\sla}[1]{/\!\!\!#1}
\newcommand{\slac}[1]{/\!\!\!\!#1}
%
% Momenta
%========
%
\newcommand{\iemom}{p_{_-}}                    % 2f incoming momenta
\newcommand{\ipmom}{p_{_+}}
\newcommand{\oemom}{q_{_-}}                    % 2f outgoing momenta
\newcommand{\opmom}{q_{_+}}
%
% Scalar product of two momenta
%==============================
%
\newcommand{\spro}[2]{{#1}\cdot{#2}}
%
% gammas
%=======
%
\newcommand{\gfour}{\gamma_4}                    
\newcommand{\gfd}{\gamma_5}                    
\newcommand{\gap}{\lpar 1+\gamma_5\rpar}
\newcommand{\gam}{\lpar 1-\gamma_5\rpar}
\newcommand{\gdp}{\gamma_+}
\newcommand{\gdm}{\gamma_-}
\newcommand{\gdpm}{\gamma_{\pm}}
\newcommand{\gad}{\gamma}
\newcommand{\gapm}{\lpar 1\pm\gamma_5\rpar}
\newcommand{\gadi}[1]{\gamma_{#1}}
\newcommand{\gadu}[1]{\gamma_{#1}}
\newcommand{\gapu}[1]{\gamma^{#1}}
\newcommand{\sigd}[2]{\sigma_{#1#2}}
\newcommand{\sumsp}{\overline{\sum_{\mbox{spins}}}}
%
% Special functions & integrals
%==============================
%
\newcommand{\li}[2]{\mathrm{Li}_{#1}\lpar\displaystyle{#2}\rpar} % polylog
\newcommand{\etaf}[2]{\eta\lpar#1,#2\rpar}
\newcommand{\lkall}[3]{\lambda\lpar#1,#2,#3\rpar}       % Kallen's lambda
\newcommand{\slkall}[3]{\lambda^{1/2}\lpar#1,#2,#3\rpar}%\sqrt
\newcommand{\segam}{\Gamma}                             % Euler's Gamma
\newcommand{\egam}[1]{\Gamma\lpar#1\rpar}               % Euler's Gamma
\newcommand{\ebe}[2]{B\lpar#1,#2\rpar}                  % Euler's beta
\newcommand{\ddel}[1]{\delta\lpar#1\rpar}               % Dirac's delta
\newcommand{\drii}[2]{\delta_{#1#2}}                    % Kronecker's delta
\newcommand{\driv}[4]{\delta_{#1#2#3#4}}                % gen.   "
\newcommand{\intmomi}[2]{\int\,d^{#1}#2}
\newcommand{\intmomii}[3]{\int\,d^{#1}#2\,\int\,d^{#1}#3}
\newcommand{\intfx}[1]{\int_{\scriptstyle 0}^{\scriptstyle 1}\,d#1}
\newcommand{\intfxy}[2]{\int_{\scriptstyle 0}^{\scriptstyle 1}\,d#1\,
                        \int_{\scriptstyle 0}^{\scriptstyle #1}\,d#2}
\newcommand{\intfxyz}[3]{\int_{\scriptstyle 0}^{\scriptstyle 1}\,d#1\,
                         \int_{\scriptstyle 0}^{\scriptstyle #1}\,d#2\,
                         \int_{\scriptstyle 0}^{\scriptstyle #2}\,d#3}
\newcommand{\Beta}[2]{{\rm{B}}\lpar #1,#2\rpar}
\newcommand{\sBeta}{\rm{B}}
\newcommand{\sign}[1]{{\rm{sign}}\lpar{#1}\rpar}
%
% Z widths
%=========
%
\newcommand{\gn}{\Gamma_{\nu}}
\newcommand{\gel}{\Gamma_{\fe}}
\newcommand{\gmu}{\Gamma_{\mu}}
\newcommand{\gff}{\Gamma_{\ff}}
\newcommand{\gt}{\Gamma_{\tau}}
\newcommand{\gl}{\Gamma_{\fl}}
\newcommand{\gq}{\Gamma_{\fq}}
\newcommand{\gu}{\Gamma_{\fu}}
\newcommand{\gd}{\Gamma_{\fd}}
\newcommand{\gc}{\Gamma_{\fc}}
\newcommand{\gs}{\Gamma_{\fs}}
\newcommand{\gbq}{\Gamma_{\ffb}}
\newcommand{\gz}{\Gamma_{_{\zb}}}
\newcommand{\gw}{\Gamma_{_{\wb}}}
\newcommand{\gh}{\Gamma_{_{h}}}
\newcommand{\ghb}{\Gamma_{_{\hb}}}
\newcommand{\gi}{\Gamma_{\rm{inv}}}
\newcommand{\gzs}{\Gamma^2_{_{\zb}}}
%
% Quantum numbers
%================
%
\newcommand{\tcie}{I^{(3)}_{\fe}}
\newcommand{\tcim}{I^{(3)}_{\flm}}
\newcommand{\tcif}{I^{(3)}_{\ff}}
\newcommand{\tciq}{I^{(3)}_{\fq}}
\newcommand{\tcib}{I^{(3)}_{\ffb}}
\newcommand{\tcih}{I^{(3)}_h}
\newcommand{\tcii}{I^{(3)}_i}
\newcommand{\tcift}{I^{(3)}_{\tilde f}}
\newcommand{\tcifp}{I^{(3)}_{f'}}
\newcommand{\wispt}[1]{I^{(3)}_{#1}}
\newcommand{\ql}{Q_l}
\newcommand{\qe}{Q_e}
\newcommand{\qu}{Q_u}
\newcommand{\qd}{Q_d}
\newcommand{\qb}{Q_b}
\newcommand{\qt}{Q_t}
\newcommand{\qup}{Q'_u}
\newcommand{\qdp}{Q'_d}
\newcommand{\qmu}{Q_{\mu}}
\newcommand{\qes}{Q^2_e}
\newcommand{\qec}{Q^3_e}
\newcommand{\qus}{Q^2_u}
\newcommand{\qds}{Q^2_d}
\newcommand{\qbs}{Q^2_b}
\newcommand{\qts}{Q^2_t}
\newcommand{\qbc}{Q^3_b}
\newcommand{\qf}{Q_f}
\newcommand{\qfs}{Q^2_f}
\newcommand{\qfc}{Q^3_f}
\newcommand{\qff}{Q^4_f}
\newcommand{\qep}{Q_{e'}}
\newcommand{\qfp}{Q_{f'}}
\newcommand{\qfps}{Q^2_{f'}}
\newcommand{\qfpc}{Q^3_{f'}}
\newcommand{\qq}{Q_q}
\newcommand{\qqs}{Q^2_q}
\newcommand{\qi}{Q_i}
\newcommand{\qis}{Q^2_i}
\newcommand{\qj}{Q_j}
\newcommand{\qjs}{Q^2_j}
\newcommand{\QW}{Q_{_\wb}}
\newcommand{\QWs}{Q^2_{_\wb}}
\newcommand{\Qd}{Q_d}
\newcommand{\Qds}{Q^2_d}
\newcommand{\Qu}{Q_u}
\newcommand{\Qus}{Q^2_u}
\newcommand{\vi}{v_i}
\newcommand{\vis}{v^2_i}
\newcommand{\ai}{a_i}
\newcommand{\ais}{a^2_i}
%
% Self-energies
%==============
%
\newcommand{\piv}{\Pi_{_V}}
\newcommand{\pia}{\Pi_{_A}}
\newcommand{\piva}{\Pi_{_{V,A}}}
\newcommand{\pivi}[1]{\Pi^{({#1})}_{_V}}
\newcommand{\piai}[1]{\Pi^{({#1})}_{_A}}
\newcommand{\pivai}[1]{\Pi^{({#1})}_{_{V,A}}}
\newcommand{\pih}{{\hat\Pi}}
\newcommand{\sgh}{{\hat\Sigma}}
\newcommand{\Pgg}{\Pi_{\ph\ph}}
\newcommand{\Ptg}{\Pi_{_{3Q}}}
\newcommand{\Ptt}{\Pi_{_{33}}}
\newcommand{\Pzg}{\Pi_{_{\zb\ab}}}
\newcommand{\Pzga}[2]{\Pi^{#1}_{_{\zb\ab}}\lpar#2\rpar}
\newcommand{\Pf}{\Pi_{_F}}
\newcommand{\Sgg}{\Sigma_{_{\ab\ab}}}
\newcommand{\Szg}{\Sigma_{_{\zb\ab}}}
\newcommand{\SVV}{\Sigma_{_{\vb\vb}}}
\newcommand{\USvv}{{\hat\Sigma}_{_{\vb\vb}}}
\newcommand{\Sww}{\Sigma_{_{\wb\wb}}}
\newcommand{\Swwg}{\Sigma^{_G}_{_{\wb\wb}}}
\newcommand{\Szz}{\Sigma_{_{\zb\zb}}}
\newcommand{\Shh}{\Sigma_{_{\hb\hb}}}
\newcommand{\Spzz}{\Sigma'_{_{\zb\zb}}}
\newcommand{\Stg}{\Sigma_{_{3Q}}}
\newcommand{\Stt}{\Sigma_{_{33}}}
\newcommand{\bSww}{{\overline\Sigma}_{_{WW}}}
\newcommand{\bStg}{{\overline\Sigma}_{_{3Q}}}
\newcommand{\bStt}{{\overline\Sigma}_{_{33}}}
\newcommand{\Sssn}{\Sigma_{_{\hkn\hkn}}}
\newcommand{\Sssc}{\Sigma_{_{\phi\phi}}}
\newcommand{\Szn}{\Sigma_{_{\zb\hkn}}}
\newcommand{\Swc}{\Sigma_{_{\wb\hkg}}}
\newcommand{\mix}[2]{{\cal{M}}^{#1}\lpar{#2}\rpar}
\newcommand{\bmix}[2]{\Pi^{{#1},F}_{_{\zb\ab}}\lpar{#2}\rpar}
\newcommand{\hPgg}[2]{{\hat{\Pi}^{{#1},F}}_{_{\ph\ph}}\lpar{#2}\rpar}
\newcommand{\hmix}[2]{{\hat{\Pi}^{{#1},F}}_{_{\zb\ab}}\lpar{#2}\rpar}
\newcommand{\Dz}[2]{{\cal{D}}_{_{\zb}}^{#1}\lpar{#2}\rpar}
\newcommand{\bDz}[2]{{\cal{D}}^{{#1},F}_{_{\zb}}\lpar{#2}\rpar}
\newcommand{\hDz}[2]{{\hat{\cal{D}}}^{{#1},F}_{_{\zb}}\lpar{#2}\rpar}
\newcommand{\Szzd}[2]{\Sigma'^{#1}_{_{\zb\zb}}\lpar{#2}\rpar}
\newcommand{\Swwd}[2]{\Sigma'^{#1}_{_{\wb\wb}}\lpar{#2}\rpar}
\newcommand{\Shhd}[2]{\Sigma'^{#1}_{_{\hb\hb}}\lpar{#2}\rpar}
\newcommand{\ZFren}[2]{{\cal{Z}}^{#1}\lpar{#2}\rpar}
\newcommand{\WFren}[2]{{\cal{W}}^{#1}\lpar{#2}\rpar}
\newcommand{\HFren}[2]{{\cal{H}}^{#1}\lpar{#2}\rpar}
\newcommand{\WI}{\cal{W}}
%
% QCD varia
%==========
%
\newcommand{\cf}{c_f}
\newcommand{\Cf}{C_{_F}}
\newcommand{\Nf}{N_f}
\newcommand{\Nc}{N_c}
\newcommand{\Ncs}{N^2_c}
\newcommand{\nf }{n_f}
\newcommand{\nfs}{n^2_f}
\newcommand{\nfc}{n^3_f}
\newcommand{\MSB}{\overline{MS}}
\newcommand{\LMSB}{\Lambda_{\overline{\mathrm{MS}}}}
\newcommand{\LMSBp}{\Lambda'_{\overline{\mathrm{MS}}}}
\newcommand{\LMSBS}{\Lambda^2_{\overline{\mathrm{MS}}}}
\newcommand{\LMSBv }{\mbox{$\Lambda^{(5)}_{\overline{\mathrm{MS}}}$}}
\newcommand{\LMSBvS}{\mbox{$\left(\Lambda^{(5)}_{\overline{\mathrm{MS}}}\right)^2$}}
\newcommand{\LMSBt }{\mbox{$\Lambda^{(3)}_{\overline{\mathrm{MS}}}$}}
\newcommand{\LMSBtS}{\mbox{$\left(\Lambda^{(3)}_{\overline{\mathrm{MS}}}\right)^2$}}
\newcommand{\LMSBf }{\mbox{$\Lambda^{(4)}_{\overline{\mathrm{MS}}}$}}
\newcommand{\LMSBfS}{\mbox{$\left(\Lambda^{(4)}_{\overline{\mathrm{MS}}}\right)^2$}}
\newcommand{\LMSBn }{\mbox{$\Lambda^{(\nf)}_{\overline{\mathrm{MS}}}$}}
\newcommand{\LMSBnS}{\mbox{$\left(\Lambda^{(\nf)}_{\overline{\mathrm{MS}}}\right)^2$}}
\newcommand{\LMSBnml }{\mbox{$\Lambda^{(\nf-1)}_{\overline{\mathrm{MS}}}$}}
\newcommand{\LMSBnmlS}{\mbox{$\left(\Lambda^{(\nf-1)}_{\overline{\mathrm{MS}}}\right)^2$}}
\newcommand{\Bnf}{\lpar\nf \rpar}
\newcommand{\Bnfm}{\lpar\nf-1 \rpar}
\newcommand{\LuM}{L_{_M}}
\newcommand{\bef}{\beta_{\ff}}
\newcommand{\befs}{\beta^2_{\ff}}
\newcommand{\befc}{\beta^3_{f}}
\newcommand{\alsp}{\alpha'_{_S}}
\newcommand{\api}{\displaystyle \frac{\als(s)}{\pi}}
\newcommand{\alss}{\alpha^2_{_S}}
\newcommand{\ztwo}{\zeta(2)}
\newcommand{\ztri}{\zeta(3)}
\newcommand{\zfor}{\zeta(4)}
\newcommand{\zfiv}{\zeta(5)}
\newcommand{\bi}[1]{b_{#1}}
\newcommand{\ci}[1]{c_{#1}}
\newcommand{\Ci}[1]{C_{#1}}
\newcommand{\bip}[1]{b'_{#1}}
\newcommand{\cip}[1]{c'_{#1}}
%
% Numerical factors
%==================
%
\newcommand{\osps}{16\,\pi^2}
\newcommand{\srt}{\sqrt{2}}
\newcommand{\ospsi}{\displaystyle{\frac{i}{16\,\pi^2}}}
%
% 2f processes
%=============
%
\newcommand{\tfpromu}{\mbox{$e^+e^-\to \mu^+\mu^-$}}
\newcommand{\tfprotau}{\mbox{$e^+e^-\to \tau^+\tau^-$}}
\newcommand{\tfproe}{\mbox{$e^+e^-\to e^+e^-$}}
\newcommand{\tfpronu}{\mbox{$e^+e^-\to \barnu\nu$}}
\newcommand{\tfproqq}{\mbox{$e^+e^-\to \barq q$}}
\newcommand{\tfprohad}{\mbox{$e^+e^-\to\,$} hadrons}
%
% brems. processes
%-----------------
%
\newcommand{\bpromu}{\mbox{$e^+e^-\to \mu^+\mu^-\ph$}}
\newcommand{\bprotau}{\mbox{$e^+e^-\to \tau^+\tau^-\ph$}}
\newcommand{\bproe}{\mbox{$e^+e^-\to e^+e^-\ph$}}
\newcommand{\bpronu}{\mbox{$e^+e^-\to \barnu\nu\ph$}}
\newcommand{\bproqq}{\mbox{$e^+e^-\to \barq q \ph$}}
%
% 2b processes
%-------------
%
\newcommand{\tbprow} {\mbox{$e^+e^-\to \wbp \wbm $}}
\newcommand{\tbproz} {\mbox{$e^+e^-\to \zb  \zb  $}}
\newcommand{\tbproh} {\mbox{$e^+e^-\to \zb  \hb  $}}
\newcommand{\tbprozg}{\mbox{$e^+e^-\to \zb  \ph  $}}
\newcommand{\tbprog} {\mbox{$e^+e^-\to \ph  \ph  $}}
%
% Line-style for propagators
%===========================
%
\newcommand{\Fermionline}[1]{
\vcenter{\hbox{
  \begin{picture}(60,20)(0,{#1})
  \SetScale{2.}
    \ArrowLine(0,5)(30,5)
  \end{picture}}}
}
\newcommand{\AntiFermionline}[1]{
\vcenter{\hbox{
  \begin{picture}(60,20)(0,{#1})
  \SetScale{2.}
    \ArrowLine(30,5)(0,5)
  \end{picture}}}
}
\newcommand{\Photonline}[1]{
\vcenter{\hbox{
  \begin{picture}(60,20)(0,{#1})
  \SetScale{2.}
    \Photon(0,5)(30,5){2}{6.5}
  \end{picture}}}
}
\newcommand{\Gluonline}[1]{
\vcenter{\hbox{
  \begin{picture}(60,20)(0,{#1})
  \SetScale{2.}
    \Gluon(0,5)(30,5){2}{6.5}
  \end{picture}}}
}
\newcommand{\Wbosline}[1]{
\vcenter{\hbox{
  \begin{picture}(60,20)(0,{#1})
  \SetScale{2.}
    \Photon(0,5)(30,5){2}{4}
    \ArrowLine(13.3,3.1)(16.9,7.2)
  \end{picture}}}
}
\newcommand{\Zbosline}[1]{
\vcenter{\hbox{
  \begin{picture}(60,20)(0,{#1})
  \SetScale{2.}
    \Photon(0,5)(30,5){2}{4}
  \end{picture}}}
}
\newcommand{\Philine}[1]{
\vcenter{\hbox{
  \begin{picture}(60,20)(0,{#1})
  \SetScale{2.}
    \DashLine(0,5)(30,5){2}
  \end{picture}}}
}
\newcommand{\Phicline}[1]{
\vcenter{\hbox{
  \begin{picture}(60,20)(0,{#1})
  \SetScale{2.}
    \DashLine(0,5)(30,5){2}
    \ArrowLine(14,5)(16,5)
  \end{picture}}}
}
\newcommand{\Ghostline}[1]{
\vcenter{\hbox{
  \begin{picture}(60,20)(0,{#1})
  \SetScale{2.}
    \DashLine(0,5)(30,5){.5}
    \ArrowLine(14,5)(16,5)
  \end{picture}}}
}
%
% SM Lagrangian in \Rxi
%======================
%
\newcommand{\gauge}{g}
\newcommand{\gpar}{\xi}
\newcommand{\gparA}{\xi_{_A}}
\newcommand{\gparZ}{\xi_{_Z}}
\newcommand{\gpari}[1]{\gpar_{#1}}
\newcommand{\gparis}[1]{\gpar^2_{#1}}
\newcommand{\gpariq}[1]{\gpar^4_{#1}}
\newcommand{\gpars}{\xi^2}
\newcommand{\dgpar}{\Delta\gpar}
\newcommand{\dgparA}{\Delta\gparA}
\newcommand{\dgparZ}{\Delta\gparZ}
\newcommand{\gparq}{\xi^4}
\newcommand{\gparAs}{\xi^2_{_A}}
\newcommand{\gparAq}{\xi^4_{_A}}
\newcommand{\gparZs}{\xi^2_{_Z}}
\newcommand{\gparZq}{\xi^4_{_Z}}
\newcommand{\Rxi}{R_{\gpar}}
\newcommand{\hxi}{\chi}
%
% Lagrangiains
%-------------
%
\newcommand{\LSM}{{\cal{L}}_{_{\rm{SM}}}}
\newcommand{\LSMr}{{\cal{L}}^{\rm{ren}}_{_{\rm{SM}}}}
\newcommand{\LYM}{{\cal{L}}_{_{YM}}}
\newcommand{\Lzer}{{\cal{L}}_{_{0}}}
\newcommand{\Lone}{{\cal{L}}^{{\bos},I}}
\newcommand{\Lpro}{{\cal{L}}_{\rm{prop}}}
\newcommand{\Ls  }{{\cal{L}}_{_{S}}}
\newcommand{\Lsi }{{\cal{L}}^{I}_{_{S}}}
\newcommand{\Lgf }{{\cal{L}}_{gf  }}
\newcommand{\Lgfi}{{\cal{L}}^{I}_{gf}}
\newcommand{\Lf  }{{\cal{L}}^{{\fer},I}_{\ssV}}
\newcommand{\LHf }{{\cal{L}}^{\fer}_{\ssS}}
\newcommand{\LHfi}{{\cal{L}}^{{\fer},I}_{\ssS}}
\newcommand{\Lren}{{\cal{L}}_{\rm{ren}}}
\newcommand{\Lct}{{\cal{L}}_{\rm{ct}}}
\newcommand{\Lcti}[1]{{\cal{L}}^{#1}_{\rm{ct}}}
\newcommand{\LctI}{{\cal{L}}^{(2)}_{\rm{ct}}}
\newcommand{\Llone}{{\cal{L}}}
\newcommand{\LQED}{{\cal{L}}_{_{\rm{QED}}}}
\newcommand{\LQEDr}{{\cal{L}}^{\rm{ren}}_{_{\rm{QED}}}}
\newcommand{\FST}[3]{F_{#1#2}^{#3}}
\newcommand{\cD}[1]{D_{#1}}
\newcommand{\pd}[1]{\partial_{#1}}
\newcommand{\tgen}[1]{\tau^{#1}}
\newcommand{\gbl}{g_1}
\newcommand{\lctt}[3]{\varepsilon_{#1#2#3}}
\newcommand{\lctf}[4]{\varepsilon_{#1#2#3#4}}
\newcommand{\lctfb}[4]{\varepsilon\lpar{#1#2#3#4}\rpar}
\newcommand{\slct}{\varepsilon}
%gauge fixing
\newcommand{\cgfi}[1]{{\cal{C}}^{#1}}
\newcommand{\cgfZ}{{\cal{C}}_{_Z}}
\newcommand{\cgfA}{{\cal{C}}_{_A}}
\newcommand{\cgfZs}{{\cal{C}}^2_{_Z}}
\newcommand{\cgfAs}{{\cal{C}}^2_{_A}}
%parameters of scalar potential
\newcommand{\hpms}{\mu^2}
\newcommand{\hpal}{\alpha_{_H}}
\newcommand{\hpals}{\alpha^2_{_H}}
\newcommand{\hpbe}{\beta_{_H}}
\newcommand{\hpbep}{\beta^{'}_{_H}}
\newcommand{\hpla}{\lambda}
\newcommand{\hpalf}{\alpha_{f}}
\newcommand{\hpbef}{\beta_{f}}
%transformation parameters
\newcommand{\tpar}[1]{\Lambda^{#1}}
%M,L-operators
\newcommand{\Mop}[2]{{\rm{M}}^{#1#2}}
\newcommand{\Lop}[2]{{\rm{L}}^{#1#2}}
\newcommand{\Lgen}[1]{T^{#1}}
\newcommand{\Rgen}[1]{t^{#1}}
\newcommand{\fpari}[1]{\lambda_{#1}}
\newcommand{\fQ}[1]{Q_{#1}}
\newcommand{\unm}{I}
\newcommand{\cDsla}{/\!\!\!\!D}
%
% A-B-C-D functions
%==================
%
\newcommand{\saff}[1]{A_{#1}}                    % A form-factors
\newcommand{\aff}[2]{A_{#1}\lpar #2\rpar}                   
\newcommand{\sbff}[1]{B_{#1}}                    % B form-factors
\newcommand{\sfbff}[1]{B^{F}_{#1}}
\newcommand{\bff}[4]{B_{#1}\lpar #2;#3,#4\rpar}             
\newcommand{\bfft}[3]{B_{#1}\lpar #2,#3\rpar}             
\newcommand{\fbff}[4]{B^{F}_{#1}\lpar #2;#3,#4\rpar}        
\newcommand{\cdbff}[4]{\Delta B_{#1}\lpar #2;#3,#4\rpar}             
\newcommand{\sdbff}[4]{\delta B_{#1}\lpar #2;#3,#4\rpar}             
\newcommand{\cdbfft}[3]{\Delta B_{#1}\lpar #2,#3\rpar}             
\newcommand{\sdbfft}[3]{\delta B_{#1}\lpar #2,#3\rpar}             
\newcommand{\scff}[1]{C_{#1}}                    % C form-factors
\newcommand{\scffo}[2]{C_{#1}\lpar{#2}\rpar}                
\newcommand{\cff}[7]{C_{#1}\lpar #2,#3,#4;#5,#6,#7\rpar}    
\newcommand{\sccff}[5]{c_{#1}\lpar #2;#3,#4,#5\rpar} 
\newcommand{\sdff}[1]{D_{#1}}                    % D form-factors
\newcommand{\dffp}[7]{D_{#1}\lpar #2,#3,#4,#5,#6,#7;}       
\newcommand{\dffm}[4]{#1,#2,#3,#4\rpar}                     
\newcommand{\bzfa}[2]{B^{F}_{_{#2}}\lpar{#1}\rpar}
\newcommand{\bzfaa}[3]{B^{F}_{_{#2#3}}\lpar{#1}\rpar}
\newcommand{\shcff}[4]{C_{_{#2#3#4}}\lpar{#1}\rpar}
\newcommand{\shdff}[6]{D_{_{#3#4#5#6}}\lpar{#1,#2}\rpar}
\newcommand{\scdff}[3]{d_{#1}\lpar #2,#3\rpar} 
\newcommand{\scaldff}[1]{{\cal{D}}^{#1}}
\newcommand{\caldff}[2]{{\cal{D}}^{#1}\lpar{#2}\rpar}
\newcommand{\caldfft}[3]{{\cal{D}}_{#1}^{#2}\lpar{#3}\rpar}
%
% a-b-c-d functions
%------------------
%
\newcommand{\slaff}[1]{a_{#1}}                        
\newcommand{\slbff}[1]{b_{#1}}                        
\newcommand{\slbffh}[1]{{\hat{b}}_{#1}}    
\newcommand{\ssldff}[1]{d_{#1}}                        
\newcommand{\sslcff}[1]{c_{#1}}                        
\newcommand{\slcff}[2]{c_{#1}^{(#2)}}                        
\newcommand{\sldff}[2]{d_{#1}^{(#2)}}                        
\newcommand{\lbff}[3]{b_{#1}\lpar #2;#3\rpar}         
\newcommand{\lbffh}[2]{{\hat{b}}_{#1}\lpar #2\rpar}   
\newcommand{\lcff}[8]{c_{#1}^{(#2)}\lpar  #3,#4,#5;#6,#7,#8\rpar}         
\newcommand{\ldffp}[8]{d_{#1}^{(#2)}\lpar #3,#4,#5,#6,#7,#8;}
\newcommand{\ldffm}[4]{#1,#2,#3,#4\rpar}                   
%
% I-J functions
%--------------
%
\newcommand{\Iff}[4]{I_{#1}\lpar #2;#3,#4 \rpar}
\newcommand{\Jff}[4]{J_{#1}\lpar #2;#3,#4 \rpar}
\newcommand{\Jds}[5]{{\bar{J}}_{#1}\lpar #2,#3;#4,#5 \rpar}
%--
% n-dimension and epsilons
%=========================
%
\newcommand{\nhmt}{\frac{n}{2}-2}
\newcommand{\nhmts}{{n}/{2}-2}
\newcommand{\omnh}{1-\frac{n}{2}}
\newcommand{\nhmo}{\frac{n}{2}-1}
\newcommand{\fmon}{4-n}
\newcommand{\lpi}{\ln\pi}
\newcommand{\lmass}[1]{\ln #1}
\newcommand{\egnh}{\egam{\frac{n}{2}}}
\newcommand{\egomnh}{\egam{1-\frac{n}{2}}}
\newcommand{\egtmnh}{\egam{2-\frac{n}{2}}}
\newcommand{\Ddr}{{\ds\frac{1}{{\bar{\varepsilon}}}}}
\newcommand{\Ddrs}{{\ds\frac{1}{{\bar{\varepsilon}^2}}}}
\newcommand{\Ddrd}{{\bar{\varepsilon}}}
\newcommand{\ept}{\hat\varepsilon}
\newcommand{\Ddrh}{{\ds\frac{1}{\hat{\varepsilon}}}}
\newcommand{\Ddrp}{{\ds\frac{1}{\varepsilon'}}}
\newcommand{\Ddrps}{\lpar{\ds{\frac{1}{\varepsilon'}}}\rpar^2}
\newcommand{\dre}{\varepsilon}
\newcommand{\drei}[1]{\varepsilon_{#1}}
\newcommand{\epp}{\varepsilon'}
\newcommand{\eps}{\varepsilon^*}
%--
%-- Im for masses and propagators
%================================
\newcommand{\ep}{\epsilon}
%--
\newcommand{\propbt}[6]{{{#1_{#2}#1_{#3}}\over{\lpar #1^2 + #4 
-\ib\ep\rpar\lpar\lpar #5\rpar^2 + #6 -\ib\ep\rpar}}}
\newcommand{\propbo}[5]{{{#1_{#2}}\over{\lpar #1^2 + #3 - \ib\ep\rpar
\lpar\lpar #4\rpar^2 + #5 -\ib\ep\rpar}}}
\newcommand{\propc}[6]{{1\over{\lpar #1^2 + #2 - \ib\ep\rpar
\lpar\lpar #3\rpar^2 + #4 -\ib\ep\rpar
\lpar\lpar #5\rpar^2 + #6 -\ib\ep\rpar}}}
%--
\newcommand{\propa}[2]{{1\over {#1^2 + #2^2 - \ib\ep}}}
\newcommand{\propb}[4]{{1\over {\lpar #1^2 + #2 - \ib\ep\rpar
\lpar\lpar #3\rpar^2 + #4 -\ib\ep\rpar}}}
\newcommand{\propbs}[4]{{1\over {\lpar\lpar #1\rpar^2 + #2 - \ib\ep\rpar
\lpar\lpar #3\rpar^2 + #4 -\ib\ep\rpar}}}
\newcommand{\propat}[4]{{#3_{#1}#3_{#2}\over {#3^2 + #4^2 - \ib\ep}}}
\newcommand{\propaf}[6]{{#5_{#1}#5_{#2}#5_{#3}#5_{#4}\over 
{#5^2 + #6^2 -\ib\ep}}}
\newcommand{\momeps}[1]{#1^2 - \ib\ep}
\newcommand{\mopeps}[1]{#1^2 + \ib\ep}
%--
\newcommand{\propz}[1]{{1\over{#1^2 + \mzs - \ib\ep}}}
\newcommand{\propw}[1]{{1\over{#1^2 + \mws - \ib\ep}}}
\newcommand{\proph}[1]{{1\over{#1^2 + \mhs - \ib\ep}}}
\newcommand{\propf}[2]{{1\over{#1^2 + #2}}}
\newcommand{\propzrg}[3]{{{\delta_{#1#2}}\over{{#3}^2 + \mzs - \ib\ep}}}
\newcommand{\propwrg}[3]{{{\delta_{#1#2}}\over{{#3}^2 + \mws - \ib\ep}}}
\newcommand{\propzug}[3]{{
      {\delta_{#1#2} + \displaystyle{{{#3}^{#1}{#3}^{#2}}\over{\mzs}}}
                         \over{{#3}^2 + \mzs - \ib\ep}}}
\newcommand{\propwug}[3]{{
      {\delta_{#1#2} + \displaystyle{{{#3}^{#1}{#3}^{#2}}\over{\mws}}}
                        \over{{#3}^2 + \mws - \ib\ep}}}
%----------------------------------------
\newcommand{\thf}[1]{\theta\lpar #1\rpar}
\newcommand{\epf}[1]{\varepsilon\lpar #1\rpar}
\newcommand{\singp}{\stackrel{sing}{\rightarrow}}
\newcommand{\aint}[3]{\int_{#1}^{#2}\,d #3}
\newcommand{\aroot}[1]{\sqrt{#1}}
\newcommand{\gramc}{\Delta_3}
\newcommand{\gramd}{\Delta_4}
\newcommand{\pinch}[2]{P^{(#1)}\lpar #2\rpar}
\newcommand{\pinchc}[2]{C^{(#1)}_{#2}}
\newcommand{\pinchd}[2]{D^{(#1)}_{#2}}
\newcommand{\loarg}[1]{\ln\lpar #1\rpar}
\newcommand{\loargr}[1]{\ln\lrbr #1\rrbr}
\newcommand{\lsoarg}[1]{\ln^2\lpar #1\rpar}
\newcommand{\ltarg}[2]{\ln\lpar #1\rpar\lpar #2\rpar}
\newcommand{\rfun}[2]{R\lpar #1,#2\rpar}
\newcommand{\pinchb}[3]{B_{#1}\lpar #2,#3\rpar}
\newcommand{\lga}{\ph}
\newcommand{\lzga}{\ssZ\ph}
%
% Auxiliary functions
%
\newcommand{\afa}[5]{A_{#1}^{#2}\lpar #3;#4,#5\rpar}
\newcommand{\bfa}[5]{B_{#1}^{#2}\lpar #3;#4,#5\rpar} 
\newcommand{\hfa}[5]{H_{#1}^{#2}\lpar #3;#4,#5\rpar}
\newcommand{\rfa}[5]{R_{#1}^{#2}\lpar #3;#4,#5\rpar}
\newcommand{\afao}[3]{A_{#1}^{#2}\lpar #3\rpar}
\newcommand{\bfao}[3]{B_{#1}^{#2}\lpar #3\rpar}
\newcommand{\hfao}[3]{H_{#1}^{#2}\lpar #3\rpar}
\newcommand{\rfao}[3]{R_{#1}^{#2}\lpar #3\rpar}
\newcommand{\afas}[2]{A_{#1}^{#2}}
\newcommand{\bfas}[2]{B_{#1}^{#2}}
\newcommand{\hfas}[2]{H_{#1}^{#2}}
\newcommand{\rfas}[2]{R_{#1}^{#2}}
\newcommand{\tfas}[2]{T_{#1}^{#2}}
\newcommand{\afaR}[6]{A_{#1}^{\gpar}\lpar #2;#3,#4,#5,#6 \rpar}
\newcommand{\bfaR}[6]{B_{#1}^{\gpar}\lpar #2;#3,#4,#5,#6 \rpar}
\newcommand{\hfaR}[6]{H_{#1}^{\gpar}\lpar #2;#3,#4,#5,#6 \rpar}
\newcommand{\shfaR}[1]{H_{#1}^{\gpar}}
\newcommand{\rfaR}[6]{R_{#1}^{\gpar}\lpar #2;#3,#4,#5,#6 \rpar}
\newcommand{\srfaR}[1]{R_{#1}^{\gpar}}
\newcommand{\afaRg}[5]{A_{#1 \lga}^{\gpar}\lpar #2;#3,#4,#5 \rpar}
\newcommand{\bfaRg}[5]{B_{#1 \lga}^{\gpar}\lpar #2;#3,#4,#5 \rpar}
\newcommand{\hfaRg}[5]{H_{#1 \lga}^{\gpar}\lpar #2;#3,#4,#5 \rpar}
\newcommand{\shfaRg}[1]{H_{#1\lga}^{\gpar}}
\newcommand{\rfaRg}[5]{R_{#1 \lga}^{\gpar}\lpar #2;#3,#4,#5 \rpar}
\newcommand{\srfaRg}[1]{R_{#1\lga}^{\gpar}}
\newcommand{\afaRt}[3]{A_{#1}^{\gpar}\lpar #2,#3 \rpar}
\newcommand{\hfaRt}[3]{H_{#1}^{\gpar}\lpar #2,#3 \rpar}
\newcommand{\hfaRf}[4]{H_{#1}^{\gpar}\lpar #2,#3,#4 \rpar}
\newcommand{\afasm}[4]{A_{#1}^{\lpar #2,#3,#4 \rpar}}
\newcommand{\bfasm}[4]{B_{#1}^{\lpar #2,#3,#4 \rpar}}
\newcommand{\color}[1]{c_{#1}}
\newcommand{\htf}[2]{H_2\lpar #1,#2\rpar}
\newcommand{\rof}[2]{R_1\lpar #1,#2\rpar}
\newcommand{\rtf}[2]{R_3\lpar #1,#2\rpar}
\newcommand{\rtrans}[2]{R_{#1}^{#2}}
\newcommand{\momf}[2]{#1^2_{#2}}
\newcommand{\Scalvert}[8][70]{
  \vcenter{\hbox{
  \SetScale{0.8}
  \begin{picture}(#1,50)(15,15)
    \Line(25,25)(50,50)      \Text(34,20)[lc]{#6} \Text(11,20)[lc]{#3}
    \Line(50,50)(25,75)      \Text(34,60)[lc]{#7} \Text(11,60)[lc]{#4}
    \Line(50,50)(90,50)      \Text(11,40)[lc]{#2} \Text(55,33)[lc]{#8}
    \GCirc(50,50){10}{1}          \Text(60,48)[lc]{#5} 
  \end{picture}}}
  }
%
% db-s additions, beware, I've modified above also
%
\newcommand{\tHs}{\mu}
\newcommand{\tHsz}{\mu_{_0}}
\newcommand{\tHss}{\mu^2}
\newcommand{\Reb}{{\rm{Re}}}
\newcommand{\Imb}{{\rm{Im}}}
%
% gp's additions 
%
\newcommand{\spd}{\partial}
\newcommand{\fun}[1]{f\lpar{#1}\rpar}
\newcommand{\ffun}[2]{F_{#1}\lpar #2\rpar}
\newcommand{\gfun}[2]{G_{#1}\lpar #2\rpar}
\newcommand{\sffun}[1]{F_{#1}}
\newcommand{\csffun}[1]{{\cal{F}}_{#1}}
\newcommand{\sgfun}[1]{G_{#1}}
\newcommand{\tpfi}{\lpar 2\pi\rpar^4\ib}
\newcommand{\ffv}{F_{_V}}
\newcommand{\fga}{G_{_A}}
\newcommand{\ffm}{F_{_M}}
\newcommand{\ffs}{F_{_S}}
\newcommand{\fgp}{G_{_P}}
\newcommand{\fge}{G_{_E}}
\newcommand{\ffa}{F_{_A}}
\newcommand{\ffps}{F_{_P}}
\newcommand{\ffe}{F_{_E}}
\newcommand{\gacom}[2]{\lpar #1 + #2\gfd\rpar}
\newcommand{\mft}{m_{\tilde f}}
\newcommand{\qft}{Q_{f'}}
\newcommand{\vft}{v_{\tilde f}}
\newcommand{\subb}[2]{b_{#1}\lpar #2 \rpar}
\newcommand{\fwfr}[5]{\Sigma\lpar #1,#2,#3;#4,#5 \rpar}
\newcommand{\slim}[2]{\lim_{#1 \to #2}}
\newcommand{\sprop}[3]{
{#1\over {\lpar q^2\rpar^2\lpar \lpar q+ #2\rpar^2+#3^2\rpar }}}
%
% roots, variables, coefficients
%
\newcommand{\xroot}[1]{x_{#1}}
\newcommand{\yroot}[1]{y_{#1}}
\newcommand{\zroot}[1]{z_{#1}}
\newcommand{\lvar}{l}
\newcommand{\rvar}{r}
\newcommand{\tvar}{t}
\newcommand{\uvar}{u}
\newcommand{\vvar}{v}
\newcommand{\xvar}{x}
\newcommand{\yvar}{y}
\newcommand{\zvar}{z}
\newcommand{\yvarp}{y'}
\newcommand{\rvars}{r^2}
\newcommand{\vvars}{v^2}
\newcommand{\xvars}{x^2}
\newcommand{\yvars}{y^2}
\newcommand{\zvars}{z^2}
\newcommand{\rvarc}{r^3}
\newcommand{\xvarc}{x^3}
\newcommand{\yvarc}{y^3}
\newcommand{\zvarc}{z^3}
\newcommand{\rvarq}{r^4}
\newcommand{\xvarq}{x^4}
\newcommand{\yvarq}{y^4}
\newcommand{\zvarq}{z^4}
\newcommand{\avar}{a}
\newcommand{\avars}{a^2}
\newcommand{\avarc}{a^3}
\newcommand{\avari}[1]{a_{#1}}
\newcommand{\avart}[2]{a_{#1}^{#2}}
\newcommand{\delvari}[1]{\delta_{#1}}
\newcommand{\rvari}[1]{r_{#1}}
\newcommand{\xvari}[1]{x_{#1}}
\newcommand{\yvari}[1]{y_{#1}}
\newcommand{\zvari}[1]{z_{#1}}
\newcommand{\rvart}[2]{r_{#1}^{#2}}
\newcommand{\xvart}[2]{x_{#1}^{#2}}
\newcommand{\yvart}[2]{y_{#1}^{#2}}
\newcommand{\zvart}[2]{z_{#1}^{#2}}
\newcommand{\rvaris}[1]{r^2_{#1}}
\newcommand{\xvaris}[1]{x^2_{#1}}
\newcommand{\yvaris}[1]{y^2_{#1}}
\newcommand{\zvaris}[1]{z^2_{#1}}
\newcommand{\Xvar}{X}
\newcommand{\Xvars}{X^2}
\newcommand{\Xvari}[1]{X_{#1}}
\newcommand{\Xvaris}[1]{X^2_{#1}}
\newcommand{\Yvar}{Y}
\newcommand{\Yvars}{Y^2}
\newcommand{\Yvari}[1]{Y_{#1}}
\newcommand{\Yvaris}[1]{Y^2_{#1}}
%---
\newcommand{\lnx}{\ln\xvar}
\newcommand{\lnz}{\ln\zvar}
\newcommand{\lnsx}{\ln^2\xvar}
\newcommand{\lnsz}{\ln^2\zvar}
\newcommand{\lncz}{\ln^3\zvar}
\newcommand{\lnomz}{\ln\lpar 1-\zvar\rpar}
\newcommand{\lnsomz}{\ln^2\lpar 1-\zvar\rpar}
\newcommand{\ccoefi}[1]{c_{#1}}
\newcommand{\ccoeft}[2]{c^{#1}_{#2}}
%
% Matrices
%
\newcommand{\Smat}{{\cal{S}}}
\newcommand{\Mmat}{{\cal{M}}}
\newcommand{\Xmat}[1]{X_{#1}}
\newcommand{\XmatI}[1]{X^{-1}_{#1}}
\newcommand{\unitmat}{I}
\newcommand{\Kmat}{{C}}
\newcommand{\Kmatc}{{C}^{\dagger}}
\newcommand{\Kmati}[1]{{C}_{#1}}
\newcommand{\Kmatci}[1]{{C}^{\dagger}_{#1}}
\newcommand{\ffac}[2]{f_{#1}^{#2}}
\newcommand{\Ffac}[1]{F_{#1}}
\newcommand{\Rvec}[2]{R^{(#1)}_{#2}}
\newcommand{\momfl}[2]{#1_{#2}}
\newcommand{\momfs}[2]{#1^2_{#2}}
\newcommand{\fpseZ}{A^{^{FP,Z}}}
\newcommand{\fpseA}{A^{^{FP,A}}}
\newcommand{\fptZ}{T^{^{FP,Z}}}
\newcommand{\fptA}{T^{^{FP,A}}}
\newcommand{\dprop}{\overline\Delta}
\newcommand{\dpropi}[1]{d_{#1}}
\newcommand{\dpropic}[1]{d^{c}_{#1}}
\newcommand{\dpropii}[2]{d_{#1}\lpar #2\rpar}
\newcommand{\dpropis}[1]{d^2_{#1}}
\newcommand{\dproppi}[1]{d'_{#1}}
\newcommand{\psf}[4]{P\lpar #1;#2,#3,#4\rpar}
\newcommand{\ssf}[5]{S^{(#1)}\lpar #2;#3,#4,#5\rpar}
\newcommand{\csf}[5]{C_{_S}^{(#1)}\lpar #2;#3,#4,#5\rpar}
%
% polarization vectors
%=====================
%
\newcommand{\lvec}{l}
\newcommand{\lvecs}{l^2}
\newcommand{\lveci}[1]{l_{#1}}
\newcommand{\mvec}{m}
\newcommand{\mvecs}{m^2}
\newcommand{\mveci}[1]{m_{#1}}
\newcommand{\nvec}{n}
\newcommand{\nvecs}{n^2}
\newcommand{\nveci}[1]{n_{#1}}
\newcommand{\epi}[1]{\epsilon_{#1}}
\newcommand{\phep}[1]{\ep_{#1}}
\newcommand{\sphep}{\ep}
\newcommand{\vbep}[1]{e_{#1}}
\newcommand{\vbepp}[1]{e^{+}_{#1}}
\newcommand{\vbepm}[1]{e^{-}_{#1}}
\newcommand{\svbep}{e}
%
% longitudinal polarizations
%===========================
%
\newcommand{\lpol}{\lambda}
\newcommand{\spol}{\sigma}
\newcommand{\rpol}{\rho  }
\newcommand{\kpol}{\kappa}
\newcommand{\lpols}{\lambda^2}
\newcommand{\spols}{\sigma^2}
\newcommand{\rpols}{\rho^2}
\newcommand{\kpols}{\kappa^2}
\newcommand{\lpoli}[1]{\lambda_{#1}}
\newcommand{\spoli}[1]{\sigma_{#1}}
\newcommand{\rpoli}[1]{\rho_{#1}}
\newcommand{\kpoli}[1]{\kappa_{#1}}
%
% some vectors
%=============
%
\newcommand{\uvec}{u}
\newcommand{\uveci}[1]{u_{#1}}
%
% Momenta:
%=========
%
\newcommand{\imom}{q}
\newcommand{\imomi}[1]{q_{#1}}
\newcommand{\imoms}{q^2}
\newcommand{\pmom}{p}
\newcommand{\pmomp}{p'}
\newcommand{\pmoms}{p^2}
\newcommand{\pmomq}{p^4}
\newcommand{\pmomx}{p^6}
\newcommand{\pmomi}[1]{p_{#1}}
\newcommand{\pmomis}[1]{p^2_{#1}}
%--
\newcommand{\Pmom}{P}
\newcommand{\Pmoms}{P^2}
\newcommand{\Pmomi}[1]{P_{#1}}
\newcommand{\Pmomis}[1]{P^2_{#1}}
\newcommand{\Kmom}{K}
\newcommand{\Kmoms}{K^2}
\newcommand{\Kmomi}[1]{K_{#1}}
\newcommand{\Kmomis}[1]{K^2_{#1}}
%--
\newcommand{\kmom}{k}
\newcommand{\kmoms}{k^2}
\newcommand{\kmomi}[1]{k_{#1}}
\newcommand{\lmom}{l}
\newcommand{\lmoms}{l^2}
\newcommand{\lmomi}[1]{l_{#1}}
\newcommand{\qmom}{q}
\newcommand{\qmoms}{q^2}
\newcommand{\qmomi}[1]{q_{#1}}
\newcommand{\qmomis}[1]{q^2_{#1}}
\newcommand{\smom}{s}
\newcommand{\smoms}{s^2}
\newcommand{\smomi}[1]{s_{#1}}
\newcommand{\tmom}{t}
\newcommand{\tmoms}{t^2}
\newcommand{\tmomi}[1]{t_{#1}}
\newcommand{\Trmom}{Q}
\newcommand{\Prmom}{P}
\newcommand{\gmv}{Q^2}
\newcommand{\Trmoms}{Q^2}
\newcommand{\Prmoms}{P^2}
\newcommand{\Ptmoms}{T^2}
\newcommand{\Pumoms}{U^2}
\newcommand{\Trmomq}{Q^4}
\newcommand{\Prmomq}{P^4}
\newcommand{\Ptmomq}{T^4}
\newcommand{\Pumomq}{U^4}
\newcommand{\Trmomx}{Q^6}
\newcommand{\Trmomi}[1]{Q_{#1}}
\newcommand{\Trmomis}[1]{Q^2_{#1}}
\newcommand{\Prmomi}[1]{P_{#1}}
\newcommand{\pone}{p_1}
\newcommand{\ptwo}{p_2}
\newcommand{\ptre}{p_3}
\newcommand{\pfor}{p_4}
\newcommand{\pones}{p_1^2}
\newcommand{\ptwos}{p_2^2}
\newcommand{\ptres}{p_3^2}
\newcommand{\pfors}{p_4^2}
\newcommand{\poneq}{p_1^4}
\newcommand{\ptwoq}{p_2^4}
\newcommand{\ptreq}{p_3^4}
\newcommand{\pforq}{p_4^4}
\newcommand{\modmom}[1]{\mid{\vec{#1}}\mid}
\newcommand{\modmomi}[2]{\mid{\vec{#1}}_{#2}\mid}
\newcommand{\vect}[1]{{\vec{#1}}}
\newcommand{\Energ}{E}
\newcommand{\Energp}{E'}
\newcommand{\Energpp}{E''}
\newcommand{\Energs}{E^2}
\newcommand{\Energc}{E^3}
\newcommand{\Energf}{E^4}
\newcommand{\Energv}{E^5}
\newcommand{\Energx}{E^6}
\newcommand{\Energi}[1]{E_{#1}}
\newcommand{\Energt}[2]{E_{#1}^{#2}}
\newcommand{\Energis}[1]{E^2_{#1}}
\newcommand{\energ}{e}
\newcommand{\energp}{e'}
\newcommand{\energpp}{e''}
\newcommand{\energs}{e^2}
\newcommand{\energi}[1]{e_{#1}}
\newcommand{\energt}[2]{e_{#1}^{#2}}
\newcommand{\energis}[1]{e^2_{#1}}
\newcommand{\wenerg}{w}
\newcommand{\wenergs}{w^2}
\newcommand{\wenergi}[1]{w_{#1}}
\newcommand{\wenergp}{w'}
\newcommand{\wenergpp}{w''}
%
% kinematical cuts
%=================
%
\newcommand{\ecut}{e}
\newcommand{\ecuts}{e^2}
\newcommand{\ecuti}[1]{e^{#1}}
\newcommand{\ccut}{c_m}
\newcommand{\ccuti}[1]{c_{#1}}
\newcommand{\ccuts}{c^2_m}
\newcommand{\scuts}{s^2_m}
\newcommand{\ccutis}[1]{c^2_{#1}}
\newcommand{\ccutic}[1]{c^3_{#1}}
\newcommand{\ccutc}{c^3_m}
\newcommand{\rcut}{\varrho}
\newcommand{\rcuts}{\varrho^2}
\newcommand{\rcuti}[1]{\varrho_{#1}}
\newcommand{\rcutu}[1]{\varrho^{#1}}
\newcommand{\Dcut}{\Delta}
%
%-----LIB_VERT_XI1.TEX--------------------------
\newcommand{\dwf}{\delta_{_{WF}}}
\newcommand{\gbar}{\overline g}
\newcommand{\PP}{\mbox{PP}}
\newcommand{\mv}{m_{_V}}
\newcommand{\bGv}{{\overline\Gamma}_{_V}}
\newcommand{\Umuv}{\hat{\mu}_\ssV}
\newcommand{\Svv}{{\Sigma}_\ssV}
\newcommand{\muv}{p_\ssV}
\newcommand{\muvb}{\mu_{\ssV_{0}}}
\newcommand{\URPvv}{{P}_\ssV}
\newcommand{\RPvv}{{P}_\ssV}
\newcommand{\Svvrem}{{\Sigma}_\ssV^{\mathrm{rem}}}
\newcommand{\USvvrem}{\hat{\Sigma}_\ssV^{\mathrm{rem}}}
\newcommand{\Gv}{\Gamma_{_V}}
%
% Renormalization 
%
\newcommand{\param}{p}
\newcommand{\parami}[1]{p^{#1}}
\newcommand{\paramb}{p_{0}}
\newcommand{\Zcon}{Z}
\newcommand{\Zconi}[1]{Z_{#1}}
\newcommand{\zconi}[1]{z_{#1}}
\newcommand{\Zconim}[1]{{Z^-_{#1}}}
\newcommand{\zconim}[1]{{z^-_{#1}}}
\newcommand{\Zcont}[2]{Z_{#1}^{#2}}
\newcommand{\zcont}[2]{z_{#1}^{#2}}
\newcommand{\zcontm}[2]{z_{#1}^{{#2}-}}
\newcommand{\sZconi}[2]{\sqrt{Z_{#1}}^{\;#2}}
\newcommand{\php}[3]{e^{#1}_{#2}\lpar #3 \rpar}
\newcommand{\gacome}[1]{\lpar #1 - \gfd\rpar}
\newcommand{\sPj}[2]{\Lambda^{#1}_{#2}}
\newcommand{\sPjs}[2]{\Lambda_{#1,#2}}
\newcommand{\amos}{\mbox{$M^2_{_1}$}}
\newcommand{\amts}{\mbox{$M^2_{_2}$}}
\newcommand{\er}{e_{_{R}}}
\newcommand{\epr}{e'_{_{R}}}
\newcommand{\ers}{e^2_{_{R}}}
\newcommand{\erc}{e^3_{_{R}}}
\newcommand{\erq}{e^4_{_{R}}}
\newcommand{\erf}{e^5_{_{R}}}
\newcommand{\sour}{J}
\newcommand{\sourb}{\overline J}
\newcommand{\lrm}{M_{_R}}
%
% db: fermionic self-energies and vertex libraries
%
\newcommand{\vlami}[1]{\lambda_{#1}}
\newcommand{\vlamis}[1]{\lambda^2_{#1}}
\newcommand{\Vvert}{V}
\newcommand{\Avert}{A}
\newcommand{\Svert}{S}
\newcommand{\Pvert}{P}
\newcommand{\vvert}{F}
\newcommand{\Cvert}{\cal{V}}
\newcommand{\Bvert}{\cal{B}}
\newcommand{\Vveri}[2]{V_{#1}^{#2}}
\newcommand{\Fveri}[1]{{\cal{F}}^{#1}}
\newcommand{\Cveri}[1]{{\cal{V}}\lpar{#1}\rpar}
\newcommand{\Bveri}[1]{{\cal{B}}\lpar{#1}\rpar}
\newcommand{\Vverti}[3]{V_{#1}^{#2}\lpar{#3}\rpar}
\newcommand{\Averti}[3]{A_{#1}^{#2}\lpar{#3}\rpar}
\newcommand{\Gverti}[3]{G_{#1}^{#2}\lpar{#3}\rpar}
\newcommand{\Zverti}[3]{Z_{#1}^{#2}\lpar{#3}\rpar}
\newcommand{\Hverti}[2]{H^{#1}\lpar{#2}\rpar}
\newcommand{\Wverti}[3]{W_{#1}^{#2}\lpar{#3}\rpar}
\newcommand{\Cverti}[2]{{\cal{V}}_{#1}^{#2}}
\newcommand{\vverti}[3]{F^{#1}_{#2}\lpar{#3}\rpar}
\newcommand{\averti}[3]{{\overline{F}}^{#1}_{#2}\lpar{#3}\rpar}
\newcommand{\fveone}[1]{f_{#1}}
\newcommand{\fvetri}[3]{f^{#1}_{#2}\lpar{#3}\rpar}
\newcommand{\gvetri}[3]{g^{#1}_{#2}\lpar{#3}\rpar}
\newcommand{\cvetri}[3]{{\cal{F}}^{#1}_{#2}\lpar{#3}\rpar}
\newcommand{\hvetri}[3]{{\hat{\cal{F}}}^{#1}_{#2}\lpar{#3}\rpar}
\newcommand{\avetri}[3]{{\overline{\cal{F}}}^{#1}_{#2}\lpar{#3}\rpar}
\newcommand{\fverti}[2]{F^{#1}_{#2}}
\newcommand{\cverti}[2]{{\cal{F}}_{#1}^{#2}}
\newcommand{\fV}{f_{_{\Vvert}}}
\newcommand{\gA}{g_{_{\Avert}}}
\newcommand{\fVi}[1]{f^{#1}_{_{\Vvert}}}
\newcommand{\seai}[1]{a_{#1}}
\newcommand{\seapi}[1]{a'_{#1}}
\newcommand{\seAi}[2]{A_{#1}^{#2}}
\newcommand{\sewi}[1]{w_{#1}}
\newcommand{\seWi}[1]{W_{#1}}
\newcommand{\seWsi}[1]{W^{*}_{#1}}
\newcommand{\seWti}[2]{W_{#1}^{#2}}
\newcommand{\sewti}[2]{w_{#1}^{#2}}
\newcommand{\seSig}[1]{\Sigma_{#1}\lpar\sla{\pmom}\rpar}
\newcommand{\ww}{w}
%
% gp: sm_renorm_oneloop
%
\newcommand{\bbff}[1]{{\overline B}_{#1}}
\newcommand{\sW}{p_{_W}}
\newcommand{\sZ}{p_{_Z}}
\newcommand{\ssp}{s_p}
\newcommand{\fW}{f_{_W}}
\newcommand{\fZ}{f_{_Z}}
\newcommand{\tabn}[1]{Tab.(\ref{#1})}
\newcommand{\subMSB}[1]{{#1}_{\mbox{$\overline{\scriptscriptstyle MS}$}}}
\newcommand{\supMSB}[1]{{#1}^{\mbox{$\overline{\scriptscriptstyle MS}$}}}
\newcommand{\redMSB}{{\mbox{$\overline{\scriptscriptstyle MS}$}}}
\newcommand{\gpbb}{g'_{0}}
\newcommand{\Zconip}[1]{Z'_{#1}}
\newcommand{\bpff}[4]{B'_{#1}\lpar #2;#3,#4\rpar}             % B' form-factor
\newcommand{\xidf}{\xi^2-1}
\newcommand{\tDdr}{1/{\bar{\varepsilon}}}
\newcommand{\cRz}{{\cal R}_{_Z}}
\newcommand{\cRg}{{\cal R}_{\gamma}}
\newcommand{\Sz}{\Sigma_{_Z}}
\newcommand{\alh}{{\hat\alpha}}
\newcommand{\alhz}{\alpha_{_Z}}
\newcommand{\Phzg}{{\hat\Pi}_{_{\zb\ab}}}
\newcommand{\fvvert}{F^{\rm vert}_{_V}}
\newcommand{\gavert}{G^{\rm vert}_{_A}}
\newcommand{\bmv}{{\overline m}_{_V}}
\newcommand{\Sgn}{\Sigma_{\gamma\hkn}}
\newcommand{\tabns}[2]{Tabs.(\ref{#1}--\ref{#2})}
\newcommand{\rmboxd}{{\rm Box}_d\lpar s,t,u;M_1,M_2,M_3,M_4\rpar}
\newcommand{\rmboxc}{{\rm Box}_c\lpar s,t,u;M_1,M_2,M_3,M_4\rpar}
%
% D-functions
%
\newcommand{\Afaci}[1]{A_{#1}}
\newcommand{\Afacis}[1]{A^2_{#1}}
\newcommand{\upar}[1]{u}
\newcommand{\upari}[1]{u_{#1}}
\newcommand{\vpari}[1]{v_{#1}}
\newcommand{\lpari}[1]{l_{#1}}
\newcommand{\Lpari}[1]{l_{#1}}
\newcommand{\Nff}[2]{N^{(#1)}_{#2}}
\newcommand{\Sff}[2]{S^{(#1)}_{#2}}
\newcommand{\sSff}{S}
\newcommand{\FQED}[2]{F_{#1#2}}
\newcommand{\fbpsif}{{\overline{\psi}_f}}
\newcommand{\fpsif}{\psi_f}
\newcommand{\etafd}[2]{\eta_d\lpar#1,#2\rpar}
\newcommand{\sigdu}[2]{\sigma_{#1#2}}
\newcommand{\scalc}[4]{c_{_0}\lpar #1;#2,#3,#4\rpar}
\newcommand{\scald}[2]{d_{_0}\lpar #1,#2\rpar}
\newcommand{\pir}[1]{\Pi^{\rm ren}\lpar #1\rpar}
\newcommand{\sigh}{\sigma_{\rm had}}
\newcommand{\dah}{\Delta\alpha^{(5)}_{\rm had}}
\newcommand{\dat}{\Delta\alpha_{\rm top}}
\newcommand{\Vqed}[3]{V_1^{\rm sub}\lpar#1;#2,#3\rpar}
\newcommand{\thetah}{{\hat\theta}}
\newcommand{\mtsix}{m^6_t}
\newcommand{\smlon}{\frac{\mlones}{s}}
\newcommand{\lntwo}{\ln 2}
\newcommand{\wmin}{w_{\rm min}}
\newcommand{\kmin}{k_{\rm min}}
\newcommand{\scaldi}[3]{d_{_0}^{#1}\lpar #2,#3\rpar}
\newcommand{\mdls}{\Big|}
\newcommand{\smf}{\frac{\mfs}{s}}
\newcommand{\bint}{\beta_{\rm int}}
\newcommand{\IRv}{V_{_{\rm IR}}}
\newcommand{\IRr}{R_{_{\rm IR}}}
\newcommand{\fssts}{\frac{s^2}{t^2}}
\newcommand{\fssus}{\frac{s^2}{u^2}}
\newcommand{\optM}{1+\frac{t}{M^2}}
\newcommand{\opuM}{1+\frac{u}{M^2}}
\newcommand{\ftM}{\lpar -\frac{t}{M^2}\rpar}
\newcommand{\fuM}{\lpar -\frac{u}{M^2}\rpar}
\newcommand{\omsM}{1-\frac{s}{M^2}}
\newcommand{\xsf}{\sigma_{_{\rm F}}}
\newcommand{\xsb}{\sigma_{_{\rm B}}}
\newcommand{\afb}{A_{_{\rm FB}}}
\newcommand{\rsoft}{\rm soft}
\newcommand{\rms}{\rm s}
\newcommand{\rsmx}{\sqrt{s_{\rm max}}}
\newcommand{\rspm}{\sqrt{s_{\pm}}}
\newcommand{\rsp}{\sqrt{s_{+}}}
\newcommand{\rsm}{\sqrt{s_{-}}}
\newcommand{\sigmx}{\sigma_{\rm max}}
\newcommand{\gG}[2]{G_{#1}^{#2}}
\newcommand{\gacomm}[2]{\lpar #1 - #2\gfd\rpar}
\newcommand{\fcsx}{\frac{1}{\ctwsix}}
\newcommand{\fcq}{\frac{1}{\ctwf}}
\newcommand{\fcs}{\frac{1}{\ctws}}
\newcommand{\affs}[2]{{\cal A}_{#1}\lpar #2\rpar}                   % A
\newcommand{\stwei}{s_{\theta}^8}
\def\mdan{\vspace{1mm}\mpar{\hfil$\downarrow$new\hfil}\vspace{-1mm}
          \ignorespaces}
\def\muan{\vspace{-1mm}\mpar{\hfil$\uparrow$new\hfil}\vspace{1mm}\ignorespaces}
\def\mlan{\vspace{-1mm}\mpar{\hfil$\rightarrow$new\hfil}\vspace{1mm}\ignorespaces}
\def\mnnew{\mpar{\hfil NEWNEW \hfil}\ignorespaces}
%
% db's of libraries
%                  
\newcommand{\boxc}[2]{{\cal{B}}_{#1}^{#2}}
\newcommand{\boxct}[3]{{\cal{B}}_{#1}^{#2}\lpar{#3}\rpar}
\newcommand{\hboxc}[3]{\hat{{\cal{B}}}_{#1}^{#2}\lpar{#3}\rpar}
\newcommand{\vev}{\langle v \rangle}
\newcommand{\vevi}[1]{\langle v_{#1}\rangle}
\newcommand{\vevs}{\langle v^2   \rangle}
\newcommand{\fwfrV}[5]{\Sigma_{_V}\lpar #1,#2,#3;#4,#5 \rpar}
\newcommand{\fwfrS}[7]{\Sigma_{_S}\lpar #1,#2,#3;#4,#5;#6,#7 \rpar}
\newcommand{\fSi}[1]{f^{#1}_{_{\Svert}}}
\newcommand{\fPi}[1]{f^{#1}_{_{\Pvert}}}
\newcommand{\mXs}{m_{_X}}
\newcommand{\mXss}{m^2_{_X}}
\newcommand{\mYs}{M^2_{_Y}}
\newcommand{\xik}{\xi_k}
\newcommand{\xiks}{\xi^2_k}
\newcommand{\mpls}{m^2_+}
\newcommand{\mmis}{m^2_-}
%
%--
\newcommand{\SN}{\Sigma_{_N}}
\newcommand{\SC}{\Sigma_{_C}}
\newcommand{\SPN}{\Sigma'_{_N}}
\newcommand{\PFf}{\Pi^{\fer}_{_F}}
\newcommand{\PFb}{\Pi^{\bos}_{_F}}
\newcommand{\dPZ}{\Delta{\hat\Pi}_{_Z}}
\newcommand{\Sfin}{\Sigma_{_F}}
\newcommand{\Sfir}{\Sigma_{_R}}
\newcommand{\Sfinh}{{\hat\Sigma}_{_F}}
\newcommand{\Sfinf}{\Sigma^{\fer}_{_F}}
\newcommand{\Sfinbh}{\Sigma^{\bos}_{_F}}
\newcommand{\alf}{\alpha^{\fer}}
\newcommand{\alhfz}{\alpha^{\fer}\lpar{\ssZ}\rpar}
\newcommand{\alhfs}{\alpha^{\fer}\lpar{\sman}\rpar}
\newcommand{\gfQ}{g^f_{_{Q}}}
\newcommand{\gfL}{g^f_{_{L}}}
\newcommand{\ccf}{\frac{\gbs}{16\,\pi^2}}
\newcommand{\chq}{{\hat c}^4}
\newcommand{\muuq}{m_{u'}}
\newcommand{\muus}{m^2_{u'}}
\newcommand{\mdd}{m_{d'}}
%--
\newcommand{\clf}[2]{\mathrm{Cli}_{_#1}\lpar\displaystyle{#2}\rpar}
\def\stes{\sin^2\theta}
\def\acal{\cal A}
\def\alr{A_{_{\rm{LR}}}}
\newcommand{\barQ}{\overline Q}
\newcommand{\Sptg}{\Sigma'_{_{3Q}}}
\newcommand{\Sptt}{\Sigma'_{_{33}}}
\newcommand{\Ppgg}{\Pi'_{\ph\ph}}
\newcommand{\Pww}{\Pi_{_{\wb\wb}}}
\newcommand{\capV}[2]{{\cal F}^{#2}_{_{#1}}}
%--
\newcommand{\bt}{\beta_t}
\newcommand{\mhsix}{M^6_{_H}}
\newcommand{\topt}{{\cal T}_{33}}
\newcommand{\topq}{{\cal T}_4}
\newcommand{\Phzgf}{{\hat\Pi}^{\fer}_{_{\zb\ab}}}
\newcommand{\Phzgb}{{\hat\Pi}^{\bos}_{_{\zb\ab}}}
\newcommand{\Sfirh}{{\hat\Sigma}_{_R}}
\newcommand{\Szgh}{{\hat\Sigma}_{_{\zb\ab}}}
\newcommand{\Szghb}{{\hat\Sigma}^{\bos}_{_{\zb\ab}}}
\newcommand{\Szghf}{{\hat\Sigma}^{\fer}_{_{\zb\ab}}}
\newcommand{\Szgb}{\Sigma^{\bos}_{_{\zb\ab}}}
\newcommand{\Szgf}{\Sigma^{\fer}_{_{\zb\ab}}}
%--
\newcommand{\chig}{\chi_{_{\ph}}}
\newcommand{\chiz}{\chi_{_{\zb}}}
\newcommand{\Sfih}{{\hat\Sigma}}
\newcommand{\Szzh}{\hat{\Sigma}_{_{\zb\zb}}}
\newcommand{\dPZf}{\Delta{\hat\Pi}^f_{_{\zb}}}
\newcommand{\khZdf}[1]{{\hat\kappa}^{#1}_{_{\zb}}}
\newcommand{\chf}{{\hat c}^4}
%---
%for process sm_ola
\newcommand{\amp}[2]{{\cal{A}}_{_{#1}}^{\rm{#2}}}
%--
\newcommand{\hatvm}[1]{{\hat v}^-_{#1}}
\newcommand{\hatvp}[1]{{\hat v}^+_{#1}}
\newcommand{\hatvpm}[1]{{\hat v}^{\pm}_{#1}}
\newcommand{\kvz}[1]{\kappa^{\zb #1}_{_V}}
%--
\newcommand{\barp}{\overline p}                
\newcommand{\delw}{\Delta_{_{\wb}}}
\newcommand{\bdelw}{{\bar{\Delta}}_{_{\wb}}}
\newcommand{\bdelf}{{\bar{\Delta}}_{\ff}}
\newcommand{\delz}{\Delta_{_\zb}}
\newcommand{\deli}[1]{\Delta\lpar{#1}\rpar}
\newcommand{\chizb}{\chi_{_\zb}}
\newcommand{\Swwp}{\Sigma'_{_{\wb\wb}}}
\newcommand{\epph}{\varepsilon'/2}
\newcommand{\sbffp}[1]{B'_{#1}}                    
\newcommand{\epss}{\varepsilon^*}
%--
\newcommand{\Ddrhs}{{\ds\frac{1}{\hat{\varepsilon}^2}}}
\newcommand{\lnmsb}{L_{_\wb}}
\newcommand{\lnsmsb}{L^2_{_\wb}}
\newcommand{\tpni}{\lpar 2\pi\rpar^n\ib}
\newcommand{\tpn}{2^n\,\pi^{n-2}}
%--
\newcommand{\cmf}{M_f}
\newcommand{\cmfs}{M^2_f}
\newcommand{\toDdr}{{\ds\frac{2}{{\bar{\varepsilon}}}}}
\newcommand{\troDdr}{{\ds\frac{3}{{\bar{\varepsilon}}}}}
\newcommand{\totDdr}{{\ds\frac{3}{{2\,\bar{\varepsilon}}}}}
\newcommand{\foDdr}{{\ds\frac{4}{{\bar{\varepsilon}}}}}
\newcommand{\smh}{m_{_H}}
\newcommand{\smhs}{m^2_{_H}}
\newcommand{\Ph}{\Pi_{_\hb}}
\newcommand{\Sphh}{\Sigma'_{_{\hb\hb}}}
\newcommand{\bh}{\beta}
\newcommand{\alsn}{\alpha^{(n_f)}_{_S}}
\newcommand{\smq}{m_q}
\newcommand{\smqp}{m_{q'}}
\newcommand{\shb}{h}
\newcommand{\hab}{A}
\newcommand{\hbpm}{H^{\pm}}
\newcommand{\hbp}{H^{+}}
\newcommand{\hbm}{H^{-}}
\newcommand{\msh}{M_h}
\newcommand{\mha}{M_{_A}}
\newcommand{\mhc}{M_{_{H^{\pm}}}}
\newcommand{\mshs}{M^2_h}
\newcommand{\mhas}{M^2_{_A}}
\newcommand{\barfp}{\overline{f'}}                
\newcommand{\chiii}{{\hat c}^3}
\newcommand{\chiv}{{\hat c}^4}
\newcommand{\chv}{{\hat c}^5}
\newcommand{\chvi}{{\hat c}^6}
\newcommand{\alsvi}{\alpha^{6}_{_S}}
\newcommand{\tww}{t_{_W}}
\newcommand{\ti}{t_{_1}}
\newcommand{\tii}{t_{_2}}
\newcommand{\tiii}{t_{_3}}
\newcommand{\tiv}{t_{_4}}
\newcommand{\psla}{\hbox{\rlap/p}}
\newcommand{\qsla}{\hbox{\rlap/q}}
\newcommand{\nsla}{\hbox{\rlap/n}}
\newcommand{\lsla}{\hbox{\rlap/l}}
\newcommand{\msla}{\hbox{\rlap/m}}
\newcommand{\cnsla}{\hbox{\rlap/N}}
\newcommand{\clsla}{\hbox{\rlap/L}}
\newcommand{\cmsla}{\hbox{\rlap/M}}
\newcommand{\blmt}{\lrbr - 3\rrbr}
\newcommand{\blfo}{\lrbr 4 1\rrbr}
\newcommand{\bltp}{\lrbr 2 +\rrbr}
%---------------------------------
% mixed QCD
\newcommand{\clitwo}[1]{{\rm{Li}}_{2}\lpar{#1}\rpar}
\newcommand{\clitri}[1]{{\rm{Li}}_{3}\lpar{#1}\rpar}
% subleadings
\newcommand{\xt}{x_{\ft}}
\newcommand{\zt}{z_{\ft}}
\newcommand{\Ht}{h_{\ft}}
\newcommand{\xts}{x^2_{\ft}}
\newcommand{\zts}{z^2_{\ft}}
\newcommand{\Hts}{h^2_{\ft}}
\newcommand{\ztc}{z^3_{\ft}}
\newcommand{\Htc}{h^3_{\ft}}
\newcommand{\ztq}{z^4_{\ft}}
\newcommand{\Htq}{h^4_{\ft}}
\newcommand{\ztv}{z^5_{\ft}}
\newcommand{\Htv}{h^5_{\ft}}
\newcommand{\ztx}{z^6_{\ft}}
\newcommand{\Htx}{h^6_{\ft}}
\newcommand{\ztz}{z^7_{\ft}}
\newcommand{\Htz}{h^7_{\ft}}
\newcommand{\sht}{\sqrt{\Ht}}
\newcommand{\atan}[1]{{\rm{arctan}}\lpar{#1}\rpar}
\newcommand{\dbff}[3]{{\hat{B}}_{_{{#2}{#3}}}\lpar{#1}\rpar}
\newcommand{\ztbs}{{\bar{z}}^{2}_{\ft}}
\newcommand{\ztb}{{\bar{z}}_{\ft}}
\newcommand{\Htbs}{{\bar{h}}^{2}_{\ft}}
\newcommand{\Htb}{{\bar{h}}_{\ft}}
\newcommand{\Hztb}{{\bar{hz}}_{\ft}}
\newcommand{\Ln}[1]{{\rm{Ln}}\lpar{#1}\rpar}
\newcommand{\Lns}[1]{{\rm{Ln}}^2\lpar{#1}\rpar}
\newcommand{\wt}{w_{\ft}}
\newcommand{\wts}{w^2_{\ft}}
\newcommand{\wtb}{\overline{w}}
%--
\newcommand{\fra}{\frac{1}{2}}
\newcommand{\frb}{\frac{1}{4}}
\newcommand{\frc}{\frac{3}{2}}
\newcommand{\frd}{\frac{3}{4}}
\newcommand{\fre}{\frac{9}{2}}
\newcommand{\frf}{\frac{9}{4}}
\newcommand{\frg}{\frac{5}{4}}
\newcommand{\frh}{\frac{5}{2}}
\newcommand{\fri}{\frac{1}{8}}
\newcommand{\frj}{\frac{7}{4}}
\newcommand{\frl}{\frac{7}{8}}
\newcommand{\Spzzh}{\hat{\Sigma}'_{_{\zb\zb}}}
\newcommand{\sss}{s\sqrt{s}}
\newcommand{\sqs}{\sqrt{s}}
\newcommand{\Rtg}{R_{_{3Q}}}
\newcommand{\Rtt}{R_{_{33}}}
\newcommand{\Rww}{R_{_{\wb\wb}}}
\newcommand{\ssZ}{{\scriptscriptstyle{\zb}}}
\newcommand{\ssW}{{\scriptscriptstyle{\wb}}}
\newcommand{\ssH}{{\scriptscriptstyle{\hb}}}
\newcommand{\ssV}{{\scriptscriptstyle{\vb}}}
\newcommand{\ssA}{{\scriptscriptstyle{A}}}
\newcommand{\ssB}{{\scriptscriptstyle{B}}}
\newcommand{\ssC}{{\scriptscriptstyle{C}}}
\newcommand{\ssD}{{\scriptscriptstyle{D}}}
\newcommand{\ssF}{{\scriptscriptstyle{F}}}
\newcommand{\ssG}{{\scriptscriptstyle{G}}}
\newcommand{\ssL}{{\scriptscriptstyle{L}}}
\newcommand{\ssM}{{\scriptscriptstyle{M}}}
\newcommand{\ssN}{{\scriptscriptstyle{N}}}
\newcommand{\ssP}{{\scriptscriptstyle{P}}}
\newcommand{\ssQ}{{\scriptscriptstyle{Q}}}
\newcommand{\ssR}{{\scriptscriptstyle{R}}}
\newcommand{\ssS}{{\scriptscriptstyle{S}}}
\newcommand{\ssT}{{\scriptscriptstyle{T}}}
\newcommand{\ssU}{{\scriptscriptstyle{U}}}
\newcommand{\ssX}{{\scriptscriptstyle{X}}}
\newcommand{\ssY}{{\scriptscriptstyle{Y}}}
\newcommand{\ssWF}{{\scriptscriptstyle{WF}}}
%--
\newcommand{\DiagramFermionToBosonFullWithMomenta}[8][70]{
  \vcenter{\hbox{
  \SetScale{0.8}
  \begin{picture}(#1,50)(15,15)
    \put(27,22){$\nearrow$}      
    \put(27,54){$\searrow$}    
    \put(59,29){$\to$}    
    \ArrowLine(25,25)(50,50)      \Text(34,20)[lc]{#6} \Text(11,20)[lc]{#3}
    \ArrowLine(50,50)(25,75)      \Text(34,60)[lc]{#7} \Text(11,60)[lc]{#4}
    \Photon(50,50)(90,50){2}{8}   \Text(80,40)[lc]{#2} \Text(55,33)[ct]{#8}
    \Vertex(50,50){2,5}          \Text(60,48)[cb]{#5} 
    \Vertex(90,50){2}
  \end{picture}}}
  }
\newcommand{\DiagramFermionToBosonPropagator}[4][85]{
  \vcenter{\hbox{
  \SetScale{0.8}
  \begin{picture}(#1,50)(15,15)
    \ArrowLine(25,25)(50,50)
    \ArrowLine(50,50)(25,75)
    \Photon(50,50)(105,50){2}{8}   \Text(90,40)[lc]{#2}
%    \GCirc(50,50){10}{0.5}         \Text(80,48)[cb]{#3}
    \Vertex(50,50){0.5}         \Text(80,48)[cb]{#3}
    \GCirc(82,50){8}{1}            \Text(55,48)[cb]{#4}
    \Vertex(105,50){2}
  \end{picture}}}
  }
\newcommand{\DiagramFermionToBosonEffective}[3][70]{
  \vcenter{\hbox{
  \SetScale{0.8}
  \begin{picture}(#1,50)(15,15)
    \ArrowLine(25,25)(50,50)
    \ArrowLine(50,50)(25,75)
    \Photon(50,50)(90,50){2}{8}   \Text(80,40)[lc]{#2}
    \BBoxc(50,50)(5,5)            \Text(55,48)[cb]{#3}
    \Vertex(90,50){2}
  \end{picture}}}
  }
\newcommand{\DiagramFermionToBosonFull}[3][70]{
  \vcenter{\hbox{
  \SetScale{0.8}
  \begin{picture}(#1,50)(15,15)
    \ArrowLine(25,25)(50,50)
    \ArrowLine(50,50)(25,75)
    \Photon(50,50)(90,50){2}{8}   \Text(80,40)[lc]{#2}
    \Vertex(50,50){2.5}          \Text(60,48)[cb]{#3}
    \Vertex(90,50){2}
  \end{picture}}}
  }
%--
\newcommand{\expgw}{\frac{\gf\mws}{2\srt\,\pi^2}}
\newcommand{\expgz}{\frac{\gf\mzs}{2\srt\,\pi^2}}
\newcommand{\Spww}{\Sigma'_{_{\wb\wb}}}
\newcommand{\shf}{{\hat s}^4}
%--
\newcommand{\acz}{\scff{0}}
\newcommand{\acoo}{\scff{11}}
\newcommand{\acod}{\scff{12}}
\newcommand{\acdo}{\scff{21}}
\newcommand{\acdd}{\scff{22}}
\newcommand{\acdt}{\scff{23}}
\newcommand{\acdq}{\scff{24}}
\newcommand{\acto}{\scff{31}}
\newcommand{\actd}{\scff{32}}
\newcommand{\actt}{\scff{33}}
\newcommand{\actq}{\scff{34}}
\newcommand{\actc}{\scff{35}}
\newcommand{\acts}{\scff{36}}
\newcommand{\acoA}{\scff{1A}}
\newcommand{\acdA}{\scff{2A}}
\newcommand{\acdB}{\scff{2B}}
\newcommand{\acdC}{\scff{2C}}
\newcommand{\acdD}{\scff{2D}}
\newcommand{\actA}{\scff{3A}}
\newcommand{\actB}{\scff{3B}}
\newcommand{\actC}{\scff{3C}}
%--
\newcommand{\ada}{\sdff{0}}
\newcommand{\adb}{\sdff{11}}
\newcommand{\adc}{\sdff{12}}
\newcommand{\add}{\sdff{13}}
\newcommand{\ade}{\sdff{21}}
\newcommand{\adf}{\sdff{22}}
\newcommand{\adg}{\sdff{23}}
\newcommand{\adh}{\sdff{24}}
\newcommand{\adi}{\sdff{25}}
\newcommand{\adj}{\sdff{26}}
\newcommand{\adl}{\sdff{27}}
\newcommand{\adm}{\sdff{31}}
\newcommand{\adn}{\sdff{32}}
\newcommand{\ado}{\sdff{33}}
\newcommand{\adp}{\sdff{34}}
\newcommand{\adq}{\sdff{35}}
\newcommand{\adr}{\sdff{36}}
\newcommand{\ads}{\sdff{37}}
\newcommand{\adt}{\sdff{38}}
\newcommand{\adu}{\sdff{39}}
\newcommand{\adw}{\sdff{310}}
\newcommand{\adv}{\sdff{311}}
\newcommand{\ady}{\sdff{312}}
\newcommand{\adz}{\sdff{313}}
%--
\newcommand{\admt}{\frac{\tman}{\sman}}
\newcommand{\admu}{\frac{\uman}{\sman}}
\newcommand{\frm}{\frac{3}{8}}
\newcommand{\frn}{\frac{5}{8}}
\newcommand{\fro}{\frac{15}{8}}
\newcommand{\frp}{\frac{3}{16}}
\newcommand{\frq}{\frac{5}{16}}
\newcommand{\frr}{\frac{1}{16}}
\newcommand{\frs}{\frac{7}{2}}
\newcommand{\frt}{\frac{7}{16}}
\newcommand{\fru}{\frac{1}{3}}
\newcommand{\frw}{\frac{2}{3}}
\newcommand{\frz}{\frac{4}{3}}
\newcommand{\fry}{\frac{13}{3}}
\newcommand{\fraa}{\frac{11}{4}}
%--
\newcommand{\bee}{\beta_{e}}
\newcommand{\beW}{\beta_{_\wb}}
%--
\newcommand{\toDdrh}{{\ds\frac{2}{{\hat{\varepsilon}}}}}
\newcommand{\bqas}{\begin{eqnarray*}}
\newcommand{\eqas}{\end{eqnarray*}}
%--
\newcommand{\mhcub}{M^3_{_H}}
%--
\newcommand{\adComA}{\sdff{A}}
\newcommand{\adComB}{\sdff{B}}
\newcommand{\adComC}{\sdff{C}}
\newcommand{\adComD}{\sdff{D}}
\newcommand{\adComE}{\sdff{E}}
\newcommand{\adComF}{\sdff{F}}
\newcommand{\adComG}{\sdff{G}}
\newcommand{\adComH}{\sdff{H}}
\newcommand{\adComI}{\sdff{I}}
\newcommand{\adComJ}{\sdff{J}}
\newcommand{\adComL}{\sdff{L}}
\newcommand{\adComM}{\sdff{M}}
\newcommand{\adComN}{\sdff{N}}
\newcommand{\adComO}{\sdff{O}}
\newcommand{\adComP}{\sdff{P}}
\newcommand{\adComQ}{\sdff{Q}}
\newcommand{\adComR}{\sdff{R}}
\newcommand{\adComS}{\sdff{S}}
\newcommand{\adComT}{\sdff{T}}
\newcommand{\adComU}{\sdff{U}}
%--
\newcommand{\adComAc}{\sdff{A}^c}
\newcommand{\adComBc}{\sdff{B}^c}
\newcommand{\adComCc}{\sdff{C}^c}
\newcommand{\adComDc}{\sdff{D}^c}
\newcommand{\adComEc}{\sdff{E}^c}
\newcommand{\adComFc}{\sdff{F}^c}
\newcommand{\adComGc}{\sdff{G}^c}
\newcommand{\adComHc}{\sdff{H}^c}
\newcommand{\adComIc}{\sdff{I}^c}
\newcommand{\adComJc}{\sdff{J}^c}
\newcommand{\adComLc}{\sdff{L}^c}
\newcommand{\adComMc}{\sdff{M}^c}
\newcommand{\adComNc}{\sdff{N}^c}
\newcommand{\adComOc}{\sdff{O}^c}
\newcommand{\adComPc}{\sdff{P}^c}
\newcommand{\adComQc}{\sdff{Q}^c}
\newcommand{\adComRc}{\sdff{R}^c}
\newcommand{\adComSc}{\sdff{S}^c}
\newcommand{\adComTc}{\sdff{T}^c}
\newcommand{\adComUc}{\sdff{U}^c}
%--
\newcommand{\adComAf}{\sdff{A}^f}
\newcommand{\adComBf}{\sdff{B}^f}
\newcommand{\adComCf}{\sdff{F}^f}
\newcommand{\adComDf}{\sdff{D}^f}
\newcommand{\adComEf}{\sdff{E}^f}
\newcommand{\adComFf}{\sdff{F}^f}
\newcommand{\adComGf}{\sdff{G}^f}
\newcommand{\adComHf}{\sdff{H}^f}
\newcommand{\adComIf}{\sdff{I}^f}
\newcommand{\adComJf}{\sdff{J}^f}
\newcommand{\adComLf}{\sdff{L}^f}
\newcommand{\adComMf}{\sdff{M}^f}
\newcommand{\adComNf}{\sdff{N}^f}
\newcommand{\adComOf}{\sdff{O}^f}
\newcommand{\adComPf}{\sdff{P}^f}
\newcommand{\adComQf}{\sdff{Q}^f}
\newcommand{\adComRf}{\sdff{R}^f}
\newcommand{\adComSf}{\sdff{S}^f}
\newcommand{\adComTf}{\sdff{T}^f}
\newcommand{\adComUf}{\sdff{U}^f}
%--
\newcommand{\adComBfc}{\sdff{B}^{fc}} 
\newcommand{\adComCfco}{\sdff{C}^{fc1}}
\newcommand{\adComCfcd}{\sdff{C}^{fc2}} 
\newcommand{\adComCfct}{\sdff{C}^{fc3}} 
\newcommand{\adComDfc}{\sdff{D}^{fc}}
\newcommand{\adComEfc}{\sdff{E}^{fc}}
\newcommand{\adComFfc}{\sdff{F}^{fc}}
\newcommand{\adComGfc}{\sdff{G}^{fc}}
\newcommand{\adComHfc}{\sdff{H}^{fc}}
\newcommand{\adComLfc}{\sdff{L}^{fc}}
%--
\newcommand{\afba}[1]{A^{#1}_{_{\rm FB}}}
\newcommand{\alra}[1]{A^{#1}_{_{\rm LR}}}
%--
\newcommand{\adComAt}{\sdff{A}^t}
\newcommand{\adComBt}{\sdff{B}^t}
\newcommand{\adComCt}{\sdff{T}^t}
\newcommand{\adComDt}{\sdff{D}^t}
\newcommand{\adComEt}{\sdff{E}^t}
\newcommand{\adComFt}{\sdff{T}^t}
\newcommand{\adComGt}{\sdff{G}^t}
\newcommand{\adComHt}{\sdff{H}^t}
\newcommand{\adComIt}{\sdff{I}^t}
\newcommand{\adComJt}{\sdff{J}^t}
\newcommand{\adComLt}{\sdff{L}^t}
\newcommand{\adComMt}{\sdff{M}^t}
\newcommand{\adComNt}{\sdff{N}^t}
\newcommand{\adComOt}{\sdff{O}^t}
\newcommand{\adComPt}{\sdff{P}^t}
\newcommand{\adComQt}{\sdff{Q}^t}
\newcommand{\adComRt}{\sdff{R}^t}
\newcommand{\adComSt}{\sdff{S}^t}
\newcommand{\adComTt}{\sdff{T}^t}
\newcommand{\adComUt}{\sdff{U}^t}
%--
\newcommand{\adComAtt}{\sdff{A}^{\tau}}
\newcommand{\adComBtt}{\sdff{B}^{\tau}}
\newcommand{\adComCtt}{\sdff{T}^{\tau}}
\newcommand{\adComDtt}{\sdff{D}^{\tau}}
\newcommand{\adComEtt}{\sdff{E}^{\tau}}
\newcommand{\adComFtt}{\sdff{T}^{\tau}}
\newcommand{\adComGtt}{\sdff{G}^{\tau}}
\newcommand{\adComHtt}{\sdff{H}^{\tau}}
\newcommand{\adComItt}{\sdff{I}^{\tau}}
\newcommand{\adComJtt}{\sdff{J}^{\tau}}
\newcommand{\adComLtt}{\sdff{L}^{\tau}}
\newcommand{\adComMtt}{\sdff{M}^{\tau}}
\newcommand{\adComNtt}{\sdff{N}^{\tau}}
\newcommand{\adComOtt}{\sdff{O}^{\tau}}
\newcommand{\adComPtt}{\sdff{P}^{\tau}}
\newcommand{\adComQtt}{\sdff{Q}^{\tau}}
\newcommand{\adComRtt}{\sdff{R}^{\tau}}
\newcommand{\adComStt}{\sdff{S}^{\tau}}
\newcommand{\adComTtt}{\sdff{T}^{\tau}}
\newcommand{\adComUtt}{\sdff{U}^{\tau}}
%--
\newcommand{\etavz}[1]{\eta^{\zb #1}_{_V}}
%--
\newcommand{\phanst}{$\hphantom{\sigma^{s+t}\ }$}
\newcommand{\phanat}{$\hphantom{A_{FB}^{s+t}\ }$}
\newcommand{\phanss}{$\hphantom{\sigma^{s}\ }$}
\newcommand{\phanas}{$\hphantom{A_{FB}^{s}\ }$} 
\newcommand{\pbb}{\,\mbox{\bf pb}}
\newcommand{\pe}{\,\%\:}
\newcommand{\pc}{\,\%}
%--
\newcommand{\temiv}{10^{-4}}
\newcommand{\temv}{10^{-5}}
\newcommand{\temvi}{10^{-6}}
\newcommand{\di}[1]{d_{#1}}
%--
\newcommand{\delip}[1]{\Delta_+\lpar{#1}\rpar}
\newcommand{\propbb}[5]{{{#1}\over {\lpar #2^2 + #3 - \ib\varepsilon\rpar
\lpar\lpar #4\rpar^2 + #5 -\ib\varepsilon\rpar}}}
\newcommand{\cfft}[5]{C_{#1}\lpar #2;#3,#4,#5\rpar}    
\newcommand{\ppl}[1]{p_{+{#1}}}
\newcommand{\pmi}[1]{p_{-{#1}}}
%--
\newcommand{\bpox}{\beta^2_{\xi}}
\newcommand{\bffdiff}[5]{B_{\rm d}\lpar #1;#2,#3;#4,#5\rpar}             
\newcommand{\cffdiff}[7]{C_{\rm d}\lpar #1;#2,#3,#4;#5,#6,#7\rpar}    
\newcommand{\affdiff}[2]{A_{\rm d}\lpar #1;#2\rpar}             
\newcommand{\Dqf}{\Delta\qf}
\newcommand{\bposx}{\beta^4_{\xi}}
\newcommand{\svverti}[3]{f^{#1}_{#2}\lpar{#3}\rpar}
\newcommand{\Mods}{\mbox{$M^2_{12}$}}
\newcommand{\Mots}{\mbox{$M^2_{13}$}}
\newcommand{\Motq}{\mbox{$M^4_{13}$}}
\newcommand{\Mdts}{\mbox{$M^2_{23}$}}
\newcommand{\Mdos}{\mbox{$M^2_{21}$}}
\newcommand{\Mtds}{\mbox{$M^2_{32}$}}
\newcommand{\dffpt}[3]{D_{#1}\lpar #2,#3;}           
\newcommand{\quu}{Q_{uu}}
\newcommand{\qdd}{Q_{dd}}
\newcommand{\qud}{Q_{ud}}
\newcommand{\qdu}{Q_{du}}
%--
\newcommand{\msPj}[6]{\Lambda^{#1#2#3}_{#4#5#6}}
%--
\newcommand{\bdiff}[4]{B_{\rm d}\lpar #1,#2;#3,#4\rpar}             
\newcommand{\bdifff}[7]{B_{\rm d}\lpar #1;#2;#3;#4,#5;#6,#7\rpar}             
\newcommand{\adiff}[3]{A_{\rm d}\lpar #1;#2;#3\rpar}  
%--
\newcommand{\aw}{a_{_\wb}}
\newcommand{\az}{a_{_\zb}}
%--
\newcommand{\sct}[1]{sect.~\ref{#1}}
%--
\newcommand{\dreim}[1]{\varepsilon^{\rm M}_{#1}}
\newcommand{\drem}{\varepsilon^{\rm M}}
%--
\newcommand{\hcapV}[2]{{\hat{\cal F}}^{#2}_{_{#1}}}
%--
\newcommand{\swww}{{\scriptscriptstyle \wb\wb\wb}}
\newcommand{\szhz}{{\scriptscriptstyle \zb\hb\zb}}
\newcommand{\shzh}{{\scriptscriptstyle \hb\zb\hb}}
\newcommand{\bwith}[3]{\beta^{#3}_{#1}\lpar #2\rpar}
\newcommand{\Shhh}{{\hat\Sigma}_{_{\hb\hb}}}
\newcommand{\Sphhh}{{\hat\Sigma}'_{_{\hb\hb}}}
%--
\newcommand{\seWilc}[1]{w_{#1}}
\newcommand{\seWtilc}[2]{w_{#1}^{#2}}
%--
\newcommand{\eilc}{\gamma}
\newcommand{\eilcs}{\gamma^2}
\newcommand{\eilcc}{\gamma^3}
\newcommand{\eilcb}{{\overline{\gamma}}}
\newcommand{\eilcbs}{{\overline{\gamma}^2}}
%--
\newcommand{\Sttww}{\Sigma_{_{33;\wb\wb}}}
\newcommand{\bSttww}{{\overline\Sigma}_{_{33;\wb\wb}}}
%--
\newcommand{\Pggtg}{\Pi_{\ph\ph;3Q}}

%
% Title
%
%%% \include{ltf_note_title}
\title{\bf Precision Calculation Project Report
\thanks{The numerical results are based on the work of {\tt TOPAZ0} and 
{\tt ZFITTER} teams.
At present, the following physicists are active members of the two teams: 
{\tt TOPAZ0}:  G. Montagna, O. Nicrosini, G. Passarino and F. Piccinini;  
{\tt ZFITTER}: D. Bardin, P. Christova, M. Jack, L. Kalinovskaya, 
A. Olshevski, S. Riemann, T. Riemann.}}

\author{
Dmitri Bardin \footnote{e-mail: {\tt Dmitri.Bardin@cern.ch}} \\
Laboratory of Nuclear Problems, JINR, Dubna, Russia \\[3mm]
and  \\[3mm]
Martin Gr\"unewald \footnote{e-mail: {\tt Martin.Grunewald@cern.ch}} \\
Institute of Physics, Humboldt University, Berlin, Germany \\[3mm]
and  \\[3mm]
Giampiero Passarino \footnote{e-mail: {\tt Giampiero@to.infn.it}}  \\
Dipartimento di Fisica Teorica, Universit\`a di Torino, Italy \\
INFN, Sezione di Torino, Italy \\[1cm]
}

\maketitle

\begin{abstract}
  \normalsize \noindent The complete list of definitions for
  quantities relevant in the analysis of SLD/LEP-1 results around the
  $\zb$-resonance is given.  The common set of conventions adopted by
  the programs {\tt TOPAZ0} and {\tt ZFITTER}, following the
  recommendations of the LEP electroweak working group, is reviewed.
  The relevance of precision calculations is discussed in detail both
  for pseudo-observables (PO) and for realistic observables (RO).  The
  model-independent approach is also discussed.  A critical assessment
  is given of the comparison between {\tt TOPAZ0} and {\tt ZFITTER}.
\end{abstract}

\clearpage

\tableofcontents

\clearpage

%
% Body
%

%%% \input{ltf_note_1}
%----------------------------------------------------------------
\section{Motivations for the Upgrading of Precision Calculations}
%----------------------------------------------------------------

The main motivation for upgrading precision calculations around the
$\zb$-reso\-nance with the programs {\tt TOPAZ0}~\cite{kn:topaz0} and
{\tt ZFITTER}~\cite{kn:zfitter} and for making public the results is a
reflection of questions frequently asked by the experimental
community:

{\em A complete definition of lineshape and asymmetry pseudo
  observables (POs), together with the residual Standard Model (SM)
  dependence in model-independent fits, is needed.  This includes a
  description on what is actually taken from the SM.
  
  Both codes calculate POs. A definition of these POs is needed,
  showing that {\tt TOPAZ0}~\cite{kn:topaz0} and {\tt
    ZFITTER}~\cite{kn:zfitter} use the same definition so that any
  discrepancy is really a measure of missing higher-order corrections.
  This should include quantities like $\mz$, $\gz$, $\gff$,
  $\sigma^0_{\had}$ and $\afba{0}$, and also $\gvf$, $\gaf$ and
  $\seffsf{\rm lept}$.  }

%---------------------------------------
\subsection{Goals of this Report}
%---------------------------------------

In 1989 the CERN Report `$\zb$ Physics at LEP1', \cite{kn:yr1989} has
provided a central documentation of the theoretical basis for the
physics analysis of the LEP results.  Although being quite
comprehensive, an update on the discussion of radiative corrections
became necessary in 1995, detailed in the CERN report `Reports of the
Working Group on Precision Calculations for the $\zb$ Resonance',
\cite{kn:ocyr95}.  The structure of the latter report was determined
by a central part describing the situation for the electroweak
observables as obtained by various independent calculations, including
the remaining theoretical uncertainties, followed by comprehensive
descriptions of the QCD aspects of electroweak $\zb$ physics.

A new step was taken in early 1998 with a note on the `Upgrading of
Precision Calculations for Electroweak
Observables'~\cite{kn:prevnote}, where one focused on the calculation
of the pseudo-observables.

It is now time to move a step forward and fully revise the comparisons
not only for POs but also for realistic observables (ROs), i.e., total
cross-sections and forward-backward asymmetries, both extrapolated and
with realistic cuts.  Our goal, therefore, has been to upgrade and to
compare critically the complete {\tt TOPAZ0} and {\tt ZFITTER}
predictions with a particular emphasis on demanding the following
criteria:
\begin{itemize}
  
\item Comparisons, after the upgrading, should be consistently better
  than what they were in earlier studies.

\item At the peak all relative deviations among total cross-sections
  and absolute deviations among asymmetries should be below $0.1$
  per-mill.
  
\item At the wings, typically $\sqrt{\sman} = \mz \pm 1.8\,\GeV$, they
  should be below $0.3$ per-mill.
  
\end{itemize}
It is important to observe that a comparison for ROs at the level of
$10^{-4}$ has never been attempted before.

The numerical results reported in this article are calculated with
{\tt TOPAZ0} version 4.4~\cite{T44source} and {\tt ZFITTER} version
5.20~\cite{Z520source}.  After a careful examination of the new
upgrading of {\tt TOPAZ0} and {\tt ZFITTER} contained in these
versions we are able to report in general a good agreement in our
comparisons.  The worst case for pseudo-observables is represented by
the $\ffb\barb$-channel.  This fact, however, was largely expected:
this particular channel is where the next-to-leading two-loop
electroweak corrections are missing and, therefore, here is where we
face a larger level of theoretical uncertainties.  We find
satisfactory agreement in all the comparisons performed for realistic
observables but one: the inclusion of initial-final QED interference
in the presence of realistic cuts, i.e., acollinearity and polar angle
cuts and energy thresholds.  This fact will be discussed in detail in
Section~\ref{sec:IFI}.

In this context we would like to emphasise that the implementation of
the next-to-leading corrections, as well as of any higher-order
corrections, makes {\em stable} all theoretical predictions: the
degree of arbitrariness of the various implementations is reduced with
the introduction of newly computed terms.

Coming back to the reason for the present upgrading, we may say that
there are additional motivations for it, which we illustrate in the
following section.

%-------------------------------------------
\subsection{List of Improved I/O Parameters}
%-------------------------------------------

For all results, if not stated otherwise we use
%--
\bq
\mz = 91.1867\,\GeV.
\eq
%--
In fixing the set of input parameters we take the lepton masses as in
PDG'98~\cite{kn:pdg98}. They are as follows:
%--
\bq
\me = 0.51099907\,\MeV, \quad
\mm = 105.658389\,\MeV, \quad
\mtau = 1.77705\, \GeV.
\eq
%--
Since all renormalization schemes use the Fermi constant, $\gf$, we
refer to a recent calculation \cite{kn:stuart}, giving an improved
value of $\gf= 1.16637(1)\times 10^{-5}\,\GeV^{-2}$ to be compared
with the old one, $\gf= 1.16639(2)\times
10^{-5}\,\GeV^{-2}$~\cite{kn:pdg98}.

An important issue concerns the evaluation of $\alpha_{\rm QED}$ at
the mass of the $\zb$.  There is an agreement in our community on
using the following strategy. Define
%--
\bq
\alpha(\mz) = \frac{\alpha(0)}{1-\Delta\alpha^{(5)}(\mz) 
-\Delta_{\rm{top}}(\mz)-\Delta^{\alpha\als}_{\rm{top}}(\mz)}\;, 
\eq
%--
where one has $\Delta\alpha^{(5)}(\mz) = \Delta\alpha_{\rm{lept}} +
\Delta\alpha^{(5)}_{\rm{had}}$.  In both codes the input parameter is
now $\Delta\alpha^{(5)}_{\rm{had}}$, as it is the contribution with
the largest uncertainty, while the calculation of the top
contributions to $\Delta\alpha$ is left for the code.  This should
become common to all codes.

The programs {\tt TOPAZ0} and {\tt ZFITTER} include the recently
computed $\ord{\alpha^3}$ terms of \cite{kn:ste} for
$\Delta\alpha_{\rm{lept}}$, and use as default
$\Delta\alpha^{(5)}_{\rm{had}} = 0.0280398$ taken from \cite{kn:ej}.
As explained above the latter parameter can be reset by the user.
Using the default one obtains $1/\alpha^{(5)}(\mz) = 128.877$, to
which one must add the $\ft\bart$ contribution and the
$\ord{\alpha\als}$ correction induced by the $\ft\bart$ loop with
gluon exchange \cite{kn:ak}.

For the improved calculation of $\Delta\alpha_{\rm{lept}}$ we find the
result as reported in \tabn{tab1}.  For the $\ft$ contribution we
derive the results listed in \tabn{tab2}.  For the $\ord{\alpha\als}$
corrections induced by the $\ft\bart$ loop we report the results in
\tabn{tab3} with the value of $\als(\mt)$ as reported in \tabn{tab3p}.

%--
\begin{table}[p]
\begin{center}
\renewcommand{\arraystretch}{1.25}
\begin{tabular}{|c||c|}
\hline
   & $10^4 \times \Delta\alpha_{\rm{lept}}$  \\
\hline
\hline
{\tt TOPAZ0}   & 314.97644    \\
\hline
{\tt ZFITTER}  & 314.97637    \\
\hline
\end{tabular}
\caption[]{
  $10^{4}\times\Delta_{\rm{lept}}(\mz)$. }
\label{tab1}
\end{center}
\end{table}
%--

%--
\begin{table}[p]
\begin{center}
\renewcommand{\arraystretch}{1.25}
\begin{tabular}{|c||c|c|c|}
\hline
$\mt\,$[GeV]  & 168.8     & 173.8     & 178.8     \\
\hline
\hline
{\tt TOPAZ0}  & -0.622230 & -0.585844 & -0.552589 \\
\hline
{\tt ZFITTER} & -0.622230 & -0.585844 & -0.552589 \\
\hline
\end{tabular}
\caption[]{
  $10^{4} \times \Delta_{\rm top}(\mz)$. }
\label{tab2}
\end{center}
\end{table}
%--

%--
\begin{table}[p]
\begin{center}
\renewcommand{\arraystretch}{1.25}
\begin{tabular}{|c||c|c|c|}
\hline
$\als(\mz)/\mt\,$[GeV] &   168.8   &   173.8   &   178.8    \\
\hline
\hline
0.116                  & -0.108440 & -0.101593 & -0.095371  \\
                       & -0.108440 & -0.101593 & -0.095371  \\
0.119                  & -0.110994 & -0.103976 & -0.097599  \\
                       & -0.110994 & -0.103962 & -0.097600  \\
0.122                  & -0.113536 & -0.106347 & -0.099816  \\
                       & -0.113536 & -0.106347 & -0.099816  \\
\hline
\end{tabular}
\caption[]{
  {\tt TOPAZ0} (first row) / {\tt ZFITTER} (second row) results for
  $10^{4} \times \Delta^{\alpha\als}_{\rm{top}}(\mz)$. }
\label{tab3}
\end{center}
\end{table}
%--

%--
\begin{table}[p]
\begin{center}
\renewcommand{\arraystretch}{1.25}
\begin{tabular}{|c||c|c|c|}
\hline
$\als(\mz)/\mt\,$[GeV] &  168.8   &  173.8   &  178.8    \\
\hline
\hline
0.116                  &  0.10631 &  0.10589 &  0.10548  \\
                       &  0.10631 &  0.10589 &  0.10548  \\
0.119                  &  0.10881 &  0.10837 &  0.10795  \\
                       &  0.10881 &  0.10837 &  0.10795  \\
0.122                  &  0.11130 &  0.11084 &  0.11040  \\
                       &  0.11130 &  0.11084 &  0.11040  \\
\hline
\end{tabular}
\caption[]{
  {\tt TOPAZ0} (first row) / {\tt ZFITTER} (second row) results for
  $\als(\mt)$.  }
\label{tab3p}
\end{center}
\end{table}
%--

%-------------------------------------
\section{Analysis of the Measurements}
%-------------------------------------

In the following we take the LEP-1 measurements of hadronic and
leptonic cross sections and leptonic forward-backward asymmetries as
an example to discuss the data analysis strategy.

%-------------------------------------
\subsection{The Experimental Strategy}
%-------------------------------------

Technically, each LEP experiment extracts POs, namely $\mz, \gz,
\sigma^0_{\had}, R_{\fe,\flm,\flt}$ and $\afba{0,\fe,\flm,\flt}$ (see
Section~\ref{sec:PO-def} for a definition), from their measured
cross-sections and asymmetries (realistic observables).  The four sets
of POs are combined, taking correlated errors between the LEP
experiments into account, in order to obtain a LEP-average set of
POs~\cite{kn:lepewwg98}. The latter is then interpreted, for example
within the frame-work of the Minimal Standard Model.

Ideally, one would like to combine the results of the LEP experiments
at the level of the measured cross-sections and asymmetries - a goal
that has never been achieved so far because of the intrinsic
complexity, given the large number of measurements with different cuts
and the complicated structure of the experimental covariance matrices
relating their errors.  As a consequence, the practical attitude of
the four LEP experiments is to stay with a {\em Model-Independent}
(MI) fit, i.e., from ROs $\to$ POs ($\,\oplus\,$ a Standard Model
remnant) for each experiment, and to average the four sets of POs.
The result of this procedure is a set of best values for POs which are
of course important quantities in their own right.  The extraction of
Lagrangian parameters, $\mz, \mt, \mh$, $\als(\mzs)$ and
$\alpha(\mzs)$, is based on the LEP-averaged POs.

There remain several questions to be answered: the main one is, to
what extend are the POs a {\em Model-Independent} (MI) description of
the measurements?  Furthermore, are they valid even in the case where
the Standard Model is not the correct theory?  Note, that many effects
are absorbed into the POs.  Since POs are determined by fitting
realistic observables (ROs), one has to clarify what is actually taken
from the SM (such as imaginary parts and parts which have been moved
to $\zb-\ph$ interference terms and photon-exchange terms) making the
MI results dependent on the SM.

The POs are unsatisfactory for many reasons but to some level of
accuracy they describe well the experimental measurements at the $\zb$
peak.  How well and what is lost in the reduction ROs $\to$ POs is
exactly the kind of question that the LEP community is trying to
answer.

In the case of the Standard Model and the measurements of hadronic and
leptonic cross sections and leptonic forward-backward asymmetries at
LEP-1, it has been tested by each LEP experiment how the results on
the SM parameters differ between a SM fit to its own measured ROs, and
a SM fit to the POs which themselves are derived in an MI fit to the
same ROs~\cite{kn:ADLO}.  In the MI fit to determine the POs, the SM
initialisation has been performed with $\mz=91.1867\,$GeV, $\mt=
175\,$GeV, $\mh=150\,$GeV, $\als(\mzs)=0.119$, and
$\Delta\alpha^{(5)}_{\rm{had}}(\mzs) = 0.02804$ ($1/\alpha^{(5)}(\mzs)
= 128.878$).  In the two SM fits (to ROs and to POs) to be compared,
in both cases $\mt= 173.8 \pm 5.0\,$GeV and
$\Delta\alpha^{(5)}_{\rm{had}}(\mzs) = 0.02804\pm0.00065$
($1/\alpha^{(5)}(\mzs) = 128.878 \pm 0.090$) are included as external
constraints.

For each experiment, the largest difference in central values,
relative to the fitted errors, is observed for the value of the SM
parameter $\mz$, up to 30\% of the fit error.  The fit error itself is
unchanged.  The shift in central value also depends on the Higgs mass
used in the SM initialisation of the MI fit to determine the POs.  One
reason for these observations is the following, but more detailed
studies are needed.  The experimentally preferred value of the
Higgs-boson mass may be different from the value of the Higgs-boson
mass used in the SM initialisation of the MI fit. In particular, the
latter affects the value of the $\zb-\ph$ interference term for the
hadronic cross-section which must be taken from the SM for the MI fit.
Since the $\zb-\ph$ interference term for the hadronic cross-section
is highly anti-correlated with $\mz$ when derived from the LEP-1 data,
the choice of $\mh$ affects the fitted value of $\mz$ in the MI fit,
and subsequently the extraction of SM parameters.  For the other four
SM parameters, the observed differences in fitted central values and
errors are usually below 10\% to 15\% of the fit error on this
parameter.

As the LEP average is a factor of two more precise, care has to be
taken in the averaging procedure.  An alternative way to extract SM
parameters from the LEP measurements, avoiding the intermediate step
of POs, is to average directly the SM parameters which have been
obtained by each experiment through a SM fit to its own ROs.  This
alternative should indeed also be pursued by the experiments.

%-------------------------------------
\subsection{The Theoretical Strategy}
%-------------------------------------

Within the context of the SM the ROs are described in terms of some set
of amplitudes
%--
\bq
A_{_{\rm SM}} = A_{\ph} + A_{_{\zb}} + \mbox{non-factorizable},
\eq
%--
where the last term is due to all those contributions that do not
factorize into the Born-like amplitude, e.g., weak boxes.  Once the
matrix element $A_{_{\rm SM}}$ is computed, squared and integrated to
obtain the cross section, a convolution with initial- and final-state
QED and final-state QCD radiation follows:
%--
\bq
\sigma(\sman) = \int d\zvar~H_{\rm in}(\zvar,\sman)\,H_{\rm fin}(\zvar,\sman)\,
{\hat\sigma}(\zvar\sman),
\eq
%--
where $H_{\rm in}(\zvar,\sman)$ and $H_{\rm fin}(\zvar,\sman)$ are
so-called, {\em radiator} or {\em flux} functions accounting for {\em
  Initial- (Final-) State Radiation, ISR (FSR)}, respectively, and
${\hat\sigma}(\zvar\sman)$ is the kernel cross-section of the hard
process, evaluated at the reduced squared centre-of-mass energy
$s'=zs$.

It is a well-known fact that the structure of the matrix element
changes after inclusion of higher-order electroweak corrections.  One
needs the introduction of complex-valued form-factors which depend on
the two Mandelstam variables $\sman$ and $\tman$.  The separation into
insertions for the $\ph$ exchange and for the $\zb$ exchange is lost.

The weak boxes are present as non-resonating insertions to the
electroweak form-factors. At the $\zb$ resonance, the one-loop weak
box terms are small, with relative contributions $\leq 10^{-4}$.  If
we neglect them, the $\tman$-dependence is turned off.  The
$\tman$-dependence would also spoil factorisation of the form-factors
into products of effective vector and axial-vector couplings.  In all
comparisons of ROs presented in this report, the weak boxes are taken
into account because we go off resonance up to $\sqrt{s}=\mz\pm3$~GeV.

Full factorisation is re-established by neglecting various terms that
are of the order $\ord{\alpha{\gz}/{\mz}}$.  The resulting effective
vector and axial-vector couplings are complex valued and dependent on
$\sman$.  The factorisation is the result of a variety of
approximations which are valid at the $\zb$ resonance to the accuracy
needed, and which are indispensable in order to relate POs to ROs.

After the above mentioned series of approximations we arrive at the so
called $\zb$-boson pole approximation, which is actually equivalent to
setting $\sman=\mzs$ in the form-factors.  After de-convoluting ROs of
QED and QCD radiation the set of approximations transform ROs into
POs.

A source of potential ambiguity is linked to the adopted strategy for
extracting MI POs from the measured ROs. Indeed, the full SM
calculation in a MI analysis is performed only once at the beginning
where one needs to specify in addition to $\mz$, which is also a PO,
the (remaining) relevant SM parameters $\mt,\mh,\alpha(\mz),\als(\mz)$
for the SM-complement of the MI parameterisation, ${\rm RO} = {\rm RO}
({\rm PO} \oplus {\overline{\rm SM}})$.  Subsequent steps in the MI
calculation then go directly via $\mz$, total and partial widths, and
$\zb\ff\barf$ couplings.

This part of the procedure is particularly cumbersome. However, one
has to live with the fact that -- for practical reasons -- the POs
will be combined among the LEP experiments and survive forever.  The
cross-section and asymmetry measurements will be published by the
experiments, but most likely no one will ever undertake the effort to
combine them.  Therefore one is left with the task of making sure that
the adopted procedure is acceptable.

%---------------------------
\section{Pseudo-Observables}
%---------------------------

There remains to be investigated the systematic errors arising from
theory and possible {\em ambiguities} in the definition of the MI fit
parameters, the POs.

%--------------------------------------------
\subsection{Definition of Pseudo-Observables}
%--------------------------------------------
\label{sec:PO-def}

Independent of the particular realization of the effective couplings
they are complex-valued functions, due to the imaginary parts of the
diagrams. In the past this fact had some relevance only for realistic
observables while for pseudo-observables they were {\em
  conventionally} defined to include only real parts.  This convention
has changed lately with the introduction of next-to-leading
corrections: imaginary parts, although not next-to-leading in a strict
sense, are sizeable two-loop effects.  These are enhanced by factors
$\pi^2$ and sometimes also by a factor $\Nf$, with $\Nf$ being the
total number of fermions (flavour$\,\otimes\,$colour) in the SM.  Once
we include the best of the two-loop terms then imaginary parts should
also come in.  The latest versions of {\tt TOPAZ0} and {\tt ZFITTER}
therefore include imaginary parts of the $\zb$-resonance form factors.

The explicit formulae for the $\zb\ff\barf$ vertex are always written
starting from a Born-like form of a pre-factor $\times$ fermionic
current, where the Born parameters are promoted to effective,
scale-dependent parameters,
%--
\bq
\rZf\gadu{\mu}\,\lrbr \lpar \tcif + \ib\,a_{\ssL}\rpar \gdp 
- 2\,\qf\kZdf{\ff} \smans + \ib\,a_{\ssQ}\rrbr
= \gadu{\mu}\,\bigl(\Gvf + \Gaf\,\gfd\bigr),
\label{prototype}
\eq
%--
where $\gdp=1+\gfd$ and $a_{\ssQ,\ssL}$ are the SM imaginary parts.
Note that imaginary parts are always factorized in {\tt ZFITTER} and
added linearly in {\tt TOPAZ0}.

By definition, the total and partial widths of the $\zb$ boson include
all corrections, also QED and QCD corrections.  The partial decay
width is therefore described by the following expression:
%--
\bq
\gff\equiv\Gamma\lpar \zb\to\ff\barf\rpar = 4\,\cf\,\Gamma_0\,\bigl(
|\Gvf|^2\,R^{\ff}_{\ssV} + |\Gaf|^2\,R^{\ff}_{\ssA}\bigr) 
+\Delta_{_{\rm EW/QCD}}\;,
\label{defgammaf}
\eq
%--
where $\cf = 1$ or $3$ for leptons or quarks $(\ff=\fl,\fq)$, and the
radiator factors $R^{\ff}_{\ssV}$ and $R^{\ff}_{\ssA}$ describe the
final state QED and QCD corrections and take into account the fermion
mass $\mf$.

There is a large body of contributions to the radiator factors in
particular for the decay $\zb\to q\barq$; both {\tt TOPAZ0} and {\tt
  ZFITTER} implement the results that have been either derived or, in
few cases, confirmed in some more general setting by the Karlsruhe
group, see for instance \cite{kn:kpeople}.  The splitting between
radiators and effective couplings follows well defined recipes that
can be found and referred to in~\cite{kn:ocyr95, kn:thebook}.  In
particular our choice has been that top-mass dependent QCD corrections
are to be considered as QCD corrections and included in the radiators
and not in the effective quark couplings.

The last term,
%--
\bq \Delta_{_{\rm EW/QCD}}= 
 \Gamma^{(2)}_{_{\rm EW/QCD}} - 
 \frac{\als}{\pi}\,\Gamma^{(1)}_{_{\rm EW}}\;, 
\eq 
%--
accounts for the non-factorizable corrections.  The standard partial
width, $\Gamma_0$, is
%--
\bq
\Gamma_0 = {{\gf\mzc}\over {24\srt\,\pi}} = 82.945(7)\,\MeV.
\eq
%--
The hadronic and leptonic pole cross-sections are defined by
%--
\bq
\sigma^0_{\had} = 12\pi\,\frac{\gel\gh}{\mzs\gzs} \qquad
\sigma^0_{\ell} = 12\pi\,\frac{\gel\gl}{\mzs\gzs} \;,
\eq
%--
where $\gz$ is the total decay width of the $\zb$ boson, i.e, the sum
of all partial decay widths. Note that the mass and total width of the
$\zb$ boson are defined based on a propagator term $\chi$ with an
$s$-dependent width:
%--
\bq
\chi^{-1}(s) = s-\mzs+is\gz/\mz\;.
\eq
The effective electroweak mixing angles ({\em effective sinuses}) are
always defined by
%--
\bq
4\,|\qf|\seffsf{\ff} = 1-\frac{\Reb\;\Gvf}{\Reb\;\Gaf} =
1-\frac{\gvf}{\gaf}
\;,
\label{defeffsin}
\eq
%--
where we define
%--
\bq
\gvf = \Reb\;\Gvf, \qquad \gaf = \Reb\;\Gaf.
\eq
%--
The forward-backward asymmetry $\afba{}$ is defined via 
%--
\bq
\afba{}=\frac{\sigma_{_{\rm{F}}}-\sigma_{_{\rm{B}}}}
             {\sigma_{_{\rm{F}}}+\sigma_{_{\rm{B}}}}\;,
\qquad
\sigma_{_{\rm{T}}}=\sigma_{_{\rm{F}}}+\sigma_{_{\rm{B}}}\;,
\eq
%--
where $\sigma_{_{\rm{F}}}$ and $\sigma_{_{\rm{B}}}$ are the cross
sections for forward and backward scattering, respectively.  Before
analysing the forward-backward asymmetries we have to describe the
inclusion of imaginary parts.  $\afba{}$ is calculated as
%--
\bq
\afba{} = \frac{3}{4}\frac{\sigma_{_{\rm VA}}}{\sigma_{_{\rm T}}}\;,
\eq
%--
where
%---
\bqa
\sigma_{_{\rm VA}} &=& \frac{\gf\mzs}{\sqrt{2}}\,\sqrt{\rhoe\rhof}\,
\qe\qf\Reb\Bigl[\alpha^*(\mzs)\Gve\Gaf\chi(\sman)
\Bigr]  
\nl{}&&
+\frac{\gf^2\mzq}{8\,\pi}\,\rhoe\rhof 
\Reb\Bigl[\Gve\lpar\Gae\rpar^*\Bigr] 
\Reb\Bigl[\Gvf\lpar\Gaf\rpar^*\Bigr] 
\sman\,|\chi(\sman)|^2.
\label{sva}
\eqa
%--
In case of quark-pair production, an additional radiator factor
multiplies $\sigma_{_{\rm VA}}$, see also \eqn{eq:afbqcd}.

This result is valid in the realization where $\rhof$ is a real
quantity, i.e., the imaginary parts are not re-summed in $\rhof$.  In
this case
%--
\bq
\Gvf = \Reb\lpar \Gvf\rpar + \ib\,\Imb \lpar \Gvf\rpar =
\gvf + \ib\,\Imb \lpar \Gvf\rpar, 
\qquad
\Gaf = \tcif + \ib\,\Imb \lpar \Gaf\rpar.
\eq
%--
Otherwise $\Gaf = \tcif$ is a real quantity but $\rhof$ is complex
valued and \eqn{sva} has to be changed accordingly, i.e., we introduce
%--
\bq
\gvf = \sqrt{\rhof}\,\vc{\ff}\;, 
\qquad 
\gaf = \sqrt{\rhof}\,\tcif\;,
\eq
%--
with
%--
\bq
\vc{\ff}=\tcif-2\qf\seffsf{\ff}\;.
\eq
%--
For the peak asymmetry, the presence of $\rho$'s is irrelevant since
they will cancel in the ratio. We have
%---
\bqa
\hat\afba{{{0}}\rm f} &=& \frac{3}{4}\,\hat A_{\fe}\,\hat A_{\ff},
\nl
\hat A_{\ff} &=&
\frac{2\,\Reb\bigl[\Gvf\lpar\Gaf\rpar^*\bigr]}
{\bigl(\big|\Gvf\big|^2 + \big|\Gaf\big|^2\bigr)}\;.
\label{defasym}
\eqa
%---
The question is what to do with imaginary parts in \eqn{defasym}.  For
partial widths, as they absorb all corrections, the convention is to
use
%--
\bq
\big|\Gvaf\big|^2 =\bigl(\Reb\Gvaf\bigr)^2 + \bigl(\Imb\Gvaf\bigr)^2.
\eq
%--
%therefore it seems logical to maintain this convention also for
%asymmetries. Consequently, for the numerator of \eqn{defasym} one
%should also keep the imaginary parts. 
On the contrary, the PO peak asymmetry $\afba{0\rm f}$ will be defined by
an analogy of equation \eqn{defasym} where {\em conventionally} imaginary
parts are not included
%---
\bqa
\afba{0\rm f}&=&\frac{3}{4}\,{\cal A}_{\fe}\,{\cal A}_{\ff},
\nl
{\cal A}_{\ff}&=&\frac{2\bigl(\gvf\gaf\bigr)}{\bigl(\gvf\bigr)^2 
+ \bigr(\gaf\bigr)^2}\;.
\label{defasymre}
\eqa
%---
We note, that \eqn{defasymre} is not an approximation of
\eqn{defasym}.  Both are POs and both could be used as the {\em
  definition}.  Numerically, they give very similar results: {\tt
  ZFITTER} calculates for the two definitions in \eqn{defasym} and
\eqn{defasymre}, $\hat\afba{0\rm l}=0.0160692$ and $\afba{0\rm
  l}=0.0160739$.  The absolute difference, $0.0000047$, is more than
two orders of magnitude smaller than the current experimental error of
$0.00096$~\cite{kn:lepewwg98}.

In contrast to POs, which are defined, it is impossible to avoid
imaginary parts for ROs without spoiling the comparison between the
theoretical prediction and the experimental measurement.  Then one has
to start with \eqn{sva}.  We will develop \eqn{sva} in the realization
where imaginary parts are added linearly.  For the $\zb\zb$ part of
the VA cross-section one derives:
%--
\bq
\Reb\Bigl[ \Gve\lpar \Gae\rpar^*\Bigr] 
\Reb\Bigl[ \Gvf\lpar \Gaf\rpar^*\Bigr].
\eq
%--
This collapses to a familiar expression if the axial-vector
coefficients are real, however one cannot factorize and simplify the
$\rho$'s especially away from the pole because of the $\zb\ph$
component.  For the $\zb\ph$ part of the $VA$ cross-section one has
the following result:
%--
\bq
\Reb\bigl[\alpha^*(\sman)\chi(\sman)\bigr]\,
\Reb\lpar \Gae\Gaf\rpar
-
\Imb\bigl[\alpha^*(\sman)\chi(\sman)\bigr]\,
\Imb\lpar \Gae\Gaf\rpar.
\eq
%--

A definition of the PO heavy quark forward-backward asymmetry
parameter which would include mass effects is
%--
\bq
{\cal A}_{\ffb} = {{2\,\gvb\gab}\over {\frac{1}{2}\,\lpar 3 - \beta^2\rpar 
(\gvb)^2 + \beta^2\,(\gab)^2}}\,\beta,
\label{ab}
\eq
%--
where $\beta$ is the $\ffb$-quark velocity.  For $\afba{0\rm b}$ {\tt
  TOPAZ0/ZFITTER} find (with $\mb=4.7$ GeV)
%---
\bqa
0.102611/0.102634 \quad &\mbox{for}& \quad \mb\ne 0, 
\nl
0.102594/0.102617 \quad &\mbox{for}& \quad \mb  = 0, 
\eqa
%--
with a $0.000017/0.000017$ difference to be compared with the
experimental error of $0.0021$~\cite{kn:lepewwg98}.
The difference is very small, due to an accidental cancellation of the
mass corrections between the numerator and denominator of \eqn{ab}.
This occurs for down-type quarks where $(\gvb)^2 \approx (\gab)^2/2$
and where
%--
\bqa
\afba{0\rm b} & \approx &
\frac{3}{4}\,\frac{2\,\gve\gae}{(\gve)^2+(\gae)^2}\,
\frac{2\,\gvb\gab}{(\gvb)^2 + (\gab)^2}\,\lpar 1 + \delta_{\rm mass}\rpar,  
\nl
\delta_{\rm mass}&\approx& 4\,\frac{\mqs}{\sman}{{(\gab)^2/2 - (\gvb)^2}
\over{(\gvb)^2 + (\gab)^2}}.
\eqa
%---
For the $\fc$-quark this difference is even bigger ($\sim 0.000025$
for $\mc=1.5$ GeV, to be compared with the experimental error of
0.0044~\cite{kn:lepewwg98}), one more example that for $\ffb$-quarks
we meet an accidental cancellation.  Note that the mass effect should
be even smaller since running quark masses seem to be the relevant
quantities instead of the pole ones.  Therefore, our definition of the
POs forward--backward asymmetry and coupling parameter will be as in
\eqn{defasymre}.

The most important upgradings in the SM calculation of POs have been
already described in \cite{kn:prevnote}.  In particular they consist
of the inclusion of higher-order QCD corrections, mixed
electroweak-QCD corrections~\cite{kn:mix}, and next-to-leading
two-loop corrections of $\ord{\alpha^2\mts}$~\cite{kn:dfs}.

In Ref.~\cite{kn:dfs} the two-loop $\ord{\alpha^2\mts}$ corrections
are incorporated in the theoretical calculation of $\mw$ and
$\seffsf{\rm{lept}}$.  More recently the complete calculation of the
decay rate of the $\zb$ has been made available to us~\cite{kn:prel}.
The only case that is not covered is the one of final $\ffb$-quarks,
because it involves non-universal $\ord{\alpha^2\mts}$ vertex
corrections.

Another development in the computation of radiative corrections to the
hadronic decay of the $\zb$ is contained in two papers, which together
provide complete corrections of $\ord{\alpha\als}$ to $\Gamma(\zb\to
\fq\barq)$ with $\fq=\fu,\fd,\fs,\fc$ and $\ffb$.  In the first
reference of~\cite{kn:mix} the decay into light quarks is treated.  In
the second one the remaining diagrams contributing to the decay into
bottom quarks are considered and thus the mixed two-loop corrections
are complete.

%------------------------------------------
\subsection{Model Independent Calculations}
%------------------------------------------
     
To summarise the MI ansatz, one starts with the SM, which introduces
complex-valued couplings, calculated to some order in perturbation
theory.  Next we define $\gvf,\gaf$ as the real parts of the effective
couplings and $\Gamma_{\ff}$ as the physical partial width absorbing
all radiative corrections including the imaginary parts of the
couplings and fermion mass effects.  Furthermore, we introduce the
ratios of partial widths
%--
\bq
R_{\fq}=\frac{\Gamma_{\fq}}{\gh}\;,
\qquad
R_{\fl}=\frac{\gh}{\Gamma_{\fl}}\;,
\eq
%--
for quarks and leptons, respectively.  

The LEP collaborations report POs for the following sets:
%--
\bq
\bigl(\mz,\gz,\sigma^{0}_{\had},R_{\ff},\afba{0,\ff}\bigr);
\quad 
\bigl(\mz,\gz,\gh,\gvf,\gaf\bigr);
\quad
\bigl(\mz,\gz,\gh,\seffsf{\ff},\rho_{\ff}\bigr).
\eq
%--
In order to extract $\gvf,\gaf$ from $\Gamma_{\ff}$ one has to get the
SM-remnant from \eqn{defgammaf}, all else is trivial.  However, the
parameter transformation cannot be completely MI, due to the residual
SM dependence appearing inside \eqn{defgammaf}.  

In conclusion, the flow of the MI calculation requested by the
experimental collaborations is:
\begin{enumerate}
  
\item Pick the Lagrangian parameters $\mt,\mh$ etc. for the explicit
  calculation of the residual SM-dependent part.
  
\item Perform the SM initialisation of everything, such as imaginary
  parts etc.  giving, among other things, the complement
  ${\overline{\rm SM}}$.
  
\item Select $\gvf,\gaf$.
  
\item Perform a SM-like calculation of $\Gamma_f$, but using arbitrary
  values for $\gvf, \gaf$, and only the rest, namely
%--
\bq
R^{\ff}_{\ssV}, \quad R^{\ff}_{\ssA}\;,  
\qquad
\Delta_{_{\rm EW/QCD}}\;,  
\qquad
\Imb\;\Gvf, \quad \Imb\;\Gaf\;,
\eq
%--
from the SM.
\end{enumerate}

An example of the parameter transformations is the following: starting
from $\mz, \gz, \sigma^{0}_{\had}, R_{\fe,\flm,\flt}$ and
$\afba{0,\fe,\flm,\flt}$ we first obtain
%--
\bqa
\gel &=& \mz\gz\,\biggl[{{\sigma^{0}_{\had}}\over{12\,\pi R_{\fe}}}\biggr]^{1/2},
\nl
\gh  &=& \mz\gz\,\biggl[{{\sigma^{0}_{\had}R_{\fe}}\over {12\,\pi}}\biggr]^{1/2}.
\eqa
%---
With 
%--
\bq
{\cal A}_{\fe} = \frac{2}{\sqrt{3}}\sqrt{\afba{0,\fe}}\;, 
\qquad 
\mbox{and} \quad \gamma = \frac{\gf\mzc}{6\,\srt\,\pi}\;,
\eq
%--
we subtract QED radiation,
%--
\bq
\gel^0 = \frac{\gel}{\ds{1+\frac{3}{4}\,\frac{\alpha(\mzs)}{\pi}}}\;,
\eq
%--
and get
%---
\bqa
\seffsf{\fe} &=& \frac{1}{4}\,
\lpar 1 + {{\sqrt{1-{\cal A}^2_{\fe}}-1}\over{{\cal A}_{\fe}}}\rpar,  
\nl
\rho_{\fe} &=& \frac{\gel^0}{\gamma}\,
\biggl[\lpar\frac{1}{2}-2\,\seffsf{\fe}\rpar^2 
+\frac{1}{4} + \lpar\Imb\;\Gve\rpar^2 + \lpar\Imb\;\Gae\rpar^2 \biggr]^{-1}.
\qquad
\eqa
%---
With
%--
\bq
{\cal A}_{\ff} = \frac{4}{3}\,{{\afba{0,\ff}}\over{{\cal A}_{\fe}}}\;,
\eq
%--
we further obtain
%---
\bqa
\seffsf{\ff} &=& \frac{1}{4\,\big|\qf\big|}\,
\lpar 1 + {{\sqrt{1 - {\cal{A}}^2_{\ff}}-1}\over{{\cal{A}}_{\ff}}}\rpar,  
\nl
\rho_{\ff} &=& \frac{\Gamma_f^0}{\gamma}\,
\biggl[\lpar\frac{1}{2}-2|\qf|\seffsf{\ff}\rpar^2 
+\frac{1}{4} + \lpar\Imb\;\Gvf\rpar^2 + \lpar\Imb\;\Gaf\rpar^2 \biggr]^{-1},
\qquad
\eqa
%---
for $\ff = \flm,\flt$.  For quarks one should remember to subtract
first non-factorizable terms and then to distinguish between
$R^{\ff}_{\ssV}$ and $R^{\ff}_{\ssA}$.

%------------------------------------------
\subsection{Results for Pseudo-Observables}
%------------------------------------------

Having established a common input parameter set (IPS) we now turn to
discussing the results for pseudo-observables (POs).  For POs we use two
reference sets of values:
%--
\bq
\mz = 91.1865\,\GeV, \quad \mt = 171.1\,\GeV, \quad \mh = 76\,\GeV,
\quad \als(\mz) = 0.119,
\eq
%--
which corresponds to the minimum of the $\chi^2$ of the summer-1998
fit to the SM, see~\cite{kn:lepewwg98}, and
%--
\bq
\mz = 91.1867\,\GeV, \quad \mt = 173.8\,\GeV, \quad \mh = 100\,\GeV,
\quad \als(\mz) = 0.119,
\eq
%--
which is our {\em preferred setup} in this report.

For the usual list of POs that enter any SM fit we derive the results
of \tabns{tab4_1}{tab4_2}. Here we take into account the updated value
for $\gf$ and compare results calculated with the old value, $\gf =
1.16639 \times 10^{-5}\,\GeV^{-2}$, and with the new value, $\gf =
1.16637 \times 10^{-5}\,\GeV^{-2}$.  For all other results presented
in this report, the new value of $\gf$ is used.

\begin{table}[p]
\begin{center}
\renewcommand{\arraystretch}{1.25}
\begin{tabular}{|c||c||c|c|c|}
\hline
Observable       & Summer 1998 &  Old $\gf$  &  New $\gf$ & Diff.   \\
\hline
$1/\alpha^{(5)}$ &    128.878  &\multicolumn{2}{c|}{128.877}  &     \\
\cline{3-4}
                 & $\pm$0.090  &\multicolumn{2}{c|}{128.877}  &     \\   
\hline
$\mz$[GeV](Input)&    91.1865  &\multicolumn{2}{c|}{91.1865}  &     \\
\hline
$\mt$[GeV](Input)&    171.1    &\multicolumn{2}{c|}{171.1}    &     \\
\hline
$\mh$[GeV](Input)&    76.0     &\multicolumn{2}{c|}{76.0}     &     \\
\hline
\hline
$\gz\,$[GeV]     &    2.4958   &   2.49543   &   2.49538  & 0.05 MeV\\
                 &$\pm$0.0024  &   2.49564   &   2.49559  &         \\
\hline                                          
$\sigma^0_{\had}\,$[nb]&41.473 &   41.4743   &   41.4743  & -       \\
                 &$\pm$0.058   &   41.4759   &   41.4759  &         \\
\hline
$\Rl$            &    20.748   &   20.7468   &   20.7467  & 0.0001  \\
                 &$\pm$0.026   &   20.7453   &   20.7452  &         \\
\hline
$\afba{0,\rm l}$ &    0.01613  &   0.0161823 &  0.0161725 & 0.00001 \\
                 &$\pm$0.00096 &   0.0161686 &  0.0161588 &         \\
\hline
${\cal A}_{\fe}$ &    0.1467   &   0.146889  &  0.146844  & 0.0005  \\
                 &$\pm$0.0017  &   0.146827  &  0.146782  &         \\
\hline
$\seffsf{\rm{lept}}$&  0.23157 &   0.231539  &  0.231544  & -0.00001\\
                 &$\pm$0.00018 &   0.231547  &  0.231552  &         \\
\hline
$\mw\,$[GeV]     &    80.37    &   80.3722   &  80.3718   & 0.4 MeV \\
                 &$\pm$0.09    &   80.3724   &  80.3721   &         \\
\hline
\end{tabular}
\caption[]{
  Table of POs, first entry is {\tt TOPAZ0}, second is {\tt ZFITTER}.
  The experimental results show the status of summer
  1998~\cite{kn:lepewwg98}.  Old $\gf$ is $\gf = 1.16639 \times
  10^{-5}\,GeV^{-2}$, new $\gf$ is $\gf = 1.16637 \times
  10^{-5}\,GeV^{-2}$.  {\rm Diff} is difference between old and new
  $\gf$. }
\label{tab4_1}
\end{center}
\end{table}
%--

%--
\begin{table}[p]
\begin{center}
\renewcommand{\arraystretch}{1.25}
\begin{tabular}{|c||c||c|c|c|}
\hline
Observable       & Summer 1998 &  Old $\gf$  &  New $\gf$ & Diff.   \\
\hline
$1/\alpha^{(5)}$ & 128.878     &\multicolumn{2}{c|}{128.877}  &     \\
                 &$\pm$ 0.0021 &\multicolumn{2}{c|}{128.877}  &     \\
\hline
$\mz$[GeV](Input)&    91.1865  &\multicolumn{2}{c|}{91.1865}  &     \\
\hline
$\mt$[GeV](Input)&    171.1    &\multicolumn{2}{c|}{171.1}    &     \\
\hline
$\mh$[GeV](Input)&    76.0     &\multicolumn{2}{c|}{76.0}     &     \\
\hline
\hline
$\Rb$            &    0.21590  &   0.215913  &  0.215913  & -       \\
                 &$\pm$0.00076 &   0.215897  &  0.215898  &         \\
\hline
$\Rc$            &    0.1722   &   0.172223  &  0.172222  & -       \\
                 &$\pm$0.0048  &   0.172224  &  0.172223  &         \\
\hline
$\afba{0,\ffb}$  &    0.1028   &   0.102912  &  0.102881  & 0.00003 \\
                 &$\pm$0.0021  &   0.102927  &  0.102895  &         \\
\hline
$\afba{0,\fc}$   &    0.0734   &   0.0735700 &  0.0735456 & 0.00002 \\
                 &$\pm$0.0045  &   0.0735365 &  0.0735121 &         \\
\hline
${\cal A}_{\ffb}$&    0.935    &   0.934724  &  0.934720  & -       \\   
                 &$\pm$0.018   &   0.934678  &  0.934674  &         \\
\hline                                                        
${\cal A}_{\fc}$ &    0.668    &   0.667806  &  0.667787  & 0.00002 \\
                 &$\pm$0.028   &   0.667784  &  0.667765  &         \\
\hline
\end{tabular}
\caption[]{
  Table of POs, first entry is {\tt TOPAZ0}, second is {\tt ZFITTER}.
  The experimental results show the status of summer
  1998~\cite{kn:lepewwg98}.  Old $\gf$ is $\gf = 1.16639 \times
  10^{-5}\,GeV^{-2}$, new $\gf$ is $\gf = 1.16637 \times
  10^{-5}\,GeV^{-2}$.  {\rm Diff} is difference between old and new
  $\gf$. }
\label{tab4_2}
\end{center}
\end{table}
%--

\clearpage

The full list contains more POs and is given in \tabn{tab5+6}, where
we include the relative and absolute difference between {\tt TOPAZ0}
and {\tt ZFITTER} in units of per-mill:
%--
\bq
\delta = 10^3 \times\,\frac{\rm T-Z}{\rm T} \qquad
\Delta = 10^3 \times\,(    {\rm T-Z}      ) \;.
\eq
%--
In \tabn{tab5+6} we report also some POs which are not usually taken
into account in fitting the experimental data.  With the exception of
$\seffsf{\ffb}$, for which we find a difference of $0.4$ per-mill, the
relative deviation is always (well) below $0.15$ per-mill.  The larger
difference in $\seffsf{\ffb}$ is perhaps not surprising since the
$\ffb$ sector did not undergo any update aimed to including
next-to-leading two-loop electroweak effects in $\mt$, which are not
available for this channel.

In \tabns{tab4bis_1}{tab4bis_2} we analyse the POs as a function of
$\mh$ in logarithmic spacing, $\mh = 10, 30, 100, 300, 1000\,$GeV,
therefore including the region of low $\mh$ (here $\mz = 91.1867$ has
been used).

The variations of POs as a function of the Higgs boson mass are an
important issue related to the theoretical uncertainty in the
determination of constraints on SM parameters.  From the most recent
study~\cite{kn:lepewwg98} one derives that the region below a Higgs
mass of about $70\,$GeV has a comparatively larger theoretical
uncertainty, although the current $90\,$GeV lower limit on the Higgs
mass from the direct search makes it less interesting.

In \tabns{tab4bis_1}{tab4bis_2} we compare a relevant set of POs in
the preferred calculational setup of {\tt TOPAZ0} and {\tt ZFITTER}.
In \tabn{tab4bis_1} we report the relative deviations, in per-mill,
between the two predictions.  Everywhere this deviation is (well)
below $0.1\,$ per-mill, even at very low values of the Higgs mass.
For the asymmetries of \tabn{tab4bis_2} we report absolute deviations
in units of $10^{-3}$.  The largest absolute deviation is found for
${\cal A}_{\fe}$ at $\mh = 10\,$GeV, giving $0.14 \times 10^{-3}$.
For $\mh = 100\,$GeV all absolute deviations in \tabn{tab4bis_2} are
below $0.08 \times 10^{-3}$.

The good agreement of POs calculated by {\tt TOPAZ0} and {\tt ZFITTER}
verifies a posteriori the consistency of definitions for POs in the
two programs, although one should understand that the agreement is
necessary but not sufficient for consistency.  We come back to this
question when we discuss realistic observables and their calculations
in terms of POs.

It is instructive to compare few examples with the old results of
\cite{kn:ocyr95}, obtained for $\mh = 300\,$GeV.  For
$\seffsf{\rm{lept}}$ the relative difference {\tt T/Z} has moved from
$0.21$ to $0.004$ per-mill at $\mh = 300\,$GeV and it is now
everywhere below $0.06$ per-mill, reached at very low Higgs masses.
For $\mw$ it was $0.087$ per-mill and it is now $0.015$ per-mill with
a maximum of $0.029$ for very large values of $\mh$.  Finally, the
absolute deviation for ${\cal A}_{\fe}$ was $0.38 \times 10^{-3}$ and
it is now $0.05 \times 10^{-3}$ with a maximum of $0.14 \times
10^{-3}$ at low values of $\mh$.

%--
\begin{table}[p]
\begin{center}
\renewcommand{\arraystretch}{1.25}
\begin{tabular}{|c||c|c|c|}
\hline
Observable &{\tt TOPAZ0}&{\tt ZFITTER}&$10^3\times\frac{T-Z}{T}$\\
\hline
\hline
$1/\alpha^{(5)}(\mz)$  &  128.877 & 128.877 &       \\
\hline
$1/\alpha(\mz)$        &  128.887 & 128.887 &       \\
\hline
$\mw\,$[GeV]           &  80.3731 & 80.3738 & -0.009\\
\hline
$\sigma^0_{\had}\,$[nb]&  41.4761 & 41.4777 & -0.04 \\
\hline
$\sigma^0_{\lep}\,$[nb]&  1.9995  & 1.9997  & -0.12 \\
\hline
$\Gamma_{\had}\,$[GeV] &  1.74211 & 1.74223 & -0.07 \\
\hline
$\gz\,$[GeV]           &  2.49549 & 2.49573 & -0.10 \\
\hline
$\gn\,$[MeV]           &  167.207 & 167.234 & -0.16 \\
$\gel\,$[MeV]          &   83.983 &  83.995 & -0.14 \\
$\gmu\,$[MeV]          &   83.983 &  83.995 & -0.14 \\
$\gt\,$[MeV]           &   83.793 &  83.805 & -0.14 \\
$\gu\,$[MeV]           &  300.129 & 300.154 & -0.08 \\
$\gd\,$[MeV]           &  382.961 & 382.996 & -0.09 \\
$\gc\,$[MeV]           &  300.069 & 300.092 & -0.08 \\
$\gbq\,$[MeV]          &  375.997 & 375.993 &  0.01 \\
$\gi\,$[GeV]           &  0.50162 & 0.50170 & -0.16 \\
\hline
$R_l$                  &  20.7435 & 20.7420  & 0.07 \\
$R^0_b$                & 0.215829 & 0.215811 & 0.08 \\
$R^0_c$                & 0.172245 & 0.172246 &-0.01 \\
\hline
$\seffsf{\rm{lept}}$   & 0.231596 & 0.231601 &-0.02 \\
$\seffsf{\ffb}$        & 0.232864 & 0.232950 &-0.37 \\
$\seffsf{\fc}$         & 0.231491 & 0.231495 &-0.02 \\
\hline
$\rhoe$                & 1.00513  & 1.00528  &-0.15 \\
$\rhoi{\ffb}$          & 0.99413  & 0.99424  &-0.11 \\
$\rhoi{\fc}$           & 1.00582  & 1.00598  &-0.16 \\
\hline
\hline
Observable & {\tt TOPAZ0} & {\tt ZFITTER}    &$10^3\times(T-Z)$\\
\hline
\hline
$\afba{0,\rm l}$       & 0.016084 & 0.016074 & 0.01 \\
$\afba{0,\ffb}$        & 0.102594 & 0.102617 &-0.02 \\
$\afba{0,\fc}$         & 0.073324 & 0.073300 & 0.02 \\
\hline
${\cal A}_{\fe}$       & 0.146440 & 0.146396 & 0.04 \\
${\cal A}_{\ffb}$      & 0.934654 & 0.934607 & 0.05 \\
${\cal A}_{\fc}$       & 0.667609 & 0.667595 & 0.01\\
\hline
\end{tabular}
\caption[]{
  Complete table of POs, from {\tt TOPAZ0} and {\tt ZFITTER}. }
\label{tab5+6}
\end{center}
\end{table}
%--

%--
\begin{table}[p]
\begin{center}
\renewcommand{\arraystretch}{1.25}
\begin{tabular}{|c||c|c|c|c|c|}
\hline
&\multicolumn{5}{|c|}{$\mh\,$[GeV]}                                      \\
\hline
Observable       & 10       &    30    &    100   &    300   &   1000    \\
\hline
\hline
$\gz\,$[GeV]     & 2.49298  & 2.49618  & 2.49549  & 2.49227  & 2.48732   \\
                 & 2.49322  & 2.49645  & 2.49573  & 2.49240  & 2.48751   \\
                 & -0.10    & -0.11    & -0.10    & -0.05    & -0.08     \\
\hline                                           
$\sigma^0_{\had}\,$[nb]
                 & 41.4739  & 41.4744  & 41.4761  & 41.4788  & 41.4830   \\
                 & 41.4748  & 41.4761  & 41.4777  & 41.4798  & 41.4831   \\
                 & -0.02    & -0.04    & -0.04    & -0.02    & -0.002    \\
\hline                                           
$\sigma^0_{\lep}\,$[nb]
                 & 1.99797  & 1.99851  & 1.99947  & 2.00062  & 2.00209   \\
                 & 1.99811  & 1.99874  & 1.99970  & 2.00074  & 2.00208   \\
                 &-0.07     &-0.12     &-0.12     &-0.06     &-0.005     \\
\hline
$R_{\fl}$        & 20.7580  & 20.7527  & 20.7435  & 20.7330  & 20.7199   \\  
                 & 20.7570  & 20.7511  & 20.7420  & 20.7322  & 20.7200   \\
                 & +0.05    & +0.08    & +0.07    & +0.04    & -0.005    \\
\hline
$\seffsf{\rm{lept}}$
                 & 0.230698 & 0.231044 & 0.231596 & 0.232175 & 0.232845  \\  
                 & 0.230712 & 0.231056 & 0.231601 & 0.232176 & 0.232838  \\
                 & -0.061   & -0.052   & -0.022   & -0.004   & +0.030    \\
\hline
$\mw\,$[GeV]     & 80.4587  & 80.4298  & 80.3731  & 80.2989  & 80.2045   \\  
                 & 80.4583  & 80.4297  & 80.3738  & 80.3001  & 80.2068   \\
                 & +0.005   & +0.001   & -0.009   & -0.015   & -0.029    \\
\hline
$\Rb$            & 0.215759 & 0.215794 & 0.215829 & 0.215839 & 0.215824  \\   
                 & 0.215763 & 0.215775 & 0.215811 & 0.215845 & 0.215857  \\
                 & -0.02    & +0.09    & +0.08    & -0.03    & -0.15     \\
\hline
$\Rc$            & 0.172305 & 0.172280 & 0.172245 & 0.172213 & 0.172184  \\   
                 & 0.172301 & 0.172281 & 0.172246 & 0.172210 & 0.172174  \\
                 & +0.02    & -0.01    & -0.01    & +0.02    & +0.06     \\
\hline
\end{tabular}
\caption[]{
  Table of POs as a function of the Higgs boson mass; first entry is
  {\tt TOPAZ0}, second is {\tt ZFITTER}, third entry is 1-T/Z in
  per-mill. }
\label{tab4bis_1}
\end{center}
\end{table}
%--

%--
\begin{table}[p]
\begin{center}
\renewcommand{\arraystretch}{1.25}
\begin{tabular}{|c||c|c|c|c|c|}
\hline
&\multicolumn{5}{|c|}{$\mh\,$[GeV]}                                      \\
\hline
Observable       & 10       &    30    &    100   &    300   &   1000    \\
\hline
\hline
$\afba{0,\rm l}$ & 0.017672 & 0.017052 & 0.016084 & 0.015098 & 0.013994  \\   
                 & 0.017647 & 0.017031 & 0.016074 & 0.015096 & 0.014006  \\
                 & +0.03    & +0.02    & +0.01    & +0.002   & -0.01     \\
\hline
$\afba{0,\ffb}$  & 0.107564 & 0.105656 & 0.102594 & 0.099373 & 0.095637  \\ 
                 & 0.107587 & 0.105665 & 0.102617 & 0.099410 & 0.095711  \\ 
                 & -0.02    & -0.01    & -0.02    & -0.04    & -0.07     \\
\hline
$\afba{0,\fc}$   & 0.077216 & 0.075714 & 0.073324 & 0.070827 & 0.067949  \\
                 & 0.077157 & 0.075663 & 0.073300 & 0.070824 & 0.067983  \\
                 & +0.06    & +0.05    & +0.02    & +0.003   & -0.03     \\
\hline
${\cal A}_{\fe}$ & 0.15350  & 0.15078  & 0.14644  & 0.14188  & 0.13660   \\
                 & 0.15340  & 0.15069  & 0.14640  & 0.14187  & 0.13666   \\
                 & +0.10    & +0.09    & +0.04    & +0.01    & -0.06     \\
\hline
${\cal A}_{\ffb}$& 0.935220 & 0.935003 & 0.934654 & 0.934283 & 0.933837  \\
                 & 0.935165 & 0.934947 & 0.934607 & 0.934251 & 0.933844  \\
                 & +0.06    & +0.06    & +0.05    & +0.03    & -0.01     \\
\hline                                                        
${\cal A}_{\fc}$ & 0.670709 & 0.669517 & 0.667609 & 0.665601 & 0.663267  \\ 
                 & 0.670666 & 0.669481 & 0.667595 & 0.665605 & 0.663302  \\
                 & +0.04    & +0.04    & +0.01    & -0.004   & -0.04     \\
\hline
\end{tabular}
\caption[]{
  Table of POs as a function of the Higgs boson mass; first entry is
  {\tt TOPAZ0}, second is {\tt ZFITTER}, third is $10^3\times(T-Z)$. }
\label{tab4bis_2}
\end{center}
\end{table}
%--

\clearpage

%---------------------------------------------------------
\subsection{Theoretical Uncertainties for Pseudo-Observables}
%---------------------------------------------------------
\label{sec:PO-err}

Here we discuss the theoretical uncertainties associated with the SM
calculation of POs.  In \tabns{tab19}{tab20} we give the central
value, the minus error and the plus error as predicted by {\tt TOPAZ0}
and compare with the current total experimental error where available.
The procedure is straightforward: both codes have a preferred
calculational setup and options to be varied, options having to do
with the remaining theoretical uncertainties and the corresponding
implementation of higher-order terms.  To give an example, we have now
LO and NLO two-loop EW corrections but we are still missing the NNLO
ones and this allows for variations in the final recipe for $\rho_f$,
etc.

{\tt TOPAZ0} has been run over all the options remaining after
implementation of NLO corrections, and all the results for POs are
collected.  We use
\begin{itemize}
  
\item[--] {\em central} for the value of the PO evaluated with the
  preferred setup;
  
\item[--] minus error for $\mbox{PO}_{\it central} - \min_{\rm
    opt}\,\mbox{PO}$;
  
\item[--] plus error for $\max_{\rm opt}\,\mbox{PO} - \mbox{PO}_{\it
    central}$.

\end{itemize}
%--

%--
\begin{table}[ht]
\begin{center}
\renewcommand{\arraystretch}{1.25}
\begin{tabular}{|c||c|c|c|c|}
\hline
Observable       & central & minus error & plus error & total exp. error \\
\hline
\hline
$1/\alpha^{(5)}(\mz)$  &  128.877 & -       &  -         & \\
$1/\alpha(\mz)$        &  128.887 & -       &  -         & \\
\hline
\hline
$\mw\,$[GeV]           &  80.3731 & 5.8 MeV &  0.3 MeV   & 64 MeV \\
$\sigma^0_{\had}\,$[nb]&  41.4761 & 1.0 pb  &  1.6 pb    & 58 pb \\
$\sigma^0_{\ell}\,$[nb]&   1.9995 & 0.17 pb &  0.26 pb   & 3.5 pb \\
\hline                                   
$\gn\,$[MeV]           &  167.207 & 0.017   &  0.001     &  \\
$\gel\,$[MeV]          &   83.983 & 0.010   &  0.0005    & 0.10$^*$ \\
$\gmu\,$[MeV]          &   83.983 & 0.010   &  0.0005    &\\
$\gt\,$[MeV]           &   83.793 & 0.010   &  0.0005    &\\
$\gu\,$[MeV]           &  300.129 & 0.047   &  0.013     &\\
$\gd\,$[MeV]           &  382.961 & 0.054   &  0.010     &\\
$\gc\,$[MeV]           &  300.069 & 0.047   &  0.013     &\\
$\gbq\,$[MeV]          &  375.997 & 0.208   &  0.077     &\\
$\Gamma_{\rm{had}}\,$[GeV]&1.74211& 0.26 MeV&  0.11 MeV  &  2.3 MeV$^*$\\
$\gi\,$[GeV]           &  0.50162 & 0.05 MeV& 0.002 MeV  &  1.9 MeV$^*$\\
$\gz\,$[GeV]           &  2.49549 & 0.34 MeV&  0.11 MeV  &  2.4 MeV    \\
\hline
\end{tabular}
\caption[]{
  Theoretical uncertainties on POs from {\tt TOPAZ0}.  The
  experimental error is that of summer 1998~\cite{kn:lepewwg98}.  $*)$
  assumes lepton universality.  }
\label{tab19}
\end{center}
\end{table}

\clearpage

%--
\begin{table}[t]
\begin{center}
\renewcommand{\arraystretch}{1.25}
\begin{tabular}{|c||c|c|c|c|}
\hline
Observable  & central & minus error & plus error & total exp. error \\
\hline
\hline
$R_l$                  &  20.7435 & 0.0020   & 0.0013    & 0.026$^*$    \\
$R^0_b$                & 0.215829 & 0.000100 & 0.000031  & 0.00074  \\
$R^0_c$                & 0.172245 & 0.000005 & 0.000024  & 0.0044   \\
\hline
$\seffsf{\rm{lept}}$   & 0.231596 & 0.000035 & 0.000033  & 0.00018$^*$  \\
%                         0.0010(0.00029$^\dagger)$       \\
$\seffsf{\ffb}$        & 0.232864 & 0.000002 & 0.000048  &          \\
$\seffsf{\fc}$         & 0.231491 & 0.000029 & 0.000033  &          \\
\hline
$\afba{0,\rm l}$       & 0.016084 & 0.000057 & 0.000060  & 0.00096$^*$  \\
$\afba{0,\ffb}$        & 0.102594 & 0.000184 & 0.000195  & 0.0021   \\
$\afba{0,\fc}$         & 0.073324 & 0.000142 & 0.000149  & 0.0044   \\
\hline
${\cal A}_{\fe}$       & 0.146440 & 0.000259 & 0.000275  & 0.0017$^*$   \\
${\cal A}_{\ffb}$      & 0.934654 & 0.000032 & 0.000005  & 0.035    \\
${\cal A}_{\fc}$       & 0.667609 & 0.000114 & 0.000103  & 0.040    \\
\hline
$\rhoe$                & 1.00513  & 0.00010  & 0.000005  & 0.0012$^*$   \\
$\rhoi{\ffb}$          & 0.99413  & 0.00048  & 0.000001  &          \\
$\rhoi{\fc}$           & 1.00582  & 0.00010  & 0.000005  &          \\
\hline
\end{tabular}
\caption[]{
  Theoretical uncertainties on POs from {\tt TOPAZ0}.  The
  experimental error is that of summer 1998~\cite{kn:lepewwg98}.  $*)$
  assumes lepton universality.  }
\label{tab20}
\end{center}
\end{table}
%--

One can see from \tabns{tab19}{tab20} that there is a sizeable
reduction of the theoretical uncertainty associated with POs compared
to the findings of \cite{kn:ocyr95}.  This is in accordance with the
work of \cite{kn:prevnote} and is mainly due to the implementation of
next-to-leading corrections in {\tt TOPAZ0} and {\tt ZFITTER}.  We do
not show any estimate for theoretical uncertainty in POs from {\tt
  ZFITTER}, since they are typically more narrow.

Within {\tt TOPAZ0} the central values are defined by the following
flags: {\tt OU0='S'} (fixed), {\tt OU1='Y'}, {\tt OU2='N'} (fixed),
{\tt OU3='Y'} (fixed), {\tt OU4='N'}, {\tt OU5='Y'}, {\tt OU6='Y'},
{\tt OU7='N'}, {\tt OU8='C'}.  Plus and minus errors are obtained by
changing the flags to the following values: {\tt OU1='N'}, {\tt
  OU4='Y'}, {\tt OU5='N'}, {\tt OU6='N'}, {\tt OU7='Y'}, {\tt OU8='L'}
or {\tt OU8='R'} (one by one).  {\tt ZFITTER}'s central values are
produced with the default flag setting.  Plus and minus errors are
obtained by changing the flags to the following values: {\tt
  SCAL=0,4}; {\tt HIGS=0,1}; {\tt SCRE=0,1,2}; {\tt EXPR=0,1,2}; {\tt
  EXPF=0,1,2}; {\tt HIG2=0,1}.

When performing a SM analysis of measured POs, several SM fits should
be performed, changing the flags as indicated above.  The differences
in fitted values are a measure of the theoretical uncertainty in
extracting SM parameters from the measured POs.

\clearpage

%%% \input{ltf_note_2}
%------------------------------
\section{Realistic Observables}
%------------------------------

The ROs, measured cross-sections and forward-backward asymmetries, are
computed in the context of the SM, see \cite{kn:thebook}.  Thus the
comparison between {\tt TOPAZ0} and {\tt ZFITTER} is mainly a SM
comparison. In addition, however, one of the goals will be to pin down

\begin{itemize}

\item the definition of POs;
  
\item the calculation of ROs in terms of the defined POs for the
  purpose of MI fits, showing that for POs with values as calculated
  in the SM, the ROs are {\em by construction} identical to the full
  SM RO calculation.

\end{itemize}

The last point requires expressing $\rho$'s and effective mixing
angles in terms of POs, assuming the validity of the SM.  After this
transformation the ROs will be given as a function of the POs at their
SM values.  This is not at all a trivial affair because of gauge
invariance and one should remember that gauge invariance at the $\zb$
pole (on-shell gauge invariance) is entirely another story from gauge
invariance at any arbitrary scale (off-shell gauge invariance).  Some
of the re-summations that are allowed at the pole and that heavily
influence the definition of effective $\zb$ couplings are not
trivially extendible to the off-shell case. Therefore, the expression
for RO=RO(PO), at arbitrary $\sman$, requires a careful examination
and should be better understood as RO=RO(PO,${\overline{\rm SM}}$),
that is, for example:
%--
\bq
\sigma_{\rm MI} = \sigma_{\rm SM}\lpar R_{\fl},\afba{0,l},\dots \to \gvf,\gaf
\to \rho_f,\seffsf{\ff}; \mbox{residual SM}\rpar.
\label{defmiapp}
\eq
%--
As long as the procedure does not violate gauge invariance and the POs
are given SM values, there is nothing wrong with the calculations.  It
is clear that in this case the SM ROs coincide with the MI ROs.  

The next question is, of course, do the ROs in MI calculations agree
for POs not having SM values - at least over a range of PO values
corresponding to current experimental errors on POs for a single
experiment, i.e., two to three times the error on the LEP-average POs?
It is clear that the present procedure, ${\overline{\rm SM}}$ fixed
and POs varying around their SM values, is wrong in principle but one
should content oneself with testing its accuracy.

There is another reason to be worried, one should avoid any
interpretative strategy such that the pattern becomes: raw data
$\,\to\,$ RO {\em decoded} into PO $\,\to\,$ Lagrangian parameters
(any Lagrangian, SM, MSSM, etc.), if only one {\em decoder} (code that
allows for MI studies) has been built for that purpose.  The glimpse
we want to have of nature should not depend on the {\em decoder}.

%-------------------------------------------
\subsection{Setup for Realistic Observables}
%-------------------------------------------

In this note we discuss ROs for $\sman$-channel processes, thus
excluding Bhabha scattering.  The ROs, cross-section and
forward-backward asymmetries, are computed for the setup specified in
\tabn{tab7}.  This setup will be referred to as the fully extrapolated
setup.

\begin{table}[ht]
\begin{center}
\renewcommand{\arraystretch}{1.25}
\begin{tabular}{|c||c|}
\hline
INPUT               & $\mz = 91.1867, \mt = 173.8, \mh = 100,
                      \als(\mz) = 0.119$  \\
\hline
\hline
$\sqrt{\sman}$      & $\mz-3,\, \mz-1.8,\, \mz,\, \mz+1.8,\, \mz+3$  \\
$\smani{0}/\sman$   & 0.01  \\
WEAK BOXES          & YES   \\
IFI                 & NO ($\to$ Section~\ref{sec:IFI}) \\
ISPP                & NO ($\to$ Section~\ref{sec:SPP}) \\
\hline
\end{tabular}
\caption[]{
  Extrapolated setup for the calculations of ROs.  Masses and
  $\sqrt{\sman}$ are in GeV.  }
\label{tab7}
\end{center}
\end{table}
%--

The choice of the energies is dictated by the fact that the most
precise SLD/LEP-1 measurements are at the pole and at $\pm 1.8\,$GeV
away from the pole.  The parameter $\smani{0}$ is usually referred to
as the $\smanp$-cut, i.e., $\smani{0}=\smanp_{\rm min}$, where the
following definition applies:

\noindent
{\em $\sqrt{\smanp}$ is the centre-of-mass energy of the $\fep\fem$
  system after initial-state radiation.}

In general the effects of initial-final QED interference (IFI) and of
initial-state pair-production are not included.  They will be
discussed separately in Sections~\ref{sec:IFI} and~\ref{sec:SPP},
respectively.

When SM parameters are varied we use
%--
\bqas
\mh = 10, 30, 100, 300, 1000\,\GeV.
\eqas
%--

%---------------------------------------------------------------------------
\subsection{Next-to-Leading and Mixed Corrections for Realistic Observables}
%---------------------------------------------------------------------------

The inclusion of mixed two-loop correction for RO, at $s\ne \mzs$, can
only represent an approximation to the real answer.  Consider the term
giving
%---
\bqa
\Delta_{_{\rm EW/QCD}}&=&
\Nc\,\frac{\als}{\pi}\frac{\alpha^2\mz}{12\pi}
\biggl\{
\biggl[\lpar\gcp{\fq}\rpar^4+\lpar\gcm{\fq}\rpar^4\biggr]
\biggl[\Cf\Lambda_2^{(1)}\lpar\frac{\sman}{\mzs}\rpar  
- \Lambda_2^{(0)}\lpar\frac{\sman}{\mzs}\rpar\biggr] 
\nl{}&&
+\frac{\gcm{\fq}}{2\stws}
\biggl( 
\gcm{\fq}\biggl[\Cf\Lambda_2^{(1)}\lpar\frac{\sman}{\mws}\rpar
-\Lambda_2^{(0)}\lpar\frac{\sman}{\mws}\rpar\biggr] 
\nl{}&&
+6\,\tciq\frac{\ctw}{\stw}
\biggl[\Cf\Lambda_3^{(1)}\lpar\frac{\sman}{\mws}\rpar
-\Lambda_3^{(0)}\lpar\frac{\sman}{\mws}\rpar\biggr]
\biggr)
\biggr\},
\eqa
%---
with $\gcpm{\fq}=\gcv{\fq}\pm\gca{\fq}$ and $\Cf = 4/3$. The functions
$\Lambda_i$ are given by an expansion in $\als$ and, in computing the
$\zb$ width one sets $\sman = \mzs$ obtaining a gauge invariant
answer. For arbitrary $\sman$ the following
happens.\footnote{A.~Czarnecki and J.~K\"uhn, private communication.}
$\Lambda^{(1)}_2(\xvar)$ is very simple, it quickly approaches its
asymptotic value $-3/8$ as $\xvar$ grows.  So it is legitimate to use
the same value as at $\xvar=1$, i.e., $-0.37 \pm 0.04$.

With $\Lambda^{(1)}_3$ the situation is a bit more complicated. Its
value can of course still be found using the formula (14) of the first
Ref. of \cite{kn:mix}, (although the error bar will be larger than at
$x=1$).  However, for a heavy off-shell $\zb$ the decay into a real
$\wb$ (plus a pair of quarks) becomes more important, and the
coefficient $\Lambda^{(1)}_3$ gives only one part of the full mixed
QCD/electroweak corrections.  An estimate of the QCD corrections to
the real W emission is given in \cite{kn:cm}.  The fact that
$\wb(\mbox{on-shell})\fq\barq$ is a genuine four-fermion event does
not help too much: the whole issue of separating two- from
four-fermion events at sufficiently high energy has not yet been
systematised.

In addition, if we are away from the $\zb$ pole, we have to take into
account $\ph$-exchange.

The strategy adopted by {\tt TOPAZ0} in this case will be the
following: the amplitude squared due to $\zb$-exchange is something
like
%--
\bq
\kappa\,{{sf(\sman)}\over {\lpar\sman-\mzs\rpar^2 + \smans\gz^2/\mzs}}\;,
\eq
%--
where $\kappa$ denotes a collection of coupling constants. This we
rewrite as
%--
\bq
\kappa\,{{\sman f(\sman) - \mzs f(\mzs)}\over 
{\lpar\sman-\mzs\rpar^2 + \smans\gzs/\mzs}} 
+\kappa\,{{\gz^2\mzs}\over 
{\lpar\sman-\mzs\rpar^2 + \smans\gzs/\mzs}}\,\frac{f(\mzs)}{\gzs}\;.
\eq
%--
The splitting is motivated by the fact that $f(\sman)$ is not gauge
invariant while $f(\mzs)$ is. Moreover, for on-shell $\zb$ bosons we
have at our disposal an improved calculation, e.g., including mixed
two-loop effects. Thus we write
%--
\bq
\kappa\,{{sf(\sman) - \mzs f(\mzs)}\over 
{\lpar\sman-\mzs\rpar^2 + \smans\gzs/\mzs}} 
+{{\gz^2\mzs}\over{\lpar\sman-\mzs\rpar^2 + \smans\gzs/\mzs}}\,F_{\rm imp}(\mzs),
\label{CKHSS}
\eq
%--
with $F = \kappa f/\gzs$, and $F_{\rm imp}$ is the two-loop corrected,
on-shell, expression.

Within {\tt ZFITTER} the implementation of CKHSS correction was done
in a simplified way. The numbers for non-factorized $\ord{\alpha\als}$
corrections for different channels $\zb\to\ff\barf$, reported in
\cite{kn:mix}, are hard-wired to the code for calculating POs.  An
analogy of \eqn{CKHSS} was used for ROs with the inclusion of,
properly normalised, non-factorized $\ord{\alpha\als}$ corrections.  A
detailed comparison of {\tt TOPAZ0/ZFITTER} numbers with/without CKHSS
corrections (DD versus DDD) was done and excellent agreement was
found.  It does not look surprising since the correction is small
($\sim 0.3$ per-mill) and its crude implementation works in practice.

The same applies for the next-to-leading, $\ord{\alpha^2\mts}$
corrections~\cite{kn:dfs}.  Here {\tt TOPAZ0} uses
%---
\bqa
\kappa\,{{sf(\sman)}\over {\lpar\sman-\mzs\rpar^2 + \smans\gz^2/\mzs}}&=&
\kappa\,{{sf_{_{\rm LO}}(\sman) - \mzs f_{_{\rm LO}}(\mzs)}\over 
{\lpar\sman-\mzs\rpar^2 + \smans\gzs/\mzs}} 
\nl&&
+{{\gzs\mzs}\over 
{\lpar\sman-\mzs\rpar^2 + \smans\gzs/\mzs}}\,F_{_{\rm NLO}}(\mzs),\quad
\eqa
%---
with LO,NLO indicating leading and next-to-leading corrections.

Implementation of NLO, two-loop EW corrections, into {\tt ZFITTER} is
very involved and cannot be described in all details here.  It makes
use of the {\tt FORTRAN} code {\tt m2tcor} \cite{kn:degpr} and is due
to common work with G. Degrassi and P. Gambino done in February~1998
\cite{kn:dbdgpg}.  The full collection of the relevant formulae is
presented in \cite{kn:thebook}.  Again, a careful comparison of {\tt
  TOPAZ0/ZFITTER} numbers with/without NLO corrections was undertaken
and good agreement was registered.  Since the implementation into {\tt
  TOPAZ0} is completely independent, the agreement is a convincing
argument to conclude that both implementations are correct.

%---------------------------------
\subsection{Final-State Radiation}
%---------------------------------

One should realize that $\smanp$ is not equivalent to the invariant
mass of the final-state $\ff\barf$ system, $\Mlones(\ff\barf)$, due to
final state QED and QCD radiation.  Furthermore, in the presence of a
$\smanp$-cut the correction for final state QED radiation is simply
%--
\bq
R^{\rm FS}_{\rm{QED}}(\sman) = 
  \frac{3}{4}\,\qfs\,\frac{\alpha(\sman)}{\pi}\;,
\eq
%--
while for a $\Mlones$-cut the correction is more complicated, see
\cite{kn:fsr}.  For full angular acceptance one derives the following
corrections:
%---
\bqa 
\sigma\lpar\sman\rpar &=&  
\frac{\alpha}{4\pi}\qfs\sigma^0\lpar{\sman}\rpar
\biggl\{-2\xvars 
+4\biggl[
\lpar\zvar+\frac{\zvars}{2}+2\ln\xvar\rpar\ln\frac{\sman}{\mfs}
\\{}&&
+\zvar\lpar 1 +\frac{\zvar}{2}\rpar\ln\zvar 
+2\ztwo-2\li{2}{\xvar}
-2\ln\xvar + \frac{5}{4}- 3\zvar - \frac{\zvars}{4}   
\biggr]
\biggr\}.
\label{imc1}
\nn
\eqa
%--
Here we have introduced
%--
\bq
\xvar=1-\zvar,\qquad
\zvar=\Mlones\lpar\ff\barf\rpar/\sman. 
\eq
%--
For an $\smanp$-cut both QED and QCD final-state radiation are
included through an inclusive correction factor.  For $\Mlones$-cut
the result remains perfectly defined for leptons, however for hadronic
final states there is a problem.  This has to do with QCD final-state
corrections. Indeed we face the following situation:

\begin{itemize}
  
\item for $\fep\fem\to\ff\barf \gamma$ the exact correction factor is
  known at $\ord{\alpha}$ even in the presence of a
  $\Mlones(\ff\barf)$ cut \cite{kn:fsr};
  
\item the complete set of final-state QCD corrections are known up to
  $\ord{\als^3}$ (see \cite{kn:rqm} and also \cite{kn:kpeople}) only
  for the fully inclusive setup, i.e., no cut on the $\ff\barf$
  invariant mass;
  
\item the mixed two-loop QED/QCD final-state corrections are also
  known only for a fully inclusive setup \cite{kn:kat};
  
\item at $\ord{\als}$ QCD final-state corrections in presence of a
  $\Mlones$-cut follow from the analogous QED calculation
  \cite{kn:abl}.

\end{itemize}

The ideal thing would be to have QED $\,\oplus\,$ QCD final-state {\em
  radiator factors} $R$, with a kinematical cut imposed on
$\Mlones(\ff\barf)$.  Missing this calculation, that would give the
correct $\ord{\alpha\als}$ factors with cuts, we have three options
%--
\bqa
R^{\rm FS} &=& 1 + R^{\rm FS}_{\rm QED, cut} + R^{\rm FS}_{\rm QCD, ext}\;,
\nl
R^{\rm FS} &=& \lpar 1 + R^{\rm FS}_{\rm QCD, ext}\rpar\,
               \lpar 1 + R^{\rm FS}_{\rm QED, cut}\rpar,  \nl
R^{\rm FS} &=& 1 + R^{\rm FS}_{\rm QED, cut} + R^{\rm FS}_{\rm QCD, cut}\;.
\label{eq44}
\eqa
%--
The (QCD,ext) corrections are understood up to $\ord{\als^3}$, while
those corresponding to the (QCD,cut) setup are only computed at
$\ord{\als}$.  The first of \eqn{eq44} is our preferred option.

%\clearpage

%--------------------------------------------
\section{De-Convoluted Realistic Observables}
%--------------------------------------------

Our goal in describing the theoretical uncertainties for realistic
observables is twofold.  First we want to discuss the effect of QED
radiation by comparing different radiators and then we have to give a
critical assessment of the theoretical uncertainty in the predicted
cross-sections, de-convoluted of QED effects, i.e., the purely weak
uncertainty.  We therefore define several levels of de-convolution:
%--
\begin{itemize}
  
\item Single-de-convolution (SD), giving the kernel cross-sections
  without initial-state QED radiation, but including all final-state
  correction factors.
  
\item Double-de-convolution (DD), giving the kernel cross-sections
  without ini\-tial- and final-state QED radiation and without any
  final-state QCD radiation.  There is an additional level, to be
  called DDD, and the difference between DD and DDD deserves a word of
  comment.  The improvement upon naive electroweak/QCD factorisation
  contains two effects, the FTJR correction~\cite{kn:ftjr} which gives
  the leading two-loop answer for the $\ffb\barb$-channel and the
  CKHSS correction~\cite{kn:mix} which gives the correct answer for
  the remaining mixed corrections in all quark channels.
  
  In DD-mode FTJR and CKHSS corrections are kept while in DDD-mode
  they are excluded.  This option allows us to keep under control the
  implementation of the new CKHSS correction.
  
\item DD, DDD with only $\zb-\zb$ exchange (DDZ, DDDZ), weak boxes are
  not included,
  
\item DDD with only $\zb \oplus \ph$ (DDZG), i.e. no $\zb-\ph$
  interference and weak boxes are not included.

\end{itemize}

Rather than comparing only the complete results for ROs, i.e.,
including initial-state QED radiation, final-state QED and QCD
radiation, initial-state pair-production and initial-final QED
interference, we do more.  The reason is given by the observation that
often an agreement on the complete result may be a consequence of
several compensations of the single components.  This is why we want
to compare component by component and the procedure will allow us to
formulate a quantitative statement on the overall theoretical
uncertainty.  In this way the errors are independent of the amount of
cancellation between the various components.

We define
%--
\bq
\sigma_{_{\rm T}}(X)\quad\mbox{and}\quad
\afba{}(X)={{\sigma_{_{\rm FB}}(X)}\over{\sigma_{_{\rm T}}(X)}}\;, 
\qquad
\mbox{with X= SD,DD,DDD,DDZ,DDZG}.
\eq
%--
For SD-quantities that contain final-state QED radiation, we further
distinguish between two series, the so-called $\Mlones$-series where a
cut on $\Mlones(\ff\barf)$ is applied and the so-called
$\smanp$-series.  For SD setup the latter implies that no cut is
applied (fully extrapolated setup). In this way also the final-state
QED correction factors with/without cuts on the final-state fermions
can be compared.

%----------------------------------------
\subsection{De-Convoluted Cross-Sections}
%----------------------------------------

By comparing SD with DD quantities we are able to disentangle the
effect of initial-state QED radiation from final-state QED
$\,\oplus\,$ QCD radiation.

The DDZ or DDZG modes are included for convenience of the reader but
deserve an additional comment. Clearly, away from the $\zb$-peak,
diagrams with $\zb$ or $\ph$ exchanges are not gauge invariant and,
therefore, we expect deviations in the result of the two codes.
However, they are useful to understand the pattern of agreement and
also to show the internal consistency of codes, for instance
%--
\bq
\sigma^{DDZ}_{\ff}(\mz) = \sigma^0_{\ff}
\lrbr 1 + \frac{3}{4}\,\frac{\alpha(\mzs)}{\pi}\rrbr^{-1}
\lrbr 1 + \frac{3}{4}\,\qfs\frac{\alpha(\mzs)}{\pi}
                            + \delta^{\rm{QCD}}\rrbr^{-1},
\label{deservinganimprovement}
\eq
%--
is an equality between RO and PO that must be satisfied.  The
comparison between DDZ and DDZG modes, moreover, gives an estimate of
the $\zb-\ph$ interference effects, before folding with QED radiation.

\begin{figure}[p]
\begin{center}
%includegraphics[width=0.75\linewidth,clip=true,bb=90 90 425 740]{fignote1.eps}
\includegraphics[width=0.75\linewidth]{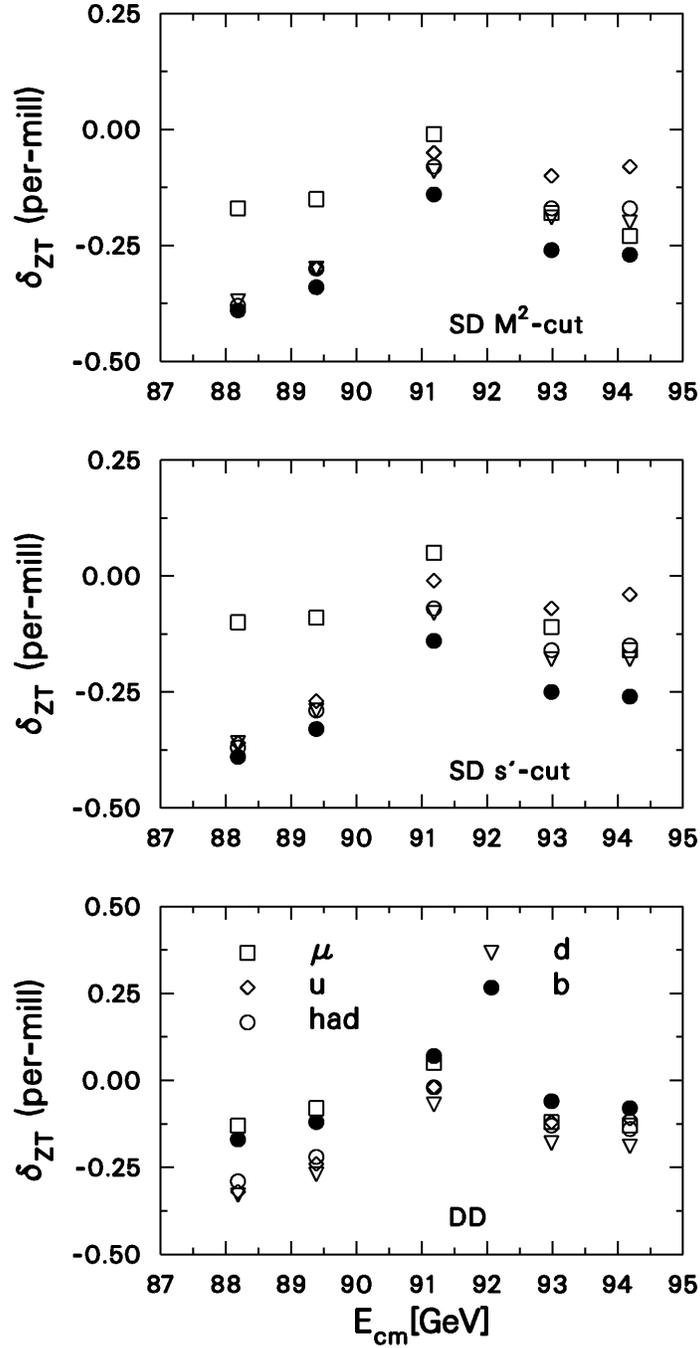}
\caption[]{
  Relative deviations between {\tt TOPAZ0} and {\tt ZFITTER} for total
  fermionic cross-section. DD-mode is shown with SD-modes
  corresponding to no-cut or a $\Mlones(\ff\barf)$-cut of
  $0.01\,\sman$. }
\label{fignote1}
\end{center}
\end{figure}

In \tabn{tab11f} we start our comparison for RO de-convoluted
observables showing results for $\sigma_{\ff}$ in DD and SD
(no-cut and $\Mlones$-cut) DDD modes. The reference point has
been fixed to $\mh = 100\,$GeV.  The relative deviations $\delta_{TZ}
=\,$T/Z-1 in per-mill are shown in \fig{fignote1}.  At $\sqrt{\sman} =
\mz$ we register $0.05$ per-mill for muons both for DD mode and for
SD, no-cut mode. For hadrons we have $-0.02$ per-mill in DD mode
and $-0.07$ per-mill in SD, no-cut mode.  There are tiny
variations when we consider the SD $\Mlones$-cut mode.

The differences between DD and SD describe different implementations
of QED final-state radiation and of QCD corrections.  Our comparison
between the two SD branches shows that final-state QED correction
factors are correctly implemented for the fully inclusive setup
(no-cut) and for a cut on the invariant mass of the $\ff\barf$
pair.  The agreement between DD and SD for hadrons shows that also
final-state QCD factors are under control.

As we have already illustrated there are other sources of final-state
$\als$-dependent corrections, due to the non-factorisation of QCD and
purely electroweak effects.  The SD mode also accounts for
non-factorizable two-loop effects and the agreement between two
calculations in SD mode is an agreement for the $R^{\ff}_{\ssV,\ssA}$
factors of \eqn{defgammaf} and for the interplay between them and the
FTJR/CKHSS effects.  From \fig{fignote1} we observe a $-0.02, -0.07$
and $+0.07$ per-mill differences for $\fu,\fd$ and $\ffb$ quarks at
the peak.  Therefore the overall agreement for the hadronic
cross-section, $-0.02$ per-mill, is also the result of some partial
compensation between contributions from up- and down-type quarks.

Note that the agreement remains very good also for $\ffb$-quarks where
next-to-leading corrections are not available and where, therefore,
one would expect larger deviations between {\tt TOPAZ0} and {\tt
  ZFITTER}.

Note the following relation between the pseudo-observable $\Rb$ and
the ratio of DD cross-sections
%--
\bq
\Rb = {{\sigma_{\ffb}}\over {\sigma_{\rm{had}}}}\mdls_{\sqrt{\sman} = \mz}
        - \left\{ \ba{ll}
        0.00146 & \mbox{{\tt TOPAZ0}} \\
        0.00146 & \mbox{{\tt ZFITTER}}\;.
        \ea\right.
\eq
%--
The difference reflects the ${\overline{SM}}$-remnant effect, since
the ratio of RO cross-sections has $\ph$-exchange, imaginary parts,
$\dots$, and (substantially negligible) weak boxes.

%--
\begin{table}[p]
\begin{center}
\renewcommand{\arraystretch}{1.1}
\begin{tabular}{|c||c|c|c|c|c|}
\hline
 & \multicolumn{5}{c|}{Centre-of-mass energy in GeV} \\
\cline{2-6}
 & $\mz - 3$ & $\mz - 1.8$ & $\mz$ & $\mz + 1.8$ & $\mz + 3$  \\
\hline 
\hline
$\sigma_{\mu}\,$[nb]  DD $\,\equiv\,$ DDD 
& 0.29999 & 0.65718 & 2.00341 & 0.65856 & 0.31047  \\
& 0.30003 & 0.65724 & 2.00331 & 0.65863 & 0.31051  \\    
$\sigma_{\mu}\,$[nb]  SD -- no-cut
& 0.30055 & 0.65839 & 2.00711 & 0.65978 & 0.31104  \\
& 0.30058 & 0.65845 & 2.00700 & 0.65985 & 0.31108  \\         
$\sigma_{\mu}\,$[nb]  SD -- $\Mlones$-cut
& 0.30047 & 0.65821 & 2.00656 & 0.65960 & 0.31095  \\
& 0.30052 & 0.65832 & 2.00659 & 0.65971 & 0.31102  \\     
\hline 
\hline
$\sigma_{\fu}\,$[nb]  DD 
& 0.99648 & 2.21803 & 6.82893 & 2.23330 & 1.04203  \\
& 0.99682 & 2.21861 & 6.82913 & 2.23355 & 1.04214  \\
$\sigma_{\fu}\,$[nb]  SD -- no-cut
& 1.04290 & 2.32118 & 7.14541 & 2.33633 & 1.08993  \\
& 1.04330 & 2.32183 & 7.14551 & 2.33647 & 1.08996  \\        
$\sigma_{\fu}\,$[nb]  SD -- $\Mlones$-cut
& 1.04277 & 2.32090 & 7.14452 & 2.33604 & 1.08979  \\
& 1.04320 & 2.32162 & 7.14486 & 2.33626 & 1.08986  \\          
\hline 
\hline
$\sigma_{\fd}\,$[nb]  DD 
& 1.26996 & 2.84741 & 8.79775 & 2.86549 & 1.32868  \\
& 1.27040 & 2.84820 & 8.79838 & 2.86600 & 1.32893  \\
$\sigma_{\fd}\,$[nb]  SD -- no-cut
& 1.31395 & 2.94596 & 9.10195 & 2.96450 & 1.37458  \\
& 1.31444 & 2.94683 & 9.10268 & 2.96502 & 1.37482  \\          
$\sigma_{\fd}\,$[nb]  SD -- $\Mlones$-cut
& 1.31391 & 2.94587 & 9.10166 & 2.96441 & 1.37453  \\
& 1.31441 & 2.94672 & 9.10249 & 2.96496 & 1.37479  \\          
\hline 
\hline
$\sigma_{\fc}\,$[nb]  DD 
& 0.99648 & 2.21803 & 6.82893 & 2.23330 & 1.04203  \\       
& 0.99682 & 2.21861 & 6.82913 & 2.23355 & 1.04214  \\
$\sigma_{\fc}\,$[nb]  SD -- no-cut
& 1.04267 & 2.32070 & 7.14397 & 2.33588 & 1.08972  \\
& 1.04307 & 2.32133 & 7.14405 & 2.33601 & 1.08975  \\
$\sigma_{\fc}\,$[nb]  SD -- $\Mlones$-cut
& 1.04262 & 2.32057 & 7.14359 & 2.33575 & 1.08967         \\
& 1.04303 & 2.32124 & 7.14375 & 2.33592 & 1.08971  \\
\hline 
\hline
$\sigma_{\ffb}\,$[nb] DD 
& 1.25204 & 2.80753 & 8.67631 & 2.82663 & 1.31091  \\
& 1.25226 & 2.80787 & 8.67577 & 2.82681 & 1.31101  \\
$\sigma_{\ffb}\,$[nb]  SD -- no-cut
& 1.28945 & 2.89170 & 8.93787 & 2.91233 & 1.35080  \\
& 1.28995 & 2.89266 & 8.93909 & 2.91306 & 1.35115  \\          
$\sigma_{\ffb}\,$[nb]  SD -- $\Mlones$-cut
& 1.28945 & 2.89168 & 8.93780 & 2.91231 & 1.35079  \\
& 1.28995 & 2.89266 & 8.93908 & 2.91306 & 1.35115  \\     
\hline 
\hline
$\sigma_{\rm{had}}\,$[nb] DD 
& 5.78492 & 12.93841 & 39.92967 & 13.02421 & 6.05233  \\
& 5.78670 & 12.94148 & 39.93079 & 13.02591 & 6.05313  \\
$\sigma_{\rm{had}}\,$[nb]  SD -- no-cut
& 6.00291 & 13.42550 & 41.43114 & 13.51353 & 6.27961  \\
& 6.00518 & 13.42948 & 41.43401 & 13.51559 & 6.28050  \\          
$\sigma_{\rm{had}}\,$[nb]  SD -- $\Mlones$-cut
& 6.00265 & 13.42490 & 41.42929 & 13.51293 & 6.27933  \\
& 6.00500 & 13.42907 & 41.43272 & 13.51517 & 6.28030  \\     
\hline 
\end{tabular}
\caption[]{{\tt TOPAZ0/ZFITTER} comparison of $\sigma_{\ff}$ 
  de-convoluted: DD and SD fully extrapolated mode and $\Mlones$-cut
  of $0.01\,\sman$. Here $\mh = 100\,$GeV. }
\label{tab11f}
\end{center}
\end{table}
%--

%--
%--

\clearpage

%-------------------------------------
\subsection{De-Convoluted Asymmetries}
%-------------------------------------
%--
In \tabn{tab11g} we show DD and SD modes for the de-convoluted muonic
and heavy quark forward-backward asymmetries.
%--
\begin{table}[ht]
\begin{center}
\renewcommand{\arraystretch}{1.1}
\begin{tabular}{|c||c|c|c|c|c|}
\hline
& \multicolumn{5}{c|}{Centre-of-mass energy in GeV} \\
  \cline{2-6}
& $\mz - 3$ & $\mz - 1.8$ & $\mz$ & $\mz + 1.8$ & $\mz + 3$  \\
\hline 
\hline
$\afba{\mu}$  DD 
& -0.26170 & -0.15037 & 0.01745 & 0.17510 & 0.27002  \\
& -0.26167 & -0.15037 & 0.01741 & 0.17502 & 0.26991  \\
$\afba{\mu}$  SD -- no-cut
& -0.26122 & -0.15010 & 0.01742 & 0.17478 & 0.26952  \\
& -0.26119 & -0.15010 & 0.01737 & 0.17469 & 0.26941  \\
$\afba{\mu}$  SD -- $\Mlones$-cut
& -0.26128 & -0.15013 & 0.01742 & 0.17481 & 0.26958  \\
& -0.26122 & -0.15011 & 0.01738 & 0.17471 & 0.26944  \\
\hline 
\hline
$\afba{\fc}$  DD 
& -0.09383 & -0.02507 & 0.07411 & 0.16636 & 0.22319  \\
& -0.09376 & -0.02506 & 0.07405 & 0.16624 & 0.22304  \\
$\afba{\fc}$  SD -- no-cut
& -0.08977 & -0.02399 & 0.07092 & 0.15922 & 0.21364  \\
& -0.08968 & -0.02396 & 0.07086 & 0.15909 & 0.21348  \\
$\afba{\fc}$  SD -- $\Mlones$-cut
& -0.08978 & -0.02399 & 0.07093 & 0.15923 & 0.21365  \\
& -0.08968 & -0.02396 & 0.07086 & 0.15910 & 0.21349  \\
\hline 
\hline
$\afba{\ffb}$ DD 
& 0.03652 & 0.06403 & 0.10311 & 0.13956 & 0.16242  \\
& 0.03615 & 0.06361 & 0.10262 & 0.13903 & 0.16187  \\
$\afba{\ffb}$ SD -- no-cut
& 0.03556 & 0.06233 & 0.10035 & 0.13579 & 0.15802  \\
& 0.03529 & 0.06209 & 0.10014 & 0.13562 & 0.15788  \\
$\afba{\ffb}$  SD -- $\Mlones$-cut 
& 0.03556 & 0.06233 & 0.10035 & 0.13580 & 0.15803  \\
& 0.03529 & 0.06209 & 0.10014 & 0.13562 & 0.15788  \\
\hline 
\end{tabular}
\caption[]{
  {\tt TOPAZ0/ZFITTER} comparison of $\afba{\mu,\fc,\ffb}$ in DD-mode
  and SD-modes. }
\label{tab11g}
\end{center}
\end{table}
%--

\begin{figure}[htbp]
\begin{center}
%includegraphics[width=0.75\linewidth,clip=true,bb=90 90 425 740]{fignote6.eps}
\includegraphics[width=0.75\linewidth]{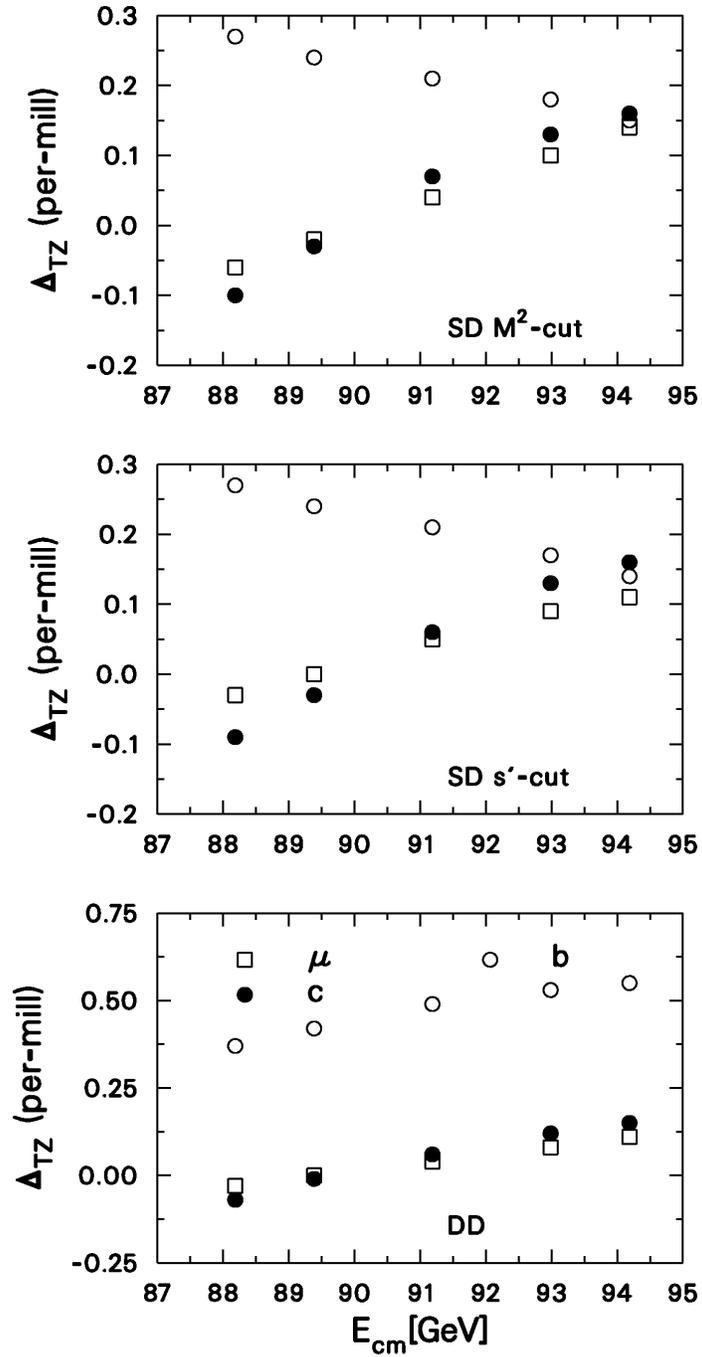}
\caption[]{
  Absolute deviations between {\tt TOPAZ0} and {\tt ZFITTER} for
  de-convoluted forward-backward asymmetries. }
\label{fignote6}
\end{center}
\end{figure}

In \fig{fignote6} we show the absolute deviation in per-mill between
{\tt TOPAZ0} and {\tt ZFITTER} predictions for the forward-backward
asymmetries in DD mode and in SD-modes.  From this figure one
understands that $\afba{\ffb}$ is indeed the RO showing the largest
deviations between the two codes.  This is hardly a surprise, given
the comparison for $\seffsf{\ffb}$ reported in \tabn{tab5+6}.  In both
cases it is the absence of next-to-leading corrections in the
$\ffb\barb$-channel (due to missing non-universal next-to-leading
terms) that stays at the root of the relatively large theoretical
uncertainty.  

For $\afba{\mu}$ the absolute deviations is always below $0.14$
per-mill ($0.04 - 0.05$ at the peak).  For $\afba{\fc}$ the agreement
is also very good, deviations below $0.16$ per-mill (and only at the
wings) and peak asymmetries differing of $0.06 - 0.07$ per-mill.  This
sort of agreement and consistency between DD-mode and SD-modes shows
that also final state QCD corrections are under control in the
$\fc$-channel.

%--
%--

\clearpage

Note that QCD corrections for the forward-backward asymmetry can well
be approximated by an expansion in the parameter
$\xvar=4\mqS/\sman=4r^2\;$,
%---
\bq
\sigma_{\rm VA}^{\fq}\to\sigma_{\rm VA}^{\fq}\,
\lpar 1 + \frac{\als}{\pi}\,f_1\rpar,
\label{eq:afbqcd}
\eq
%--
with a correction factor, $f_1$, which we write as \cite{kn:abl}
%--
\bq
f_1=\frac{8}{3}{\xvar}^{\frac{1}{2}}
+\lpar\frac{7}{3}+\frac{\pi^2}{18}-\frac{2}{3}l_r+\frac{1}{3}l_r^2\rpar\xvar
-\frac{40}{27}{\xvar}^{\frac{3}{2}}
+\lpar\frac{55}{24}+\frac{\pi^2}{12}-\frac{19}{12}l_r+\frac{1}{2}l_r^2\rpar
\xvars 
+\cdots,
%+\ord{{\xvar}^{\frac{5}{2}}},
\eq
%--
where $l_r=\ln r$. These first order corrections vanish in the
massless limit.  The asymmetry changes as
%--
\bq
\afba{\fq}\to\afba{\fq}\,\lrbr 1 - \frac{\als}{\pi}\,
\lpar 1 - f_1\rpar\rrbr.
\eq
%--
For $\ffb$-quarks the inclusion of QCD final state correction improves
the {\tt TOPAZ0}-{\tt ZFITTER} agreement.  This fact is not completely
satisfactory, signalling some difference (and some uncertainty) in the
implementation of electroweak/QCD radiative corrections for the
$\ffb\barb$-channel.

%--------------------------------------------------------------
\subsection{Higgs-Mass Dependence of De-Convoluted Observables}
%--------------------------------------------------------------

\begin{figure}[htbp]
\begin{center}
%includegraphics[width=0.75\linewidth,clip=true,bb=5 90 425 740]{fignote2.eps}
\includegraphics[width=0.95\linewidth]{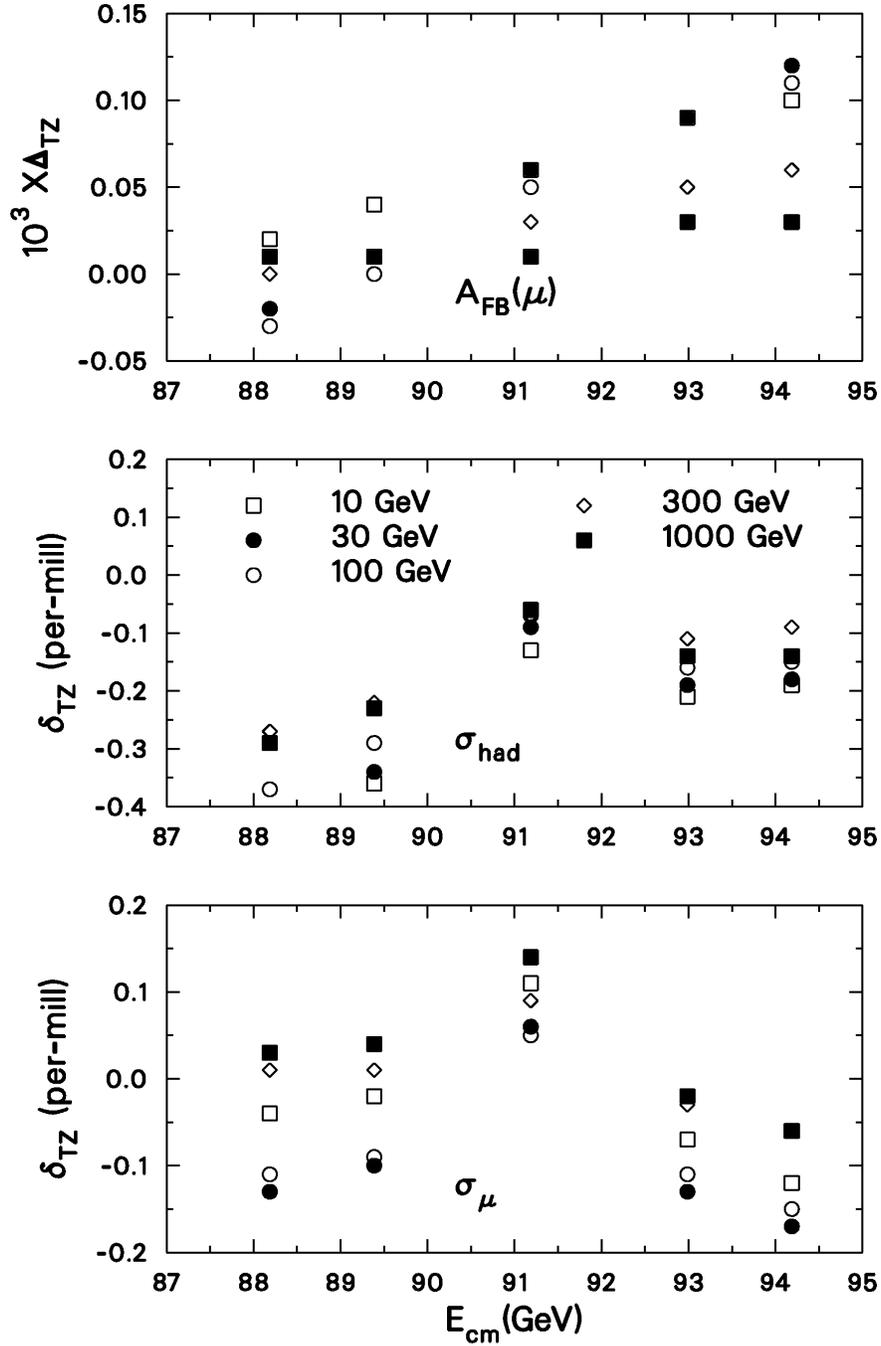}
\caption[]{
  Relative deviations between {\tt TOPAZ0} and {\tt ZFITTER} for
  $\sigma(\mu)$, $\afba{\mu}$ and $\sigma(\rm had)$ for different
  values of the Higgs boson mass in SD no-cut mode.}
\label{fignote2}
\end{center}
\end{figure}

Our comparison for de-convoluted ROs has to be extended to a wide
range of values for the Higgs boson mass: differences indicate
theoretical uncertainties in using the measurements to constrain the
mass of the Higgs boson.  In \tabns{tab11a}{tab11aa} we present
cross-sections and asymmetries in the Higgs-mass range $\mh$ range of
$10-1000\,$ GeV.  In \fig{fignote2} we show the relative deviations
between the {\tt TOPAZ0} and {\tt ZFITTER} prediction for
$\sigma(\mu)$, $\sigma(\rm had)$ and $\afba{\mu}$ in SD no-cut mode as
a function of the Higgs mass ranging from $10\,$GeV to $1000\,$GeV.

The figure confirms the good agreement between the two sets of
predictions: differences in the peak muonic cross-sections are
everywhere below $0.14$ per-mill, reached only at the boundaries of
the interval in $\mh$ ($0.05$ per-mill at $\mh = 100\,$GeV).
Deviations in the peak hadronic cross-sections vary from $-0.13$
per-mill at very low values of the Higgs boson mass to $-0.07$
per-mill at $\mh = 100\,$GeV and stay practically constant for higher
values of $\mh$.  The variations in relative differences for
$\sigma(\mu)$ are $0.09$ per-mill at peak and $0.16, 0.11$ per mill at
the wings ($\mz \pm 3$).  For $\sigma_{\rm{had}}$ we have $0.07$
per-mill at peak and $0.17, 0.10$ per-mill at the wings.  For
$\afba{\mu}$ we observe absolute differences which are everywhere
below 0.00012 and at the peak below 0.00006.

%--
\begin{table}[htp]
\begin{center}
\renewcommand{\arraystretch}{1.1}
\begin{tabular}{|c||c|c|c|c|c|}
\hline
   \multicolumn{6}{|c|}{$\sigma(\mu)$~[nb]} \\
\hline
  & \multicolumn{5}{c|}{$\mh$in GeV} \\
  \cline{2-6}
$\sqrt{\sman}$[GeV]  & 10 & 30 & 100 & 300 & 1000  \\
\hline 
\hline
$\mz-3$   &  0.30000 & 0.30055 & 0.30055 & 0.30011 & 0.29938\\
          &  0.30001 & 0.30059 & 0.30058 & 0.30011 & 0.29937\\
\hline
$\mz-1.8$ &  0.65724 & 0.65832 & 0.65839 & 0.65766 & 0.65643\\
          &  0.65726 & 0.65839 & 0.65845 & 0.65766 & 0.65641\\
\hline
$\mz$     &  2.00558 & 2.00613 & 2.00711 & 2.00827 & 0.20098\\
          &  2.00537 & 2.00601 & 2.00700 & 2.00809 & 0.20095\\
\hline
$\mz+1.8$ &  0.65810 & 0.65968 & 0.65978 & 0.65900 & 0.65772\\
          &  0.65815 & 0.65976 & 0.65985 & 0.65902 & 0.65773\\
\hline
$\mz+3$   &  0.31003 & 0.31102 & 0.31104 & 0.31053 & 0.30971\\
          &  0.31007 & 0.31107 & 0.31108 & 0.31054 & 0.30973\\

\hline 
\end{tabular}
\caption[]{
  {\tt TOPAZ0/ZFITTER} comparison for the muonic cross-section in SD
  fully extrapolated mode for different values of the Higgs boson
  mass. }
\label{tab11a}
\end{center}
\end{table}

%--
\begin{table}[htp]
\begin{center}
\renewcommand{\arraystretch}{1.1}
\begin{tabular}{|c||c|c|c|c|c|}
\hline
   \multicolumn{6}{|c|}{$\sigma(\rm had)$~[nb]} \\
\hline
  & \multicolumn{5}{c|}{$\mh$in GeV} \\
  \cline{2-6}
$\sqrt{\sman}$[GeV]  & 10 & 30 & 100 & 300 & 1000  \\
\hline 
\hline
$\mz-3$    &   5.99467 &   6.00515 &   6.00291 &   5.99098 &   5.97237\\
           &   5.99740 &   6.00774 &   6.00518 &   5.99267 &   5.97414\\
\hline
$\mz-1.8$  &  13.40951 &  13.42945 &  13.42550 &  13.40383 &  13.37020\\
           &  13.41443 &  13.43405 &  13.42948 &  13.40683 &  13.37339\\
\hline
$\mz$      &  41.42574 &  41.42837 &  41.43114 &  41.43364 &  41.43777\\
           &  41.43106 &  41.43209 &  41.43401 &  41.43645 &  41.44034\\
\hline
$\mz+1.8$  &  13.48804 &  13.51762 &  13.51353 &  13.49007 &  13.45456\\
           &  13.49074 &  13.52007 &  13.51559 &  13.49146 &  13.45640\\
\hline
$\mz+3$    &   6.26327 &   6.28209 &   6.27961 &   6.26553 &   6.24431\\
           &   6.26441 &   6.28316 &   6.28050 &   6.26606 &   6.24513\\
\hline 
\end{tabular}
\caption[]{
  {\tt TOPAZ0/ZFITTER} comparison for the hadronic cross-section in SD
  fully extrapolated mode for different values of the Higgs boson
  mass. }
\label{tab11b}
\end{center}
\end{table}
%----------

%--
%--

\clearpage

%--
\begin{table}[ht]
\begin{center}
\renewcommand{\arraystretch}{1.1}
\begin{tabular}{|c||c|c|c|c|c|}
\hline
   \multicolumn{6}{|c|}{$\afba{\mu}$} \\
\hline
  & \multicolumn{5}{c|}{$\mh$in GeV} \\
  \cline{2-6}
$\sqrt{\sman}$[GeV]  & 10 & 30 & 100 & 300 & 1000  \\
\hline 
\hline
$\mz-3$   & -0.25984 &-0.26021 &-0.26122 &-0.26246 &-0.26398\\
          & -0.25986 &-0.26019 &-0.26119 &-0.26246 &-0.26399\\
\hline
$\mz-1.8$ & -0.14864 &-0.14909 &-0.15010 &-0.15126 &-0.15264\\
          & -0.14868 &-0.14910 &-0.15010 &-0.15127 &-0.15265\\
\hline
$\mz$     &  0.01899 & 0.01838 & 0.01742 & 0.01644 & 0.01534\\
          &  0.01893 & 0.01832 & 0.01737 & 0.01641 & 0.01533\\
\hline
$\mz+1.8$ &  0.17648 & 0.17566 & 0.17478 & 0.17401 & 0.17324\\
          &  0.17639 & 0.17557 & 0.17469 & 0.17396 & 0.17321\\
\hline
$\mz+3$   &  0.27130 & 0.27035 & 0.26952 & 0.26890 & 0.26833\\
          &  0.27120 & 0.27023 & 0.26941 & 0.26884 & 0.26830\\
\hline 
\end{tabular}
\caption[]{
  {\tt TOPAZ0/ZFITTER} comparison for the muonic forward-backward
  asymmetry in SD fully extrapolated mode for different values of the
  Higgs boson mass. }
\label{tab11aa}
\end{center}
\end{table}

%-----------------------------------
\subsection{Standard Model Remnants}
%-----------------------------------

POs are determined by fitting ROs, but actually something is still
taken from the SM (imaginary parts, parts which have been moved to
interference terms and photon-exchange terms) making the
model-independent results dependent on the SM.  How complicated is
such a description?  Within the codes we consider sub-contributions to
the DD de-convoluted quantities, 1) total DD, 2) DD with $\zb$
exchange only, 3) DD with $\zb \oplus \ph$ without interference.  For
instance we construct the relative and absolute differences
%---
\bqa
\delta^{\rm int}\sigma &=&
{{\sigma^{\rm DD}_{\rm T}}\over{\sigma^{\rm DD}_{\zb+\ph}}} - 1
\qquad \mbox{in percent},  
\nl
\Delta^{\rm int}\afba{\flm} &=& 
\afba{\mu,\rm DD} - \lpar\afba{\mu,\rm DD}\rpar_{\zb+\ph}.
\eqa
%--
They are reported in \tabn{tab12}.  The effect of the $\zb - \ph$
interference is negative below the peak, vanishingly small around it
and turning positive and large above it. The effect of $\zb - \ph$
interference is particularly important for the forward-backward
asymmetry, as its energy dependence is governed by this interference.

%--
\begin{table}[ht]
\begin{center}
\renewcommand{\arraystretch}{1.1}
\begin{tabular}{|c||c|c|c|c|c|}
  \hline
  & \multicolumn{5}{c|}{Centre-of-mass energy in GeV} \\
  \cline{2-6}
  & $\mz - 3$ & $\mz - 1.8$ & $\mz$ & $\mz + 1.8$ & $\mz + 3$  \\
  \hline \hline
$\delta^{\rm int}\sigma_{\flm}$  & $-0.209\,\%$ & $-0.136\,\%$ & $-0.028\,\%$ & 
                                  $+0.072\,\%$ & $+0.132\,\%$  \\
\hline
$\delta^{\rm int}\sigma_{\rm had}$  & $-0.492\,\%$ & $-0.301\,\%$ & $-0.029\,\%
$ & $+0.226\,\%$ & $+0.384\,\%$  \\
\hline
$\Delta^{\rm int}\afba{\flm}$  & -0.27703 & -0.16611 & 0.00145 &
                                 0.15923 &  0.25441  \\
  \hline 
\end{tabular}
\caption[]{
  {\tt TOPAZ0} relative differences $\delta^{\rm int}\sigma$ and
  absolute differences $\Delta^{\rm int}\afba{\flm}$. }
\label{tab12}
\end{center}
%\end{table}
%--
%--
%\begin{table}[ht]
\begin{center}
\renewcommand{\arraystretch}{1.1}
\begin{tabular}{|c||c|c|c|c|c|}
\hline
  & \multicolumn{5}{c|}{Centre-of-mass energy in GeV} \\
  \cline{2-6}
  & $\mz - 3$ & $\mz - 1.8$ & $\mz$ & $\mz + 1.8$ & $\mz + 3$  \\
\hline \hline
$\sigma_{\flm}\,$[nb]  No Ims 
& 0.29996 & 0.65713 & 2.00343 & 0.65855 & 0.31045 \\
$\sigma_{\flm}\,$[nb]  
& 0.29999 & 0.65718 & 2.00341 & 0.65856 & 0.31047 \\
\hline                        
Diff.[pb]   
& +0.03   & +0.05   & -0.02   & -0.01   & +0.02  \\
\hline \hline
$\sigma_{\rm had}\,$[nb]  No Ims
& 5.78583 & 12.94061 & 39.93848 & 13.02635 & 6.05322  \\
$\sigma_{\rm had}\,$[nb]  
& 5.78492 & 12.93841 & 39.92967 & 13.02421 & 6.05233  \\
\hline                            
Diff.[pb]    
& -0.91   & -2.20    & -8.81    & -2.14    & -0.89  \\
\hline \hline
$\afba{\flm}$ No Ims 
& -0.26311 & -0.15181 & 0.01598 & 0.17364 & 0.26858  \\
$\afba{\flm}$ 
& -0.26170 & -0.15037 & 0.01745 & 0.17510 & 0.27002  \\
\hline                            
Diff.    
& -0.00141 & -0.00144 &-0.00147 &+0.00146 & -0.00144 \\
\hline
\end{tabular}
\caption[]{
  {\tt TOPAZ0} comparison of DD (completely de-convoluted) RO
  with/without imaginary parts (Ims) in couplings and form factors. }
\label{tab13}
\end{center}
\end{table}
%--

Among the de-convoluted quantities the most relevant are those
computed at $s = \mzs$, which have an obvious counterpart in the PO,
that we have already computed, i.e., $\sigma^0_{\ell}\,$,
$\sigma^0_{\had}\,$, and $\afba{0,\rm l}$.

There is however a noticeable difference between the two sets,
represented by the interference of the $\zb-\ph$ $s$-channel diagrams,
including the imaginary parts in $\alpha(\sman)$ and in the
form-factors, the latter being particularly relevant for the leptonic
forward-backward asymmetry. This effect is illustrated in
\tabn{tab13}.

%---------------------------------------------------------
\subsection{The $\zb-\ph$ Interference for Cross-Sections}
%---------------------------------------------------------

One must evaluate the residual SM dependence of the so-called
model-inde\-pen\-dent parameters; one simple source of such a SM
dependence is due to the interference terms for cross-sections, which
are governed by the value of $\seffsf{\rm lept}$ and therefore depend
on the values of $\mt$ and $\mh$ chosen.\footnote{The LEP community
  has agreed on a set of numbers, $\mt=175\,$GeV and $\mh =
  150\,$GeV.}  Note that for leptonic final states, one can use the
POs to express the interference terms, at least up to imaginary parts
which must be taken from the SM as usual.  This is possible because
the interference terms are proportional to the effective couplings
which can be derived from the POs $R_{\ell}$ and
$\afba{0,\ell}$.\footnote{This is realised for MI calculations with
  {\tt TOPAZ0}.  For MI calculations with {\tt ZFITTER}, it is
  realised for the effective-couplings interfaces, but not for the
  partial-width interface.} However, for the inclusive hadronic final
state, which is a sum over the five light quark flavours, the
interference terms must be taken completely from the SM.\footnote{This
  is avoided in the S-Matrix ansatz~\cite{kn:smatrix-1,kn:smatrix-2},
  which treats also the $\zb-\ph$ interference terms for cross
  sections and asymmetries as free and independent parameters to be
  determined from the data.  The experimental measurements are also
  analysed within this extended MI ansatz.  Combined LEP results are
  given in~\cite{kn:lepewwg98}.}

\clearpage

%--
\begin{table}[ht]
\begin{center}
\renewcommand{\arraystretch}{1.1}
\begin{tabular}{|c||c|c|c|c|c|}
  \hline
  & \multicolumn{5}{c|}{Centre-of-mass energy in GeV} \\
  \cline{2-6}
$\mh\,$[GeV]  & $\mz - 3$ & $\mz - 1.8$ & $\mz$ & $\mz + 1.8$ & $\mz + 3$  \\
  \hline \hline
  \multicolumn{6}{|c|}{$\delta^{\rm int}\sigma_{\flm}$} \\
  \hline \hline
10   & $-0.229\,\%$ & $-0.150\,\%$ & $-0.030\,\%$ & 
       $+0.082\,\%$ & $+0.149\,\%$  \\
100  & $-0.209\,\%$ & $-0.136\,\%$ & $-0.028\,\%$ & 
       $+0.072\,\%$ & $+0.132\,\%$  \\
1000 & $-0.181\,\%$ & $-0.119\,\%$ & $-0.026\,\%$ & 
       $+0.060\,\%$ & $+0.111\,\%$  \\
  \hline \hline
  \multicolumn{6}{|c|}{$\delta^{\rm int}\sigma_{\rm had}$} \\
  \hline \hline
10   & $-0.518\,\%$ & $-0.317\,\%$ & $-0.029\,\%$ & 
       $+0.240\,\%$ & $+0.407\,\%$  \\
100  & $-0.492\,\%$ & $-0.301\,\%$ & $-0.029\,\%$ & 
       $+0.226\,\%$ & $+0.384\,\%$  \\
1000 & $-0.457\,\%$ & $-0.281\,\%$ & $-0.028\,\%$ & 
       $+0.207\,\%$ & $+0.353\,\%$  \\
  \hline 
\end{tabular}
\caption[]{
  {\tt TOPAZ0} relative differences $\delta^{\rm int}\sigma(\mh) =
  \sigma^{\rm DD}/\sigma^{\rm DD}_{\ssZ+\ph} - 1$ as a function of the
  Higgs boson mass in DD-mode. }
\label{tab22}
\end{center}
\end{table}
%--

In \tabn{tab22} we show the relative deviation of excluding/including
the $\zb-\ph$ interference as a function of the Higgs boson mass in
DD-mode.  As observed before the $\zb - \ph$ interference is negative
below the peak and changes sign above it. It is vanishingly small at
the resonance for all values of $\mh$, approximately $-0.03\%$, and
can be sizeable at the wings, $-0.2\%(-0.5\%)$ at the left wing and
$+0.1\%(+0.4\%)$ at the right wing for the muonic (hadronic)
cross-section.  The Higgs-mass dependence of the $\zb - \ph$
interference is rather large, up to 20\% of the interference itself.

%%% \input{ltf_note_3}
%-----------------------------------------
\section{Convoluted Realistic Observables}
%-----------------------------------------
\label{sec:CONV-RO}

%---------------------------------------------
\subsection{Comparison for Extrapolated Setup}
%---------------------------------------------

Having discussed the status of our comparisons before the introduction
of initial-state QED radiation we now proceed to comparing the
convoluted quantities and the effect of convolution.

The default of {\tt TOPAZ0/ZFITTER} is to account for initial-state
QED radiation through a so-called additive formulation of the QED
radiator (flux-function), which is a mixture of leading-logarithms
(LL) and finite-order results.  In \cite{kn:yfsj} a proof is given
that the $\beta\xvar^{\beta-1}$ term should be factorized in front of,
at least, the LL component.  This result is obtained up to third order
LL, and there are good indications from fourth and fifth orders that
it is true to infinite order.\footnote{ S.~Jadach, private
  communication.}  Recently explicit $\alpha^3\LC$ terms became known
\cite{kn:yfsj} and also \cite{kn:yfsalso}.  For higher orders we refer
to \cite{kn:hoj} and \cite{kn:przybyc}.  In {\tt TOPAZ0/ZFITTER} the
$\ord{\alpha^3}$ radiator is implemented according to
\cite{kn:addrad}.  Recently {\tt TOPAZ0} and {\tt ZFITTER} have
implemented the order $\alpha^3$ factorized (YFS) radiator as reported
in \cite{kn:radfact}.

There is a pattern of convolution that we want to compare in our
step-by-step procedure.

\begin{itemize}
  
\item[CA3] Complete RO, with QED initial state radiation implemented
  through an additive $\ord{\alpha^3}$ radiator \cite{kn:addrad}.
  
\item[CF3] Complete RO, with QED initial state radiation implemented
  through a factorized $\ord{\alpha^3}$ radiator \cite{kn:radfact}.

\end{itemize}

The following two equations define cross-sections and forward-backward 
asym\-me\-tries {\em convoluted} with ISR:
%--
\bq
\sigma_{_{\rm T}}\lpar\sman\rpar = \int^1_{\zvari{_0}}\,d\zvar 
\Fluxf{\zvar}{\sman}{\hat\sigma}_{_{\rm T}}\bigl(\zvar\sman\bigr),
\label{fluxtotal}
\eq
%--
where $\zvari{_0}=\smani{0}/\sman$ and 
%--
\bq
\afb\lpar\sman\rpar=\frac{\pi\alpha^2\qes\qfs}{\sigma_{\rm tot}}\,
\int^1_{\zvari{_0}}\,d\zvar
\frac{1}{\lpar 1+\zvar\rpar^2}\,H_{_{\rm FB}}\lpar\zvar;\sman\rpar\,
{\hat\sigma}_{_{\rm FB}}\bigl(\zvar\sman\bigr), 
\label{fluxafb}
\eq
%--
Note that the so-called {\em radiator} (or {\em flux function}),
$\Fluxf{\zvar}{\sman}$, is known up to terms of order $\alpha^3$ while
$H_{_{\rm FB}}$ is only known up to terms of order $\alpha^2$.

The kernel cross-sections ${\hat\sigma}_{_{\rm T,FB}}$ should be
understood as the improved Born approximation (IBA), including
imaginary parts and corrected with all electroweak and possibly all
FSR (QED $\,\otimes\,$QCD) corrections where all coupling constants
and effective vector and axial weak couplings are {\em running}, i.e.,
they depend on $\smanp=\zvar\sman$ under the convolution integrals in
\eqn{fluxtotal} and \eqn{fluxafb}.  In practice this takes a lot of
CPU time and for this reason some time-saving options are foreseen in
the codes.  For instance, one may calculate effective weak couplings
only once at $\sman$ rather than at $\smanp$ thereby saving a
conspicuous amount of CPU time.

We study the accuracy of such approximations with ${\tt ZFITTER}$.  In
\tbn{tab10bisconv} we report ROs calculated with no convolution at all
(all couplings evaluated at $\sman$), convolution of $\alpha$ only
($\alpha(\sman)\to\alpha(\smanp)$), and full convolution of all
electroweak radiative corrections; corresponding to the {\tt ZFITTER}
flag {\tt CONV} with values {\tt CONV=0,1,2}, respectively.  The bulk
of the running-couplings effect is given by the $\alpha$ convolution,
in particular below the wing.  The remaining effect is totally
negligible at the resonance and below, growing to $\sim -0.15$
per-mill above the resonance for cross-sections and remaining
negligible for the asymmetries.  This study proves that one may avoid
using the CPU-time consuming full convolution of all electroweak
radiative corrections and that it is sufficient to keep the $\alpha$
convolution only.  This is the default used for the {\tt ZFITTER}
numbers reported in this article.

In {\tt TOPAZ0} all universal electroweak corrections and final-state
QCD corrections are put in convolution with initial-state QED
radiation and therefore the couplings, $\alpha$, and $\als$ are
evaluated at the scale $\smanp$.  Weak boxes, vertices and expanded
bosonic self-energy corrections are added linearly, evaluated at the
nominal energy. The latter is also true for IFI.

\begin{figure}[p]
\begin{center}
%includegraphics[width=0.75\linewidth,clip=true,bb=90 90 425 740]{fignote3.eps}
\includegraphics[width=0.75\linewidth]{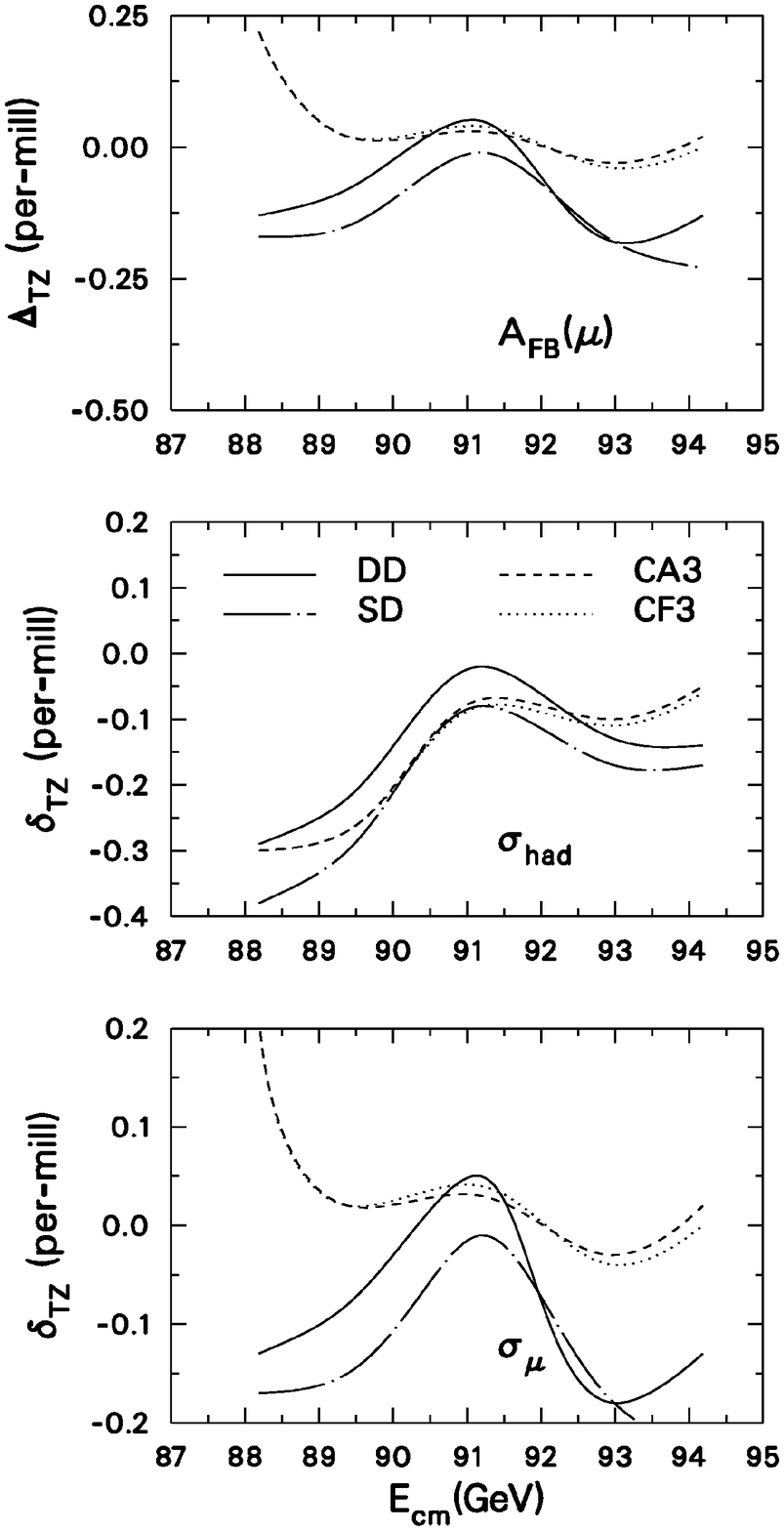}
\caption[]{
  Relative deviations between {\tt TOPAZ0} and {\tt ZFITTER} for
  muonic and hadronic cross-section in DD, SD, CA3 and CF3 modes. The
  last three are corresponding to a cut $\Mlones(\ff\barf)$ of
  $0.01\,\sman$. }
\label{fignote3}
\end{center}
\end{figure}

\begin{figure}[p]
\begin{center}
%includegraphics[width=0.75\linewidth,clip=true,bb=90  90 425 740]{fignote3s.eps}
\includegraphics[width=0.75\linewidth]{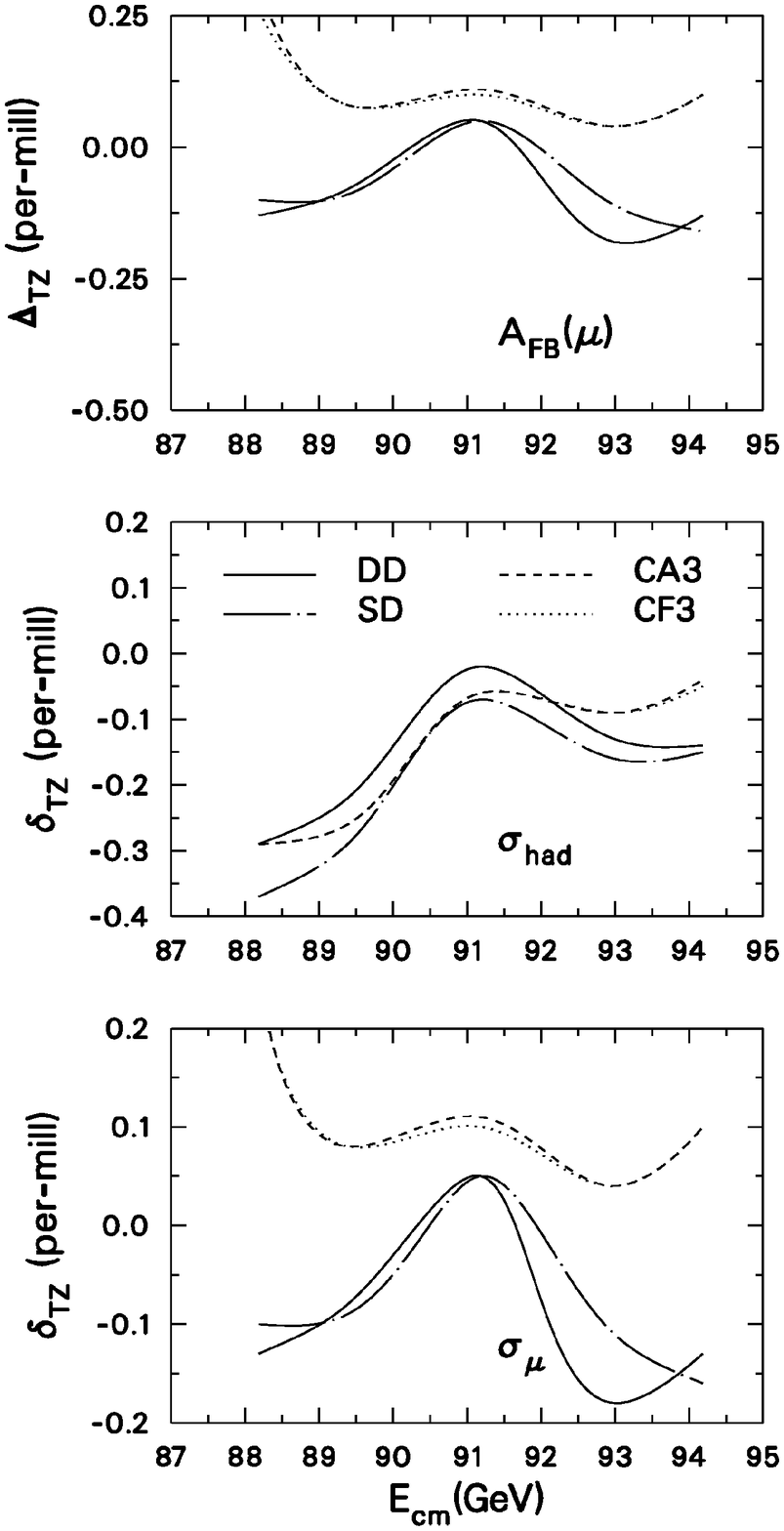}
\caption[]{
  Relative deviations between {\tt TOPAZ0} and {\tt ZFITTER} for
  muonic and hadronic cross-section in DD, SD, CA3 and CF3 modes. The
  last three are corresponding to a cut on $\smanp$ of $0.01\,\sman$.
  }
\label{fignote3s}
\end{center}
\end{figure}

Results are shown in \tabn{tab10} where we include all steps in the
process de-convoluted $\,\to\,$ convoluted, i.e., DD, SD, CA3 and CF3.
The reported results refer to an $\Mlones(\ff\barf)$ cut of
$0.01\,\sman$.  The relative deviations between {\tt TOPAZ0} and {\tt
  ZFITTER} are shown in \fig{fignote3}.  The numbers of \tabn{tab10}
are produced with a $\Mlones(\ff\barf)$-cut of $0.01\,\sman$. It is
instructive to compare with similar results obtained by imposing the
an $\smanp$-cut, e.g. $\smanp \ge 0.01\,\sman$.  The comparison is
shown in \tabn{tab10bis} and \fig{fignote3s}.  From \tabn{tab10bis}
one sees that the differences between the two cuts are of
$0.43\,(1.38)\,$pb for the muonic (hadronic) peak cross-section.
There is really no problem as long as the procedure is fully
specified.

It emerges from this comparisons that the agreement for the muonic
cross-section for energies below the peak is quite reasonable but less
satisfactory than for hadrons.  Given the agreement at the level of
de-convoluted cross-sections and once we have observed that the
de-convolution is satisfactory for hadrons, we come to the conclusion
that for muons the low-$\qmoms$ region, where the Coulomb pole and
mass effects (for very loose $\smanp$-cuts) may become relevant, gives
the dominant difference in $\sigma_{\flm}$.

%--
\begin{table}[t]
\begin{center}
\renewcommand{\arraystretch}{1.1}
\begin{tabular}{|c||c|c|c|c|c|}
\hline
& \multicolumn{5}{c|}{Centre-of-mass energy in GeV} \\
\cline{2-6}
& $\mz - 3$ & $\mz - 1.8$ & $\mz$ & $\mz + 1.8$ & $\mz + 3$  \\
\hline 
\hline
$\sigma_{\flm}\,$[nb] 
& 0.22862 & 0.47672 &  1.48012 &  0.69526 & 0.40655 \\
& 0.22843 & 0.47653 &  1.47995 &  0.69509 & 0.40638 \\
& 0.22842 & 0.47653 &  1.47992 &  0.69505 & 0.40633 \\
&-0.04    & 0       & -0.02    & -0.06    &-0.12    \\      
\hline
$\sigma_{\rm had}\,$[nb]
& 4.45219 & 9.60235 & 30.43892 & 14.18454 & 8.19986 \\
& 4.45146 & 9.60165 & 30.43824 & 14.18391 & 8.19923 \\
& 4.45130 & 9.60139 & 30.43753 & 14.18270 & 8.19791 \\
&-0.04    &-0.03    & -0.02    & -0.09    &-0.16    \\ 
\hline
$\afba{\flm}$
&-0.28308 &-0.16979 & -0.00066 &  0.11177 & 0.15451 \\
&-0.28330 &-0.16985 & -0.00066 &  0.11182 & 0.15461 \\
&-0.28330 &-0.16985 & -0.00066 &  0.11183 & 0.15464 \\
& 0       & 0       &  0       & +0.01    &+0.03    \\
\hline
\end{tabular}
\caption[]{
  {\tt ZFITTER} illustration of the effect of convolution of
  electroweak radiative corrections in CA3 mode with a cut of
  $\smanp>0.01\sman$.  First line is no convolution, second line is
  $\alpha(s)\to\alpha(s')$ convolution, third line is full convolution
  of all electroweak radiative corrections.  Fourth line is difference
  between third and second line. }
\label{tab10bisconv}
\end{center}
\end{table}
%--

\clearpage

%--
\begin{table}[p]
\begin{center}
\renewcommand{\arraystretch}{1.1}
\begin{tabular}{|c||c|c|c|c|c|}
\hline
& \multicolumn{5}{c|}{Centre-of-mass energy in GeV} \\
\cline{2-6}
& $\mz - 3$ & $\mz - 1.8$ & $\mz$ & $\mz + 1.8$ & $\mz + 3$  \\
\hline 
\hline
$\sigma_{\flm}\,$[nb]  DD 
&  0.29999 &  0.65718 &  2.00341 &  0.65856 & 0.31047  \\
&  0.30003 &  0.65724 &  2.00331 &  0.65863 & 0.31051  \\      
$\sigma_{\flm}\,$[nb]  SD     
&  0.30047 &  0.65821 &  2.00656 &  0.65960 & 0.31095  \\
&  0.30052 &  0.65832 &  2.00659 &  0.65971 & 0.31102  \\       
$\sigma_{\flm}\,$[nb]  CA3
&  0.22840 &  0.47642 &  1.47967 &  0.69490 & 0.40628  \\
&  0.22836 &  0.47641 &  1.47962 &  0.69492 & 0.40627  \\         
$\sigma_{\flm}\,$[nb]  CF3
&  0.22841 &  0.47645 &  1.47977 &  0.69495 & 0.40630  \\ 
&  0.22837 &  0.47644 &  1.47971 &  0.69497 & 0.40629  \\        
\hline 
\hline
$\sigma_{\rm had}\,$[nb]  DD 
&  5.78492 & 12.93841 & 39.92967 & 13.02421 & 6.05233  \\
&  5.78670 & 12.94148 & 39.93079 & 13.02591 & 6.05313  \\       
$\sigma_{\rm had}\,$[nb]  SD 
&  6.00265 & 13.42490 & 41.42929 & 13.51293 & 6.27933  \\
&  6.00500 & 13.42901 & 41.43272 & 13.51517 & 6.28030  \\         
$\sigma_{\rm had}\,$[nb]  CA3
&  4.44990 &  9.59865 & 30.43501 & 14.18203 & 8.19853  \\    
&  4.45129 &  9.60132 & 30.43725 & 14.18342 & 8.19894  \\         
$\sigma_{\rm had}\,$[nb]  CF3
&  4.45016 &  9.59921 & 30.43696 & 14.18307 & 8.19901  \\     
&  4.45157 &  9.60191 & 30.43929 & 14.18451 & 8.19945  \\       
\hline 
\hline
$\afba{\flm}$  DD   
& -0.26170 & -0.15037 &  0.01745 &  0.17510 & 0.27002  \\
& -0.26167 & -0.15037 &  0.01741 &  0.17502 & 0.26991  \\
$\afba{\flm}$  SD  
& -0.26128 & -0.15013 &  0.01742 &  0.17481 & 0.26958  \\
& -0.26122 & -0.15011 &  0.01738 &  0.17471 & 0.26944  \\
$\afba{\flm}$ CA3  
& -0.28321 & -0.16981 & -0.00062 &  0.11189 & 0.15470  \\  
& -0.28336 & -0.16988 & -0.00066 &  0.11184 & 0.15464  \\
$\afba{\flm}$ CF3  
& -0.28320 & -0.16980 & -0.00062 &  0.11189 & 0.15469  \\ 
& -0.28340 & -0.16990 & -0.00066 &  0.11185 & 0.15465  \\
\hline 
\end{tabular}
\caption[]{
  {\tt TOPAZ0/ZFITTER} comparison of complete RO (CA3 and CF3) with DD
  and SD ($\Mlones$-cut) modes. }
\label{tab10}
\end{center}
\end{table}
%--

%--
%--

%--
\begin{table}[p]
\begin{center}
\renewcommand{\arraystretch}{1.1}
\begin{tabular}{|c||c|c|c|c|c|}
\hline
& \multicolumn{5}{c|}{Centre-of-mass energy in GeV} \\
\cline{2-6}
& $\mz - 3$ & $\mz - 1.8$ & $\mz$ & $\mz + 1.8$ & $\mz + 3$  \\
\hline 
\hline
$\sigma_{\flm}\,$[nb]  DD 
& 0.29999 &  0.65718 &  2.00341 &  0.65856 & 0.31047  \\
& 0.30003 &  0.65724 &  2.00331 &  0.65863 & 0.31051  \\      
$\sigma_{\flm}\,$[nb]  SD     
& 0.30055 &  0.65839 &  2.00711 &  0.65978 & 0.31104  \\
& 0.30058 &  0.65844 &  2.00700 &  0.65985 & 0.31108  \\          
$\sigma_{\flm}\,$[nb]  CA3
& 0.22849 &  0.47657 &  1.48010 &  0.69512 & 0.40642  \\     
& 0.22843 &  0.47653 &  1.47995 &  0.69509 & 0.40638  \\          
$\sigma_{\flm}\,$[nb]  CF3
& 0.22850 &  0.47660 &  1.48019 &  0.69517 & 0.40644  \\     
& 0.22844 &  0.47656 &  1.48004 &  0.69514 & 0.40640  \\         
\hline 
\hline
$\sigma_{\rm had}\,$[nb]  DD 
& 5.78492 & 12.93841 & 39.92967 & 13.02421 & 6.05233  \\
& 5.78670 & 12.94148 & 39.93079 & 13.02591 & 6.05313  \\       
$\sigma_{\rm had}\,$[nb]  SD 
& 6.00291 & 13.42550 & 41.43114 & 13.51353 & 6.27961  \\
& 6.00518 & 13.42948 & 41.43401 & 13.51559 & 6.28050  \\          
$\sigma_{\rm had}\,$[nb]  CA3
& 4.45012 &  9.59910 & 30.43639 & 14.18269 & 8.19892  \\     
& 4.45146 &  9.60165 & 30.43824 & 14.18391 & 8.19923  \\         
$\sigma_{\rm had}\,$[nb]  CF3
& 4.45038 &  9.59966 & 30.43834 & 14.18373 & 8.19940  \\     
& 4.45174 &  9.60225 & 30.44028 & 14.18499 & 8.19974  \\        
\hline 
\hline
$\afba{\flm}$  DD   
& -0.26170 & -0.15037 &  0.01745 & 0.17510 & 0.27002  \\
& -0.26167 & -0.15037 &  0.01741 & 0.17502 & 0.26991  \\
$\afba{\flm}$  SD  
& -0.26122 & -0.15010 &  0.01742 & 0.17478 & 0.26952  \\
& -0.26119 & -0.15010 &  0.01737 & 0.17469 & 0.26941  \\
$\afba{\flm}$ CA3  
& -0.28312 & -0.16977 & -0.00062 & 0.11186 & 0.15466  \\     
& -0.28330 & -0.16985 & -0.00066 & 0.11182 & 0.15461  \\
$\afba{\flm}$ CF3  
& -0.28311 & -0.16976 & -0.00062 & 0.11186 & 0.15465  \\     
& -0.28333 & -0.16987 & -0.00066 & 0.11183 & 0.15462  \\
\hline 
\end{tabular}
\caption[]{
  {\tt TOPAZ0/ZFITTER} comparison of complete RO (CA3 and CF3) with DD
  and SD ($\smanp$-cut) modes. }
\label{tab10bis}
\end{center}
\end{table}
%--

%--
%--

\clearpage

\begin{figure}[htbp]
\begin{center}
%includegraphics[width=\linewidth,clip=true,bb=15 90 425 580]{fignotesp.eps}
\includegraphics[width=\linewidth]{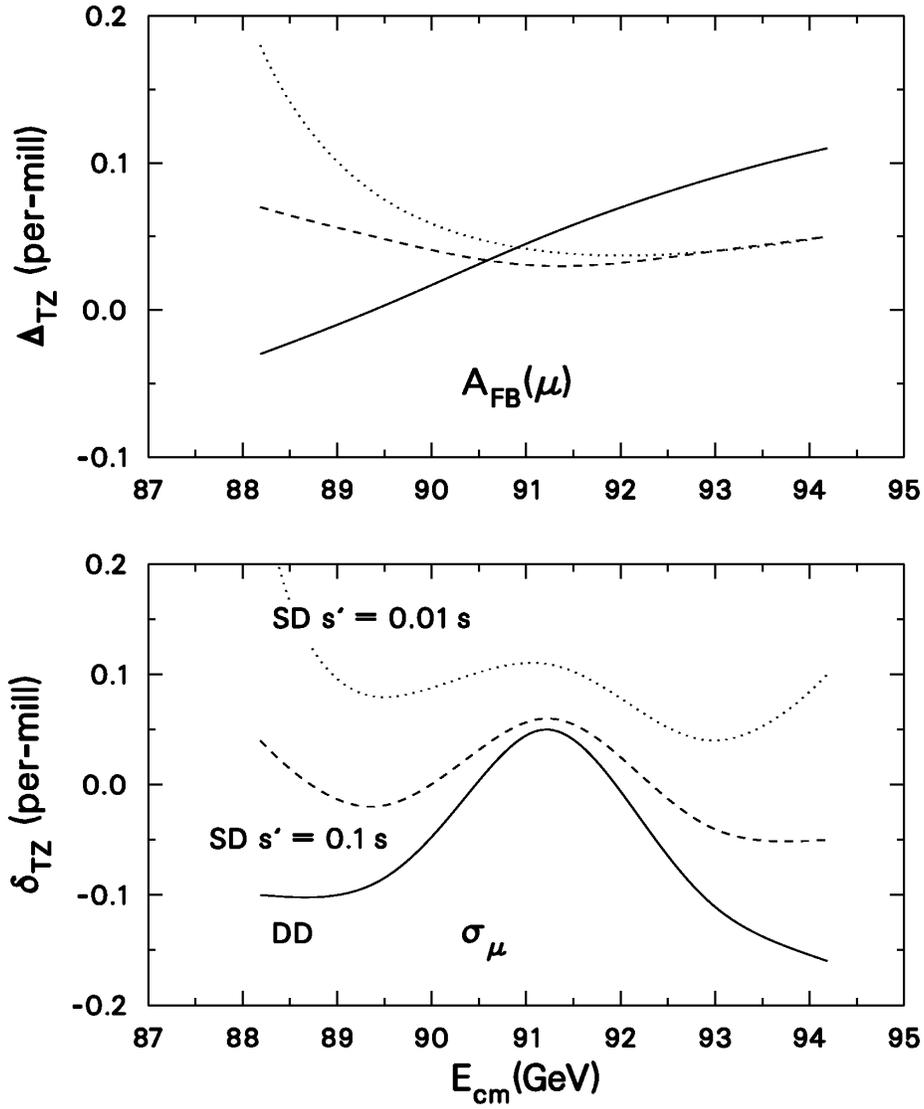}
\caption[]{
  Relative deviations between {\tt TOPAZ0} and {\tt ZFITTER} for
  muonic total cross-section and forward-backward asymmetry in SD-mode
  and CA3-mode with two set of $\smanp$-cuts. }
\label{fignotesp}
\end{center}
\end{figure}

The effect of moving the $\smanp$-cut from $0.01\,\sman$ to
$0.1\,\sman$ is shown in \tabn{tab10tris}.  As a result the agreement
becomes much better, especially on the lower tail of the resonance.
This fact is also shown in \fig{fignotesp}.

%--
\begin{table}[hp]
\begin{center}
\renewcommand{\arraystretch}{1.1}
\begin{tabular}{|c||c|c|c|c|c|}
\hline
& \multicolumn{5}{c|}{Centre-of-mass energy in GeV} \\
\cline{2-6}
& $\mz - 3$ & $\mz - 1.8$ & $\mz$ & $\mz + 1.8$ & $\mz + 3$  \\
\hline 
\hline
$\sigma_{\flm}\,$[nb]  SD     
&  0.30055 &  0.65839 &  2.00711 & 0.65978 & 0.31104  \\
&  0.30058 &  0.65845 &  2.00700 & 0.65985 & 0.31108  \\
\cline{2-6}          
$\sigma_{\flm}\,$[nb]  CA3
&  0.22849 &  0.47657 &  1.48010 & 0.69512 & 0.40642  \\     
&  0.22843 &  0.47653 &  1.47995 & 0.69509 & 0.40638  \\          
&  0.22674 &  0.47487 &  1.47845 & 0.69353 & 0.40487  \\    
&  0.22674 &  0.47489 &  1.47836 & 0.69356 & 0.40489  \\            
\hline 
\hline
$\afba{\flm}$  SD  
& -0.26122 & -0.15010 &  0.01742 & 0.17478 & 0.26952  \\
& -0.26119 & -0.15010 &  0.01737 & 0.17469 & 0.26941  \\
\cline{2-6}
$\afba{\flm}$ CA3  
& -0.28312 & -0.16977 & -0.00062 & 0.11186 & 0.15466  \\     
& -0.28330 & -0.16985 & -0.00066 & 0.11182 & 0.15461  \\
& -0.28526 & -0.17035 & -0.00061 & 0.11214 & 0.15528  \\ 
& -0.28532 & -0.17040 & -0.00064 & 0.11210 & 0.15523  \\
\hline 
\end{tabular}
\caption[]{
  {\tt TOPAZ0/ZFITTER} comparison of complete RO (CA3-mode) with SD
  ($\smanp$-cut) mode. For CA3-mode first (second) row correspond to
  {\tt T(Z)} $\smanp > 0.01\,\sman$.  Third (fourth) row give instead
  $\smanp > 0.1\,\sman$. }
\label{tab10tris}
\end{center}
\end{table}

In \tabn{tab10four} we show the comparison for heavy quark
forward-backward asymmetries, including initial-state QED radiation.
As we have observed before the $\ffb$-channel shows larger deviations.
Even though the agreement for the convoluted $\afba{\ffb}$ is
satisfactory, especially at the peak, one should not forget that a
similar comparison for DD-de-convoluted $\afba{\ffb}$ is considerably
worse so that the result of \tabn{tab10four} is also a consequence of
accidental compensations.

%--
\begin{table}[hp]
\begin{center}
\renewcommand{\arraystretch}{1.1}
\begin{tabular}{|c||c|c|c|c|c|}
\hline
& \multicolumn{5}{c|}{Centre-of-mass energy in GeV} \\
\cline{2-6}
& $\mz - 3$ & $\mz - 1.8$ & $\mz$ & $\mz + 1.8$ & $\mz + 3$  \\
\hline 
\hline
$\afba{\fc}$                                   
& -0.10600 & -0.03625 &  0.06068 &  0.12386 &  0.14840  \\     
& -0.10598 & -0.03625 &  0.06065 &  0.12377 &  0.14827  \\
& -0.02    &  0.00    & +0.03    & +0.09    & +0.13     \\
\hline
$\afba{\ffb}$
&  0.028131 &  0.05705 &  0.09611 &  0.12135 &  0.13105  \\    
&  0.028078 &  0.05701 &  0.09612 &  0.12161 &  0.13169  \\
& +0.05     & +0.04    & -0.01    & -0.26    & -0.64     \\
\hline 
\end{tabular}
\caption[]{
  {\tt TOPAZ0/ZFITTER} comparison of heavy quark forward-backward
  asymmetries in CA3-mode. Third row is the absolute deviation in
  per-mill. }
\label{tab10four}
\end{center}
\end{table}
%--

%--
%--

\clearpage

%------------------------------------------------------
\subsection{Comparison with Realistic Kinematical Cuts}
%------------------------------------------------------

We have also devoted an effort in order to present the most up-to-date
analysis for ROs with realistic kinematical cuts.  Therefore, we go
beyond the fully extrapolated set-up for muonic channel (with the
inclusion of a $\smanp/\Mlones$-cut) by considering

\begin{itemize}

\item $e^+e^- \to \flmp\flmm$  for $\theta_{\rm acc} < \theta_- < \pi -
\theta_{\rm acc}$
($\theta_{\rm acc} = 0^\circ, 20^\circ$ and $40^\circ$), \\
$\theta_{\rm acoll} < 10^\circ,~25^\circ$, and $E_{th}(\mu^{\pm}) = 1\,$GeV,

\end{itemize}
where $\theta_-$ is the final-state fermion scattering angle and
$\theta_{\rm acoll}$ the acollinearity between the final-state
fermions.  The results are shown in \tabns{tab10acol}{tab25acol}.

%--
\begin{table}[ht]
\begin{center}
\renewcommand{\arraystretch}{1.1}
\begin{tabular}{|c||c||c|c|c|c|c|}
\hline
&$\theta_{\rm acc}$&\multicolumn{5}{c|}{Centre-of-mass energy in GeV} \\
\cline{3-7}
& & $\mz - 3$ & $\mz - 1.8$ & $\mz$ & $\mz + 1.8$ & $\mz + 3$  \\
\hline\hline
$\sigma_{\flm}\,$[nb]
&$0^\circ$ & 0.21932 & 0.46287 & 1.44795 & 0.67725 & 0.39366 \\         
&      & 0.21928 & 0.46285 & 1.44781 & 0.67722 & 0.39361 \\
&      & +0.18   & +0.04   & +0.10   & +0.04   & +0.13   \\
\cline{2-7}
&$20^\circ$& 0.19990 & 0.42207 & 1.32066 & 0.61759 & 0.35886 \\
&      & 0.19987 & 0.42205 & 1.32053 & 0.61756 & 0.35881 \\
&      & +0.15   & +0.05   & +0.10   & +0.05   & +0.14   \\
\cline{2-7}
&$40^\circ$& 0.15034 & 0.31762 & 0.99428 & 0.46479 & 0.26989 \\  
&      & 0.15032 & 0.31760 & 0.99415 & 0.46474 & 0.26983 \\
&      & +0.13   & +0.06   & +0.13   & +0.11   & +0.22   \\
\hline\hline
$\afba{\flm}$                     
&$0^\circ$ &-0.28450 &-0.16914 & 0.00033 & 0.11512 & 0.16107 \\
&      &-0.28453 &-0.16911 & 0.00025 & 0.11486 & 0.16071 \\
&      & +0.03   & -0.03   & +0.08   &  +0.26  & +0.36   \\
\cline{2-7}
&$20^\circ$&-0.27509 &-0.16352 & 0.00042 & 0.11171 & 0.15645 \\
&      &-0.27506 &-0.16347 & 0.00035 & 0.11148 & 0.15616 \\
&      &-0.03    &-0.05    &+0.07    &+0.23    &+0.29    \\
\cline{2-7}
&$40^\circ$&-0.24219 &-0.14396 & 0.00054 & 0.09906 & 0.13903 \\ 
&      &-0.24207 &-0.14386 & 0.00050 & 0.09893 & 0.13891 \\
&      &-0.12    &-0.10    &+0.04    &+0.13    &+0.12    \\
\hline 
\end{tabular}
\caption[]{
  {\tt TOPAZ0/ZFITTER} comparison for muonic total cross-section and
  forward-backward asymmetry of complete RO (CA3-mode) with the
  angular acceptance ($\theta_{\rm acc}=0^\circ,10^\circ,20^\circ$)
  and acollinearity ($\theta_{\rm acol}<10^\circ$) cuts.  First row
  {\tt TOPAZ0}, second row {\tt ZFITTER}, third row relative
  (absolute) deviations in per-mill. }
\label{tab10acol}
\end{center}
\end{table}
%--

We register an agreement comparable with the one obtained with
$\smanp/\Mlones$-cut, perhaps deteriorating a little for $\afba{\flm}$
at the wings.  In conclusion the agreement between {\tt TOPAZ0} and
{\tt ZFITTER} remains rather remarkable even when the geometrical
acceptance is constrained and also final-state energies and the
acollinearity angle are bounded.

%--
\begin{table}[ht]
\begin{center}
\renewcommand{\arraystretch}{1.1}
\begin{tabular}{|c||c||c|c|c|c|c|}
\hline
&$\theta_{\rm acc}$&\multicolumn{5}{c|}{Centre-of-mass energy in GeV} \\
\cline{3-7}
& & $\mz - 3$ & $\mz - 1.8$ & $\mz$ & $\mz + 1.8$ & $\mz + 3$  \\
\hline\hline
$\sigma_{\flm}\,$[nb]
&$0^\circ$ & 0.22333 & 0.46971 & 1.46611 & 0.68690 & 0.40034 \\    
&      & 0.22328 & 0.46968 & 1.46598 & 0.68688 & 0.40031 \\
&      &+0.22    &+0.06    &+0.09    &+0.03    &+0.075   \\
\cline{2-7}
&$20^\circ$& 0.20359 & 0.42835 & 1.33731 & 0.62648 & 0.36507 \\
&      & 0.20357 & 0.42833 & 1.33718 & 0.62647 & 0.36505 \\
&      &+0.10    &+0.05    &+0.10    &+0.02    &+0.055   \\ 
\cline{2-7}
&$40^\circ$& 0.15320 & 0.32245 & 1.00698 & 0.47167 & 0.27479 \\
&      & 0.15318 & 0.32243 & 1.00682 & 0.47164 & 0.27477 \\
&      &+0.13    &+0.06    &+0.16    &+0.06    &+0.07    \\  
\hline\hline
$\afba{\flm}$
&$0^\circ$ &-0.28617 &-0.17037 &-0.00032 & 0.11324 & 0.15730 \\     
&      &-0.28647 &-0.17049 &-0.00043 & 0.11293 & 0.15682 \\
&      &+0.30    &+0.12    &+0.11    &+0.31    &+0.48    \\
\cline{2-7}
&$20^\circ$&-0.27695 &-0.16485 &-0.00026 & 0.10974 & 0.15250 \\
&      &-0.27722 &-0.16497 &-0.00037 & 0.10944 & 0.15204 \\
&      &+0.27    &+0.12    &+0.11    &+0.30    &+0.46    \\
\cline{2-7}
&$40^\circ$&-0.24423 &-0.14536 &-0.00016 & 0.09703 & 0.13492 \\ 
&      &-0.24445 &-0.14545 &-0.00026 & 0.09678 & 0.13454 \\
&      &+0.22    &+0.09    &+0.10    &+0.25    &+0.38    \\
\hline 
\end{tabular}
\caption[]{
  The same as in \tbn{tab10acol} but for the acollinearity cut
  $\theta_{\rm acol}<25^\circ$. }
\label{tab25acol}
\end{center}
\end{table}

We note that the coding in {\tt ZFITTER}, for the part involving
realistic cuts, is based on some old work \cite{kn:mbas}.  A recent
study, presented in \cite{kn:cjr}, shows that the approximations made
in the former reference ensure sufficient technical precision of the
treatment of ISR, $\ord{10^{-4}}$, at SLD/LEP-1 energies. (See
Section~\ref{sec:IFI} for the situation concerning initial-final QED
interference).  Coding in {\tt TOPAZ0} is always based on the work of
\cite{kn:fsr}.

%------------------------------------------
\subsection{Uncertainty on QED Convolution}
%------------------------------------------

We now return to a detailed analysis of initial-state QED radiation by
defining convolution factors for each realistic observable $O$, giving
the net effect of initial-state QED radiation at the various energies.
%--
\bq
\delta^{\rm dec}\lpar O\rpar = \frac{O}{O^{\rm SD}} - 1, \qquad
\Delta^{\rm dec}\lpar O\rpar = O - O^{\rm SD}. 
\eq
%--
For convenience of the reader we reproduce in \tabn{tab8} the results
for the CA3 and CF3 mode.  From \tabn{tab8} we derive the absolute
differences, for $\afba{\flm}$, and the relative ones, for
cross-sections, between additive and factorized versions of the QED
radiators. They are shown in \tabn{tab9}.

\clearpage

%--
\begin{table}[p]
\begin{center}
\renewcommand{\arraystretch}{1.1}
\begin{tabular}{|c||c|c|c|c|c|}
\hline
& \multicolumn{5}{c|}{Centre-of-mass energy in GeV} \\
\cline{2-6}
& $\mz - 3$ & $\mz - 1.8$ & $\mz$ & $\mz + 1.8$ & $\mz + 3$  \\
\hline 
\hline
$\delta^{\rm dec}(\sigma^{\rm CA3}_{\flm})$ {\tt T}  &
    -23.976  &   -27.616  &   -26.257  &     5.356  &    30.665  \\
$\delta^{\rm dec}(\sigma^{\rm CA3}_{\flm})$ {\tt Z}  &
    -24.007  &   -27.629  &   -26.261  &     5.341  &    30.631  \\
$\delta^{\rm dec}(\sigma^{\rm CF3}_{\flm})$ {\tt T}  &
    -23.973  &   -27.611  &   -26.253  &     5.364  &    30.671  \\
$\delta^{\rm dec}(\sigma^{\rm CF3}_{\flm})$ {\tt Z}  &
    -24.000  &   -27.624  &   -26.256  &     5.348  &    30.637  \\
\hline \hline
$\delta^{\rm dec}(\sigma^{\rm CA3}_{\rm had})$ {\tt T}  &
    -25.867  &   -28.501  &   -26.537  &     4.952  &    30.564  \\
$\delta^{\rm dec}(\sigma^{\rm CA3}_{\rm had})$ {\tt Z} &
    -25.873  &   -28.503  &   -26.538  &     4.945  &    30.550  \\
$\delta^{\rm dec}(\sigma^{\rm CF3}_{\rm had})$ {\tt T}  &
    -25.863  &   -28.497  &   -26.533  &     4.959  &    30.572  \\
$\delta^{\rm dec}(\sigma^{\rm CF3}_{\rm had})$ {\tt Z} &
    -25.869  &   -28.499  &   -26.533  &     4.953  &    30.558  \\
\hline \hline
$\Delta^{\rm dec}(\afba{\mu\rm CA3})$ {\tt T}  &
     -2.190  &    -1.967  &    -1.804  &    -6.292  &   -11.486  \\
$\Delta^{\rm dec}(\afba{\mu\rm CA3})$ {\tt Z} &
     -2.211  &    -1.975  &    -1.803  &    -6.287  &   -11.480  \\
$\Delta^{\rm dec}(\afba{\mu\rm CF3})$ {\tt T}  &
     -2.189  &    -1.966  &    -1.804  &    -6.292  &   -11.487  \\
$\Delta^{\rm dec}(\afba{\mu\rm CF3})$ {\tt Z} &
     -2.215  &    -1.977  &    -1.803  &    -6.286  &   -11.479  \\
\hline
\end{tabular}
\caption[]{
  RO: the effect in $\%$ of initial state QED radiation for CA3 and
  CF3 modes. }
\label{tab8}
\end{center}
\end{table}
%--

%--
\begin{table}[p]
\begin{center}
\renewcommand{\arraystretch}{1.1}
\begin{tabular}{|c||c|c|c|c|c|}
\hline
& \multicolumn{5}{c|}{Centre-of-mass energy in GeV} \\
\cline{2-6}
& $\mz - 3$ & $\mz - 1$ & $\mz$ & $\mz + 1$ & $\mz + 3$  \\
\hline \hline
\multicolumn{6}{|c|}{$10^{4} \times\,$(fact/add-1)} \\
\hline \hline
$\sigma_{\flm}$
&  0.44 & 0.63 & 0.61 & 0.72 & 0.49 \\
&  0.88 & 0.63 & 0.68 & 0.72 & 0.49 \\
\hline
$\sigma_{\rm had}$  
&  0.58 & 0.58 & 0.64 & 0.73 & 0.59 \\
&  0.61 & 0.62 & 0.67 & 0.76 & 0.62 \\
\hline \hline
\multicolumn{6}{|c|}{fact-add [pb]} \\
\hline \hline
$\sigma_{\flm}$
&  0.01 &   0.03  &   0.09  &   0.05  &   0.02 \\
&  0.02 &   0.03  &   0.10  &   0.05  &   0.02 \\
\hline
$\sigma_{\rm had}$  
&  0.26 &   0.56  &   1.95  &   1.04  &   0.48 \\
&  0.27 &   0.60  &   2.04  &   1.08  &   0.51 \\
\hline \hline
\multicolumn{6}{|c|}{$10^{5} \times\,$(fact-add)} \\
\hline \hline
$\afba{\flm}$  
&  1.00 &  1.00 & 0.00 & 0.00 & -1.00 \\
& -4.00 & -2.00 & 0.00 & 1.00 &  1.00 \\
\hline 
\end{tabular}
\caption[]{
  Absolute and relative differences in {\tt TOPAZ0} and in {\tt
    ZFITTER} for additive and factorized radiators. }
\label{tab9}
\end{center}
\end{table}
%--

\clearpage

To give an example of the developments in the treatment of QED
initial-state radiation we recall that in {\tt TOPAZ0} the following
steps have occurred:

\begin{itemize}
  
\item the leading $\ord{\alpha^2}$ result was considered in version
  1.0,
  
\item the complete $\ord{\alpha^2}$ result, i.e., leading plus NLO
  $\ord{\alpha^2\LL}$ and NNLO $\ord{\alpha^2\LL^0}$ was added in
  version 2.0 \cite{kn:t20},
  
\item the complete $\ord{\alpha^2}$ plus leading $\ord{\alpha^3\LC}$
  result as been included after version 4.0 \cite{kn:t40},

\end{itemize}
%--
where $\LL = \ln(\sman/\mes)$.  Inserting the $\ord{\alpha^3\LC}$
terms into the additive radiator leads to a negative shift of $-0.59$
per-mill in the peak hadronic cross-section. The complete shift
leading-$\alpha^2$ to leading-$\alpha^3$ is dominated by the leading
$\ord{\alpha^3}$ terms with very little influence by the NLO
$\ord{\alpha^2\LL}$ terms.

Sometimes the forward and backward cross-sections $\sigma_{_{\rm
    F,B}}$, are actually used to calculate $\afba{}$.  In \tabn{tab9c}
we present results for the total/F/B cross-sections obtained with the
additive radiator (CA3), showing the effect due to initial-state
radiation for the forward and backward cross-section separately.
Comparing results obtained with the additive and factorized radiators,
we estimate the corresponding initial-state QED uncertainty as
reported in \tbn{tab9d}.

%--
\begin{table}[ht]
\begin{center}
\renewcommand{\arraystretch}{1.1}
\begin{tabular}{|c||c|c|c|c|c|}
  \hline
$\sigma\,$[nb]  & \multicolumn{5}{c|}{Centre-of-mass energy in GeV} \\
  \cline{2-6}
  & $\mz - 3$ & $\mz - 1.8$ & $\mz$ & $\mz + 1.8$ & $\mz + 3$  \\
  \hline \hline
$\sigma_{\flm}\,$  
& 0.22849 & 0.47657 & 1.48010 & 0.69512 & 0.40642 \\
$\sigma^{\rm F}_{\flm}\,$
& 0.08190 & 0.19783 & 0.73959 & 0.38644 & 0.23464  \\
& -26.229 & -26.661 & -26.707 & -22.655 & -25.570  \\
$\sigma^{\rm B}_{\flm}\,$
& 0.14659 & 0.27874 & 0.74050 & 0.30868 & 0.17178  \\
& -22.655 & -25.570 & -24.260 & +13.388 & +51.210 \\
\hline
$\sigma_{\ffb}\,$ 
& 0.95245 & 2.06397 & 6.56297 & 3.05537 & 1.76374  \\
$\sigma^{\rm F}_{\ffb}\,$
& 0.48962 & 1.09086 & 3.59687 & 1.71307 & 0.99744  \\
& -29.290 & -28.973 & -29.175 & -26.377 & -28.229  \\
$\sigma^{\rm B}_{\ffb}\,$
& 0.46283 & 0.97311 & 2.96610 & 1.34229 & 0.76630  \\
& -26.377 & -28.229 & -27.413 & +6.642  & +34.719  \\
\hline
$\sigma_{\fc}\,$ 
& 0.77811 & 1.66456 & 5.25261 & 2.45308 & 1.42235 \\
$\sigma^{\rm F}_{\fc}\,$
& 0.34781 & 0.80211 & 2.78568 & 1.37846 & 0.81671  \\
& -27.564 & -26.845 & -27.176 & -24.904 & -26.237  \\
$\sigma^{\rm B}_{\fc}\,$
& 0.43029 & 0.86245 & 2.46693 & 1.07461 & 0.60563  \\
& -24.904 & -26.237 & -25.666 & +9.428  & +41.341  \\
\hline
\end{tabular}
\caption[]{
  Total, forward and backward $\sigma_{\ff}$ from {\tt TOPAZ0} (CA3
  mode): first entry is the cross-section, second entry is the effect
  in $\%$ of the initial state QED convolution. }
\label{tab9c}
\end{center}
\end{table}
%--

\clearpage

%--
\begin{table}[ht]
\begin{center}
\renewcommand{\arraystretch}{1.1}
\begin{tabular}{|c||c|c|c|c|c|}
\hline
$\sigma\,$[pb]  & \multicolumn{5}{c|}{Centre-of-mass energy in GeV} \\
  \cline{2-6}
  & $\mz - 3$ & $\mz - 1.8$ & $\mz$ & $\mz + 1.8$ & $\mz + 3$  \\
  \hline \hline
$\sigma_{\flm}\,$  
& 0.01 &   0.03 &     0.09  &    0.05 &     0.02 \\
& 0.02 &   0.03 &     0.10  &    0.05 &     0.02 \\
$\sigma^{\rm F}_{\flm}\,$
  & 0.01 & 0.01 & 0.05 & 0.02 & 0.01\\
  & 0.00 & 0.01 & 0.05 & 0.03 & 0.02\\
$\sigma^{\rm B}_{\flm}\,$
  & 0.01 & 0.01 & 0.05 & 0.02 & 0.01\\
  & 0.01 & 0.02 & 0.05 & 0.02 & 0.01\\
\hline
$\sigma_{\ffb}\,$
  & 0.06 & 0.12 & 0.42 & 0.22 & 0.10\\
  & 0.06 & 0.13 & 0.44 & 0.24 & 0.11\\
$\sigma^{\rm F}_{\ffb}\,$
  & 0.03 & 0.06 & 0.21 & 0.11 & 0.05\\
  & 0.03 & 0.07 & 0.28 & 0.15 & 0.07\\
$\sigma^{\rm B}_{\ffb}\,$
  & 0.03 & 0.06 & 0.21 & 0.11 & 0.05\\
  & 0.03 & 0.05 & 0.17 & 0.09 & 0.04\\
\hline
$\sigma_{\fc}\,$
  & 0.04 & 0.10 & 0.33 & 0.18 & 0.08\\
  & 0.05 & 0.10 & 0.35 & 0.19 & 0.09\\
$\sigma^{\rm F}_{\fc}\,$
  & 0.02 & 0.05 & 0.17 & 0.09 & 0.04\\
  & 0.01 & 0.05 & 0.21 & 0.13 & 0.06\\
$\sigma^{\rm B}_{\fc}\,$
  & 0.02 & 0.05 & 0.17 & 0.09 & 0.04\\
  & 0.03 & 0.06 & 0.15 & 0.06 & 0.02\\
\hline
\end{tabular}
\caption[]{
  QED IS uncertainties, CF3-CA3, in pb, for total, forward and
  backward $\sigma_{\ff}$ from {\tt TOPAZ0}, {\tt ZFITTER}. }
\label{tab9d}
\end{center}
\end{table}
%--

%-----------------------------------------------------------
\subsection{Higgs-Mass Dependence of Convoluted Observables}
%-----------------------------------------------------------

\begin{figure}[p]
\begin{center}
%includegraphics[width=0.85\linewidth,clip=true,bb=95 90 480 740]{fignote5.eps}
\includegraphics[width=0.85\linewidth]{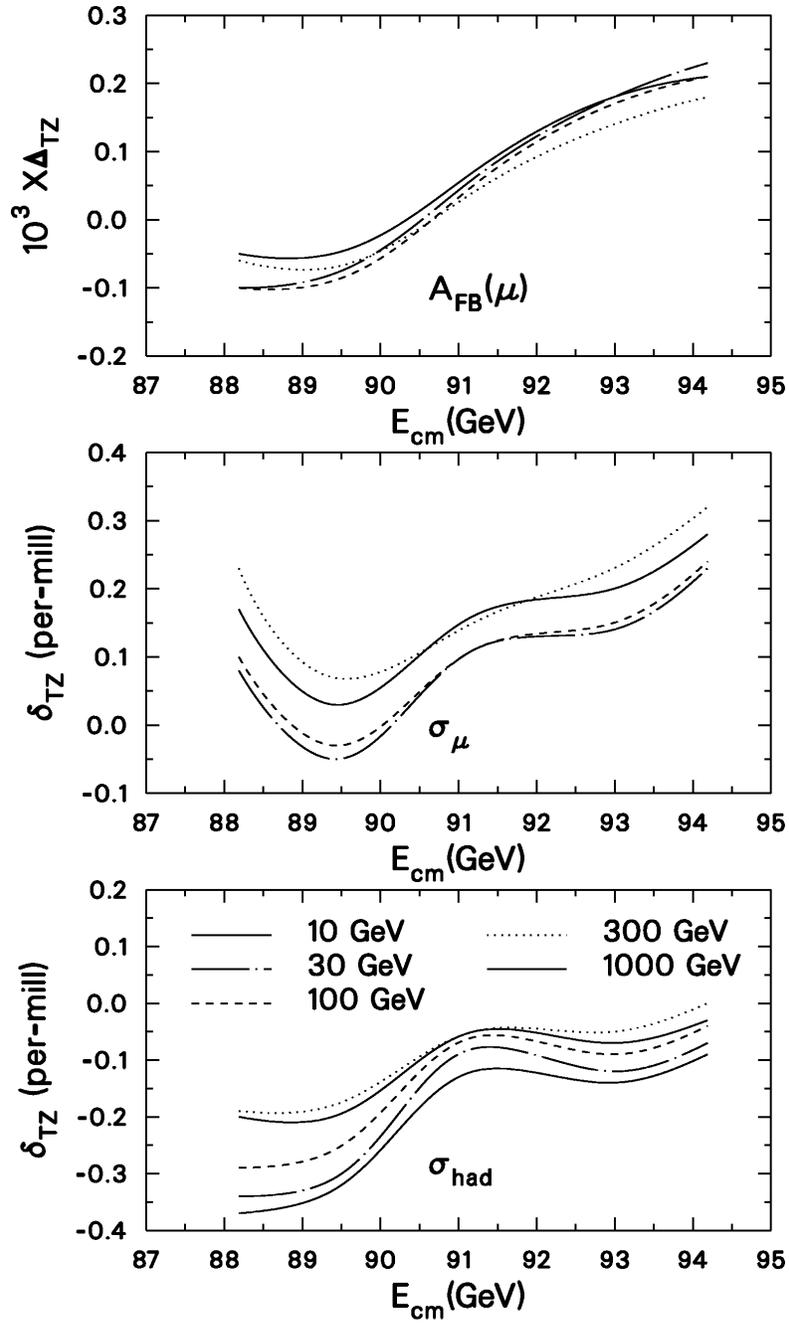}
\caption[]{
  Relative deviations between {\tt TOPAZ0} and {\tt ZFITTER} for
  hadronic cross-section in CA3 mode as a function of the Higgs boson
  mass. The curves correspond to a cut $\smanp > 0.01\,\sman$. }
\label{fignote5}
\end{center}
\end{figure}

Comparisons for Higgs masses in the range from $10\,\GeV$ to $1\,\TeV$
are needed.  For the error determination on $\mh$ in SM fits, for
example, the $\mh$ variation extends to that region, thus the
calculation must be reliable there.  Also, in principle, all the other
SM parameters should be varied to make sure the (dis)agreement is not
too dependent on actual central values.  In \fig{fignote5} we repeat a
comparison for different values of the Higgs boson mass, this time
including initial-state QED radiation in CA3 mode with an $\smanp$-cut
of $0.01\,\sman$.

For the hadronic cross-section the relative difference {\tt TOPAZ0} --
{\tt ZFITTER} varies of 0.07 per-mill at the peak and over the whole
interval $10\,\GeV - 1\,\TeV$ being -0.06 per-mill at $\mh =
100\,$GeV.  At the boundaries of the interval in $\mh$ the difference
register a variation of 0.18 per-mill at $\mh = 10\,$GeV (being -0.29
per-mill at $\mh = 100\,$GeV) and of 0.09 per mill at $\mh = 1\,$TeV
(being -0.04 per mill at $\mh = 100\,$GeV).

The results show that there is a tiny $\mh$-dependence in the {\tt
  TOPAZ0-ZFITTER} comparison at a fixed energy and that the
differences are comparatively larger below the peak and for very low
or very large values of $\mh$.

%--
%--

\clearpage

%---------------------------------------
\section{Initial-Final QED Interference}
%---------------------------------------
\label{sec:IFI}

If initial-final QED (ISR-FSR) interference (IFI) is included in a
calculation there is a conceptual problem with the meaning of the
$\smanp$-cut.  In this case the definition of the variable $\smanp$ is
unnatural since one does no longer know the origin of the radiative
photon (ISR or FSR).  Only a cut on the invariant mass of the
final-state $\ff\barf$ system, $\Mlone(\ff\barf)$, makes sense.

There is another option: to select events with little initial-state
radiation one can use a cut on the acollinearity angle $\theta_{\rm
  acol}$, between the outgoing fermion and anti-fermion.  A cut on
$\theta_{\rm acol}$ is roughly equivalent to a cut on the invariant
mass of the $\ff\barf$ system, indeed one may write
%--
\bq
\frac{\Mlones}{\sman}\approx\zvari{\rm eff} =  
{{1 - \sin(\theta_{\rm acol}/2)}\over{1 + \sin(\theta_{\rm acol}/2)}}\;,
\label{xeff}
\eq
%--
therefore a cut of $\theta_{\rm acol} < 10^{\circ}$ is roughly
corresponding to the request that $\Mlones/\sman > 0.84$.

The inclusion of initial-final QED interference in {\tt TOPAZ0} and in
{\tt ZFITTER} is done at $\ord{\alpha}$, i.e., the $\ord{\alpha}$
initial-final interference term is added linearly to the
cross-sections without entering the convolution with ISR and without
cross-talk to FSR.  For loose cuts the induced uncertainty is rather
small and the effect of the interference itself is minute, as shown
below.  On the contrary, when we select a tight acollinearity cut the
resulting limit on the energy of the emitted photon becomes more
stringent and the effect of interference grows.  The corresponding
theoretical uncertainty, due to missing higher-order corrections, is
therefore expected to be larger.

A pragmatic point of view to improve upon the pure $\ord{\alpha}$
inclusion of the interference, which however is not implemented in the
codes, would be the following: let us consider corrections not
belonging to the two classes of ISR and FSR. These non-factorizable
corrections correspond to both interference terms and QED box
diagrams.  We have, therefore, three classes of contributions for the
total cross-section: ISR, FSR and interference.

If we denote by $\sigma_{_{\rm K}}$ the sum of factorizable (Born)
plus non-factorizable cross-sections the totally radiatively corrected
process can be effectively described in terms of structure functions.
In this case $\smanp$ is still the square of the invariant mass
available after initial-state QED radiation and $\smanp \ge
\Mlones(\ff\barf)\ge\smani{0}$.  Therefore we can write
%--
\bq
\sigma(\sman) = 
\int_{\zvari{_0}}^{1} d\zvar H_{\rm in}(\zvar,\sman)\,
H_{\rm fin}(\zvar-\zvari{_0},\shat)\,
{\hat\sigma}_{_{\rm K}}\bigl(\shat;\Mlones(\ff\barf)\ge\smani{0}\bigr),
\label{monster}
\eq
%--
where $H_{\rm in,fin}$ is the initial- and final-state QED radiator
and $\zvari{_0}= \Mlones(\ff\barf)/\sman$.  Moreover
${\hat\sigma}_{_{\rm K}}$ is the kernel cross-section evaluated with
the constraint $\Mlones(\ff\barf)\ge\smani{0}$ and $\shat =
\zvar\,\sman$.  The accuracy obtainable with \eqn{monster} depends on
the various contributions included within the three factors, the
initial-state radiator $H_{\rm in}$, the final-state radiator $H_{\rm
  fin}$ and the kernel cross-section including factorizable and
non-factorizable parts.

%---------------------------------------------
\subsection{Comparison for Extrapolated Setup}
%---------------------------------------------

The effect of including or excluding initial-final state QED
interference is shown in \tabn{tab16}.  As seen from the Table, the
level of agreement between {\tt TOPAZ0} and {\tt ZFITTER} does not
deteriorate after the inclusion of initial-final state QED
interference for loose $\Mlones$-cuts.  For this setup the effect of
$\ord{\alpha}$ IFI is under control.\footnote{IFI interference
  contributions are generally small, as we know from pure
  $\ord{\alpha}$.  However there is a recent analysis by S.~Jadach and
  collaborators (private communication) which seems to indicate that
  exponentiation is quite important for the magnitude of interference
  contribution in cross-sections and asymmetries.  We have been told
  that the difference between pure $\ord{\alpha}$ and exponentiation
  is quite often a factor 2.}

%--
\begin{table}[b]
\begin{center}
\renewcommand{\arraystretch}{1.1}
\begin{tabular}{|c||c|c|c|c|c|}
\hline
& \multicolumn{5}{c|}{Centre-of-mass energy in GeV} \\
\cline{2-6}
& $\mz - 3$ & $\mz - 1.8$ & $\mz$ & $\mz + 1.8$ & $\mz + 3$  \\
\hline 
\hline
\multicolumn{6}{|c|}{No ISR/FSR interference, $\smanp \ge 0.01\,\sman$} \\
\hline 
\hline 
$\sigma_{\flm}\,$[nb]  
& 0.22849 & 0.47657 & 1.48010 & 0.69512 & 0.40642  \\    
& 0.22843 & 0.47653 & 1.47995 & 0.69509 & 0.40638  \\         
\cline{2-6}
& +0.26   &+0.08    &+0.10    &+0.04    &+0.10     \\ 
\hline 
$\sigma_{\rm had}\,$[nb]
& 4.45012 &  9.59910 & 30.43639 & 14.18269 & 8.19892  \\     
& 4.45146 &  9.60165 & 30.43824 & 14.18391 & 8.19923  \\         
\cline{2-6}
& -0.30   &  -0.27   & -0.06    & -0.09    & -0.04    \\ 
\hline
$\afba{\flm}$ 
&-0.28312 &-0.16977 &-0.00062 & 0.11186 & 0.15466  \\ 
&-0.28330 &-0.16985 &-0.00066 & 0.11182 & 0.15461  \\
\cline{2-6}
&+0.18    &+0.08    &+0.04    &+0.04    &+0.05     \\
\hline
\hline
\multicolumn{6}{|c|}{No ISR/FSR interference, $\smanp\ge\Mlones(\flmp\flmm) 
\ge 0.01\,\sman$} \\
\hline 
\hline
$\sigma_{\flm}\,$[nb]  
& 0.22840 & 0.47642 & 1.47967 & 0.69490 & 0.40628  \\    
& 0.22836 & 0.47641 & 1.47962 & 0.69492 & 0.40627  \\      
\cline{2-6}
&+0.18    &+0.02    &+0.03    &-0.03    &+0.02     \\
\hline 
$\sigma_{\rm had}\,$[nb]
&  4.44990 &  9.59865 & 30.43501 & 14.18203 & 8.19853  \\    
&  4.45129 &  9.60132 & 30.43725 & 14.18342 & 8.19894  \\         
\cline{2-6}
& -0.31    & -0.29   &  -0.07   &  -0.10    & -0.05   \\ 
\hline
$\afba{\flm}$
&-0.28321 &-0.16981 &-0.00062 &0.11189  & 0.15470  \\
&-0.28336 &-0.16988 &-0.00066 &0.11184  & 0.15464  \\
\cline{2-6}
&+0.15    &+0.07    &+0.04    &+0.05    &+0.06     \\
\hline 
\hline
  \multicolumn{6}{|c|}{ISR/FSR interference, $\smanp\ge\Mlones(\flmp\flmm) 
\ge 0.01\,\sman$} \\
  \hline \hline
$\sigma_{\flm}\,$[nb]  
& 0.22783 & 0.47566 & 1.47970 & 0.69563 & 0.40684  \\     
& 0.22779 & 0.47566 & 1.47967 & 0.69570 & 0.40686  \\
\cline{2-6}
&+0.18    & 0.0     &+0.02    &-0.10    &-0.05     \\  
\hline 
$\sigma_{\rm had}\,$[nb]
& 4.45083 & 9.59990 & 30.43495 & 14.18082 & 8.19761  \\
& 4.45221 & 9.60254 & 30.43717 & 14.18215 & 8.19797  \\
\cline{2-6}
& -0.31   & -0.27   & -0.07    &  -0.09   &   -0.04   \\ 
\hline 
$\afba{\flm}$
&-0.28242 &-0.16898 &-0.00031 & 0.11177 & 0.15467  \\ 
&-0.28287 &-0.16934 &-0.00031 & 0.11193 & 0.15473  \\
\cline{2-6}
&+0.45    &+0.36    &+0.00    &-0.16    &-0.06     \\
\hline 
\end{tabular}
\caption[]{
  Effects of ISR/FSR QED interference for $\sigma_{\flm}$,
  $\sigma_{\rm had}$ and $\afba{\flm}$ (in CA3-mode).  First row {\tt
    TOPAZ0}, second row {\tt ZFITTER} third row relative (absolute)
  deviations in per-mill.  The scale $s$ is used for the running
  $\alpha$. }
\label{tab16}
\end{center}
\end{table}
%--

%--
\begin{table}[t]
\begin{center}
\renewcommand{\arraystretch}{1.1}
\begin{tabular}{|c||c|c|c|c|c|c|}
\hline
& \multicolumn{6}{c|}{Centre-of-mass energy in GeV} \\
\cline{2-7}
&  & $\mz - 3$ & $\mz - 1.8$ & $\mz$ & $\mz + 1.8$ & $\mz + 3$  \\
\hline 
\hline
$\sigma_{\flm}\,$[nb]
&{\tt T}&  0.22311 & 0.47122 & 1.47473 & 0.68963 & 0.40076  \\
&       &  0.22153 & 0.46914 & 1.47464 & 0.69138 & 0.40204  \\
&       & -7.13    &-4.43    &-0.06    &+2.53    &+3.18     \\
\cline{2-7}
&{\tt Z}&  0.22315 & 0.47127 & 1.47466 & 0.68967 & 0.40079  \\
&       &  0.22157 & 0.46922 & 1.47463 & 0.69148 & 0.40211  \\
&       & -7.13    &-4.37    &-0.02    &+2.62    & +3.28    \\
\hline
$\afba{\flm}$
&{\tt T}& -0.28347 &-0.16849 & 0.00050 & 0.11539 & 0.16169  \\ 
&       & -0.28074 &-0.16613 & 0.00103 & 0.11425 & 0.16018  \\
&       & +2.73    &+2.36    &+0.53    &-1.14    &-1.51     \\
\cline{2-7}
&{\tt Z}& -0.28344 &-0.16849 & 0.00047 & 0.11535 & 0.16166  \\
&       & -0.28084 &-0.16634 & 0.00104 & 0.11440 & 0.16025  \\
&       & +2.60    &+2.15    &+0.57    &-1.15    &-1.41     \\
\hline 
\end{tabular}
\caption[]{
  {\tt TOPAZ0/ZFITTER} comparison of IFI effect for $\Mlones >
  0.8\,\sman$. First row is without IFI, second row is with IFI and
  third row is the net effect in per-mill.  The scale $0$ is used for
  the running $\alpha$. }
\label{tabifi08}
\end{center}
%\end{table}
%--
%
%--
%\begin{table}[b]
\begin{center}
\renewcommand{\arraystretch}{1.1}
\begin{tabular}{|c||c|c|c|c|c|c|}
\hline
& \multicolumn{6}{c|}{Centre-of-mass energy in GeV} \\
\cline{2-7}
& & $\mz - 3$ & $\mz - 1.8$ & $\mz$ & $\mz + 1.8$ & $\mz + 3$  \\
\hline 
\hline
$\sigma_{\flm}\,$[nb]& 
{\tt T} &  0.21932 & 0.46287 & 1.44794 & 0.67725 & 0.39366\\
&       &  0.21775 & 0.46081 & 1.44785 & 0.67896 & 0.39492\\
&       & -7.21    &-4.47    &-0.06    &+2.52    &+3.19   \\
\hline
$\afba{\flm}$        &
{\tt T} & -0.28450 &-0.16914 & 0.00033 & 0.11512 & 0.16106\\     
&       & -0.28163 &-0.16668 & 0.00087 & 0.11386 & 0.15937\\ 
&       & +2.87    &+2.46    &+0.54    &-1.26    &-1.69   \\
\hline 
\end{tabular}
\caption[]{
  {\tt TOPAZ0} evaluation of IFI effect for $\Mlones(\flmp\flmm) >
  13.83997\,$GeV and $\theta_{\rm acol} < 10^\circ$. First row is
  without IFI, second row is with IFI and third row is the net effect
  in per-mill. The scale $0$ is used for the running $\alpha$. }
\label{tabifit}
\end{center}
\end{table}
%--

We also perform other comparisons to understand the effect of IFI for
various setups. First of all we use an extrapolated setup with a tight
$\Mlones$-cut of $0.8\,\sman$ (it was $0.01\,\sman$ before).  Note
that such a cut is, not exactly, but approximately equivalent to
$\theta_{\rm acol} = 10^\circ$ (note that an $\Mlones$-cut takes into
account only the isotropic part of the photon phase space). The
result, reported in \tabn{tabifi08}, shows that the two codes nicely
agree for the net IFI effect also in case of tight $\Mlones$-cuts.
Note that the effect of QED initial-final interference is quite
sizeable, differently from what happens with loose cuts.

A further test, performed with {\tt TOPAZ0}, selects another setup.
Note that $E_{\rm th} = 1\,$ GeV at $\sqrt{\sman} = \mz-3$ is
equivalent to an $\Mlones$-cut of $M^2(\barf f) \approx 2\,E_{\rm
  th}\sqrt{s}\approx 13.83997\,$GeV if we consider the radiative
photon collinear with one of the final-state fermions.  In this
example we apply an $\Mlones$-cut as above and a cut of $10^\circ$ on
the acollinearity so that the photon phase space is not isotropic
anymore.  The result is shown in \tabn{tabifit}.

The cut on acollinearity, superimposed on the $\Mlones$-cut, hardly
changes the size of IFI effect, at least as seen by {\tt TOPAZ0}.
Roughly speaking, one computes IFI with a cut on $\Mlones$ simulated
by the acollinearity and has to add a small portion of the photon
phase space, subtracting, at the same time, the region not allowed by
acol-cut.  To understand which portion of photon phase-space is added
we introduce variables
%--
\bq
\Energi{+} =\frac{1}{2}\,\lpar 1 - \xvari{1} + \xvari{2}\rpar\,\sqrt{\sman}\;,
\quad
\Energi{-} = \frac{1}{2}\,\xvari{1}\,\sqrt{\sman}\;,
\quad
\Energi{\ph} = \frac{1}{2}\,\lpar 1 - \xvari{2}\rpar\,\sqrt{\sman}\;.
\label{hardbremenergies}
\eq
%--
and observe that this parametrisation has the advantage that in the
limit where we neglect $\mf$ the boundaries of the phase space are of
a triangular form
%--
\bq
\xvari{1} = \xvari{2}\;,         \\
\xvari{1} = 1 \cup \xvari{2} = 0 \;.
\eq
%--
The boundary of the phase space corresponding to the selection
criterion $\delta = \pi - \theta_{\rm ac}$ is, in the limit $\mf = 0$,
represented by
%--
\bq
\xvaris{1} - \lpar 1 + \xvari{2}\rpar\,\xvari{1} 
+ {2\over {1+\cos\theta_{\rm ac}}}\,\xvari{2} = 0.
\eq
Therefore, if we require some acollinearity cut and $M^2(\barf f) \ge s_0$,
the limits of integration are:
%--
\bqa
x^{\rm acol}_2 &\le& x_2 \le 1, \qquad x_2 \le x_1 \le 1,  \nl
\frac{s_0}{s} \le x_2 \le x^{\rm acol}_2, \qquad
x_2 &\le& x_1 \le x^-_1 \quad \cup \quad x^+_1 \le x_1 \le 1.
\eqa
%--
with
%--
\bqa
x^{\rm acol}_2 &=& {{1 - \sin(\theta_{\rm acol}/2)}\over
{1 + \sin(\theta_{\rm acol}/2)}},  \nl
x^{\pm}_1 &=& \frac{1}{2}\,\Bigl[ 1 + x_2 \pm \sqrt{x^2_2 +2\,\lpar 1-2\,\rho
\rpar\,x_2 + 1}\Bigr],  \nl
\rho &=& \frac{2}{1 + \cos\theta_{\rm acol}}
\eqa
%--

%--
\begin{table}[hp]
\begin{center}
\renewcommand{\arraystretch}{1.1}
\begin{tabular}{|c||c|c|c|c|c|}
\hline
\multicolumn{6}{|c|}{$\sigma_{\rm had}\,$[nb] } \\
\hline
& \multicolumn{5}{c|}{Centre-of-mass energy in GeV} \\
\cline{2-6}
& $\mz - 3$ & $\mz - 1.8$ & $\mz$ & $\mz + 1.8$ & $\mz + 3$  \\
\hline 
\hline
{\tt T}
& 4.44990 & 9.59865 & 30.43501 & 14.18203 & 8.19853  \\     
$\Mlones \ge 0.01\,\sman$ 
& 4.45083 & 9.59990 & 30.43495 & 14.18082 & 8.19761  \\
&+0.21    &+0.13    & -0.002   & -0.09    &-0.11     \\
\hline
{\tt Z}
& 4.45129 & 9.60132 & 30.43725 & 14.18342 & 8.19894  \\
$\Mlones \ge 0.01\,\sman$  
& 4.45221 & 9.60254 & 30.43717 & 14.18215 & 8.19797  \\
&+0.21    &+0.13    & -0.003   & -0.09    &-0.12     \\
\hline
{\tt T}
& 4.36901 & 9.45847 & 30.05724 & 13.97471 & 8.05056  \\    
$\Mlones \ge 0.8\,\sman$
& 4.37172 & 9.46205 & 30.05739 & 13.97167 & 8.04831  \\
$\alpha(\sman)$
&+0.62    &+0.38    & +0.005   & -0.22    &-0.28     \\ 
\hline
{\tt Z}
& 4.36943 & 9.45848 & 30.04915 & 13.97099 & 8.04804  \\
$\Mlones \ge 0.8\,\sman$ 
& 4.37214 & 9.46201 & 30.04923 & 13.96786 & 8.04573  \\
$\alpha(\sman)$
&+0.62    &+0.37    & +0.003   & -0.22    &-0.29     \\
\hline
{\tt T}
& 4.37186 & 9.46463 & 30.07686 & 13.98432 & 8.05646  \\ 
$\Mlones \ge 0.8\,\sman$ 
& 4.37458 & 9.46821 & 30.07701 & 13.98127 & 8.05421  \\
$\alpha(0)$ 
& +0.62   & +0.38   & +0.005   & -0.22    & -0.28    \\
\hline
\end{tabular}
\caption[]{
  {\tt TOPAZ0/ZFITTER} comparison of IFI inclusion for hadrons with
  $\Mlones$-cut. First entry is without IFI, second entry is with IFI,
  third entry is the IFI effect in per-mill. For the 0.8 cut there are
  two entries, corresponding to two different scales: $\alpha(\sman)$
  and $\alpha(0)$. }
\label{tab-ifi-had}
\end{center}
\end{table}
%--

The cut $\Mlones \ge 0.8\,\sman$ deserves a comment. For a fully
extrapolated setup we know from a complete calculation~\cite{kn:kat}
that the correct scale in the coupling is $\sman$, i.e.,
$\alpha(\sman)$.  Therefore the two codes adopt the following scales:
loose $\Mlones$ and $\smanp$ cuts (up to $0.1\sman$) with
$\alpha(\sman)$, tight cuts with $\alpha(0)$.  Now, however, we are
using a tight $\Mlones$-cut and there is an uncertainty, if one uses
$\alpha(\sman)$ then partial higher corrections are seen, as expected.
From \tabn{tab-ifi-had} we see, however, that the net effect of IFI
remains unchanged.  The uncertainty associated with the choice of the
scale, which becomes relevant only for high values of
$\smanp(M^2)$-cuts, is common to all final states, including leptons.

Note that the {\tt T/Z} agreement slightly deteriorates at such a
tight cut (up to 0.3 per-mill at resonance).  This is not at all
surprising: for such tight cuts some common exponentiation of ISR and
FSR is mandatory.  On the contrary, one should conclude that the {\tt
  T/Z} agreement remains remarkable even for very tight $\Mlones$
cuts.

In \tabn{tab10ter} we show the effect of changing the scale $\alpha(s)
\to \alpha(0)$ in the muonic cross-section and asymmetry with
$M^2$-cuts of $0.1\,s, 0.5\,s$ and $0.8\,s$.  The variations in
cross-sections range from $\approx 0.03$ per-mill at $M^2 = 0.1\,s$ up
to $\approx 4$ per-mill at $M^2 = 0.8\,s$.  The effect of a scale
change on the asymmetry is negligible.

For the hadronic cross-section at $M^2 = 0.8\,$ the variation induced
by $\alpha(s) \to \alpha(0)$ is, at the peak, a $0.65$ per-mill to be
compared with $3.75$ per-mill for muons; the sizeable difference is
due to mass effects in QED FSR, $\mm \ll m_q$.

%--
\begin{table}[t]
\begin{center}
\renewcommand{\arraystretch}{1.1}
\begin{tabular}{|c||c|c|c|c|c|}
\hline
& \multicolumn{5}{c|}{Centre-of-mass energy in GeV} \\
\cline{2-6}
& $\mz - 3$ & $\mz - 1.8$ & $\mz$ & $\mz + 1.8$ & $\mz + 3$  \\
\hline 
\hline
\multicolumn{6}{|c|}{$\sigma_{\mu}\,$[nb] } \\
\hline
{\tt T} $\Mlones \ge 0.1\,s\,\alpha(s)$ 
& 0.22615 & 0.47369 & 1.47484 & 0.69179 & 0.40382 \\
{\tt T} $\Mlones \ge 0.1\,s\,\alpha(0)$ 
& 0.22616 & 0.47370 & 1.47489 & 0.69182 & 0.40384 \\
Diff.
&  0.04   &  0.02   &  0.03   &  0.04   &  0.05   \\
\hline
{\tt T} $\Mlones \ge 0.5\,s\,\alpha(s)$ 
& 0.22026 & 0.46327 & 1.44595 & 0.67712 & 0.39442 \\
{\tt T} $\Mlones \ge 0.5\,s\,\alpha(0)$ 
& 0.22051 & 0.46380 & 1.44762 & 0.67792 & 0.39490 \\
Diff.      
&  1.13   &  1.14   &  1.08   &  1.18   &  1.22   \\
\hline
{\tt T} $\Mlones \ge 0.8\,s\,\alpha(s)$ 
& 0.20926 & 0.44215 & 1.38382 & 0.64513 & 0.37335 \\
{\tt T} $\Mlones \ge 0.8\,s\,\alpha(0)$ 
& 0.21005 & 0.44381 & 1.38903 & 0.64769 & 0.37493 \\  
Diff.      
&  3.76   &  3.74   &  3.75   &  3.95   &  4.21   \\
\hline
\hline 
\multicolumn{6}{|c|}{$\afba{\flm}$} \\
\hline
{\tt T} $\Mlones \ge 0.1\,s\,\alpha(s)$ 
& -0.28503 & -0.16986 & -0.00029 &  0.11224 &  0.15557 \\
{\tt T} $\Mlones \ge 0.1\,s\,\alpha(0)$ 
& -0.28505 & -0.16987 & -0.00029 &  0.11225 &  0.15558 \\
Diff.
&   0.02   &   0.01   &   0      &  -0.01   &  -0.01   \\
\hline
{\tt T} $\Mlones \ge 0.5\,s\,\alpha(s)$ 
& -0.28603 & -0.16981 &  0.00001 &  0.11312 &  0.15727 \\
{\tt T} $\Mlones \ge 0.5\,s\,\alpha(0)$ 
& -0.28605 & -0.16982 &  0.00001 &  0.11311 &  0.15726 \\
Diff.      
&   0.02   &   0.01   &   0      &   0.01   &   0.01   \\
\hline
{\tt T} $\Mlones \ge 0.8\,s\,\alpha(s)$ 
& -0.28086 & -0.16613 &  0.00130 &  0.11534 &  0.16236 \\
{\tt T} $\Mlones \ge 0.8\,s\,\alpha(0)$ 
& -0.28091 & -0.16616 &  0.00128 &  0.11530 &  0.16225 \\  
Diff.      
&   0.05   &   0.03   &   0.02   &   0.04   &   0.09   \\
\hline
\end{tabular}
\caption[]{
  $\sigma_{\mu}$and $\afba{\flm}$, including IFI, for various
  $M^2$-cuts.  First entry is with scale = $s$, second entry is with
  scale = $0$, third entry is the relative (absolute) difference in
  per-mill. }
\label{tab10ter}
\end{center}
\end{table}
%--

\clearpage

%------------------------------------------------------
\subsection{Comparison with Realistic Kinematical Cuts}
%------------------------------------------------------

Now we move to the classic setup, superimposing an acollinearity cut,
a cut of 1 GeV on the energy of both outgoing fermions and a cut on
the polar angular range of the outgoing fermion.  The photon phase
space is a little more complicated but not much.  An acollinearity cut
of $10^\circ$ or $25^\circ$ corresponds to $\Mlones=0.84\,\sman$ and
$\Mlones=0.64\,\sman$, and is thus regarded as a tight cut.
Therefore, the scale for $\alpha$ is $\alpha(0)$.  The comparison is
shown in \tabns{tab10acolifi1}{tab25acolifi2}.

%--
\begin{table}[b]
\begin{center}
\renewcommand{\arraystretch}{1.1}
\begin{tabular}{|c||c|c|c|c|c|}
\hline
\multicolumn{6}{|c|}{$\sigma_{\flm}\,$[nb] with $\theta_{\rm acol}<10^\circ$} \\
\hline
&\multicolumn{5}{c|}{Centre-of-mass energy in GeV} \\
\hline
$\theta_{\rm acc}$& $\mz - 3$ & $\mz - 1.8$ & $\mz$ & $\mz + 1.8$ & $\mz + 3$  \\
\hline\hline
{\tt T} \hfill $0^\circ$  
  & 0.21932  & 0.46287  & 1.44795  & 0.67725  & 0.39366 \\
  & 0.21776  & 0.46083  & 1.44785  & 0.67894  & 0.39491 \\
  & -7.16    &-4.43     &-0.07     &+2.49     &+3.17    \\
\hline
{\tt Z} \hfill $0^\circ$  
  & 0.21928  & 0.46285  & 1.44781  & 0.67722  & 0.39361 \\
  & 0.21852  & 0.46186  & 1.44782  & 0.67814  & 0.39429 \\
  &-3.48     &-2.14     &+0.01     &+1.36     &+1.72    \\
\hline
{\tt T} \hfill $20^\circ$ 
  & 0.19990  & 0.42207  & 1.32066  & 0.61759  & 0.35886 \\
  & 0.19873  & 0.42049  & 1.32027  & 0.61873  & 0.35973 \\
  &-5.89     &-3.76     &-0.30     &+1.84     &+2.42    \\
\hline
{\tt Z} \hfill $20^\circ$ 
  & 0.19987  & 0.42205  & 1.32053  & 0.61756  & 0.35881 \\
  & 0.19892  & 0.42075  & 1.32021  & 0.61857  & 0.35959 \\
  &-4.78     &-3.09     &-0.24     &+1.63     &+2.17    \\
\hline
{\tt T} \hfill $40^\circ$ 
  & 0.15034  & 0.31762  & 0.99428  & 0.46479  & 0.26989 \\
  & 0.14979  & 0.31680  & 0.99360  & 0.46514  & 0.27021 \\
  &-3.67     &-2.59     &-0.68     &+0.75     &+1.18    \\
\hline
{\tt Z} \hfill $40^\circ$ 
  & 0.15032  & 0.31760  & 0.99415  & 0.46474  & 0.26983 \\
  & 0.14978  & 0.31680  & 0.99350  & 0.46511  & 0.27016 \\
  &-3.61     &-2.53     &-0.65     &+0.80     &+1.22    \\
\hline 
\end{tabular}
\caption[]{
  {\tt TOPAZ0/ZFITTER} comparison for the muonic cross-section of
  complete RO (CA3-mode) with the angular acceptance ($\theta_{\rm
    acc}=0,20,40^\circ$) and acollinearity ($\theta_{\rm
    acol}<10^\circ$) cuts. First row is without IFI, second row with
  IFI, third row is the relative (per-mill) effect of IFI.  }
\label{tab10acolifi1}
\end{center}
\end{table}
%--
\begin{table}[ht]
\begin{center}
\renewcommand{\arraystretch}{1.1}
\begin{tabular}{|c||c|c|c|c|c|}
\hline
\multicolumn{6}{|c|}{$\afba{\flm}$ with $\theta_{\rm acol}<10^\circ$} \\
\hline
&\multicolumn{5}{c|}{Centre-of-mass energy in GeV} \\
\hline
$\theta_{\rm acc}$& $\mz - 3$ & $\mz - 1.8$ & $\mz$ & $\mz + 1.8$ & $\mz + 3$  \\
\hline\hline
{\tt T} \hfill $0^\circ$  
  & -0.28450 & -0.16914  & 0.00033  & 0.11512  & 0.16107 \\
  & -0.28158 & -0.16665  & 0.00088  & 0.11385  & 0.15936 \\
  & +2.92    & +2.49     &+0.55     &-1.27     &-1.71    \\
\hline
{\tt Z} \hfill $0^\circ$  
  & -0.28453 & -0.16911  & 0.00025  & 0.11486  & 0.16071 \\
  & -0.28282 & -0.16783  & 0.00070  & 0.11475  & 0.16059 \\
  & +1.71    & +1.28     &+0.45     &-0.11     &-0.12    \\
\hline
{\tt T} \hfill $20^\circ$ 
  & -0.27509 & -0.16352  & 0.00042  & 0.11171  & 0.15645 \\
  & -0.27259 & -0.16146  & 0.00084  & 0.11064  & 0.15499 \\
  & +2.50    & +2.06     &+0.42     &-1.07     &-1.46    \\
\hline
{\tt Z} \hfill $20^\circ$  
  & -0.27506 & -0.16347  & 0.00035  & 0.11148  & 0.15616 \\
  & -0.27408 & -0.16261  & 0.00070  & 0.11133  & 0.15594 \\
  & +0.98    & +0.86     &+0.35     &-0.15     &-0.22    \\
\hline
{\tt T} \hfill $40^\circ$ 
  & -0.24219 & -0.14396  & 0.00054  & 0.09906  & 0.13903 \\
  & -0.24041 & -0.14262  & 0.00078  & 0.09837  & 0.13809 \\
  & +1.78    & +1.34     &+0.24     &-0.69     &-0.94    \\
\hline
{\tt Z} \hfill $40^\circ$ 
  & -0.24207 & -0.14386  & 0.00050  & 0.09893  & 0.13891 \\
  & -0.24151 & -0.14343  & 0.00069  & 0.09890  & 0.13888 \\
  & +0.56    & +0.43     &+0.19     &-0.03     &-0.03    \\
\hline 
\end{tabular}
\caption[]{
  {\tt TOPAZ0/ZFITTER} comparison for the muonic forward-backward
  asymmetry of complete RO (CA3-mode) with the angular acceptance
  ($\theta_{\rm acc}=0,20,40^\circ$) and acollinearity ($\theta_{\rm
    acol}<10^\circ$) cuts.  First row is without IFI, second row with
  IFI, third row is the absolute effect of IFI (per-mill). }
\label{tab10acolifi2}
\end{center}
\end{table}
%--
\begin{table}[ht]
\begin{center}
\renewcommand{\arraystretch}{1.1}
\begin{tabular}{|c||c||c|c|c|c|c|}
\hline
\multicolumn{6}{|c|}{$\sigma_{\flm}\,$[nb] with $\theta_{\rm acol}<25^\circ$} \\
\hline
&\multicolumn{5}{c|}{Centre-of-mass energy in GeV} \\
\hline
$\theta_{\rm acc}$& $\mz - 3$ & $\mz - 1.8$ & $\mz$ & $\mz + 1.8$ & $\mz + 3$  \\
\hline\hline
{\tt T} \hfill $0^\circ$ 
  & 0.22333  & 0.46971  & 1.46611  & 0.68690  & 0.40034 \\
  & 0.22233  & 0.46838  & 1.46611  & 0.68812  & 0.40127 \\
  &-4.50     &-2.84     & 0.00     &+1.77     &+2.32    \\
\hline
{\tt Z} \hfill $0^\circ$ 
  & 0.22328  & 0.46968  & 1.46598  & 0.68688  & 0.40031 \\
  & 0.22281  & 0.46905  & 1.46603  & 0.68754  & 0.40081 \\
  &-2.11     &-1.34     &+0.03     &+0.96     &+1.25    \\
\hline
{\tt T} \hfill $20^\circ$ 
  & 0.20359  & 0.42835  & 1.33731  & 0.62648  & 0.36507 \\
  & 0.20286  & 0.42733  & 1.33700  & 0.62725  & 0.36569 \\
  &-3.60     &-2.39     &-0.23     &+1.23     &+1.70    \\
\hline
{\tt Z} \hfill $20^\circ$ 
  & 0.20357  & 0.42833  & 1.33718  & 0.62647  & 0.36505 \\
  & 0.20321  & 0.42781  & 1.33689  & 0.62684  & 0.36536 \\
  &-1.77     &-1.22     &-0.22     &+0.59     &+0.85    \\
\hline
{\tt T} \hfill $40^\circ$ 
  & 0.15320  & 0.32245  & 1.00698  & 0.47167  & 0.27479 \\
  & 0.15286  & 0.32190  & 1.00635  & 0.47185  & 0.27500 \\
  &-2.22     &-1.71     &-0.63     &+0.38     &+0.76    \\
\hline
{\tt Z} \hfill $40^\circ$ 
  & 0.15318  & 0.32243  & 1.00682  & 0.47164  & 0.27477 \\
  & 0.15287  & 0.32192  & 1.00619  & 0.47180  & 0.27496 \\
  &-2.03     &-1.58     &-0.63     &+0.34     &+0.69    \\
\hline 
\end{tabular}
\caption[]{
  {\tt TOPAZ0/ZFITTER} comparison for the muonic cross-section of
  complete RO (CA3-mode) with the angular acceptance ($\theta_{\rm
    acc}=0,20,40^\circ$) and acollinearity ($\theta_{\rm
    acol}<25^\circ$) cuts. First row is without IFI, second row with
  IFI, third row is the relative (per-mill) effect of IFI. }
\label{tab25acolifi1}
\end{center}
\end{table}
%--
\begin{table}[ht]
\begin{center}
\renewcommand{\arraystretch}{1.1}
\begin{tabular}{|c||c||c|c|c|c|c|}
\hline
\multicolumn{6}{|c|}{$\afba{\flm}$ with $\theta_{\rm acol}<25^\circ$} \\
\hline
&\multicolumn{5}{c|}{Centre-of-mass energy in GeV} \\
\hline
$\theta_{\rm acc}$& $\mz - 3$ & $\mz - 1.8$ & $\mz$ & $\mz + 1.8$ & $\mz + 3$  \\
\hline\hline
{\tt T} \hfill $0^\circ$ 
  & -0.28617 & -0.17037 & -0.00032  & 0.11324  & 0.15730 \\
  & -0.28501 & -0.16923 &  0.00006  & 0.11293  & 0.15703 \\
  & +1.16    & +1.14    & +0.38     &-0.31     &-0.27    \\
\hline
{\tt Z} \hfill $0^\circ$ 
  & -0.28647 & -0.17049 & -0.00043  & 0.11293  & 0.15682 \\
  & -0.28555 & -0.16975 & -0.00005  & 0.11307  & 0.15701 \\
  & +0.92    & +0.74    & +0.48     &+0.14     &+0.19    \\
\hline
{\tt T} \hfill $20^\circ$ 
  & -0.27695 & -0.16485 & -0.00026  & 0.10974  & 0.15250 \\
  & -0.27611 & -0.16404 &  0.00001  & 0.10955  & 0.15236 \\
  & +0.84    & +0.81    & +0.27     &-0.19     &-0.14    \\
\hline
{\tt Z} \hfill $20^\circ$ 
  & -0.27722 & -0.16497 & -0.00037  & 0.10944  & 0.15204 \\
  & -0.27657 & -0.16447 & -0.00009  & 0.10963  & 0.15229 \\
  & +0.65    & +0.50    & +0.28     &+0.19     &+0.25    \\
\hline
{\tt T} \hfill $40^\circ$ 
  & -0.24423 & -0.14536 & -0.00016  & 0.09703  & 0.13492 \\
  & -0.24381 & -0.14501 & -0.00004  & 0.09702  & 0.13500 \\
  & +0.42    & +0.35    & +0.12     &-0.01     &+0.08    \\
\hline
{\tt Z} \hfill $40^\circ$ 
  & -0.24445 & -0.14545 & -0.00026  & 0.09678  & 0.13454 \\
  & -0.24444 & -0.14542 & -0.00011  & 0.09700  & 0.13483 \\
  & +0.01    & +0.03    & +0.15     &+0.22     &+0.29    \\
\hline 
\end{tabular}
\caption[]{
  {\tt TOPAZ0/ZFITTER} comparison for the muonic forward-backward
  asymmetry of complete RO (CA3-mode) with the angular acceptance
  ($\theta_{\rm acc}=0,20,40^\circ$) and acollinearity ($\theta_{\rm
    acol}<25^\circ$) cuts. First row is without IFI, second row with
  IFI, third row is the absolute effect of IFI (per-mill). }
\label{tab25acolifi2}
\end{center}
\end{table}
%--

There are several points to be discussed here. Although for the peak
energy, we register generally a good agreement between {\tt TOPAZ0
  4.4} and {\tt ZFITTER 5.20}, the situation for the off-peak points
is signalling a disagreement.  Here {\tt TOPAZ0 4.4} confirms the same
size of IFI effect of previous situations, while {\tt ZFITTER 5.20}
sees half of the effect.  This is a somewhat unique situation, the
only one where {\tt TOPAZ0 4.4} and {\tt ZFITTER 5.20} register a
substantial disagreement.  This disagreement should not be taken as a
measure of the real theoretical uncertainty.

Recent work of the Zeuthen group~\cite{kn:cjr-prep} has contributed
substantially in understanding the origin of IFI-discrepancy, the IFI
effect being so different in the two codes when realistic cuts are
imposed.  For the case of full angular acceptance, the work of
\cite{kn:cjr-prep} contains a full list of updated results.  It
recalculates photonic corrections with acollinearity cuts, having in
mind applications to {\tt ZFITTER}.  The conclusions of this work are
that after inclusion of the new calculation the agreement with {\tt
  TOPAZ0 4.4} is 0.1 per-mill (at the wings) or better (on resonance)
for cross-sections.  Work is in progress to update the inclusion of
angular cuts.  Thanks to the important contribution of the Zeuthen
group, the IFI discrepancy with realistic cuts is on its way to be
fully solved.

As stated above, the introduction of angular cuts will represent the
next step in Zeuthen's program.  For the moment, therefore, we have at
our disposal only the comparison between {\tt TOPAZ0 4.4} and {\tt
  ZFITTER 5.20}.  This section contains many IFI numbers based on
these two programs, so that it should become clear under which
conditions the discrepancies are large or small.  Based on this the
experiments will then have to derive an adequate solution for ROs with
realistic cuts, concerning the magnitude of the effect and its
theoretical uncertainty.  Some possibilities are:

\begin{enumerate}
\item Use current IFI as implemented in {\tt TOPAZ0 4.4/ZFITTER 5.20}
  and assign overall uncertainty from the comparison,
\item Use $\ord{\alpha}$ IFI of, e.g., {\tt KORALZ} to remove IFI
  effects from the ROs and fit the ROs with {\tt TOPAZ0/ZFITTER}
  having switched off IFI.  Of course, in this case the {\tt KORALZ}
  implementation of IFI must be evaluated to obtain the theory
  uncertainty.
\item Wait for an updated {\tt ZFITTER} code and repeat the {\tt
    TOPAZ0/ZFITTER} comparison to estimate the theoretical
  uncertainty.
\end{enumerate}

%------------------------------------------------------------------
\subsection{Experimental Aspects of Initial-Final QED Interference}
%------------------------------------------------------------------

As a last remark we observe that also the treatment of initial-final
QED interference effects by the experiments in arriving at quoted ROs
has to be taken into account.  For example, if the Monte Carlo
generators used to correct for efficiency and to extrapolate for
geometrical acceptance do not contain interference (as is the case,
for example, in the current versions of {\tt JETSET/PYTHIA}, or {\tt
  KORALZ} used in multi-photon mode), then the extrapolated and quoted
ROs somehow miss interference.

One main question is therefore: do QED interference effects enter only
in the extrapolation step, e.g., $\cos(\theta) \to 1$?  If so, then at
least the fits are under control since the effect of missing IFI
cancels out when {\tt TOPAZ0/ZFITTER} are also run without
interference.  However, if interference effects already show up within
the accepted region, then not only the quoted ROs are ill-defined but
there are also problems in fitting them.

It is of course a matter of the size of the effect.  From our study we
can draw some conclusion, despite the present disagreement in case
realistic cuts are imposed.  If we stay at the $\zb$-peak where our
predictions show agreement we observe a $-0.07,-0.30,-0.68$
$(+0.01,$$-0.24,-0.65)$ per-mill effect in {\tt TOPAZ0 (ZFITTER)} on
the cross-section for $\theta_{\rm acol} < 10^\circ$ and $\theta_{\rm
  acc}= 0^\circ, 20^\circ$ and $40^\circ$.  The effect of interference
grows when reducing the angular acceptance from $0^\circ$ to
$40^\circ$.  At the wings the effect is becoming smaller for reduced
angular acceptance, but the rate of decreasing is not fast; for {\tt
  TOPAZ0/ZFITTER} the effect of IFI is still $-3.67/-3.61~(1.18/1.22)$
per-mill at $\theta_{\rm acc} = 40^\circ$ and $\sqrt{\sman} = \mz
-3~(+3)$.  Note that around $\theta_{\rm acc} = 40^\circ$ {\tt TOPAZ0}
and {\tt ZFITTER} start to agree well, so the previous statement is
not affected by differences in the codes.  For $\theta_{\rm acol} <
25^\circ$ the effect at the peak goes from zero at $\theta_{\rm acc} =
0^\circ$ to $-0.63$ per-mill at $\theta_{\rm acc} = 40^\circ$.  At the
wings the effect is $-2$ per-mill and $+0.7$ per-mill respectively for
$\theta_{\rm acc} = 40^\circ$ (with good agreement between the codes).
For asymmetries, the interference effect decreases in magnitude with
reduced angular acceptance at and below the peak energy.  This is also
the case for {\tt TOPAZ0} above the peak, while a different behaviour
is observed for {\tt ZFITTER}.

Our conclusion is that the size of the IFI effect is at the level of
few per-mill even within a fiducial volume cut.  In order to minimise
the effect we recommend to quote results, e.g., asymmetries, within a
fiducial volume, at least at the wings of the $\zb$ resonance.

\clearpage

%----------------------------------------------------------------
\section{Realistic Observables in the Model Independent Approach}
%----------------------------------------------------------------
\label{sec:comp-MI-approach}

As we have shown in \eqn{defmiapp} the main emphasis in the
MI-approach is to organise the calculation of ROs in terms of POs for
MI fits.  One has to show the internal consistency of the procedure.
First, one has to show that for POs with values as calculated in the
SM, the ROs are {\em by construction} identical to the full SM RO
calculation.  Next one has to make sure that the ROs computed by
different codes also agree for MI calculations when the POs do not
have SM values, but are varied over a range of PO values corresponding
to at least the current experimental errors on POs for a single
experiment.
 
It is one of the main goals of our study to perform such comparison
with {\tt TOPAZ0} and {\tt ZFITTER}.  Of course it will not be
possible to present here the full comparison for arbitrary ROs but we
will compare few significant examples, e.g., few ROs like
$\sigma_{\flm}, \sigma_{\had}$ and $\afba{\flm}$, at fixed
$\mz,\mt,\mh,$ $\alpha(\mzs),\als(\mzs)$, as a function of $\gz,
\sigma^{0}_{\had}, R_{\fl}$ and $\afba{0,\rm l}$.  Note that we do not
vary $\mz$, both because its experimental error is so small and
because it is also a SM parameter.  We vary the remaining POs, one by
one.  Assuming the experimental errors to be twice the current LEP
errors~\cite{kn:lepewwg98}, we obtain:
%--
\bqa
\gz &=& 2.4958 \pm 0.0048\,\GeV,  
\nl
\sigma^{0}_{\had} &=& 41.473 \pm 0.116\,\nb,  
\nl
\Rl &=& 20.748 \pm 0.052,
\nl
\afba{0,\rm l} &=& 0.01613 \pm 0.00192\,.
\label{eq:PO-LEP}
\eqa
%--
We vary the POs by twice their experimental errors.  The corresponding
deviations between the two codes will give an estimate of the
technical precision of the procedure.  For the SM initialisation, our
preferred setup of SM input parameters is used.  The effects of IFI
and ISPP are not included.

\begin{figure}[htbp]
\begin{center}
\includegraphics[width=0.7\linewidth]{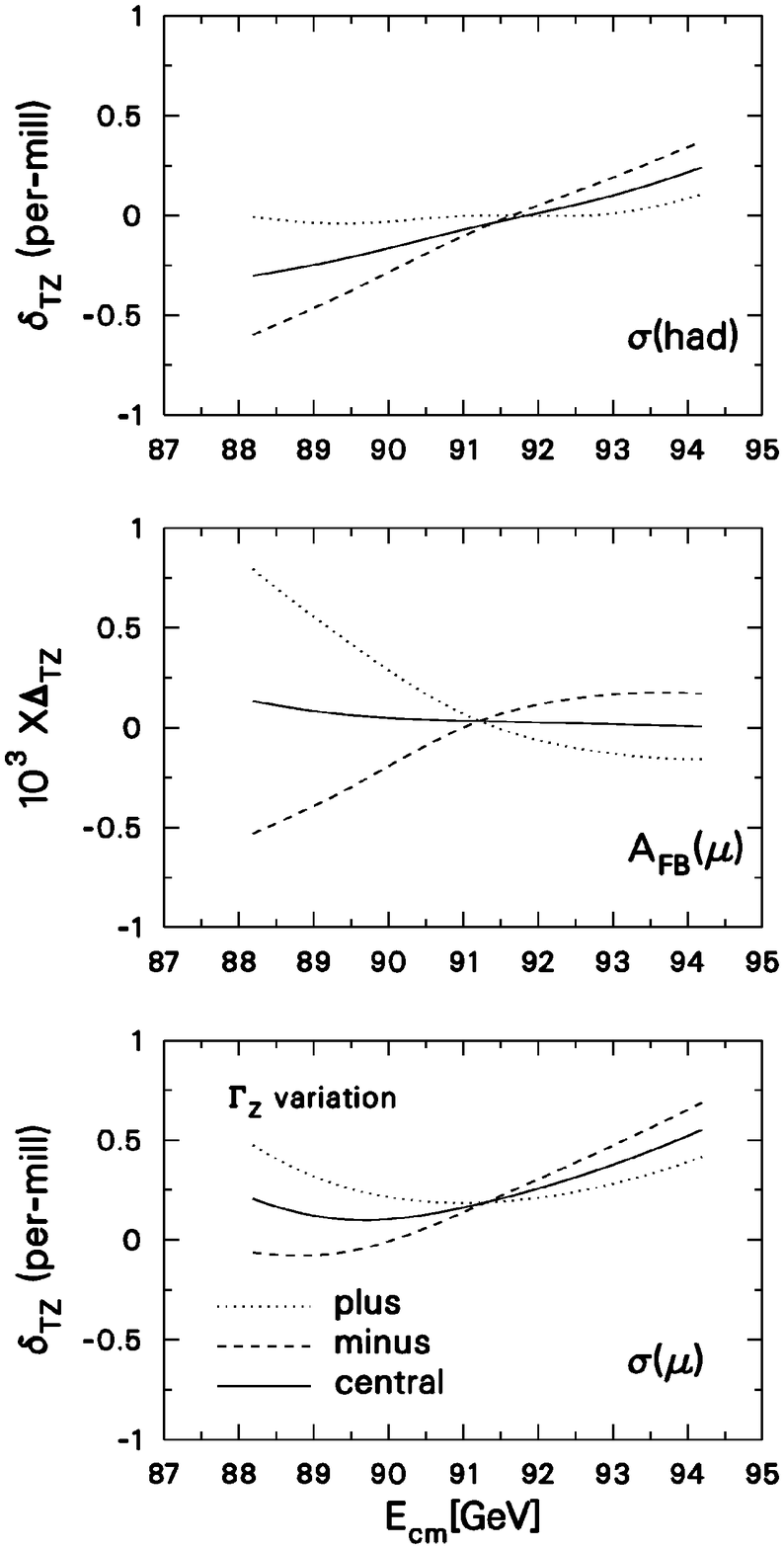}
\caption[]{
  Relative deviations between {\tt TOPAZ0} and {\tt ZFITTER} for
  muonic and hadronic cross-section and absolute deviations for muonic
  forward-backward asymmetry in CA3 mode and for $\smanp >
  0.01\,\sman$.  The solid curve gives deviations corresponding to the
  current experimental value of $\gz$, the dotted (dashed) curve
  corresponds to PO$\,\pm\, 2\,\times$ the experimental error.}
\label{fnotemi1}
\end{center}
\end{figure}

We show the MI comparisons in \figs{fnotemi1}{fnotemirc4}.  The solid
curve gives deviations corresponding to the current experimental value
of the relative PO, the dotted (dashed) curve corresponds to PO$\,\pm
\,2\,\times$ the experimental errors listed in \eqn{eq:PO-LEP}.  For
some of the figures the three curves become indistinguishable,
typically the hadronic cross section has a constant deviation between
the two MI-calculations if we vary $\Rl$ or $\afba{0,l}$.

\begin{figure}[htbp]
\begin{center}
\includegraphics[width=0.7\linewidth]{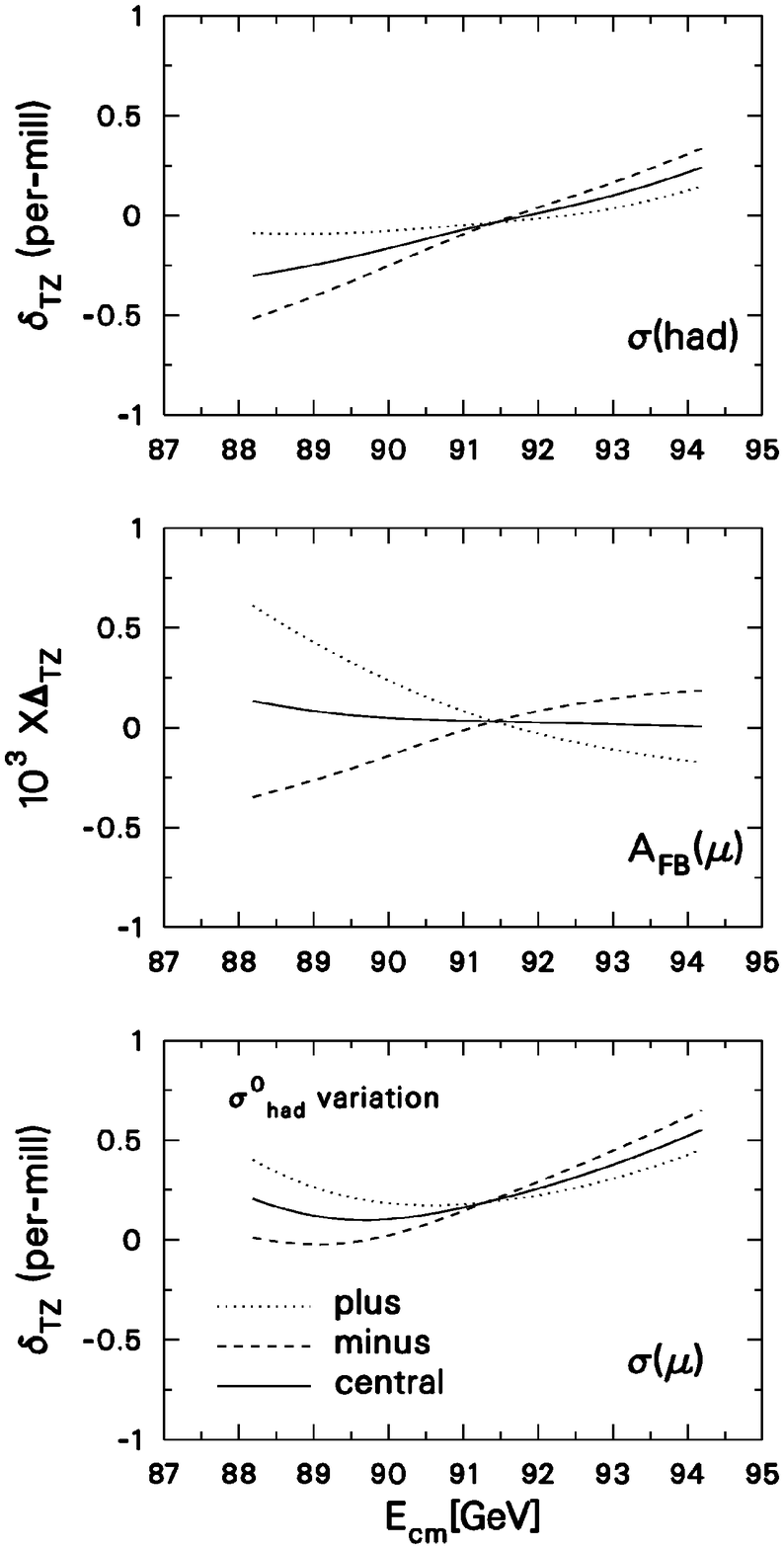}
\caption[]{
  Relative deviations between {\tt TOPAZ0} and {\tt ZFITTER} for
  muonic and hadronic cross-section and absolute deviations for muonic
  forward-backward asymmetry in CA3 mode and for $\smanp >
  0.01\,\sman$.  The solid curve gives deviations corresponding to the
  current experimental value of $\sigma^0_{\rm had}$, the
  dotted (dashed) curve corresponds to PO$\,\pm\,2\,\times$ the
  experimental error.}
\label{fnotemi2}
\end{center}
\end{figure}

\begin{figure}[htbp]
\begin{center}
\includegraphics[width=0.7\linewidth]{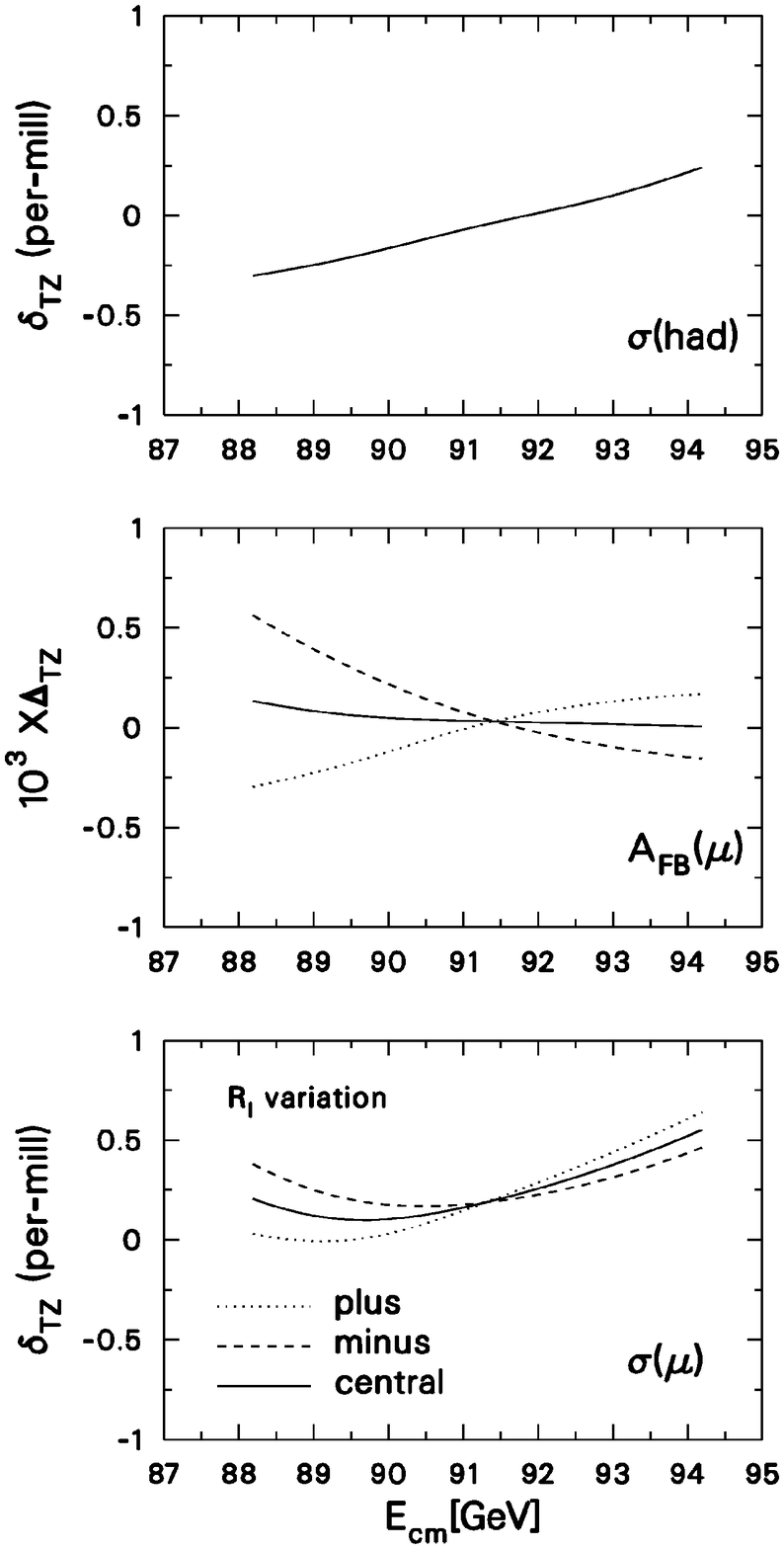}
\caption[]{
  Relative deviations between {\tt TOPAZ0} and {\tt ZFITTER} for
  muonic and hadronic cross-section and absolute deviations for muonic
  forward-backward asymmetry in CA3 mode and for $\smanp >
  0.01\,\sman$.  The solid curve gives deviations corresponding to the
  current experimental value of $\Rl$, the dotted (dashed) curve
  corresponds to PO$\,\pm\,2\,\times$ the experimental error.}
\label{fnotemi3}
\end{center}
\end{figure}

\begin{figure}[htbp]
\begin{center}
\includegraphics[width=0.7\linewidth]{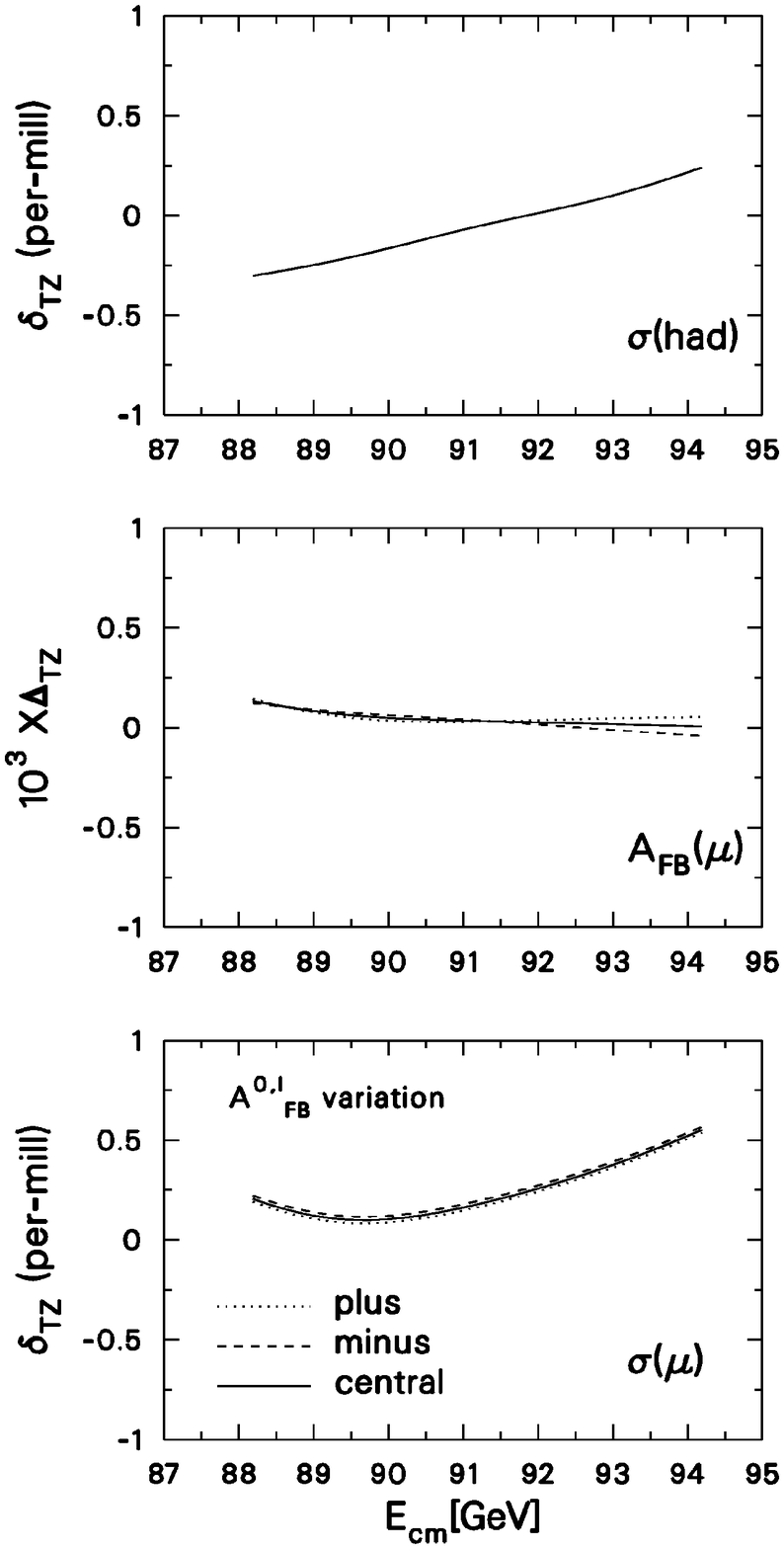}
\caption[]{
  Relative deviations between {\tt TOPAZ0} and {\tt ZFITTER} for
  muonic and hadronic cross-section and absolute deviations for muonic
  forward-backward asymmetry in CA3 mode and for $\smanp >
  0.01\,\sman$.  The solid curve gives deviations corresponding to the
  current experimental value of $\afba{0,l}$, the dotted (dashed)
  curve corresponds to PO$\,\pm\,2\,\times$ the experimental error.}
\label{fnotemi4}
\end{center}
\end{figure}

\begin{figure}[htbp]
\begin{center}
\includegraphics[width=0.85\linewidth]{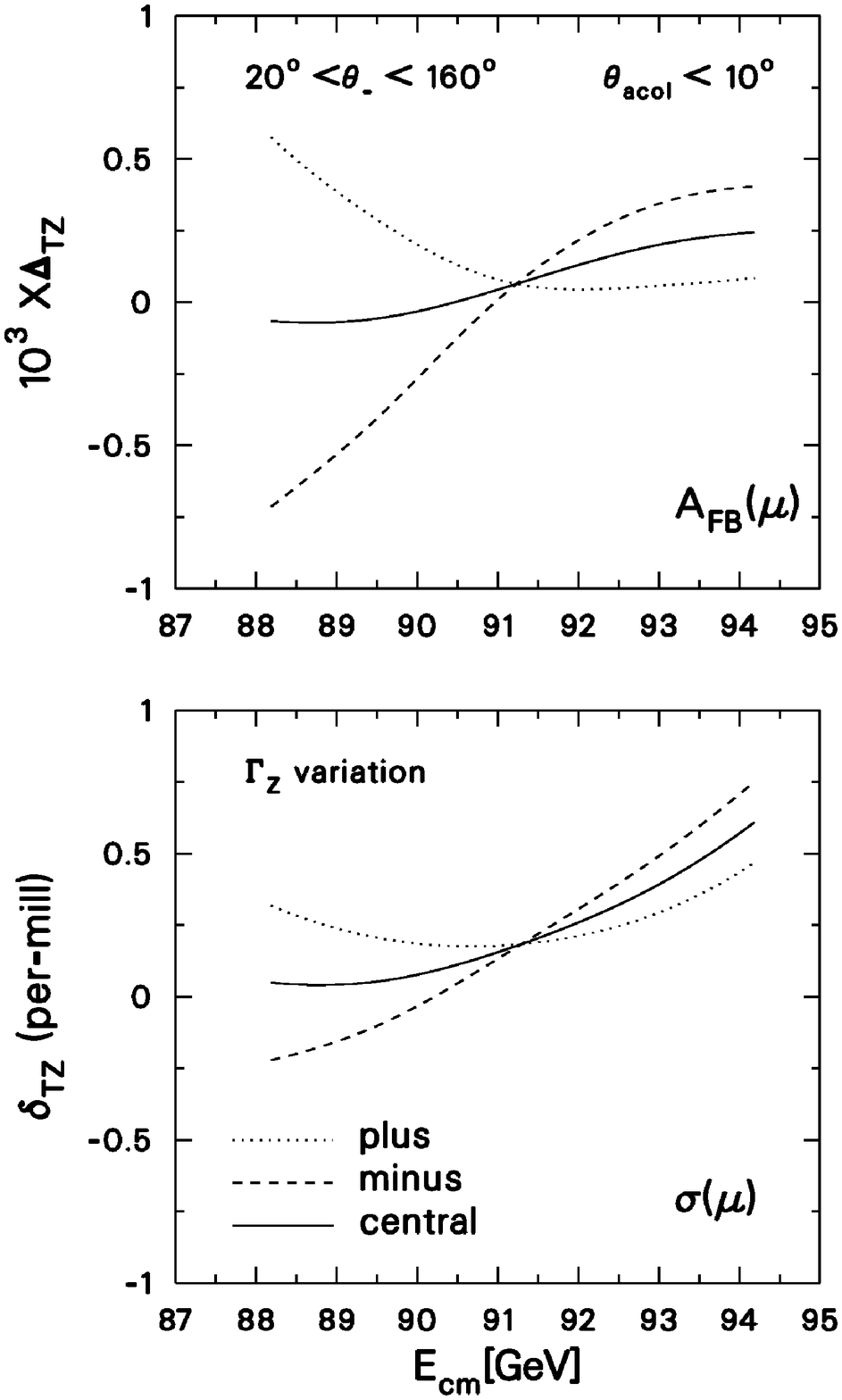}
\caption[]{
  Relative deviations between {\tt TOPAZ0} and {\tt ZFITTER} for
  muonic cross-section and absolute deviations for muonic
  forward-backward asymmetry in CA3 mode and for realistic cuts.  The
  solid curve gives deviations corresponding to the current
  experimental value of $\gz$, the dotted (dashed) curve corresponds
  to PO$\,\pm\,2\,\times$ the experimental error.}
\label{fnotemirc1}
\end{center}
\end{figure}

From \fig{fnotemi1} to \fig{fnotemi4} we use an extrapolated setup
with an $\smanp$-cut of $0.01\,\sman$ and report the relative (in
per-mill) {\tt TOPAZ0-ZFITTER} deviations for $\sigma_{\flm},
\sigma_{\had}$ and the absolute deviations for $\afba{\flm}$.  In
\figs{fnotemirc1}{fnotemirc4} we show similar comparisons for the
muonic ROs with realistic cuts.  We choose $20^\circ < \theta_- <
160^\circ$, $\theta_{\rm acoll} < 10^\circ$, and $E_{th}(\mu^{\pm}) =
1\,$GeV.  There is no appreciable difference in behaviour with respect
to the $\smanp$ case shown before.

%--
%--

%--
%--

%--
%--

%--
%--

%--
%--

%--
\begin{figure}[htbp]
\begin{center}
\includegraphics[width=0.85\linewidth]{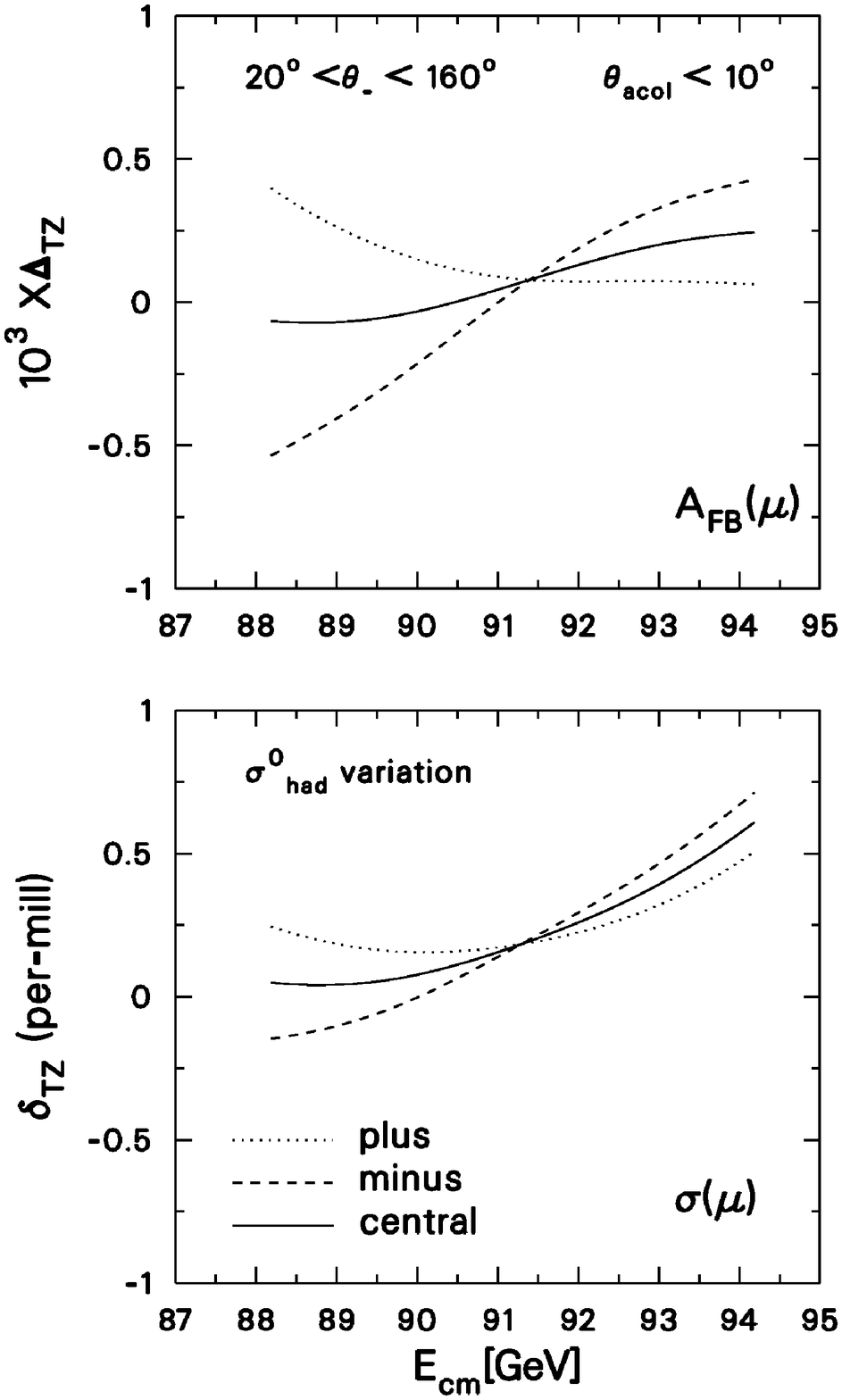}
\caption[]{
  Relative deviations between {\tt TOPAZ0} and {\tt ZFITTER} for
  muonic cross-section and absolute deviations for muonic
  forward-backward asymmetry in CA3 mode and for realistic cuts.  The
  solid curve gives deviations corresponding to the current
  experimental value of $\sigma^0_{\rm had}$, the dotted (dashed)
  curve corresponds to PO$\,\pm\,2\,\times$ the experimental error.}
\label{fnotemirc2}
\end{center}
\end{figure}

%--
\begin{figure}[htbp]
\begin{center}
\includegraphics[width=0.85\linewidth]{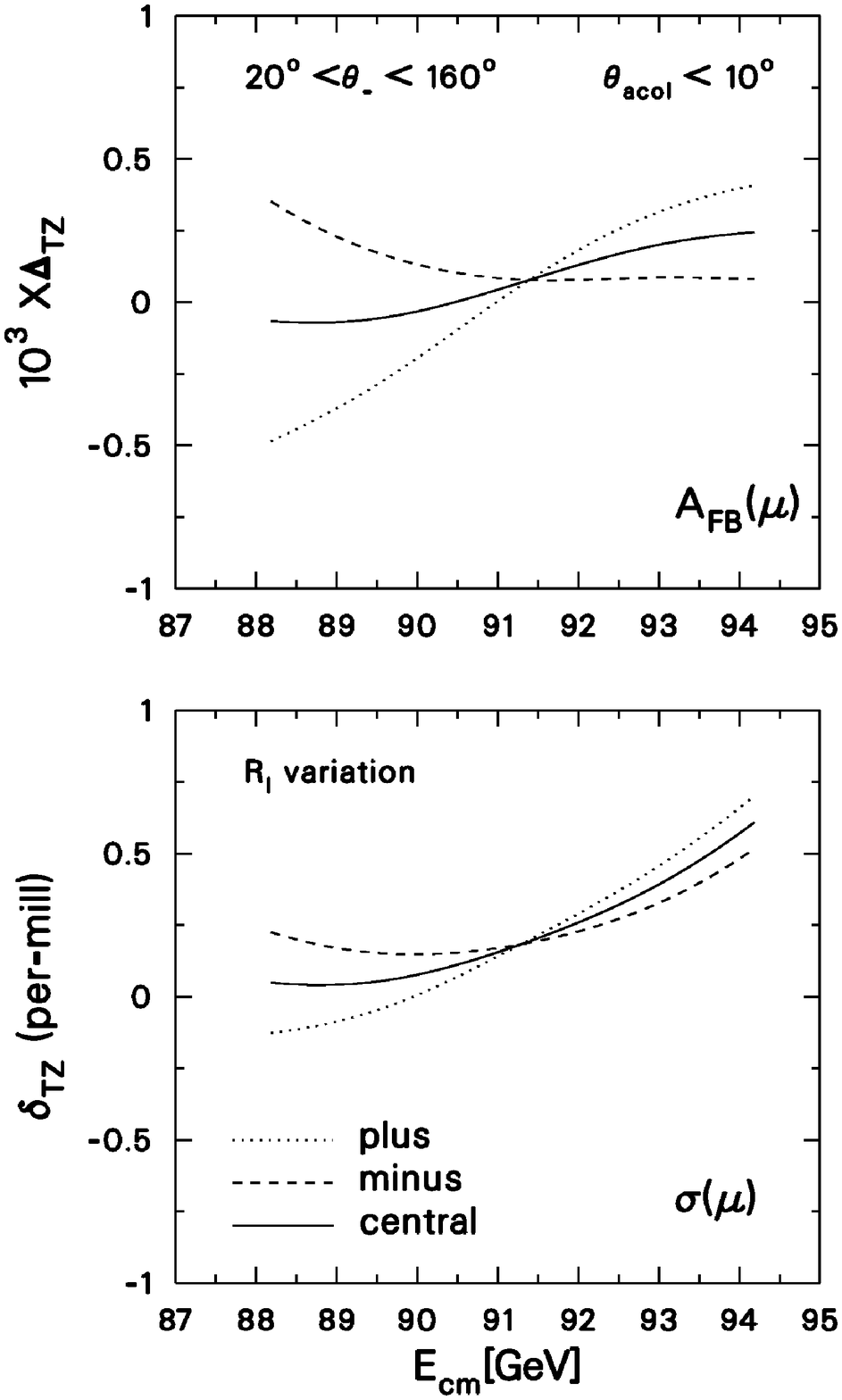}
\caption[]{
  Relative deviations between {\tt TOPAZ0} and {\tt ZFITTER} for
  muonic cross-section and absolute deviations for muonic
  forward-backward asymmetry in CA3 mode and for realistic cuts.  The
  solid curve gives deviations corresponding to the current
  experimental value of $\Rl$, the dotted (dashed) curve corresponds
  to PO$\,\pm\,2\,\times$ the experimental error.}
\label{fnotemirc3}
\end{center}
\end{figure}
%--

%--
\begin{figure}[htbp]
\begin{center}
\includegraphics[width=0.85\linewidth]{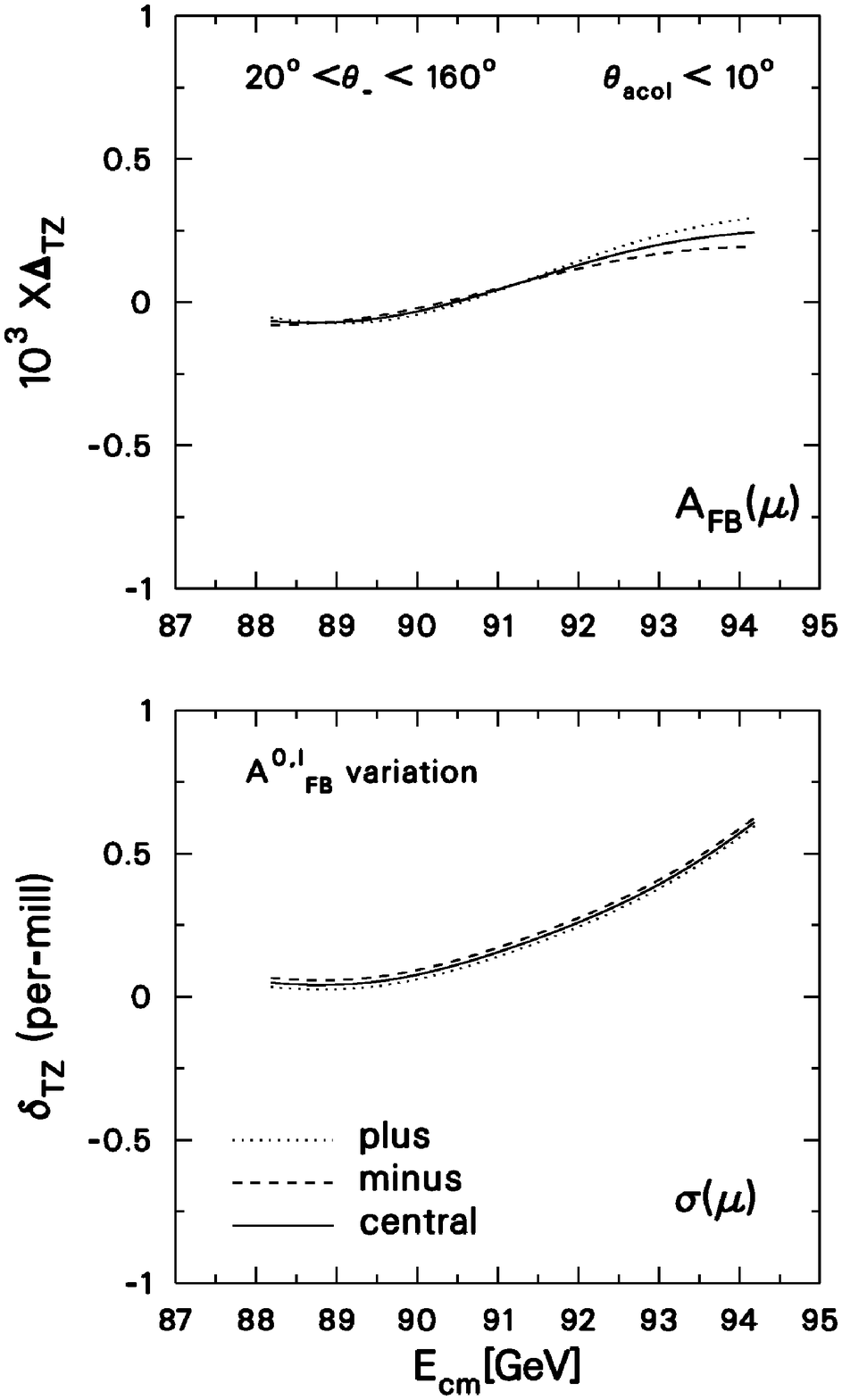}
\caption[]{
  Relative deviations between {\tt TOPAZ0} and {\tt ZFITTER} for
  muonic cross-section and absolute deviations for muonic
  forward-backward asymmetry in CA3 mode and for realistic cuts.  The
  solid curve gives deviations corresponding to the current
  experimental value of $\afba{0,l}$, the dotted (dashed) curve
  corresponds to PO$\,\pm\,2\,\times$ the experimental error.}
\label{fnotemirc4}
\end{center}
\end{figure}

\clearpage

Most of the plots in \figs{fnotemi1}{fnotemirc4} show a crossing of
the three curves (central, plus and minus) at $s = \mzs$.  This means
that at the $\zb$-pole the MI implementations of the two code are
fully equivalent while the MI treatment of the off-resonance terms is
somewhat different.  This fact is largely expected since
MI-implementations or the MI-SM splitting is far from unique.
However, the rather satisfactory level of agreement is telling us that
the associated theoretical uncertainty in extracting POs from ROs is
not substantially different or badly deteriorated with respect to the
one that we have shown in SM comparisons.

Another typical effect is that {\tt TOPAZ0} ROs tend to be lower than
{\tt ZFITTER} ROs at PO$\,-\,2\,\times\,$ experimental error and on
the low-energy side of the resonance.  With increasing energy the
curves tend to cross at the $\zb$-peak and to reverse their sign on
the high energy side.  On the contrary, for PO$\,+\,2\,\times\,$
experimental error, {\tt TOPAZ0} is higher below the resonance and
lower above it.

\begin{figure}[htbp]
\begin{center}
\includegraphics[width=0.7\linewidth]{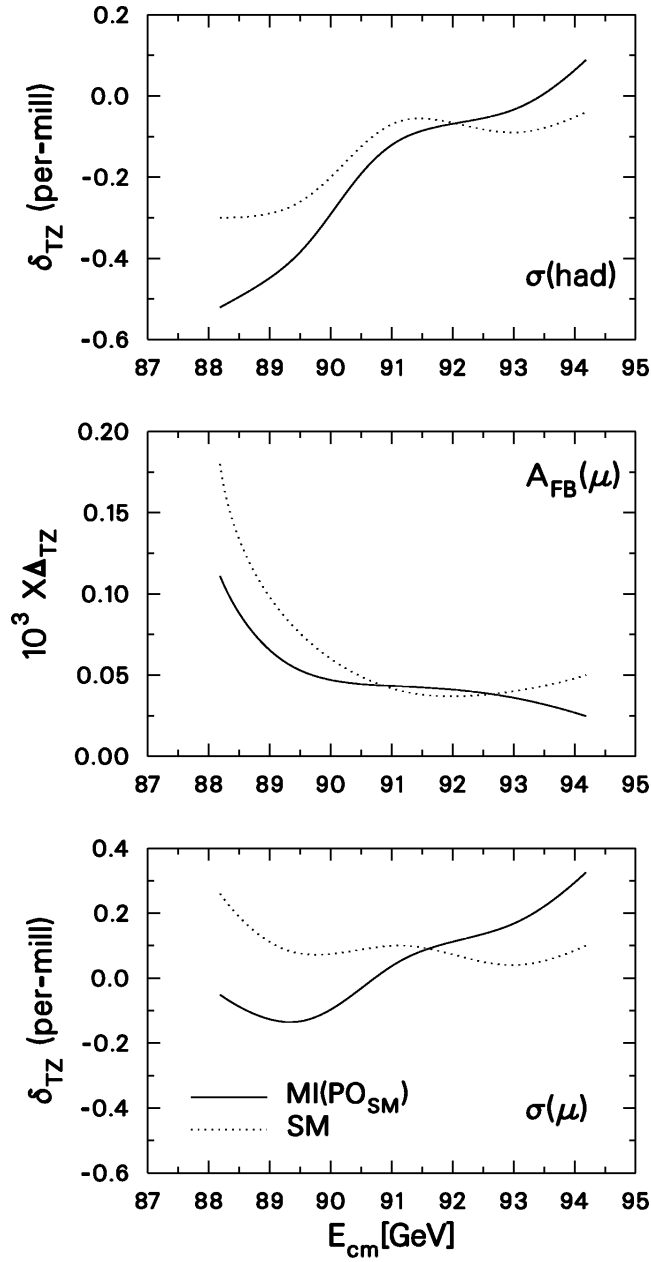}
\caption[]{
  Relative deviations between {\tt TOPAZ0} and {\tt ZFITTER} for
  muonic and hadronic cross-section and absolute deviations for muonic
  forward-backward asymmetry in CA3 mode and for $\smanp > 0.01\,s$.
  The solid curve gives deviations corresponding to SM predictions,
  the dotted curve corresponds to a MI prediction where PO are
  initialised to their SM values.}
\label{fnotemism}
\end{center}
\end{figure}

In \fig{fnotemism} we compare {\tt TOPAZ0} and {\tt ZFITTER}
predictions by computing $\sigma_{\mu}, \sigma_{\had}$ and
$\afba{\mu}$ with and $\smanp$-cut of $0.01\,s$.  The solid line gives
deviations in the SM predictions initialised with our preferred setup.
The dotted line gives deviations from the two MI predictions in case
the ROs are computed in term of POs evaluated at their (code
dependent) SM values.  From comparing the two sets of curves we see
that no serious degradation arises in the transition RO(SM,{\tt T,Z})
$\,\to\,$ RO(PO$_{\rm SM},{\tt T,Z}$).  In particular we continue to
register a very good agreement around the peak.

From these figures it emerges that, compared to the SM comparison, the
agreement is still reasonable.  Much more interesting than the
deviations shown here would be to compare differences between MI fits
performed with {\tt TOPAZ0} and {\tt ZFITTER} using the same RO data
set, clearly a task to be done by the experimental collaborations.
The differences shown in the figures will cause differences in MI-fit
results between {\tt TOPAZ0} and {\tt ZFITTER} for the same input set
of ROs.  However, the effects have to be seen compared to the fit
(experimental) errors.  Approximately, the differences in ROs seen at
the pole centre-of-mass energy will be those observed in the
corresponding POs.

%--
%--

\clearpage

%------------------------------------------------------------
\section{Theoretical Uncertainties for Realistic Observables}
%------------------------------------------------------------

When varying options in the calculation of ROs based on POs or SM
parameters, the theoretical uncertainties on ROs are obtained.  This
is discussed in the following.  However, when the experimental
collaborations analyse their measurements of ROs, this procedure is
inverted by fitting POs or SM parameters to the measured ROs: Varying
options, the changes in fitted parameters reflect the theoretical
uncertainties associated with the calculation of ROs.  This way the
theoretical uncertainties on ROs are propagated back to the fitted
parameters. Care must be taken not to double-count theoretical
uncertainties in the SM calculation of POs which propagate to ROs,
when SM parameters are determined in a fit to POs themselves
determined in a MI fit to ROs.

\subsection{Uncertainties in Standard Model Calculations}

We now propagate the various electroweak options in SM calculations
from $\mathrm{PO=PO(SM)}$ to $\mathrm{RO=RO(SM)}$.  These options are
described in Section~\ref{sec:PO-err} and are those used in describing
the theoretical uncertainties at the level of PO.  In addition there
are other uncertainties, which we have already discussed, like those
associated to different treatments of initial-state QED radiation.
The corresponding flag values in {\tt TOPAZ0/ZFITTER} are discussed in
the following.

In {\tt TOPAZ0} the flag {\tt OHC='Y'} selects next-to-leading and
higher-order hard-photon contributions. Therefore this flag should not
be varied.  Once {\tt OHC='Y'} is initialised {\tt TOPAZ0}'s flag {\tt
  ORAD} will select the type of next-to-leading and higher-order
hard-photon contributions to be included, {\tt (A,D,E,F,Y)}.  {\tt
  F}~\cite{kn:addrad} is the {\em recommended} choice; {\tt
  Y}~\cite{kn:radfact} has been implemented after version 4.3 and uses
the order $\alpha^3$ YFS radiator.  The other choices, {\tt A,D,E},
remain for compatibility tests with previous versions and are not to
be included in the estimate of the theoretical uncertainty.  {\tt
  OFS='D','Z'} selects the treatment of higher-order final-state QED
corrections, see below.  For the remaining {\tt TOPAZ0}'s flags we
observe the following.  With {\tt OWBOX} weak boxes are included {\tt
  'Y'} or not {\tt 'N'}.  The correct choice is always {\tt 'Y'}.  At
the $\zb$-resonance, but only there, weak boxes can be neglected
having a relative contribution $\le 10^{-4}$.  To give an example, for
the muonic cross-section the effect of weak boxes is $-0.01$ per-mill
at $\sqrt{s} = \mz$ and $+0.17~(-0.15)$ per-mill at the left (right)
wing.  Flags {\tt OAAS, OWEAK} must be kept equal to {\tt 'N', 'R'},
respectively.

In {\tt ZFITTER}, the flag {\tt FOT2} controls which radiator is used
for inital-state QED radiation, {\tt FOT2=-1,0,1,2,3,4,5}.  Values
{\tt FOT2=3,5} switch between the additive (default) and the
factorized order $\alpha^3$ radiator, respectively.  The remaining
{\tt FOT2} flag values are kept for compatibility tests with previous
versions and are not to be included in the estimate of the theoretical
uncertainty.  Weak boxes are switched off/on with the flag {\tt
  BOXD=0,1}, respectively, and must always be on. The flag {\tt CONV}
controls the convolution of electroweak couplings and is discussed in
Section~\ref{sec:CONV-RO}.  This flag must be kept fixed at a value
larger than zero.

For the muonic and hadronic cross-sections and for the muonic
forward-backward asymmetry we find the results reported in
\tabn{tab21}.  These results are obtained by running {\tt TOPAZ0}
under different options.  As for POs, we show the {\em central} value
for ROs evaluated at the preferred setup; the minus error for
$\mbox{RO}_{\it central} - \min_{\rm opt}\,\mbox{RO}$ and the plus
error for $\max_{\rm opt}\,\mbox{RO} - \mbox{RO}_{\it central}$.  For
cross-section the errors are reported in pb and, when available, we
also show the absolute differences {\tt T-Z}.  From \tabn{tab21} we
observe that the difference between {\tt TOPAZ0} and {\tt ZFITTER} RO
is fully compatible with the {\tt TOPAZ0}-estimated theoretical
uncertainty.

%--
\begin{table}[htbp]
\begin{center}
\renewcommand{\arraystretch}{1.1}
\begin{tabular}{|c|c|c|c|c|}
\hline
$\sqrt{\sman}\,$[GeV]   & central & minus error & plus error & T-Z \\
\hline \hline
\multicolumn{5}{|c|}{$\sigma_{\flm}$} \\
\hline \hline
$\mz-3$    &   $0.22849\,\nb$ &    $0.04\,\pb$  &   $\le 0.01\,\pb$ &
              $0.07\,\pb$ \\
$\mz-1.8$  &   $0.47657\,\nb$ &    $0.08\,\pb$  &   $0.01\,\pb$     &
              $0.04\,\pb$ \\
$\mz$      &   $1.48010\,\nb$ &    $0.09\,\pb$  &   $0.20\,\pb$     &
              $0.16\,\pb$ \\
$\mz+1.8$  &   $0.69512\,\nb$ &    $0.08\,\pb$  &   $0.06\,\pb$     &
              $0.03\,\pb$ \\
$\mz+3$    &   $0.40642\,\nb$ &    $0.06\,\pb$  &   $0.03\,\pb$     &
              $0.04\,\pb$ \\
  \hline \hline
  \multicolumn{5}{|c|}{$\afba{\flm}$} \\
  \hline \hline
$\mz-3$    &  $-0.28312$  &  $0.00009$  &  $0.00001$ & 0.00018 \\
$\mz-1.8$  &  $-0.16977$  &  $0.00008$  &  $0.00004$ & 0.00008 \\
$\mz$      &  $-0.00062$  &  $0.00006$  &  $0.00009$ & 0.00004 \\
$\mz+1.8$  &  $0.11186 $  &  $0.00004$  &  $0.00012$ & 0.00004 \\
$\mz+3$    &  $0.15466 $  &  $0.00004$  &  $0.00012$ & 0.00005 \\
  \hline \hline
  \multicolumn{5}{|c|}{$\sigma^{\flm}_{_{\rm F}}$} \\
  \hline \hline
$\mz-3$    &  $0.08190\,\nb$ & $0.03\,\pb$  &   $\le 0.01\,\pb$ & \\
$\mz-1.8$  &  $0.19783\,\nb$ & $0.05\,\pb$  &   $0.01\,\pb$ & \\
$\mz$      &  $0.73959\,\nb$ & $0.04\,\pb$  &   $0.17\,\pb$ & \\
$\mz+1.8$  &  $0.38644\,\nb$ & $0.06\,\pb$  &   $0.08\,\pb$ & \\
$\mz+3$    &  $0.23464\,\nb$ & $0.04\,\pb$  &   $0.04\,\pb$ & \\
  \hline \hline
  \multicolumn{5}{|c|}{$\sigma^{\flm}_{_{\rm B}}$} \\
  \hline \hline
$\mz-3$    &  $0.14659\,\nb$ & $0.02\,\pb$  &   $\le 0.01\,\pb$ & \\
$\mz-1.8$  &  $0.27874\,\nb$ & $0.03\,\pb$  &   $\le 0.01\,\pb$ & \\
$\mz$      &  $0.74051\,\nb$ & $0.04\,\pb$  &   $0.04\,\pb$ & \\
$\mz+1.8$  &  $0.30868\,\nb$ & $0.02\,\pb$  &   $\le 0.01\,\pb$ & \\
$\mz+3$    &  $0.17178\,\nb$ & $0.02\,\pb$  &   $\le 0.01\,\pb$ & \\
\hline 
\hline
\multicolumn{5}{|c|}{$\sigma_{\rm had}$} \\
\hline 
\hline
$\mz-3$    &  $4.45012\,\nb$ & $0.99\,\pb$  &   $1.40\,\pb$ &
              $-1.29\,\pb$ \\
$\mz-1.8$  &  $9.59909\,\nb$ & $1.81\,\pb$  &   $3.41\,\pb$ &
              $-2.49\,\pb$ \\
$\mz$      &  $30.43639\,\nb$ & $1.85\,\pb$ &   $14.27\,\pb$&
              $-11.83\,\pb$ \\
$\mz+1.8$  &  $14.18269\,\nb$ & $2.14\,\pb$ &   $6.01\,\pb$ &
              $-1.27\,\pb$ \\
$\mz+3$    &  $8.19892\,\nb$ & $1.46\,\pb$  &   $3.38\,\pb$ &
              $-0.36\,\pb$ \\
\hline
\end{tabular}
\caption[]{
  Theoretical uncertainties for $\sigma_{\flm}, \afba{\flm},
  \sigma^{\flm}_{_{\rm F,B}}$ and for $\sigma_{\rm had}$ from {\tt
    TOPAZ0}. }
\label{tab21}
\end{center}
\end{table}
%--

Finally we illustrate the effect of different treatments of
final-state QED radiation in the presence of severe kinematical cuts.
A possible source of theoretical uncertainty can be introduced when
cuts are present, due to a different treatment of higher-order
final-state QED effects: it can lead to differences which in general
depend on the experimental cuts required and that may grow for
particularly severe cuts.  It was already shown~\cite{kn:topaz0} that
two possible prescriptions,

\begin{itemize}

\item completely factorized final-state QED correction versus 

\item factorized leading-terms and non-leading contributions summed up, 

\end{itemize}

can lead to substantial differences.  In {\tt TOPAZ0} the flag {\tt
  OFS} selects the treatment of higher-order final-state QED
corrections; {\tt (D)} or {\tt (Z)}, respectively.  In {\tt ZFITTER},
the first option, completely factorized final-state QED correction, is
implemented.  {\tt TOPAZ0} predicts for $\sigma^{\mu}$ and $20^\circ <
\theta_- < 120^\circ, E_{\rm th} > 15\,$GeV, $\theta_{\rm acol} <
10^\circ$ an uncertainty of 1.2 per-mill at the five energy points.
On the contrary for a loose cut of $E_{\rm th} > 1\,$GeV we obtain a
reduction to a mere 0.1 per-mill.

\subsection{Uncertainties in Model Independent Calculations}

Having discussed the theoretical uncertainties associated with the SM
calculations of ROs, we also need to address the question of
uncertainties for the MI-calculations, $\mathrm{RO=RO(PO)}$.  The
result is that one has all errors already quoted for RO(SM),
\tabn{tab21}, $\,\oplus\,$ those derived from varying PO away from
their SM values and obtained by comparing {\tt TOPAZ0} with {\tt
  ZFITTER}, \figs{fnotemi1}{fnotemirc4}.

Note that in the SM there are relations like ${\rm PO}_1 = f({\rm
  PO}_2) = F(\mz,\mt,\dots)$.  With PO = PO(SM) one has to make sure
that, no matter how the MI-structure is built, MI({\tt T}) must be
equal to MI({\tt Z}). But when we break the SM-relations then several
possibilities arise: one may write everywhere PO$_1$ or $f({\rm
  PO}_2)$ and the difference is a measure of the associated
uncertainty.

Alternatively one can estimate this uncertainties internally to each
code by running in MI mode with all the relevant electroweak and QED
flag variations as discussed before.  The structure of the MI
calculations is such that, given the decomposition PO $\,\oplus\,$ SM
remnant, electroweak flag changing will effect the latter (the SM
complement) and leave unaltered the PO (MI) component.  To give an
example, the whole $\zb-\ph$ interference in the quark sector is taken
from the SM remnant and, therefore, influenced by flag setting.

%--------------------------------------
\section{Production of Secondary Pairs}
%--------------------------------------
\label{sec:SPP}

Finally we come to the inclusion of pair-production in the calculation
of realistic observables.  Radiative photons from the initial- or
final-state fermions may convert, leading to additional (soft)
$\ff\barf$ pairs besides the primary pair. This leads to the problem
of the signal definition, i.e., what is considered as (radiative
correction to) fermion-pair production, and what is considered as
genuine four-fermion production.  

Note that most Monte Carlo event generators used for fermion pair
production do not include the radiative production of secondary pairs.
In case of final-state pair production, visible in the detector, this
may bias efficiency and acceptance calculations, in particular if the
primary pair is a lepton pair.

%-----------------------------------------
\subsection{Initial-State Pair Production}
%-----------------------------------------

A fermionic pair of four-momentum $\qmoms$ radiated from the $e^+$ or
$e^-$ line gives a correction which is computed in \cite{kn:kkks}.
Also for this term there are different treatments, i.e., we can
exponentiate the pair-production according to the YFS
formalism~\cite{kn:jsm} or the same pairs can be included at
$\ord{\alpha^2}$.  The physical uncertainty on the pair correction is
given by that on the contribution of light-quark pairs, and is
estimated to be $1.8\cdot10^{-4}$ for cross-sections~\cite{kn:jsm}.

Both {\tt TOPAZ0} and {\tt ZFITTER} adopt a hybrid solution which
gives a remarkable agreement around the peak with the results of
\cite{kn:jsm}.  {\tt ZFITTER} includes the radiation of
$\fe,\flm,\flt$-pairs and hadronic pairs.  {\tt TOPAZ0} does not
include radiation of $\tau$-pairs, their contribution is below the
accuracy requirement of $\ord{10^{-4}}$.  For technical details we
refer to \cite{kn:ocyr95} since no further upgrading has been
performed in the area of pair production.

A cut was selected so that $\zvari{\rm min}\sman
=\smanp=\Mlones(\ff\barf) > 0.25\,\sman$, where $\ff\barf$ denotes the
{\em primary pair}, i.e., $z_{\rm min}$ is the minimum fraction of
squared invariant mass of the final state (primary pair) after
radiation of the additional initial-state pair (secondary pair).
Since we neglect terms coming from ISPP $\,\otimes\,$ FSR there will
be no difference between $\smanp$-cuts and $\Mlones$-cuts as far as
ISPP corrections are concerned.  The soft-hard separator $\Delta$ has
been fixed in the region where we see a plateau of stability.

The effect of including leptonic as well as hadronic pairs radiated
from the initial state is summarised in \tabn{tab23}.  From
\tabn{tab23} we observe that the inclusion of initial-state
pair-production lowers the cross-sections up to an energy of $\mz+3$
where the effect becomes positive.  At the peak pair production
modifies $\sigma_{\flm}$ by $-2.55~(-2.53)$ per-mill according to {\tt
  TOPAZ0} ({\tt ZFITTER}) and $\sigma_{\rm had}$ by -2.55 per-mill for
both codes.

%--
\begin{table}[t]
\begin{center}
\renewcommand{\arraystretch}{1.1}
\begin{tabular}{|c||c|c|c|c|c|}
  \hline
  & \multicolumn{5}{c|}{Centre-of-mass energy in GeV} \\
  \cline{2-6}
  & $\mz - 3$ & $\mz - 1.8$ & $\mz$ & $\mz + 1.8$ & $\mz + 3$  \\
  \hline \hline
  \multicolumn{6}{|c|}{$\sigma_{\flm}$} \\
  \hline \hline
T &  0.22849 &  0.47657 &  1.48010 &  0.69512 &  0.40642 \\     
  &  0.22796 &  0.47534 &  1.47633 &  0.69480 &  0.40713 \\
  & -2.32    & -2.58    & -2.55    & -0.46    &  1.75    \\
Z &  0.22843 &  0.47653 &  1.47995 &  0.69509 &  0.40638 \\    
  &  0.22790 &  0.47532 &  1.47621 &  0.69478 &  0.40708 \\         
  & -2.33    & -2.55    & -2.53    & -0.45    &  1.73    \\
  \hline \hline
  \multicolumn{6}{|c|}{$\afba{\flm}$} \\
  \hline \hline
T & -0.28312 & -0.16977 & -0.00062 & 0.11186 &  0.15466  \\     
  & -0.28377 & -0.17020 & -0.00062 & 0.11192 &  0.15439  \\ 
  & -0.65    & -0.43    &  0.00    & +0.06   & -0.27     \\
Z & -0.28330 & -0.16985 & -0.00066 & 0.11182 &  0.15461  \\
  & -0.28395 & -0.17028 & -0.00066 & 0.11187 &  0.15434  \\
  & -0.65    & -0.43    &  0.00    & +0.05   & -0.27     \\
  \hline \hline
  \multicolumn{6}{|c|}{$\sigma_{\rm had}$} \\
  \hline \hline
T &  4.45012 &  9.59910 & 30.43639 & 14.18269 &  8.19892 \\     
  &  4.43937 &  9.57410 & 30.35866 & 14.17587 &  8.21336 \\ 
  & -2.42    & -2.60    &-2.55     &-0.48     &  1.76    \\
Z &  4.45146 &  9.60165 & 30.43824 & 14.18391 &  8.19923 \\        
  &  4.44070 &  9.57659 & 30.36069 & 14.17709 &  8.21345 \\         
  & -2.42    & -2.62    & -2.55    & -0.48    &  1.73    \\
  \hline 
\end{tabular}
\caption[]{
  The effect of including initial state pair production in {\tt
    TOPAZ0} and {\tt ZFITTER} in CA3-mode, $\smanp>0.01\,\sman$.
  First (fourth) entry is without, second (fifth) entry is with pair
  production. Third (sixth) entry is the net effect in per-mill of the
  inclusion. }
\label{tab23}
\end{center}
\end{table}
%--

%---------------------------------------
\subsection{Final-State Pair Production}
%---------------------------------------

The current versions of {\tt TOPAZ0} (4.4) and {\tt ZFITTER} (5.20) do
not include effects of final-state pair production.  However, virtual
and real corrections due to final-state pair production cancel to a
large extent and the remaining effect is mostly absorbed in the
running electromagnetic coupling $\alpha(s)$ entering the correction
factor for FSR~\cite{kn:FSPP}.  Therefore, the experimental event
selections should not discriminate against additional (soft) pairs.
Otherwise, a correction has to be applied before {\tt TOPAZ0/ZFITTER}
calculations can be compared with the measurements.

\clearpage

%--------------------
\section{Conclusions}
%--------------------

In \tabn{tab21} we show an estimate of the theoretical error for
realistic observables as computed internally by {\tt TOPAZ0}.  Another
piece of information is given by the differences {\tt TOPAZ0} - {\tt
  ZFITTER} among the (theoretical) central values for each quantity:
these differences are basically (even though not totally) a measure of
the effect induced by a variation in the renormalization scheme.

Consider, however, the complete hadronic cross-section (in CA3
$\smanp$-mode): the differences (in per-mill) are $-0.30, -0.27,
-0.06, -0.09,-0.04$ for the five centre-of-mass energies.  There are
two different origins for them, differences already present in the
de-convoluted cross-sections (SD-mode) and differences due to
convolution with initial-state QED radiation. For the former we find
(in per-mill)
%--
\bq
-0.38,\; -0.30,\; -0.07,\; -0.15,\; -0.14. 
\eq
%--
The effect of convolution is found to be (in per-mill) 
%--
\bq
+0.06,\; +0.02,\; +0.01,\; +0.07,\; +0.14.
\eq
%--
Especially on the high-energy side of the resonance we observe a
partial compensation, leading to a small overall uncertainty.  A more
conservative attitude consists in adopting some rough approximation
thus defining
%--
\bq
\sigma(\sman) = \sigma^{\rm SD}\,\lpar 1 + \delta^{\rm dec}\rpar,
\eq
%--
and adding the errors in quadrature: 
%--
\bqa
\Delta\sigma &=& \sigma^{\rm{T}}  - \sigma^{\rm{Z}},  \quad
\Delta\sigma^{\rm SD} = \sigma^{\rm{SD,T}}  - \sigma^{\rm{SD,Z}},  
\nl
\lpar\Delta\delta^{\rm dec}\rpar^2 &=& 
\biggl[\frac{\Delta\sigma}{\sigma^{\rm SD}}\biggr]^2 + 
\biggl[\frac{\sigma\Delta\sigma^{\rm SD}}{\lpar\sigma^{\rm SD}\rpar^2}
\biggr]^2.
\eqa
%--
In this way we end up with an estimate of the theoretical error of
%--
\bqa
{}&{}& 2.73\,\pb,\; 4.76\,\pb,\; 3.51\,\pb,\; 3.23\,\pb,\; 1.71\,\pb, 
\quad \mbox{or} \nl
{}&{}&
0.061\%,\;\; 0.050\%,\;\; 0.012\%,\;\; 0.023\%,\;\; 0.021\%,
\eqa
%--
for the complete hadronic cross-section. Although quite conservative,
we consider the above as a safe estimate of the theoretical error.  A
comparison with the results of \tabn{tab21} shows a substantial
agreement with the estimate made internally by {\tt TOPAZ0}. For the
complete muonic cross-section we find
%--
\bqa
{}&{}& 
0.07\,\pb,\; 0.06\,\pb,\; 0.19\,\pb,\; 0.11\,\pb,\; 0.08\,\pb
\quad \mbox{or} \nl
{}&{}&
0.030\%,\;\; 0.014\%,\;\; 0.013\%,\;\; 0.016\%,\;\; 0.021\%.
\eqa
%--
In both cases, the uncertainty arising due to the uncertainty on the
ISPP contribution has to be added.  

For the muonic forward-backward asymmetry we find results which are
practically indistinguishable from those already reported in
\tabn{tab21}.

%-------------------------
\section{Acknowledgements}
%-------------------------

The present documentation is the result of an intensive collaboration
with the experimentalists of the LEP electroweak working group.  We
are obliged to all of them for many fruitful discussions.  We are
particularly thankful to T.~Kawamoto, A.~Olshevski and G.~Quast.  We
would like to express deep thanks to physicists who joined the {\tt
  TOPAZ0} and {\tt ZFITTER} teams in course of many years:
G.~Montagna, O.~Nicrosini, F.~Piccinini, and R.~Pittau {\tt and}
M.~Bilenky, A.~Chizhov, P.~Christova, M.~Jack, L.~Kalinovskaya,
A.~Olshevski, S.~Riemann, T.~Riemann, M.~Sachwitz, A.~Sazonov,
Yu.~Sedykh, and I.~Sheer.  Without their contributions the two
programs would not be what they are.

\clearpage
%
%%% \include{ltf_note_biblio}

%===================
\end{document}